\documentclass[linenumbers]{aastex631}

\newcommand{\teff}{$T_{\text{eff}}$}
\newcommand{\vmic}{$v_{\text{mic}}$}
\newcommand{\logg}{$\log{g}$}
\usepackage{multirow}
\usepackage{hyperref}
\usepackage[T1]{fontenc}

\begin{document}
\nolinenumbers

\title{Chemical Abundances of the Bioessential Elements C, O and S, and the Refractory Elements Fe and Ni, in Solar-type Exoplanet-hosting Stars from HARPS North and South}

\author[0000-0002-1549-626X]{Ellen Costa-Almeida}
\correspondingauthor{Ellen Costa-Almeida}
\email{ellenalmeida@on.br}
\affiliation{Observatório Nacional, Rua General José Cristino, 77, 20921-400 São Cristóvão, Rio de Janeiro, RJ, Brazil}

\author[0000-0002-9089-0136]{Luan Ghezzi}
\affiliation{Universidade Federal do Rio de Janeiro, Observatório do Valongo, Ladeira do Pedro Antônio, 43, Rio de Janeiro, RJ 20080-090, Brazil}
\affiliation{Observatório Nacional, Rua General José Cristino, 77, 20921-400 São Cristóvão, Rio de Janeiro, RJ, Brazil}

\author[0000-0001-6476-0576]{Katia Cunha}
\affiliation{Steward Observatory, University of Arizona, 933 North Cherry Avenue, Tucson, AZ 85721, USA}
\affiliation{Observatório Nacional, Rua General José Cristino, 77, 20921-400 São Cristóvão, Rio de Janeiro, RJ, Brazil}

\author[0000-0002-0134-2024]{Verne V. Smith}
\affiliation{Noirlab, 950 North Cherry Avenue, Tucson, AZ 85721, USA}

\author[0000-0002-9843-4354]{Jonathan J. Fortney}
\affiliation{Department of Astronomy and Astrophysics, University of California, Santa Cruz, CA 95064, USA}


\begin{abstract}
\nolinenumbers
We determined atmospheric and evolutionary parameters, along with chemical abundances of C, O, S, Fe, and Ni for 290 solar-type exoplanet hosting stars using high-resolution HARPS-North and HARPS-South spectra, and radii for 373 exoplanets using literature transit depths. We find that stars hosting giant exoplanets (R$_{\text{pl}}>4\ \text{R}_{\oplus}$) show enhanced [X/H] abundances compared to small exoplanet hosts for all elements analyzed. When considering only exoplanets with $P_{\text{orb}}\leq30$ days, there is a statistically significant anti-correlation between host star [Fe/H] and $P_{\text{orb}}$. However, [Fe/H] does not continue to decline as the orbital period increases, but rather rises again for exoplanets with larger orbital periods. Stars hosting only small exoplanets or hosting at least one sub-Saturn show significant differences between the populations of hot and warm exoplanets for all elements. In contrast, stars hosting at least one Jupiter-sized planet show no abundance differences. The host star C/O ratios obtained vary from 0.17 to 0.95, with giant exoplanet hosts exhibiting the lowest median C/O ratios (0.43$^{+0.02}_{-0.03}$), while the 3 -- 4 R$_\oplus$ sub-Neptune hosts in our sample exhibit the highest median C/O ratios (0.55$^{+0.05}_{-0.01}$). Our sample has 199 exoplanets with estimated masses and we find correlations between host star [O/H] and [S/H] and $\log(M_{\text{pl}}/\text{M}_\oplus)$. When segregating the sample into hot and warm exoplanet hosts, these trends are only found for warm exoplanets. Dividing the sample between low- (91 exoplanets) and high-[$\alpha$/Fe] (20 exoplanets) stars, there are trends between host star [O/H], [S/H], [Fe/H] and [Ni/H] and $\log(M_{\text{pl}}/\text{M}_\oplus)$ only for the low-[$\alpha$/Fe] sample.
\end{abstract}


\section{Introduction} 
\label{sec:intro}
Since the discovery of 51 Pegasi b, the first exoplanet orbiting a Sun-like star (\citealt{Mayor1995}), significant progress has been made through a variety of techniques and instruments used to detect exoplanets. Consequently, we have now surpassed the milestone of 6000 detected exoplanets (NASA Exoplanet Archive\footnote{\url{https://exoplanetarchive.ipac.caltech.edu/}}). These discoveries, along with the associated detailed characterizations of both exoplanets and their host stars, have enabled more comprehensive investigations into planetary formation, evolution, and habitability.

Initially suggested by \cite{Gonzalez1997} and \cite{Santos2000} and further investigated by, e.g., \cite{Fischer2005}, the planet-metallicity\footnote{Metallicity ([Fe/H]) is commonly used as a general indicator of the overall abundance of metals in the star though it is based on the iron abundance. $\text{[Fe/H]} = \log_{10}\left(N_{\text{Fe}}/N_{\text{H}}\right)_\star - \log_{10}\left(N_{\text{Fe}}/N_{\text{H}}\right)_\odot$.} correlation has revealed that the presence of short-period giant exoplanets is favored around metal-rich stars (e.g., \citealt{Mortier2013}; \citealt{Ghezzi2018}). This correlation, to this day, remains the strongest connection between exoplanets and their host stars, directly impacting our view on planet formation by providing support for the core accretion mechanism (e.g., \citealt{Pollack1996}; \citealt{Ida2004}). However, it is still unknown to what extent this trend, which weakens with decreasing exoplanetary size (\citealt{Teske2024}), is valid for smaller and terrestrial exoplanets (e.g., \citealt{Buchhave2014}; \citealt{Wang2015}; \citealt{Petigura2018b}; \citealt{Adibekyan2019}; \citealt{Wilson2022}; \citealt{Wanderley2025}). In particular, \cite{Petigura2018b} found that the occurrence of warm Super-Earths ($10<P_{\text{orb}}\leq100$ days and $1\leq R_{\text{pl}}<1.7\ R_\oplus$) remains nearly constant across metallicities ranging from -0.4 to +0.4 dex, with an occurrence rate of 20 warm Super-Earths per 100 stars. For warm Sub-Neptunes ($1.7 \leq R_{\text{pl}}<4\ R_\oplus$), this number doubles to 40 per 100 stars. However, for hot super-Earths ($P_{\text{orb}}\leq10$ days), they found an exoplanet occurrence of 4\% for stars with [Fe/H] between -0.4 dex and -0.2 dex, growing to 10\% for more metal-rich stars, with [Fe/H] between +0.2 dex and +0.4 dex. Also, for hot sub-Neptunes, the occurrence rates increase from 1\% to 8\%, respectively, when the same metallicity intervals are considered. This was also observed by \cite{Mulders2016}, who found that the occurrence rate of hot small exoplanets ($R_{\text{pl}}<4\ \text{R}_\oplus$) around host stars with super-solar metallicities is three times higher compared to host stars with subsolar metallicities.

With the increasing number of stellar abundance studies due to the availability of high-quality spectra, especially from the large spectroscopic surveys, we can now analyze the planet-metallicity correlation in more detail, particularly for smaller exoplanets detected by the Kepler and TESS missions. \cite{Wilson2022}, using abundance data from the Apache Point Observatory Galactic Evolution Experiment (APOGEE, \citealt{Majewski2017}), explored the star-planet connection for 10 elements (C, Mg, Al, Si, S, K, Ca, Mn, Fe, and Ni). They confirmed the results from \cite{Petigura2018b} regarding exoplanet occurrence rate with Fe and found that hot exoplanet occurrence rates increase by $\sim$20\% for Super-Earths and $\sim$60\% for Sub-Neptunes with an enhancement of 0.1 dex for all studied elements. However, for warm exoplanet occurrence rates, these correlations were found to be weaker. \cite{Ghezzi2026} determined abundances of 13 refractory elements (Na, Mg, Al, Si, Ca, Sc, Ti, V, Cr, Mn, Co, Ni and Cu) for 561 exoplanet-hosting stars of the California-Kepler Survey (\citealt{Petigura2017}) using high-resolution Keck/HIRES spectra and found that [X/H] abundances for Al, Si, Ca, Fe, Co, and Ni are statistically higher for systems with only hot planets relative to systems with only warm exoplanets. They also found that stars with only large exoplanets have significantly larger [X/H] abundances for Al, Si, Sc, Ti, Cr, Fe, and Co) relative to stars with only small exoplanets.

Concerning previous abundance results for the bioessential elements C, O, and S in samples of exoplanet hosts, \cite{CostaSilva2020} determined sulfur abundances for 719 FGK stars of the HARPS-GTO sample and reported no differences in the sulfur distributions between stars with and without detected exoplanets for metallicities [Fe/H]$>$-0.3 dex. They also found that lower-mass exoplanets (super-Earths and Neptune-size) are more likely to orbit stars with higher [S/Fe]\footnote{[X/Fe] = [X/H] - [Fe/H]}. \cite{DelgadoMena2021}, analyzing stars from the same HARPS sample with [Fe/H]$\gtrsim$-0.2 dex, found no significant differences in the distribution of [C/Fe] for populations of stars hosting exoplanets above and below 30 M$_\oplus$. The earlier study by \cite{Brugamyer2011} found no significant probability that the exoplanet detection rate depends on the oxygen abundance of the host star ($\sim$45\%).

Elemental abundance ratios can provide important information with respect to the distribution of chemical species in protoplanetary disks, serving as probes of the relative location of planet formation (e.g., \citealt{Thiabaud2015a}, \citeyear{Thiabaud2015b}). C/O\footnote{C/O$=10^{\text{A(C)}-\text{A(O)}}$, where A(X) is the absolute abundance of X, i.e., A(X)$=\log(N_{\text{X}})$.} was the first elemental ratio proposed to trace giant planet formation (e.g., \citealt{Seager2005}; \citealt{"O2011}), as the availability of C and O in the protoplanetary disk governs where the planet accretes most of its mass (e.g., \citealt{Thiabaud2015b}). Probing stellar C/O ratios of a sample of 16 hot Jupiters, \cite{Teske2014} found no significant trends for the host star C/O ratio as a function of exoplanet equilibrium temperature and radius. \cite{Sua2018} investigated C/O versus exoplanet mass in a sample of 99 solar-type planet-hosting stars of the HARPS-GTO sample and found that 86\% of the stars hosting high-mass exoplanets (M$_{\text{pl}}>30\ \text{M}_\oplus$) presented C/O ratios between 0.4 -- 0.8, while 14\% had C/O below 0.4, obtaining average C/O ratios of 0.46$\pm$0.11 and 0.50$\pm$0.10 for stars hosting low- and high-mass exoplanets, respectively. Also, \cite{Sharma2024} determined C, N, O, Mg, Si and Fe abundances for a sample of 149 F-, G- and K-dwarfs and giant planet-hosting stars using high-resolution VUES (Vilnius University Echelle Spectrograph) spectra and found no relationship for stellar [C/Fe], [O/Fe] and C/O with exoplanetary mass.

There is still a significant number of stars, especially planet-hosting ones, that have no abundance determinations for a variety of elements besides Fe. For instance, $\sim$24\% of the stars in the Hypatia catalog\footnote{\url{hypatiacatalog.com}} (\citealt{Hinkel2014}) do not have results for carbon abundances, while $\sim$25\% and $\sim$57\% do not have O and S abundances, respectively. Moreover, those stars for which these abundance results are available may be accompanied by large uncertainties. Since these elements are important not only for planet formation, but also for exoplanet habitability, one of the fundamental contributions that can be made in this topic is increasing the availability of homogeneous abundances and, furthermore, improving their precision -- especially for elements that are biologically and geologically essential for life on Earth, since they can provide valuable insights into internal composition and, consequently, planetary habitability (e.g., \citealt{Hinkel2018}). Habitability is, nowadays, a robust area that permeates various fields of knowledge and incorporates models of planetary formation and composition (e.g., \citealt{Thiabaud2015a}, \citeyear{Thiabaud2015b}), climate models (e.g., \citealt{Shields2019}), ecology models (e.g., \citealt{Me2021}), geological cycles, and plate tectonics (e.g., \citealt{Ehlmann2016}), among others. Due to these advances, the concept of habitability itself has broadened, considering not only the presence of liquid water on the surface of the planet, but also the availability of elements essential for life on Earth (CHNOPS) over a long timescale, in order for life to thrive and survive (e.g., \citealt{Krijt2023}). 

In this work, using publicly available high-resolution HARPS-South and HARPS-North spectra, we contribute to the increase of homogeneous chemical abundance results for the bioessential elements C, O and S, as well as the refractory elements Fe and Ni, in solar-type planet-hosting stars. Ensuring a homogeneous determination of stellar atmospheric parameters (\teff, \logg, [Fe/H]), evolutionary parameters ($R_{\star}$, $M_{\star}$ and age), and chemical abundances is essential for minimizing systematic uncertainties and discrepancies introduced by varying analysis methods. A uniform approach helps preserve the reliability of comparisons and any detected correlations. Efforts such as the one presented in this work, of homogeneously characterizing planet-hosting stars using high-resolution spectra, are increasingly important as we search for star-planet connections and chemical abundances are now being obtained for exoplanet atmospheres. For example, observations from the James Webb Space Telescope are already exploring a new diversity of molecules in the atmosphere of giant exoplanets that could not be observed with the Hubble Space Telescope, such as the detection of SO$_2$ in the atmosphere of exoplanets (e.g., \citealt{Alderson2023}; \citealt{Gressier2025}). In this context, here we will investigate the star-planet connection by probing the relationships between detailed chemical abundances of host stars and exoplanet properties such as radius, orbital period, mass, and equilibrium temperature. The results from such efforts may contribute to future refinement of planet formation theories, exploring deeply how the distribution of molecules evolve through the plotoplanetary disk and how similar is the final composition of exoplanets in comparison to their host stars. 

This paper is organized as follows. The stellar and exoplanetary samples are described in Section \ref{sec:sample_data}. Stellar parameters and chemical abundances are determined in Section \ref{sec:atm_par} and \ref{sec:abs}, respectively, and the exoplanetary radii in Section \ref{sec:exoplanet_rad_determ}. In Section \ref{sec:results}, we discuss the chemical evolution of our sample of exoplanet hosts. In Section \ref{sec:star_planet_connection}, we discuss the star-planet connection. Lastly, our concluding remarks are presented in Section \ref{sec:conclusion}.

\section{Sample and Data} 
\label{sec:sample_data}
\label{sec:2_sample}
We selected confirmed host stars from the NASA Exoplanet Archive (\citealt{ps}\footnote{Accessed on 2025-09-12 at 08:55, returning 38778 rows. This dataset or service is made available by the NASA Exoplanet Science Institute at IPAC, which is operated by the California Institute of Technology under contract with the National Aeronautics and Space Administration.}, \citealt{Akeson2013}) according to the following criteria: 5000 K $\leq$ \teff\ $\leq$ 6500 K and \logg\ $\geq$ 4.0 dex, and with exoplanets having transit depth determinations. Among the 1399 initially selected stars, we obtained good quality high-resolution spectra for 308 stars. This sample is composed mostly of F- and G-dwarfs with some K-type stars and subgiants. The coordinates and the identification of the spectra used are presented in Appendix \ref{ap:spectra}.

The data used for this work consist of publicly available high resolution HARPS-South (\citealt{Mayor2003}) and HARPS-North (\citealt{Cosentino2012}) spectra from ESO Science Archive Facility\footnote{\url{https://archive.eso.org/wdb/wdb/adp/phase3_spectral/form}} and the Italian Center for Astronomical Archive (IA2)\footnote{\url{http://archives.ia2.inaf.it/tng/}}, respectively. The HARPS (High Accuracy Radial velocity Planet Searcher, \citealt{Mayor2003}) spectrographs are twins, have a resolution of R$\sim$115,000 and cover the wavelength range 3780 -- 6910 \AA. The southern spectrograph is at the ESO 3.6 m telescope, in La Silla, Chile, and the northern spectrograph is at the 3.6 m Telescopio Nazionale Galileo (TNG), in La Palma, Canary Islands, Spain. The spectra obtained were already reduced with their instrument pipelines. We corrected them for radial velocity using PyRAF\footnote{PyRAF is a command language for Tody1986 based on the Python scripting language. It was developed at the Space Telescope Science Institute (STScI).} \texttt{fxcor} task with the Ceres 2009-02-08 solar spectrum\footnote{HARPS Solar Spectra Collection (\url{https://www.eso.org/sci/facilities/lasilla/instruments/harps/inst/monitoring/sun.html}} as a template between 6040 -- 6200 \AA. We combined the spectra of each star to improve SNR using PyRAF \texttt{scombine} task. For the final combined spectra, the median SNR is 145 with $\sim$33\% having SNR $<$ 100, $\sim$28\% having 100 $\leq$ SNR $<$ 200, $\sim$14\% having 200 $\leq$ SNR $<$ 300 and $\sim$25\% having SNR $\geq$ 300. The final SNR and the number of spectra combined are listed in Table \ref{tab:sample}. The SNR value used for each individual spectrum is the one available in the header of the fits file and the final SNR of the combined spectrum is calculated as $\text{SNR}_{\text{combined}}=\sqrt{\sum_i \text{SNR}_{i}^{2}}$.

\begin{table}
    \centering
    \begin{splittabular}{lccccccBcccccccBccccccccc}
    \hline
    \hline
    Star & G & G$_{\text{BP-RP}}$ & $\pi$ & V & A$_{\text{V}}$ & DR & $T_{\text{eff}}$ & $\log g$ & [Fe/H] & $v_{\text{mic}}$ & Age & R$_{\star}$ & M$_{\star}$ & A(C) & A(O) & A(S) & A(Ni) & SNR 5052 & SNR 5380 & SNR 6300 & SNR 6743-6758 & SNR\\
    & (mag) & (mag) & (mas) & (mag) & (mag) & & (K) & & & (km s$^{-1}$) & (Gyr) & (R$_{\odot}$) & (M$_{\odot}$) & & & & & & & & & \\
    \hline
    55 Cnc & 5.7144$\pm$0.0028 & 1.0135$\pm$0.0048 & 79.4274$\pm$0.0777 & 5.9169$\pm$0.0460 & 0.4184 & 2 & 5283$\pm$49 & 4.44$\pm$0.13 & 0.40$\pm$0.03 & 0.730$\pm$0.080 & 10.638$\pm$0.930 & 1.091$\pm$0.013 & 0.946$\pm$0.015 & 8.80$\pm$0.07 & 8.86$\pm$0.25 & 7.47$\pm$0.05 & 6.70$\pm$0.04 & 539 & 621 & 522 & 476 & 486\\
    BD+20 594 & 10.8665$\pm$0.0028 & 0.8733$\pm$0.0047 & 5.6384$\pm$0.0143 & 11.0361$\pm$0.0303 & 0.2418 & E3 & 5690$\pm$13 & 4.34$\pm$0.04 & -0.16$\pm$0.01 & 0.979$\pm$0.030 & 11.085$\pm$0.517 & 1.134$\pm$0.006 & 0.913$\pm$0.023 & 8.41$\pm$0.03 & 8.71$\pm$0.03 & 7.06$\pm$0.06 & 6.06$\pm$0.02 & 168 & 170 & 197 & 201 & 150\\
    CoRoT-1 & 13.4454$\pm$0.0028 & 0.7692$\pm$0.0049 & 1.3000$\pm$0.0167 & 13.5825$\pm$0.0303 & 0.3345 & E3 & 6337$\pm$30 & 4.42$\pm$0.06 & -0.05$\pm$0.02 & 1.463$\pm$0.070 & 2.325$\pm$0.408 & 1.263$\pm$0.016 & 1.191$\pm$0.024 & 8.34$\pm$0.04 & - & 7.17$\pm$0.06 & 6.10$\pm$0.04 & 94 & 98 & 79 & 114 & 87\\
    CoRoT-4 & 13.5490$\pm$0.0028 & 0.7600$\pm$0.0049 & 1.3623$\pm$0.0157 & 13.6835$\pm$0.0303 & 0.1365 & E3 & 6295$\pm$38 & 4.62$\pm$0.09 & 0.11$\pm$0.02 & 1.426$\pm$0.070 & 0.288$\pm$0.147 & 1.123$\pm$0.009 & 1.205$\pm$0.026 & 8.46$\pm$0.05 & - & 7.18$\pm$0.04 & 6.27$\pm$0.03 & 82 & 85 & 82 & 99 & 75\\
    \hline
    \end{splittabular}
    \caption{Sample of exoplanet-hosting stars. Column 1 shows the stellar identification. Columns 2 -- 4 show Gaia's G magnitude, G$_{\text{BP}-\text{RP}}$ color and parallax, respectively. Column 5 shows calculated V magnitude. Columns 6 and 7 shows StarHorse's extinction and which Gaia Data Release is the A$_{\text{V}}$ based on. Columns 8 -- 11 show the atmospheric parameters. Columns 12 -- 14 show age, radius and mass, respectively. Columns 15 -- 18 show the absolute abundances of C, O, S and Ni, respectively. Columns 19 -- 22 show the SNR of the spectral orders that contain the C I line at 5052~\AA, C I line at 5380~\AA, [O I] line at 6300~\AA~and all S I lines between 6743 -- 6758~\AA. Column 23 shows the total SNR of the combined spectra.}
    \label{tab:sample}
    \tablecomments{This table is published in its entirety in the machine-readable format. A portion is shown here for guidance regarding its form and content.}
\end{table}

\section{Stellar Parameters} 
\label{sec:atm_par}
\subsection{Atmospheric Parameters}
\label{sec:atm_par_determination}
We determined the effective temperatures (\teff), surface gravities (\logg), microturbulence velocities (\vmic) and metallicities ([Fe/H]), along with their uncertainties, using the methodology described in \citeauthor{Ghezzi2018} (\citeyear{Ghezzi2018}; \citealt{Martinez2019}; \citeyear{Ghezzi2021}) (hereafter, G18 and G21, respectively), which employs a classical spectroscopic method based on the excitation and ionization equilibria of Fe I and Fe II lines in an automated Python pipeline. Briefly, it uses: (i) a line list containing atomic parameters for 158 Fe I and 18 Fe II lines between 5023 -- 6862 \AA~(see Table C.2 of G18); (ii) equivalent widths (EWs) of the iron lines; (iii) Kurucz ATLAS9 ODFNEW 1-D plane parallel model atmospheres (\citealt{Castelli2003}); (iv) driver \texttt{abfind} of the Local Thermodynamic Equilibrium (LTE) line analysis code MOOG\footnote{\url{http://www.as.utexas.edu/ chris/moog.html}} (\citealt{Sneden1973}), version FEB2017. The equivalent widths were measured using the ARES v2 code\footnote{\url{https://github.com/sousasag/ARES}} (\citealt{Sousa2007}, \citeyear{Sousa2015}) adopting the parameters \textit{smoothder} = 4, \textit{space} = 2.0 m\AA, \textit{lineresol} = 0.1 m\AA, \textit{miniline} = 5 and \textit{rejt} as the SNR value of the spectrum. 

\begin{figure}[!ht]
    \centering
    \includegraphics[width=0.65\linewidth]{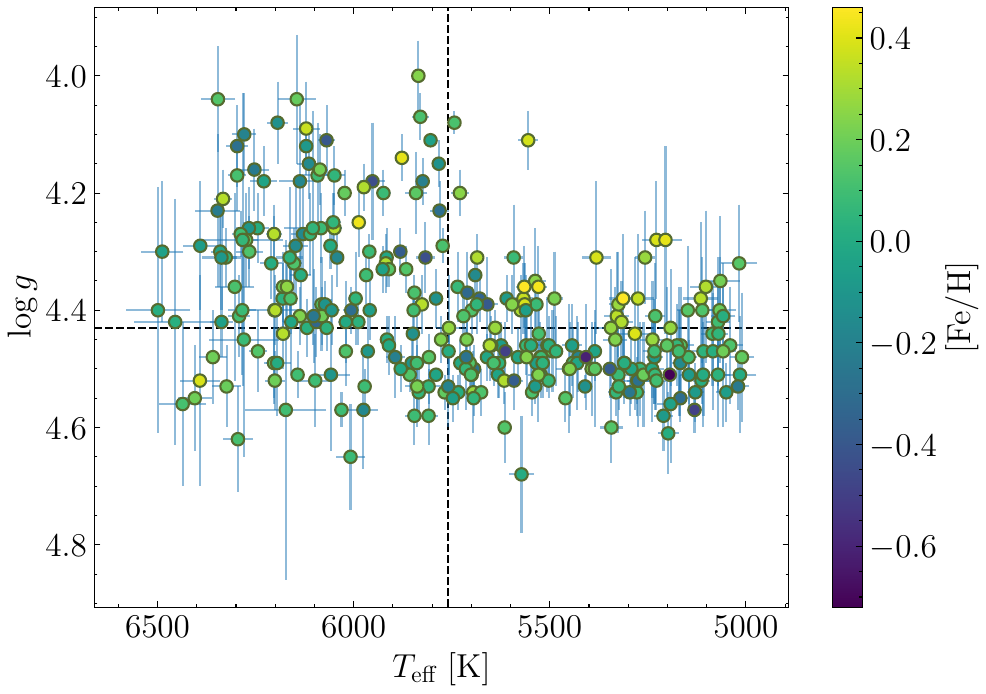}
    \caption{Kiel diagram for 290 stars with 5000 K $\leq$ \teff\ $\leq$ 6500 K and \logg\ $\geq$ 4.0 dex analyzed in this work. The dashed black lines present the solar values obtained for the Ceres 2009-02-08 solar spectrum, \teff\ = 5759$\pm$18 K, \logg\ = 4.43$\pm$0.05 dex and [Fe/H] = 0.00$\pm$0.01 dex.}
    \label{fig:03_KielDiagram}
\end{figure}

The final atmospheric parameters for 308 stars and are listed in Table \ref{tab:sample}. The median values for the uncertainties of the 308 stars are 22 K for \teff, 0.02 dex for [Fe/H], 0.06 dex for \logg\ and 0.040 km s$^{-1}$ for \vmic. Finally, as a consistency check, we determined solar atmospheric parameters using the Ceres 2009-02-08 solar spectrum (see Section \ref{sec:sample_data}) and found \teff\ = 5759$\pm$18 K, [Fe/H] = 0.00$\pm$0.01 dex where A(Fe)$_\odot$ = 7.50 dex (\citealt{Asplund2009}), \logg\ = 4.43$\pm$0.05 dex and \vmic\ = 0.936$\pm$0.030 km s$^{-1}$. Therefore, they are in excellent agreement with canonical solar values. We note that 18 stars were outside of our selection criteria of 5000 K $\leq$ \teff\ $\leq$ 6500 K and \logg\ $\geq$ 4.0 dex. Finally, we show the Kiel diagram of the 290 stars in Figure \ref{fig:03_KielDiagram}.

\subsection{Evolutionary Parameters}
\label{c2:evolutionaryparameters}
We determined stellar ages, radii, masses and surface gravities using the isochrone method with PARAM code v1.3\footnote{PARAM web interface is mantained by Léo Girardi at the Observatorio Astronomico di Padova (\url{https://stev.oapd.inaf.it/cgi-bin/param_1.3}).} (\citealt{daSilva2006}), which consists in comparing the positions of stars with theoretical isochrones in a color-magnitude diagram. The code performs a Bayesian estimation of stellar evolutionary parameters using a grid of PARSEC isochrones (\citealt{Bressan2012}), stellar observables as input parameters (\teff, [Fe/H], parallax and V magnitude) and the default Bayesian priors, which are lognormal IMFs from \cite{Chabrier2001}, constant stellar formation rate (SFR) and an age interval between 0.1 -- 12 Gyr. The effective temperatures and metallicities were spectroscopically determined in this work (see Section \ref{sec:atm_par_determination}). The parallaxes, G magnitudes and $\text{G}_{\text{BP}}-\text{G}_{\text{RP}}$ colors were obtained from Gaia DR2 (\citealt{GaiaCollaboration2018}) and DR3 data (\citealt{GaiaCollaboration2021}). In order to calculate the V magnitudes, we used the photometric relationships from Gaia DR2 Documentation Release 1.2\footnote{\url{https://gea.esac.esa.int/archive/documentation/GDR2/}} and Gaia DR3 Documentation Release 1.3\footnote{\url{https://gea.esac.esa.int/archive/documentation/GDR3/}} between Gaia and Johnson-Cousins system. For Gaia DR2 (see Chapter 5.3.7, Table 5.8),
\begin{equation}
    \text{G-V} = -0.017600 - 0.006860(\text{G}_{\text{BP}}-\text{G}_{\text{RP}}) - 0.173200(\text{G}_{\text{BP}}-\text{G}_{\text{RP}})^2,
\end{equation}
with a calibration uncertainty of 0.045858 mag associated with the photometric relationship. Thus, the final uncertainty for V is calculated in the form
\begin{equation}
    \sigma_{\text{V}} = \sqrt{(\sigma_{\text{G}})^2 + ((0.006860 + 0.346400(\text{G}_{\text{BP}}-\text{G}_{\text{RP}}))\sigma_{\text{G}_{\text{BP}}-\text{G}_{\text{RP}}})^2 + 0.045858^2}.
\end{equation}
For Gaia DR3 (Chapter 5.5.1, Table 5.9),
\begin{equation}
    \text{G-V} = -0.02704 + 0.01424(\text{G}_{\text{BP}}-\text{G}_{\text{RP}}) - 0.2156(\text{G}_{\text{BP}}-\text{G}_{\text{RP}})^2 + 0.01426(\text{G}_{\text{BP}}-\text{G}_{\text{RP}})^3,
\end{equation}
with a calibration uncertainty of 0.03017 mag associated with the photometric relationship. Thus, the final uncertainty for V is calculated in the form
\begin{equation}
    \sigma_{\text{V}} = \sqrt{(\sigma_{\text{G}})^2 + (-0.01424 + 0.4312(\text{G}_{\text{BP}}-\text{G}_{\text{RP}}) - 0.04278(\text{G}_{\text{BP}}-\text{G}_{\text{RP}})^2)\sigma_{\text{G}_{\text{BP}}-\text{G}_{\text{RP}}})^2 + 0.03017^2}.
\end{equation}

We corrected the V magnitudes for extinction using A$_{\text{V}}$ values calculated with the \texttt{StarHorse} code (\citealt{Queiroz2018}), which is an isochrone-fitting code that compares observed quantities to stellar evolutionary models to derive distances, extinctions, ages, masses, effective temperatures, metallicities and surface gravities for field stars. The code has updated versions for better performance with Gaia EDR3 data (\citealt{Anders2022}), using extinction maps of \cite{Green2019} and \cite{Drimmel2003}, and Gaia DR2 data (\citealt{Anders2019}), using its own StarHorse-derived extinction map. For 11 stars, we used DR2 data for magnitudes, parallaxes and A$_{\text{V}}$ since no A$_{\text{V}}$ correction was available in StarHorse EDR3. For 1 star, TOI-238, there is no parallax available in Gaia. Finally, for 2 stars, Kepler-449 and Kepler-477, no A$_{\text{V}}$ was available in StarHorse DR2/EDR3 so we did not determine their evolutionary parameters.

\begin{figure}[!ht]
    \centering
    \includegraphics[width=.85\linewidth]{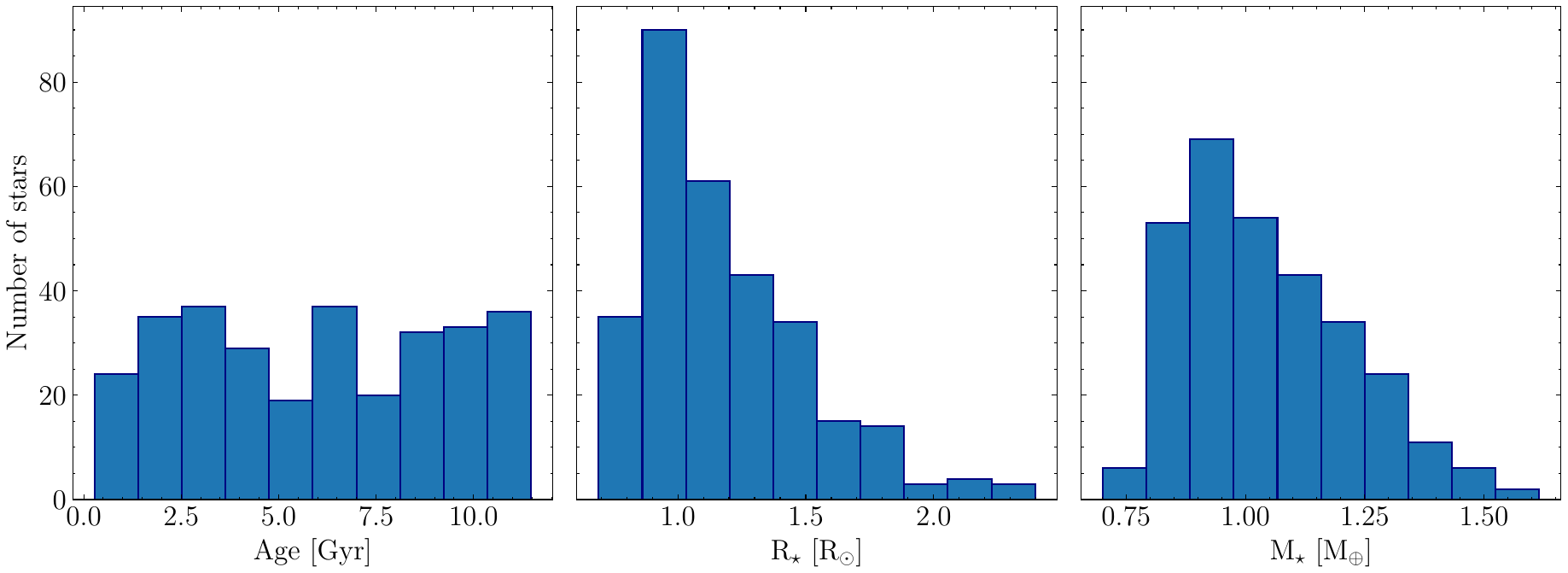}
    \caption{Evolutionary parameters determined for 300 stars using PARAM code v1.3 (\citealt{daSilva2006}).}
    \label{fig:evol_hists}
\end{figure}

We obtained evolutionary parameters for 300 stars (see Figure \ref{fig:evol_hists}). We note that, for five stars, PARAM did not converge (K2-173, K2-180, Kepler-444 , Kepler-990 and TOI-561). The evolutionary parameters determined for the stars, as well as their G and V magnitudes, G$_{\text{BP}-\text{RP}}$ color, A$_{\text{V}}$ and parallaxes, are listed in Table \ref{tab:sample}. The median values for the uncertainties are 0.64 Gyr for age, 1.3\% for $R_\star$ and 1.3\% for $M_\star$.

Finally, the 18 stars outside of our selection criteria of 5000 K $\leq$ \teff\ $\leq$ 6500 K and \logg\ $\geq$ 4.0 dex have not been included in the following abundance analysis. Comparisons with other results from the literature for atmospheric parameters (\citealt{Soubiran2022}) and stellar radii (\citealt{Stassun2017}; \citealt{Petigura2018a}; \citealt{Kruse2019}; \citealt{Loaiza-Tacuri2024}) for stars in common with this study are presented in Appendix \ref{ap:comparasions}.

\section{Chemical Abundances}
\label{sec:abs}
The atomic parameters for the C I, Ni I, and S I lines were taken from the Vienna Atomic Line Database\footnote{\url{http://vald.astro.uu.se/}} (VALD, \citealt{Ryabchikova2015}), with the exception of their $\log gf$ values for which we used solar $\log gf$ values -- following the procedure described in \cite{Ghezzi2026}. Although Ni is not a bioessential element, its inclusion was needed for the oxygen determination (see Section \ref{sec:oxygen}). For the region around the [O I] line, the atomic parameters of [O I] and nearby lines were also taken from VALD, but we used the $\log gf$ value from \cite{AllendePrieto2001} for [O I], and, for the two Ni I lines blended with the [O I], we used the weighted $\log gf$ values from \cite{Bensby2004}, which are based on the laboratory measurements of \cite{Johansson2003}. The line lists can be found in Table \ref{tab:line_list}.

\begin{table}
    \centering
    \begin{tabular}{lcrccr}
    \hline
    \hline
    ID & $\lambda_{\text{central}}$ & Ion & $\chi$ & $\log gf$ & $\log\Gamma_{\text{w}}$\\
     & (\AA) &  & (eV) &  &    (4$\pi sN_{\text{H}}$)$^{-1}$  \\
    \hline
    C I & 5052.167 & 6.0 & 7.685 & -1.318 & 0.000\\
    C I & 5380.337 & 6.0 & 7.685 & -1.608 & 0.000\\
    \hline
    S I$^8$ & 6743.483 & 16.0 & 7.866 & -1.555 & -7.160\\
    S I$^8$ & 6743.540 & 16.0 & 7.866 & -1.104 & -7.160\\
    S I$^8$ & 6743.580 & 16.0 & 7.866 & -1.004 & -7.160\\
    S I$^8$ & 6743.580 & 16.0 & 7.866 & -1.354 &  0.000\\
    S I$^8$ & 6743.640 & 16.0 & 7.866 & -1.184 &  0.000\\    
    S I$^8$ & 6756.750 & 16.0 & 7.870 & -1.915 & -7.160\\
    S I$^8$ & 6756.851 & 16.0 & 7.870 & -2.005 &  0.000\\
    S I$^8$ & 6756.960 & 16.0 & 7.870 & -1.075 & -7.160\\
    S I$^8$ & 6757.150 & 16.0 & 7.870 & -0.485 & -7.160\\
    \hline
    Ni I & 5010.938 & 28.0 & 3.635 & -0.918 & -7.225 \\
     \hline
    \end{tabular}
    \caption{List of atomic parameters of the C, S and Ni lines. The S I lines are highlighted with their multiplet numbering from \cite{Moore1945}. The columns list, respectively, the line identification, central wavelength, ionization stage following the format used by MOOG (0 means neutral), excitation potential, $\log gf$ and the logarithm of the van der Waals damping constant.}
    \label{tab:line_list}
    \tablecomments{This table is published in its entirety in the machine-readable format. A portion is shown here for guidance regarding its form and content.}
\end{table}

We measured the EWs of the C I, Ni I and S I lines using ARES v2 with the same parameters used for Fe I and Fe II lines (see Section \ref{sec:atm_par}), except for C I, for which we used \textit{lineresol} = 0.2 m\AA~as it yielded EWs more consistent with manual measurements. Also, for the C I and S I lines, the SNR adopted for the \textit{rejt} parameter is the SNR of the echelle order that includes each spectral line\footnote{The wavelength coverage of each spectral order of HARPS is available at  \url{https://www.eso.org/sci/facilities/lasilla/instruments/harps/inst/spec_form.html} and, for HARPS-N, at \url{https://www.tng.iac.es/instruments/harps/data/usermanv3.1.pdf}}. For the Ni I lines, we adopted the median SNR of the spectrum -- as it was the case for the Fe I and Fe II lines (see Section \ref{sec:atm_par_determination}). Finally, for some stars, C I or/and S I EWs provided unrealistic abundances. In such cases, they were measured using the \texttt{splot} task from PyRAF. If this did not resolve the issue, the lines were removed.

\subsection{Carbon, Nickel and Sulfur}
\label{sec:4_CNiS}

Carbon, whose dominant isotope is $^{12}$C, is formed by stars primarily during hydrostatic helium burning via the triple-$\alpha$ process. The stars that contribute most to carbon enrichment are of low- and intermediate-mass (0.8 $\text{M}_\odot$ $\leq\text{M}_\star\leq$ 8 $\text{M}_\odot$, \citealt{Matteucci2016}), as high-mass stars process a large part of their C through the $^{12}{\text{C}}(\alpha,\gamma)^{16}{\text{O}}$ reaction. The main optical indicators used to determine C abundances in FGK stars are atomic lines of C I (e.g., 5052 \AA, 5380 \AA~and 6580 \AA) and the molecular band of CH around 4300 \AA, but with only reasonable agreement between atomic and molecular abundances (e.g., \citealt{DelgadoMena2021}). The atomic lines listed above have high excitation potentials, around 8 eV, and become weak in cool stars. On the other hand, the CH band is strong and shapes the continuum around it. In this work, carbon abundances were determined based on 2 lines of C I, 5052 \AA~and 5380 \AA, using the MOOG driver \texttt{abfind} (see Table \ref{tab:line_list}).

Sulfur, with the dominant isotope $^{32}$S comprising 95\% of solar system S, is formed by high-mass stars during oxygen burning, which occurs in the final stages of evolution in massive stars (e.g., \citealt{Limongi2003}). The main indicators used historically to determine S abundances are the multiplets of S I at 6046 \AA, 6052 \AA, 6743 \AA, 6757 \AA, 8694 \AA, 9228 \AA~and 10456 \AA~(e.g., \citealt{Caffau2005}; \citealt{Spite2011}; \citealt{CostaSilva2020}) and the forbidden line [S I] at 10821 \AA~(e.g., \citealt{Matrozis2013}). Therefore, the determinations of sulfur abundances require more robust techniques due to blending with its own lines, in the case of multiplets, increasing the associated uncertainties. As a result, sulfur is an element that has historically been neglected in studies of $\alpha$-element abundances, being overshadowed by Si and Ca (\citealt{Caffau2005}). Thus, sulfur abundances are not easily found in the literature, with notable works by \cite{Luck2005}, for stars in the northern hemisphere, and \cite{CostaSilva2020}, for stars in the southern hemisphere. In this work, sulfur abundances were determined based on 9 lines from multiplet 8 of S I\footnote{Multiplet numbering from \cite{Moore1945}.}, between 6743 -- 6758 \AA, using the MOOG driver \texttt{blends} (see Table \ref{tab:line_list}).

Nickel is formed mainly in the very final stage of the life of massive stars and during type Ia supernovae (e.g., \citealt{Woosley1995}; \citealt{Limongi2003}). For massive stars, before core collapse, $^{56}\text{Fe}$ and $^{56}\text{Ni}$ are formed by successive fusion of quasi-equilibrium reactions ($\alpha$-particle capture) during explosive burning of silicon (\citealt{Arnett1996}). During a supernova Ia explosion, explosive thermonuclear nucleosynthesis further enhances Fe and Ni production, particularly of $^{56}\text{Ni}$ that decays into iron, contributing to $\sim$2/3 of total iron of the Galaxy (\citealt{Mishurov2019}). Also, due to their similar nucleosynthetic paths, the evolution of iron-peak elements (e.g., Ni, V, Cr, Mn, Co) tracks Fe. The dominant isotopes of iron and nickel are $^{56}\text{Fe}$ and $^{58}\text{Ni}$. The main indicators used to determine Ni abundances are atomic lines of Ni I in the optical region (e.g., \citealt{Adibekyan2012b}; \citealt{Ghezzi2026}; \citealt{Kirby2018}). In this work, nickel abundances were determined based on 39 lines of Ni I (\citealt{Ghezzi2026}), between 5010 -- 6843 \AA, using the MOOG driver \texttt{abfind} (see Table \ref{tab:line_list}). Finally, nickel abundances are determined as they are required for determining oxygen abundances (see Section \ref{sec:oxygen}).

\subsection{Oxygen}
\label{sec:oxygen}
Oxygen-16, the dominant O-isotope, is formed by high-mass stars during the hydrostatic burning of helium via $^{12}{\text{C}}(\alpha,\gamma)^{16}{\text{O}}$ and is ejected into the interstellar medium through supernova type II explosions. The main indicators used to determine O abundances in FGK stars are the O I line at 6158 \AA~(e.g., \citealt{BertrandeLis2015}), the forbidden [O I] lines at 6300 \AA~and 6363 \AA~(e.g., \citealt{AllendePrieto2001}; \citealt{BertrandeLis2015}), and the O I triplet at 7774 \AA~(e.g., \citealt{Steffen2015}; \citealt{Amarsi2019}; \citealt{Bergemann2021}). However, the 6158 \AA~and 6363 \AA~lines are not widely used due to their small equivalent widths, while the forbidden line at 6300 \AA~is strongly blended with 2 Ni I lines, and the triplet suffers from significant non-LTE deviations (e.g., \citealt{Amarsi2019}; \citealt{Bergemann2021}). Moreover, the lines at 6158 \AA~and 7774 \AA~have high excitation potentials, $\sim$10.7 eV and $\sim$9.2 eV, respectively, i.e., they become weak for cooler stars. Lastly, just like carbon, there can be discrepancies between the abundances based on different indicators (e.g., \citealt{AllendePrieto2001}; \citealt{Bergemann2021}). Even for the Sun, we find discrepancies between different methods and its measured oxygen abundance has changed considerably over the past 50 years: 8.92 dex by \cite{Lambert1978}; 8.66 $\pm$ 0.05 dex by \cite{Asplund2004}; 8.76 $\pm$ 0.07 dex by \cite{Caffau2008}; 8.69 $\pm$ 0.05 dex by \cite{Asplund2009}; 8.80 $\pm$ 0.03 dex by \cite{CubasArmas2020}; 8.75 $\pm$ 0.03 dex by \cite{Bergemann2021}; 8.69 $\pm$ 0.04 dex by \cite{Asplund2021}.

\begin{figure}[!ht]
    \centering
    \includegraphics[width=.65\linewidth]{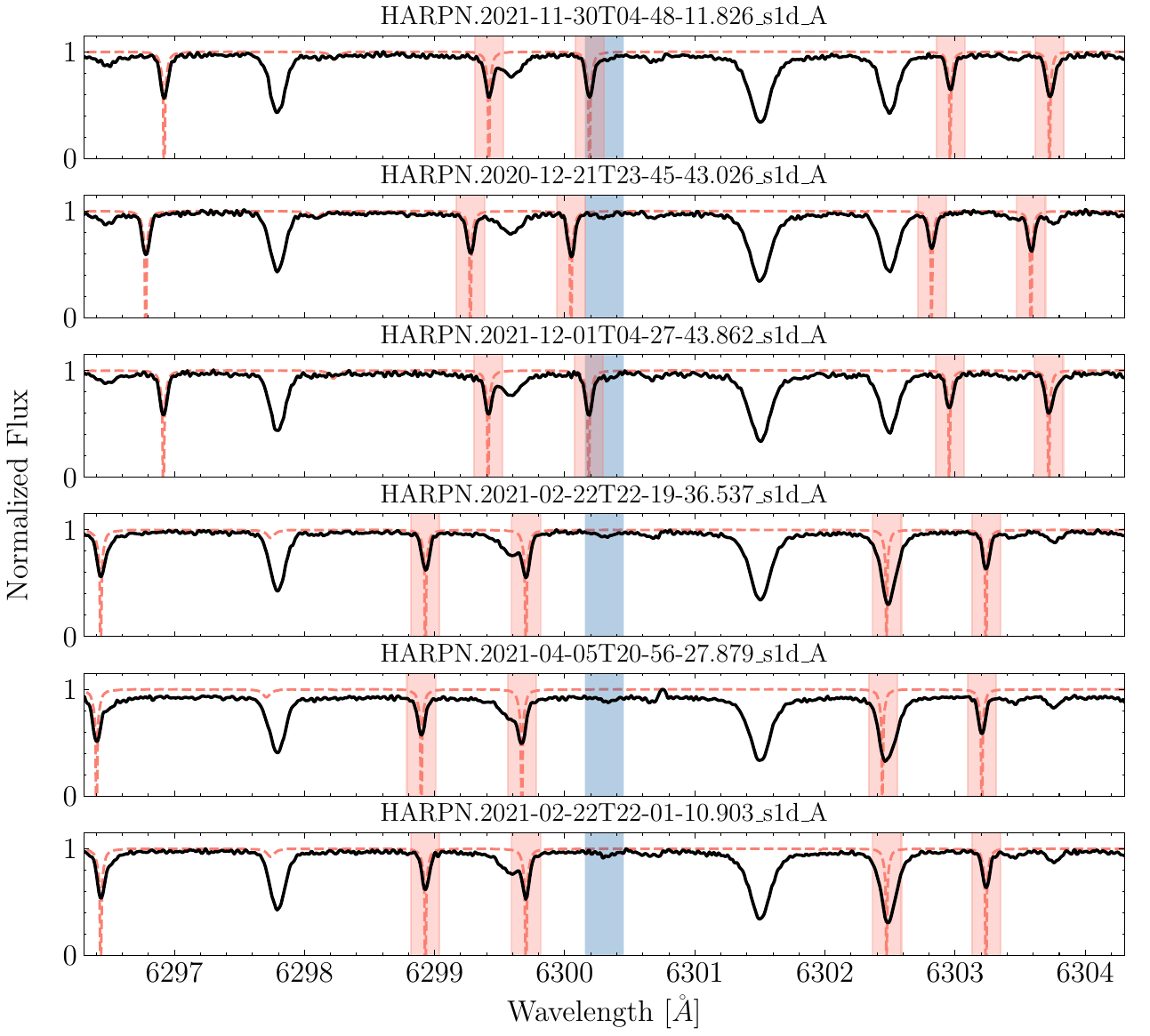}
    \caption{Telluric features around the oxygen forbidden line at 6300.304 \AA~on the spectra of HD 220197. HD 220197 spectra are shown in black and the shifted telluric spectra in dashed pink. The blue filled regions represent the range of $\pm$0.15\AA~around the [O I] line and the pink filled regions are the regions contaminated by the telluric features. The upper three panels show examples of contamination on the [O I] line, while the lower three have no contamination.}
    \label{fig:telluric}
\end{figure}

Oxygen abundances here were determined based on the forbidden line [O I] at 6300.304 \AA, using the MOOG driver \texttt{synth}, with \textit{damping} parameter set to 1 and a Gaussian broadening for the line, i.e., \textit{g} option. There are telluric features around the [O I] line, so in order to avoid possible contamination, we determined their position by using the telluric spectrum from \cite{Wallace2011} and Doppler shifting it according to every individual stellar spectrum's barycentric Earth radial velocity (available in the header of the fits file) and radial velocity (determined in this work, see Section \ref{sec:sample_data}). We used only the spectra in which a region of 0.15 \AA~around the center of the [O I] line was clean of any telluric features. In Figure \ref{fig:telluric}, we show an example of this procedure. For the star HD 220197, we see the superposition of the regions that represent the [O I] line (blue) and the telluric features (pink) in the upper three panels. Hence, these 3 spectra were not considered for our analysis. For the clean spectra, we manually removed the ones containing any sort of contamination, e.g., cosmic rays and/or deformations. Finally, the remaining spectra of each star were combined, to improve SNR, and normalized around the region including the [O I] line, between 6295 -- 6305 \AA, and also between 6235 -- 6245 \AA. This second region contains a Fe I line at 6240.646 \AA~that we use for the FWHM determination of the spectra (see below). From the 290 stars, 230 of them had clean spectra in the [O I] region.

Initial simple syntheses were performed using \textit{g} = 0.055 \AA~based on the resolution of 115000 of HARPS' twin spectrographs (FWHM$_{\text{inst}}$). Based on these synthetic spectra, we adjusted the central wavelength of the lines by shifting their wavelengths in 2 rounds: (i) within $\pm$ 0.10 \AA~with steps of 0.01 \AA; (ii) $w$ $\pm$0.01 \AA~with steps of 0.001 \AA, where $w$ is the best match found for the first round. The best match between the observed and the synthetic spectra was calculated through the minimization of the reduced $\chi^2$ (e.g., \citealt{Ghezzi2009}, \citeyear{Ghezzi2010b}, \citeyear{Ghezzi2018}),

\begin{equation}
    \chi^2_r = \frac{1}{(d-1)}\sum^n_{i = 1} \frac{(\text{O}_i - \text{S}_i)^2}{\sigma^2},
\end{equation}
where $d$ is the number of degrees of freedom and is defined as $d = n - p$, $n$ being the number of points in the observed spectra and $p$ the number of free parameters (in the wavelength adjustment, $p=1$); $\text{O}_i$ and $\text{S}_i$ are, respectively, the normalized observed and synthetic fluxes at point $i$ and $\sigma$ is the uncertainty of the observed continuum, defined as $\sigma = (\text{SNR})^{-1}$. For the Fe I line, we considered the 6240.50 -- 6240.78 \AA~range for the reduced $\chi^2$ analysis and, for [O I], 6300.10 -- 6300.50 \AA. 

To adjust the local continuum around the Fe I line, a similar reduced $\chi^2$ procedure was applied. In this case, we shifted the continuum within $\pm$0.050 of the original continuum with steps of 0.001 for only one round. For the [O I] line, due to its weakness and the presence of strong lines nearby (3 Si I lines at $\sim$6299.5 \AA~and 2 Fe I lines at $\sim$6301.6 \AA), noise was an issue. Thus, we smoothed the observed spectra using the moving average filter, which consists in expressing the data as a series of averages of different sections of itself, removing noisy fluctuations and retaining the true signal. For this, we applied a window size of 5. Finally, we adjusted the maximum of the smoothed local continuum to the maximum of the synthetic spectra between 6300.00 -- 6300.55 \AA. 

To determine the FWHM of the spectral lines, we used the Fe I line at 6240.646 \AA\footnote{Excitation potential = 2.223 eV, $\log gf$ = -3.337 and logarithm of the van der Waals damping constant = -7.661 $(4\pi s N_{\text{H}})^{-1}$.}\ because it is the closest moderately strong line ($\chi_{\text{e}}$=2.223 eV) that is isolated and has a clear local continuum. The FWHM adjustment was performed in 3 rounds, using $p = 1$: (i) FWHM varying within 0.05 -- 0.25 \AA~with steps of 0.05 \AA; (ii) FWHM varying within $s\pm$0.05 \AA~with steps of 0.01 \AA, where $s$ is the best match found for the first round; (iii) FWHM varying within $s\pm0.01$ \AA~with steps of 0.001 \AA, where $s$ is the best match found for the second round. In all of these rounds, we set the [Fe/H] as the abundance determined for the 6240 \AA~Fe I line during the atmospheric parameters determination using the \texttt{abfind} driver.

\begin{figure}[!ht]
    \centering
    \includegraphics[width=0.65\linewidth]{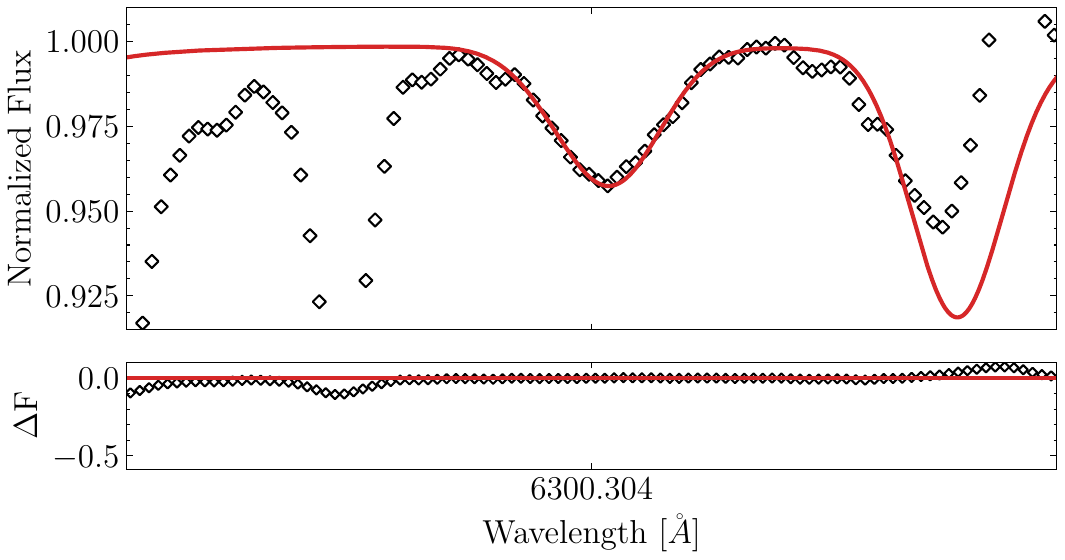}
    \caption{Example of the best fit for TOI-1710. The black diamonds represent the observed spectrum and the red line represent the synthetic spectrum. In the upper panel, we show the spectra and, in the lower panel, the residuals ($\Delta\text{F}=\text{F}_{\text{obs}}-\text{F}_{\text{synth}}$).}
    \label{fig:synthetic_spectra}
\end{figure}

The [O I] line is blended with 2 Ni I lines -- up to 60\% of the flux of the [O I] line may be attributed to the Ni I contribution (\citealt{BertrandeLis2015}) -- and weakly contaminated by a CN line, besides the telluric contamination mentioned above (e.g., \citealt{AllendePrieto2001}; \citealt{Teske2013}). Accordingly, we performed the oxygen abundance determination in 2 rounds, using $p = 2$: (i) A(O) varying within $\pm$0.50 dex with steps of 0.10 dex and A(Ni) fixed to the value determined for the star (see Section \ref{sec:4_CNiS}); (ii) A(O) varying within $o\pm0.10$ dex with steps of 0.01 dex, where $o$ is the best match found for the first round, and A(Ni) fixed to the value determined for the star. We stress that these values mentioned above are the ones from the \textit{abundances} parameter of MOOG, having the final absolute abundance for the element A(O) = 8.69 + $o$ + [Fe/H] dex, where $o$ is the best match for the second round and 8.69 is the solar absolute abundance from \cite{Asplund2009} (hereafter, A09). All best fits were inspected visually and, if needed, corrected manually. In Figure \ref{fig:synthetic_spectra}, we show an example of the best fit for TOI-1710.

As a consistency check, we determined the abundances for the Ceres 2009-02-08 solar spectrum using the same methodology described above and found A(C)$_\odot$=8.42$\pm$0.06 dex, A(O)$_\odot$=8.70$\pm$0.05 dex, A(S)$_\odot$=7.15$\pm$0.02 dex and A(Ni)$_\odot$=6.23$\pm$0.02 dex in addition to A(Fe)$_\odot$=7.50$\pm$0.01 dex. Comparing with values from A09, where A(C)$_\odot$=8.43$\pm$0.05 dex, A(O)$_\odot$=8.69$\pm$0.05 dex, A(S)$_\odot$=7.12$\pm$0.03 dex, A(Ni)$_\odot$=6.22$\pm$0.04 dex and A(Fe)$_\odot$=7.50$\pm$0.04 dex, we find an excellent agreement. This is also true for the more recent values from \cite{Asplund2021} (hereafter, A21), where A(C)$_\odot$=8.46$\pm$0.04 dex, A(O)$_\odot$=8.69$\pm$0.04 dex, A(S)$_\odot$=7.12$\pm$0.03 dex, A(Ni)$_\odot$=6.20$\pm$0.04 dex and A(Fe)$_\odot$=7.46$\pm$0.04 dex. Solar reference values from other authors can be found in Section \ref{sec:cap5_oxygen}. 

\subsection{Uncertainties}
Given that the atmospheric parameters have uncertainties associated, we estimated the change in the determined abundances by varying each atmospheric parameter by its uncertainty (plus and minus) and assumed the higher abundance difference as the uncertainty associated with that specific parameter. Lastly, we calculated the final uncertainty through a sum of squares of the uncertainties related to the atmospheric parameters and to the standard deviation of the mean of the abundances calculated for each specific element,
\begin{equation}
    \sigma_{\text{X}} = \sqrt{\sigma_{\text{X,Teff}}^2 + \sigma_{\text{X,[Fe/H]}}^2 + \sigma_{\text{X,logg}}^2 +\sigma_{\text{X,vmic}}^2 +\sigma_{\text{lines}}}.
\end{equation}
We estimated the change in the oxygen abundance caused by the uncertainty associated with the Ni abundance ($\pm\sigma_{\text{Ni}}$) and we performed the same process described for the atmospheric parameters, but with Ni, and added $\sigma_{\text{O,Ni}}$ (uncertainty on O related to the variation of the Ni abundance) to the sum of squares.

The uncertainties of C, O, S and Ni have median values of 0.04 dex, 0.03 dex, 0.06 dex and 0.02 dex, respectively, having approximately 93\%, 91\%, 83\% and 96\% of the uncertainties below 0.10 dex. We note that the stars with high uncertainties can be split into 3 separate groups: i) SNR$<$85; ii) $v_{\sin i}>$10 km s$^{-1}$; iii) [Fe/H]$>$ 0.25 dex. Finally, the determined abundances are listed in Table \ref{tab:sample}. In total, we determined C, O, S and Ni abundances for 303, 230, 272 and 307 stars, respectively, for which 289, 225, 269 and 290 stars are from the sample with 5000 K $\leq$ \teff\ $\leq$ 6500 K and \logg$\geq$4.0 dex. Also, based on abundance data available on Vizier\footnote{VizieR catalogue access tool, operated at CDS, Strasbourg, France (\url{https://vizier.cds.unistra.fr}).} and Hypatia catalog, as of December 2025, we report, for the first time, carbon abundances for 75 stars, oxygen abundances for 63 stars, sulfur abundances for 115 stars and nickel abundances for 69 stars.

\subsection{Trends of Abundances with the Effective Temperature}
\label{sec:abundances_teff}
To investigate if there are any trends between the derived abundances and the effective temperature, we performed a linear regression\footnote{The linear fits mentioned in this and the following sections were performed using the Ordinary Linear Regression (OLS) method from Python \textit{statsmodels} package.} between the absolute abundances of C, O, S and Ni with $T_{\text{eff}}$, and obtained R$^2$ values of 0.127, 0.112, 0.172 and 0.010 for C, O, S and Ni, respectively. For the angular coefficients, we obtained $t$-values of -6.461, 5.311, -7.439 and -1.720 and Pearson $p$-values of 0.000, 0.000, 0.000 and 0.086. Also, we performed Spearman tests, to investigate non-linear correlations, and obtained $p$-values of 0.000, 0.000, 0.000 and 0.042. In this work, we consider a statistically significant correlation if $p$-value$<$0.001. Furthermore, in order to better account for possible fluctuations and to improve the robustness of the statistical analysis, we bootstrapped the samples 1000 times, performed the same tests and obtained median R$^2$ values of 0.127, 0.114, 0.171 and 0.011, median $t$-values of -6.464, 5.356, -7.429 and -1.745, median $p$-values of 0.000, 0.000, 0.000 and 0.080 and median Spearman $p$-values of 0.000, 0.000, 0.000 and 0.042 for C, O, S and Ni, respectively. In summary, there is a correlation between \teff\ and C, O and S abundances mainly caused by the cooler stars.

\begin{figure}[!ht]
    \centering
    \includegraphics[width=.8\linewidth]{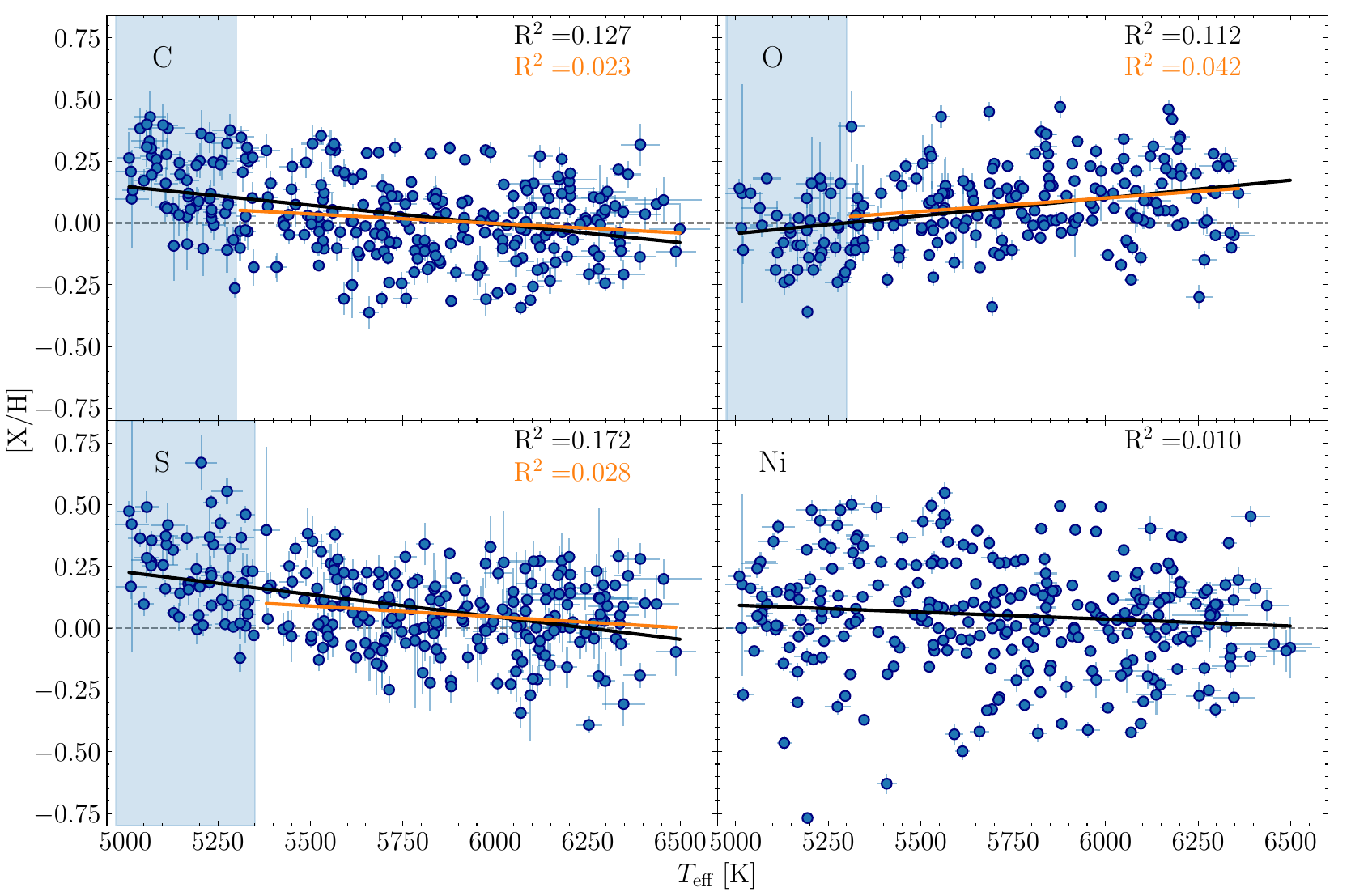}
    \caption{Distribution of [X/H] with \teff. The black line represents the linear fit for the whole sample. The orange line represents the linear fit for the stars outside the blue region. In the upper right corner, we show the R$^2$ statistics of each OLS regression. The gray dashed line represents [X/H] = 0.00 dex.}
    \label{fig:abundances_teff}
\end{figure}

To verify the location of a transition in \teff\ where the Pearson and Spearman tests do not result in a statistically significant trend, we calculated the statistics for subsamples with a minimum \teff\ varying from 5000 K to 5600 K in steps of 1 K. For C, O, and S, the \teff\ in which Pearson and Spearman $p$-values$>$0.001 and $|t$-value$|<$3 are 5229 K, 5279 K and 5284 K. For Ni, the values are 0.087, 0.042 and -1.720 for Pearson and Spearman $p$-values and angular coefficient $t$-value, respectively, at 5000 K. We performed the same tests using bootstrap with steps of 50 K and found no trends for subsamples with \teff\ above 5300 K, 5300 K and 5350 K for C, O, and S, respectively. In Figure \ref{fig:abundances_teff}, we show the distribution of [X/H] with effective temperature and the OLS regressions of the complete  (black lines) and temperature restricted (orange lines) samples.
Below, we explore the \teff\ limitations for each element.

For the C I lines, close to the C I 5052 \AA~line there is a strong Fe I line at 5051.6 \AA~that develops pronounced wings as the effective temperature decreases -- blending with the C I line (\citealt{Luck2006}). For the C I 5380 \AA~line, there is an increasing blending with an unknown feature as the effective temperature decreases (\citealt{Luck2006}). According to \cite{Luck2006}, carbon abundances are unreliable below \teff$\sim$5500 K and 5250 K for the C I lines at 5380 \AA~and 5052 \AA, respectively. In this work, we find a good agreement between the abundances of the two carbon lines for stars with \teff$\geq$5300 K. By performing a linear regression for the derived abundances for stars with \teff$\geq$5300 K, we obtain no trend with R$^2$=0.023 (median value for the distribution of the 1000 bootstrapped samples = 0.024), $t$-value=-2.327 (-2.400), Pearson $p$-value=0.021 (0.017) and Spearman $p$-value=0.054 (0.047). We also observed no trend between [C/H] and \logg, finding R$^2$=0.006 (0.006), $t$-value=-1.213 (-1.140), Pearson $p$-value=0.227 (0.238) and Spearman $p$-value=0.101 (0.112). Finally, we have 233 carbon determinations for stars with \teff$\geq$5300 K.

For the [O I] line at 6300 \AA, we notice that below 5300 K the oxygen abundances start to decrease with decreasing \teff. This behavior was also noted by \cite{BertrandeLis2015} for stars with \teff$\lesssim$5600 K for the [O I] 6300 \AA~line and \teff$\lesssim$5400 K for the O I line at 6158 \AA~but, for the latter, the abundances increase with \teff. By performing a linear regression for the derived abundances for stars with \teff$\geq$5300 K, we obtain no trend with R$^2$=0.042 (0.044), $t$-value=2.828 (2.884), Pearson $p$-value=0.005 (0.004) and Spearman $p$-value=0.002 (0.002). We also observed no trend between [O/H] and \logg, finding R$^2$=0.018 (0.019), $t$-value=-1.820 (-1.863), Pearson $p$-value=0.070 (0.064) and Spearman $p$-value=0.037 (0.035). Finally, we have 183 oxygen determinations for stars with \teff$\geq$5300 K.

For the S I lines, we notice that for \teff$<$5350 K the sulfur abundances start to increase with decreasing \teff. This happens because, as the effective temperature decreases, the blending of the S I lines with unknown lines becomes more substantial (e.g., \citealt{Takeda2016}; \citealt{Takada-Hidai2002}). \cite{Luck2015} shows a strong trend between [S/Fe] and \teff\ for stars with \teff$\sim$5500 K (see their Figure 6). By performing the linear regression for the derived abundances for stars with \teff$\geq$5350 K, we obtain no trend with R$^2$=0.028 (0.028), $t$-value=-2.441 (-2.442), Pearson $p$-value=0.016 (0.015) and Spearman $p$-value=0.074 (0.077). We also observed no trend between [S/H] and \logg, finding finding R$^2$=0.019 (0.020), $t$-value=2.007 (2.053), Pearson $p$-value=0.046 (0.041) and Spearman $p$-value=0.196 (0.172). Finally, we have 212 sulfur determinations for stars with \teff$\geq$5300 K.

The Ni I lines do not have restrictions regarding \teff\ for stars in our sample. Also, we find no trend between [Ni/H] and \logg, with R$^2$=0.006 (0.006), $t$-value=-1.300 (-1.306), Pearson $p$-value=0.195 (0.88) and Spearman $p$-value=0.024 (0.023).

Comparisons with other results from the literature for C (\citealt{Hinkel2014}; \citealt{Brewer2016}; \citealt{Biazzo2022}), O (\citealt{Brewer2016}; \citealt{Biazzo2022}), S (\citealt{Perdigon2021}; \citealt{Biazzo2022}) and Ni (\citealt{Brewer2016}; \citealt{Biazzo2022}) abundances for stars in common with this study are presented in Appendix \ref{ap:comparasions}.

\section{Exoplanetary Radii}
\label{sec:exoplanet_rad_determ}
We determined the radii for the exoplanets in our sample using the stellar radii derived in this study and transit depths from the literature. We found transit depth values for 373 exoplanets in the NASA Exoplanet Archive, totalizing 285 host stars. We determined the exoplanetary radii using the inferred stellar radii and the transit depth ($\Delta F$), which is the maximum fraction of stellar flux that is blocked during the transit of the exoplanet, by using the relation from \cite{Seager2003}:
\begin{equation}
    R_{\text{pl}} = 109.1979 \times \sqrt{\Delta F \times 10^{-6}} \times R_\star,
\end{equation}
where $R_{\text{pl}}$ is the exoplanetary radius in $\text{R}_\oplus$, $\Delta F$ is in ppm and $R_\star$ is the stellar radius in $\text{R}_\odot$. The associated uncertainties were calculated through error propagation.

In our analysis, we only considered the $\Delta F$ values of confirmed exoplanets. We obtained the $\Delta F$ values from \cite{Borde2010} (1 exoplanet), \cite{Deeg2010} (1 exoplanet), \cite{Mullally2015} (5 exoplanets), \cite{Addison2016} (1 exoplanet), \cite{Barros2016} (1 exoplanet), \cite{Pope2016} (2 exoplanets), \cite{Vanderburg2016} (5 exoplanets), \cite{Christiansen2017} (2 exoplanets), \cite{Stassun2017} (56 exoplanets), \cite{Boufleur2018} (1 exoplanet), \cite{Livingston2018} (7 exoplanets), \cite{Thompson2018} (44 exoplanets), \cite{Yu2018} (2 exoplanets), \cite{Becker2019} (1 exoplanet), \cite{Kruse2019} (32 exoplanets), \cite{Vanderburg2019} (3 exoplanets), \cite{Bonfanti2021} (1 exoplanet), \cite{AzevedoSilva2022} (1 exoplanet) and  \cite{Vivien2024} (3 exoplanets). For the rest of the 204 exoplanets, we used values ExoFOP\footnote{The Exoplanet Follow-up Observing Program website (\url{https://exofop.ipac.caltech.edu/tess/view_toi.php}) is designed to optimize resources and facilitate collaboration in follow-up studies of exoplanet candidates. ExoFOP serves as a repository for project- and community-gathered data by allowing upload and display of data and derived astrophysical parameters. ExoFOP contains stellar parameters from the TESS Input Catalog (TIC), which is served by the Mikulski Archive for Space Telescopes (MAST), and planet parameters from the NASA Exoplanet Archive.}. The exoplanetary radii determined here are listed in Table \ref{tab:planet_radii}, as well as their orbital period ($P_{\text{orb}}$) and mass ($M_{\text{pl}}$) taken from the NASA Exoplanet Archive. The median uncertainties for the exoplanetary parameters are 2\% for $\Delta F$ and $R_{\text{pl}}$, 0.0003\% for $P_{\text{orb}}$ and 12\% for $M_{\text{pl}}$.

\begin{table}
   \centering
   \begin{splittabular}{lcccccBcc}
   \hline
   \hline
   Star & Exoplanet letter & $\Delta F$ & $R_{\text{pl}}$ & $P_{\text{orb}}$ & $\Delta F$ and $P_{\text{orb}}$ reference & $M_{\text{pl}}$ & $M_{\text{pl}}$ reference\\
    & & & (R$_{\oplus}$) & (days) & & (M$_{\oplus}$) & \\
    \hline
   55 Cnc & e & 0.03847$\pm$0.00042 & 2.33660$\pm$0.03063 & 0.73655$\pm$0.00000 & ExoFOP & 7.99000$\pm$0.32000 & 1\\
    BD+20 594 & b & 0.04900$\pm$0.00250 & 2.74110$\pm$0.07231 & 41.68550$\pm$0.00300 & 2 & 22.24810$\pm$9.53490 & 2\\
    CoRoT-1 & b & 1.31400$\pm$0.00104 & 15.80940$\pm$0.20037 & 1.50896$\pm$0.00001 & ExoFOP & 327.35000$\pm$38.14000 & 3\\
    CoRoT-4 & b & 1.38123$\pm$0.04210 & 14.41210$\pm$0.24967 & 9.20165$\pm$0.00003 & ExoFOP & 228.83000$\pm$25.43000 & 4\\
    CoRoT-5 & b & 1.52427$\pm$0.11997 & 16.50163$\pm$0.69285 & 4.03790$\pm$0.00001 & ExoFOP & 148.42000$\pm$14.93700 & 5\\
    \hline
   \end{splittabular}
   \caption{Exoplanetary parameters. Column 1 shows the host-star identification. Column 2 shows the exoplanet letter. Column 3 shows the transit depth ($\Delta F$). Column 4 shows the calculated exoplanetary radii. Column 5 shows the exoplanet orbital period. Column 6 shows the reference for the transit depth and orbital period values of the exoplanet. Column 7 and 8 show the exoplanet mass and its reference.}
   \label{tab:planet_radii}
   \tablecomments{This table is published in its entirety in the machine-readable format. A portion is shown here for guidance regarding its form and content. References: [1] \cite{Bourrier2018}, [2] \cite{Stassun2017}, [3] \cite{Barge2008}, [4] \cite{Moutou2008}, [5] \cite{Rauer2009}, \cite{Borde2010}, \cite{Deeg2010}, \cite{Bonomo2017}, \cite{Boufleur2018}, \cite{Livingston2018}, \cite{Livingston2024}, \cite{Kruse2019}, \cite{Guenther2024}, \cite{Bonomo2023}, \cite{Nikolov2014}, \cite{Torres2008}, \cite{Buchhave2010}, \cite{Brown2012}, \cite{Chakrabarty2019}, \cite{Hartman2014}, \cite{Brahm2016}, \cite{Jorda2020}, \cite{Christiansen2017}, \cite{Sozzetti2021}, \cite{Polanski2024}, \cite{Rosa2024}, \cite{Kane2023}, \cite{Desidera2023}, \cite{Murgas2022}, \cite{Dransfield2022}, \cite{Egger2024}, \cite{Teske2020}, \cite{Di2020}, \cite{Ellis2021}, \cite{Bonfanti2021}, \cite{Osborn2021}, \cite{Zhang2024}, \cite{Delrez2021}, \cite{AzevedoSilva2022}, \cite{Nicholson2024}, \cite{Bonfanti2023}, \cite{Orell-Miquel2023}, \cite{Espinoza2020}, \cite{Murphy2025}, \cite{Howard2025}, \cite{Becker2019}, \cite{Vanderburg2019}, \cite{Bonfanti2025}, \cite{Petigura2020}, \cite{Vanderburg2016}, \cite{Santerne2016}, \cite{Johnson2016}, \cite{Grziwa2016}, \cite{Lillo-Box2020}, \cite{Barraga2019}, \cite{Vivien2024}, \cite{Lopez2019}, \cite{Thygesen2023}, \cite{Barros2016}, \cite{Pope2016}, \cite{Brahm2018}, \cite{Palle2019}, \cite{Kosiarek2019}, \cite{Yu2018}, \cite{Eastman2016}, \cite{Kuhn2016}, \cite{Thompson2018}, \cite{Shaw2025}, \cite{Borsato2019}, \cite{Mullally2015}, \cite{Mills2019}, \cite{Brinkman2025}, \cite{Dalba2024}, \cite{Dalba2021}, \cite{Jenkins2020}, \cite{Gill2020}, \cite{Bryant2020}, \cite{Smith2021}, \cite{Battley2024}, \cite{Schulte2025}, \cite{Nielsen2020}, \cite{Hobson2023}, \cite{Hoyer2021}, \cite{Hobson2024}, \cite{Osborn2023}, \cite{Krenn2024}, \cite{Rodriguez2021}, \cite{Georgieva2023}, \cite{Alqasim2024}, \cite{Fridlund2020}, \cite{Hawthorn2023}, \cite{Sha2021}, \cite{Armstrong2023}, \cite{Otegi2021}, \cite{Lockley2025}, \cite{Beard2024}, \cite{Turtelboom2022}, \cite{Tran2022}, \cite{Heidari2025}, \cite{Sha2023}, \cite{Kaba2022}, \cite{Persson2022}, \cite{Rice2023}, \cite{Yee2022}, \cite{TalaPinto2025}, \cite{Brahm2023}, \cite{Gill2024}, \cite{vS2025}, \cite{Frame2023}, \cite{Jones2024}, \cite{Psaridi2023}, \cite{Mantovan2024b}, \cite{Hacker2024}, \cite{Nabbie2024}, \cite{Chaturvedi2025}, \cite{Rodri2025}, \cite{Castro-Gonza2024}, \cite{Ulmer-Moll2022}, \cite{Mantovan2024a}, \cite{Grieves2022}, \cite{Ulmer-Moll2023}, \cite{Manni2025}, \cite{Eberhardt2025}, \cite{Maciejewski2014}, \cite{Addison2019}, \cite{Gillon2009}, \cite{McGruder2023}, \cite{Corte2020}, \cite{Anderson2015}, \cite{Bouchy2010}, \cite{Triaud2011}, \cite{Mancini2016}, \cite{Mancini2018}, \cite{Southworth2016}, \cite{Ciceri2016}, \cite{Nascimbeni2023}, \cite{He2013}, \cite{Southworth2015}, \cite{Anderson2014}, \cite{Ehrenreich2020}, \cite{Noguer2024}, \cite{Triaud2017}, \cite{Maciejewski2023}, \cite{Addison2016}, \cite{Hellier2015}, \cite{Gillon2014}, \cite{Anderson2017}, \cite{Wright2023}, \cite{Turner2016}, \cite{Seidel2020}, \cite{Maxted2016}, \cite{Hellier2017}, \cite{Doyle2023}, \cite{Almenara2022}, \cite{Hellier2019}, \cite{Nielsen2019}, \cite{Yee2025}, \cite{Crouzet2012} and \cite{Smith2015}.}
\end{table}

Comparisons with other results from the literature for exoplanetary radii (\citealt{Stassun2017}; \citealt{Petigura2018a}; \citealt{Kruse2019}; \citealt{Loaiza-Tacuri2024}) for exoplanets in common with this study are presented in Appendix \ref{ap:comparasions}.

\section{Chemical Evolution of Exoplanet Hosts}
\label{sec:results}
\label{sec:05}
\label{sec:tinsley}
\begin{figure}[!ht]
    \centering
    \includegraphics[width=\linewidth]{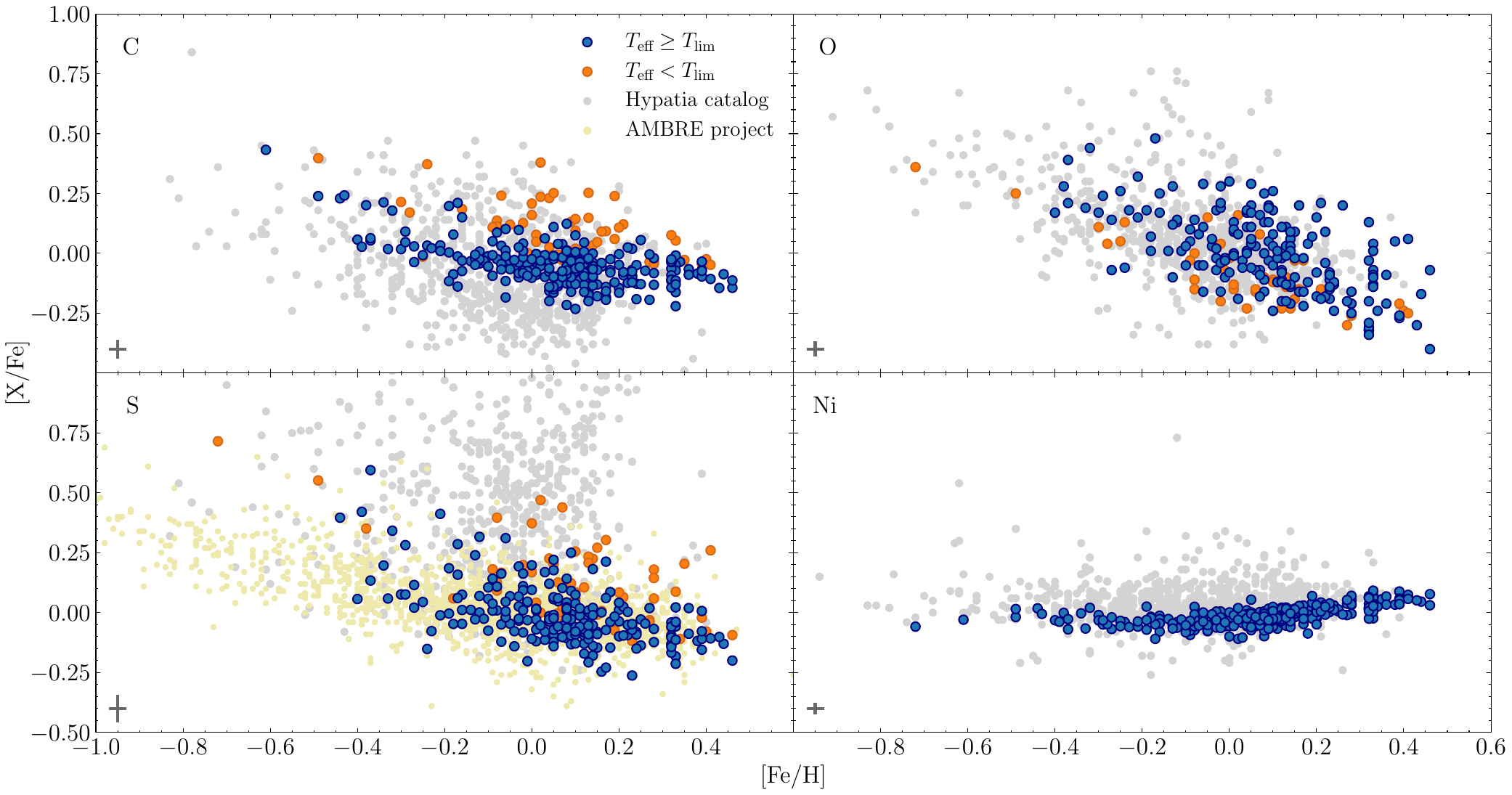}%
    \vspace{0cm}
    \begin{minipage}{\linewidth}
    \centering
    \begin{minipage}{.45\textwidth}
        \centering
        \includegraphics[width=\linewidth]{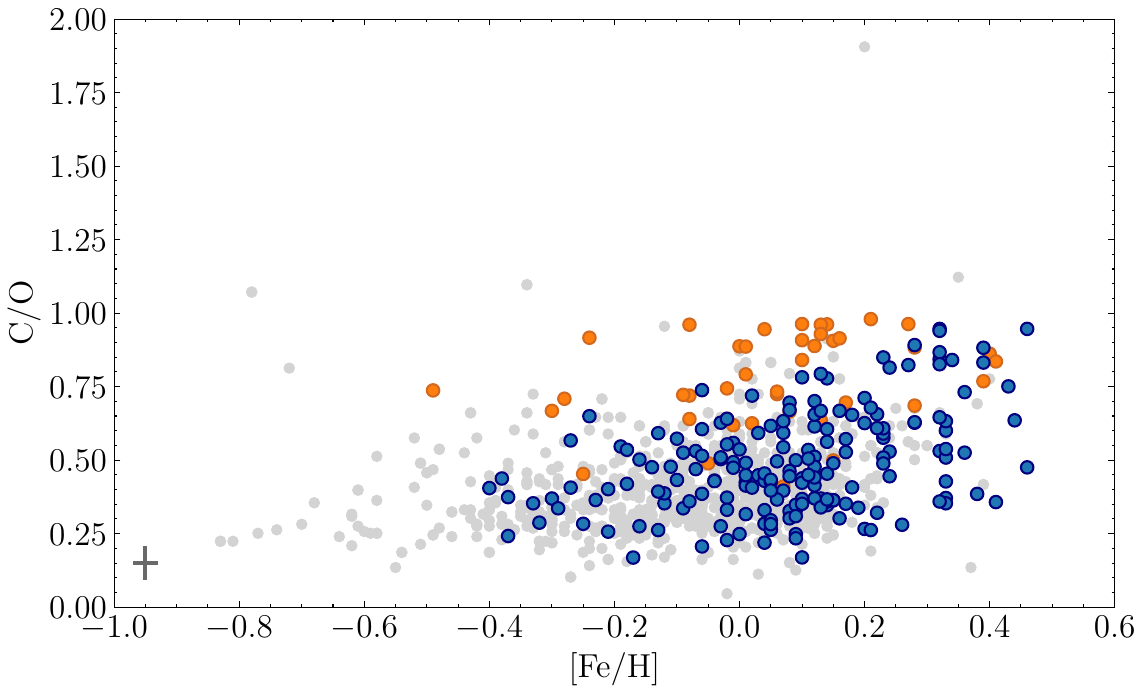}
    \end{minipage}
    \begin{minipage}{.45\textwidth}
        \centering
        \includegraphics[width=\linewidth]{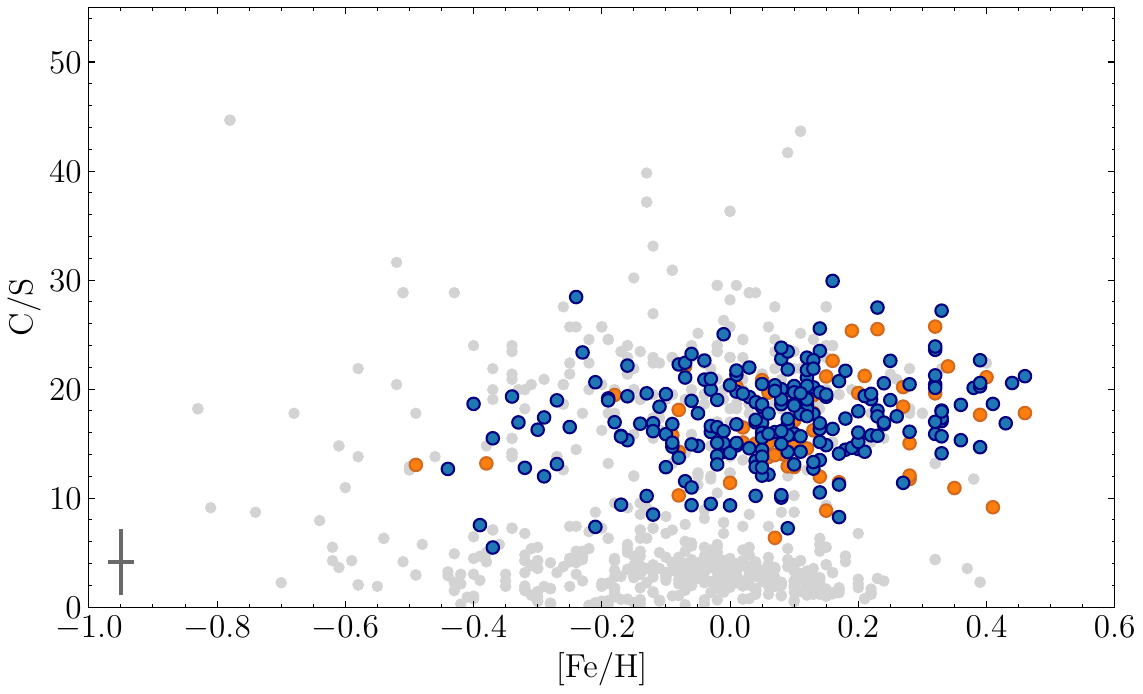}
    \end{minipage}      
    \end{minipage}%
    \vspace{0cm}%
    \begin{minipage}{\linewidth}
        \centering
    \begin{minipage}{.45\textwidth}
        \centering
        \includegraphics[width=\linewidth]{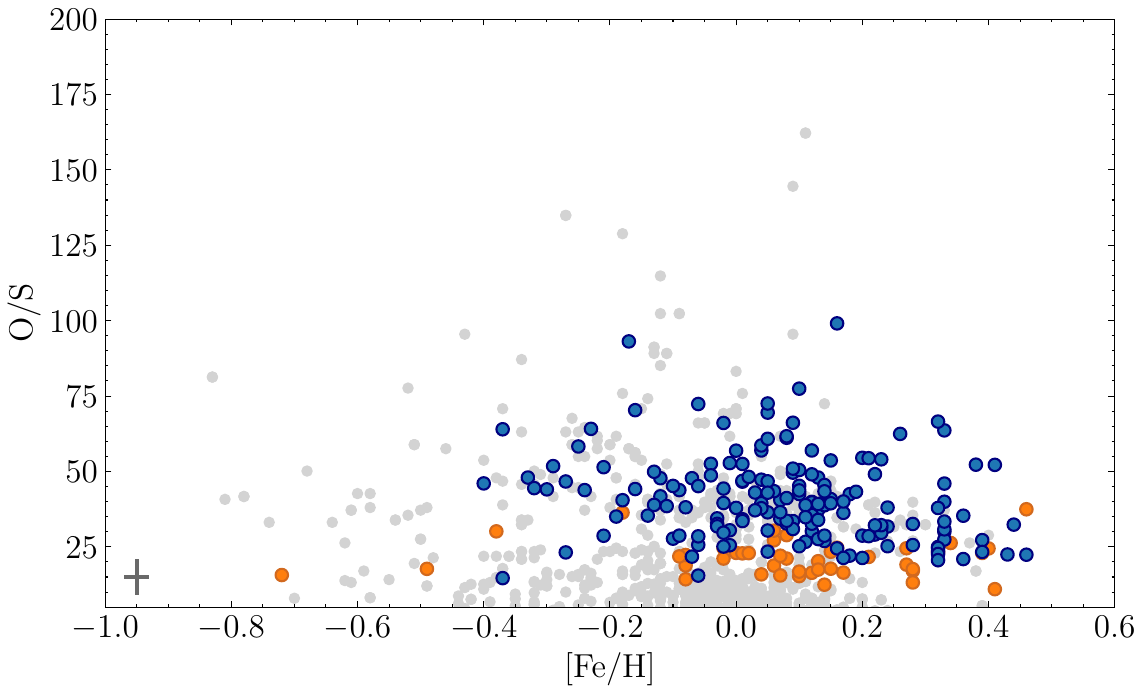}
    \end{minipage}
    \begin{minipage}{.45\textwidth}
        \centering
        \includegraphics[width=\linewidth]{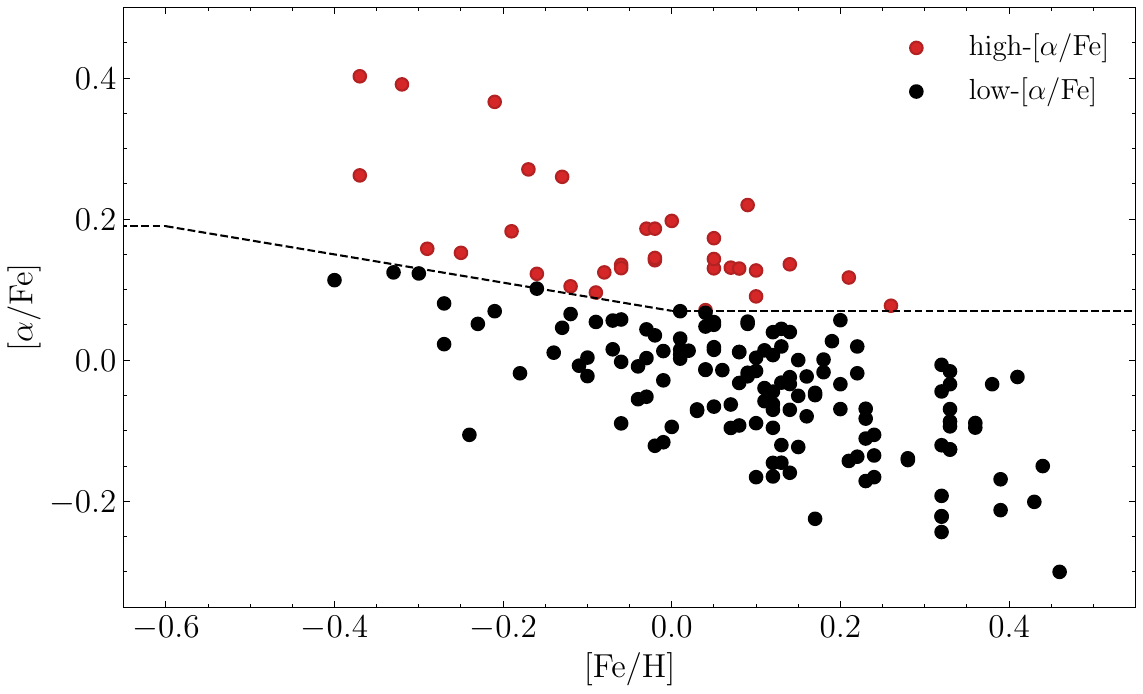}
    \end{minipage}      
    \end{minipage}%
    \caption{Distribution of [X/Fe] of C (left panel of first row), O (right panel of first row), S (left panel of second row) and Ni (right panel of second row), C/O (left panel of third row), C/S (right panel of third row) and O/S (left panel of last row) ratios  and, also, [$\alpha$/Fe] (right panel of last row) with metallicity. The blue and orange circles represent, respectively, abundances for stars having effective temperatures above and below 5300 K, 5300 K and 5350 K for C, O and S abundances, respectively. The gray and light yellow dots represent the same distributions for the stars in the Hypatia catalog (\citealt{Hinkel2014}) and in the AMBRE Project (\citealt{Perdigon2021}), respectively. In the lower left of each panel, we show the median uncertainties in gray. The red and black circles represent, respectively, stars on the high-[$\alpha$/Fe] and low-[$\alpha$/Fe] sequences.}
    \label{fig:cniso}
\end{figure}

In Figure \ref{fig:cniso}, we show the Tinsley-Wallerstein diagram, which is the canonical diagram of the chemical evolution of the Galaxy that presents the distribution of [X/Fe] versus [Fe/H], for our sample of planet hosting stars. In general, our distributions successfully reproduce the shapes expected by the time-delay model (e.g., \citealt{Matteucci2016}), in which we see the contribution of different stellar yields, low- (C, Fe, Ni) and high-mass (O, S) stars. Also, our results have significantly lower scatter when compared to non-homogeneous abundances, which is the case for those collected in the Hypatia database. Carbon shows roughly constant solar values, except in the low-metallicity regime for [Fe/H]$\lesssim$-0.2. Oxygen and sulfur show the typical behavior for $\alpha$-elements, decreasing [X/Fe] with increasing [Fe/H]. Finally, nickel closely follows Fe, showing almost a straight line on [Ni/Fe] = 0.00. We also show the distributions of the abundance ratios of C/O, C/S and O/S as functions of [Fe/H] in Figure \ref{fig:cniso}. The C/O and C/S ratios show a correlation with the metallicity as expected from the behavior of [C/Fe], [O/Fe], and [S/Fe]. The correlation of a modest increase in C/O with metallicity was also seen in \cite{Nissen2013}, \cite{Teske2014} and \cite{DelgadoMena2021}. The O/S exhibits a tendency towards an anti-correlation with [Fe/H]. 

Figure \ref{fig:cniso} also shows the distributions of [X/Fe], C/O, C/S and O/S for stars that have effective temperatures \teff$<$5300 K for C and O, and \teff$<$5350 K for S (orange filled circles). The addition of these cooler stars generally increases the scatter in the abundance results. We can see that the abundances for the cooler stars fall systematically on the upper parts of the distributions for [C/Fe] and [S/Fe], and also for the C/O ratios. On the contrary, they tend to fall below for the [O/Fe] distribution and are systematically lower for O/S. For C/S, the cooler stars are not systematically distributed.
In the following, we will only consider the abundances of stars with \teff$\geq$5300 K for C and O and \teff$\geq$5350 K for S. For Ni, we will consider all abundances determined since they do not have any limitations regarding effective temperature.

In the bottom right panel of Figure \ref{fig:cniso} we show the distribution of the [alpha/Fe] versus [Fe/H] for our sample (stars with \teff$>$5300 K for O and \teff$>$5350 K for S), where we calculate $[\alpha/\text{Fe}]$ as the average of oxygen and sulfur, the two $\alpha$-elements studied here. Following \cite{Ghezzi2026}, we used the separation estimated from Figure 1 of \cite{Adibekyan2011} to divide the host stars sample into low- and high-[$\alpha$/Fe] sequences, in which \textit{i}) $[\alpha/\text{Fe}]=0.19$ for $\text{[Fe/H]}<-0.60$; \textit{ii}) $[\alpha/\text{Fe}]=-0.20 \times\text{[Fe/H]}+0.07$ for $-0.06\leq\text{[Fe/H]}\leq0.00$; \textit{iii}) $[\alpha/\text{Fe}]=0.07$ for $\text{[Fe/H]}>0.00$, noting that in \cite{Adibekyan2011}, $[\alpha/\text{Fe}]$ is the average of [Mg/Fe], [Si/Fe] and [Ti/Fe]. The majority of our sample belongs to the low-[$\alpha$] sequence population. In total, we have 168 stars with both oxygen and sulfur abundances determined, for which 32 fall on the high-[$\alpha$/Fe] sequence (corresponding to the chemical thick disk) and 136 on the low-[$\alpha$/Fe] sequence (thin disk).
 
\section{Discussion: Star-Planet Connection}
\label{sec:star_planet_connection}
In this section, we investigate relationships between stellar abundances and planetary radii ($R_{\text{pl}}$) determined in this work, as well as orbital periods ($P_{\text{orb}}$) and masses ($M_{\text{pl}}$) taken from the NASA Exoplanet Archive (see Table \ref{tab:planet_radii}). As mentioned in Section \ref{sec:05}, we are only considering abundances of C, O and S for stars with \teff$\geq$5300 K, \teff$\geq$5300 K and \teff$\geq$5350 K, respectively. The exoplanets were divided following the boundaries of \cite{Wilson2022}. For planet size classes: (i) Sub-Earths (sE), $R_{\text{pl}}<1\ R_\oplus$; (ii) Super-Earths (SE), $1\ R_\oplus \leq R_{\text{pl}}<1.9\ R_\oplus$; (iii) Sub-Neptunes (sN), $1.9\ R_\oplus \leq R_{\text{pl}}<4\ R_\oplus$; (iv) Sub-Saturns (sS), $4\ R_\oplus \leq R_{\text{pl}}<8\ R_\oplus$; (v) Jupiters (J), $8\ R_\oplus \leq R_{\text{pl}}<23\ R_\oplus$. Also, for period classes: (i) hot, $P_{\text{orb}} \leq 10$ days; (ii) warm, $10<P_{\text{orb}} \leq 100$ days; (iii) cool, $100<P_{\text{orb}} \leq 300$ days. In total, we have 373 exoplanets with radii and orbital periods determined, 3 sE (hot), 45 SE (35 hot, 9 warm and 1 cool), 124 sN (68 hot, 54 warm and 2 cool), 44 sS (23 hot, 18 warm, 1 cool and 2 with $P_{\text{orb}}>300$ days) and 157 J (115 hot, 32 warm, 6 cool and 6 with $P_{\text{orb}}>300$ days). Given the small number of cool exoplanets and exoplanets with $P_{\text{orb}}>300$ days in our sample, these are not considered in our analysis.

\subsection{Single and Multi-planetary Systems}
\label{sec:single-vs-multi}
\begin{figure}[!ht]
    \centering
    \begin{minipage}{\linewidth}
    \centering
    \begin{minipage}{.32\textwidth}
        \centering
        \includegraphics[width=\linewidth]{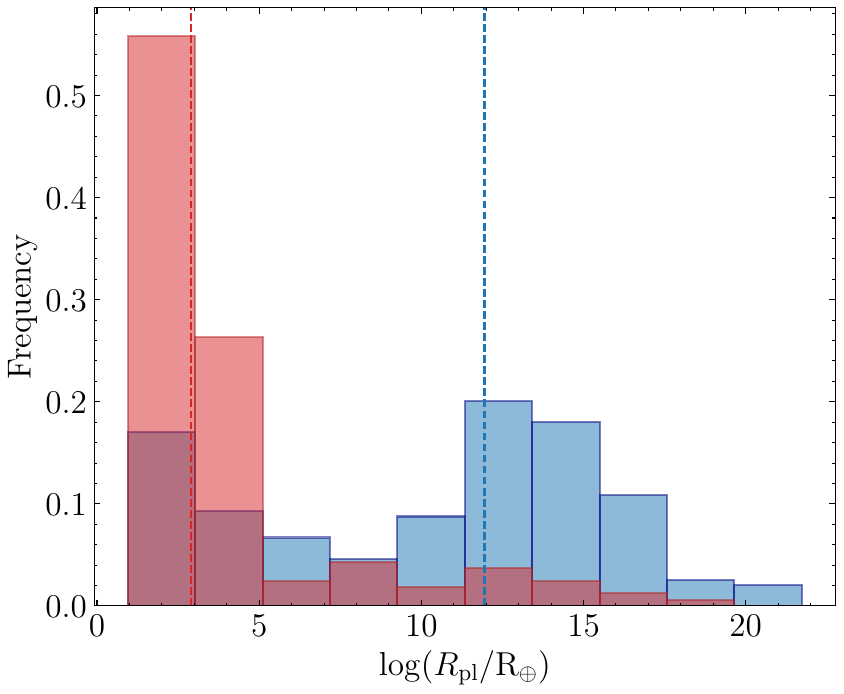}
    \end{minipage}
    \begin{minipage}{.32\textwidth}
        \centering
        \includegraphics[width=\linewidth]{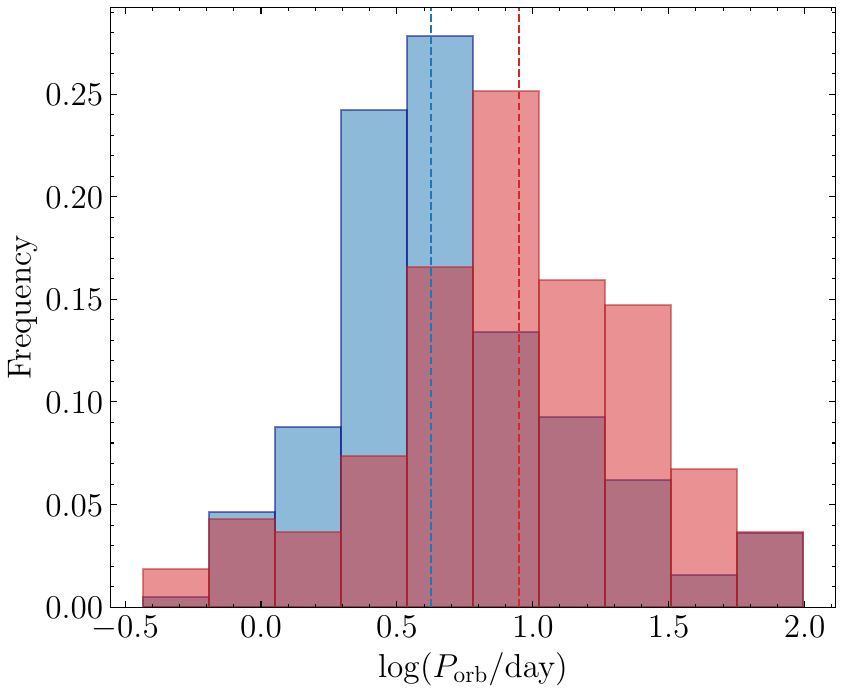}
    \end{minipage}
    \begin{minipage}{.32\textwidth}
        \centering
        \includegraphics[width=\linewidth]{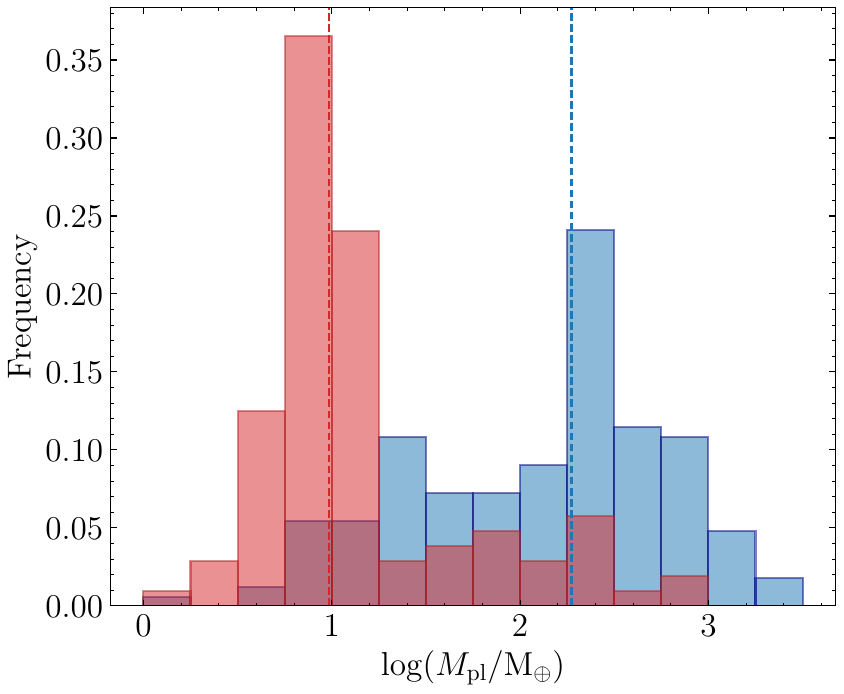}
    \end{minipage}
    \end{minipage}%
    \includegraphics[width=\linewidth]{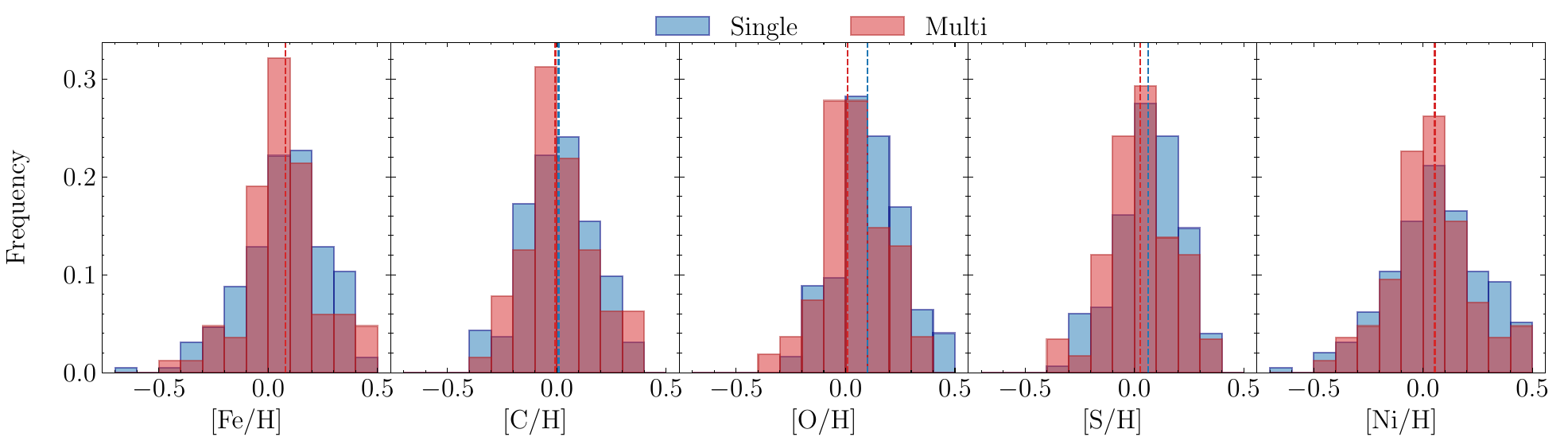}%
    \vspace{0cm}
    \includegraphics[width=.6\linewidth]{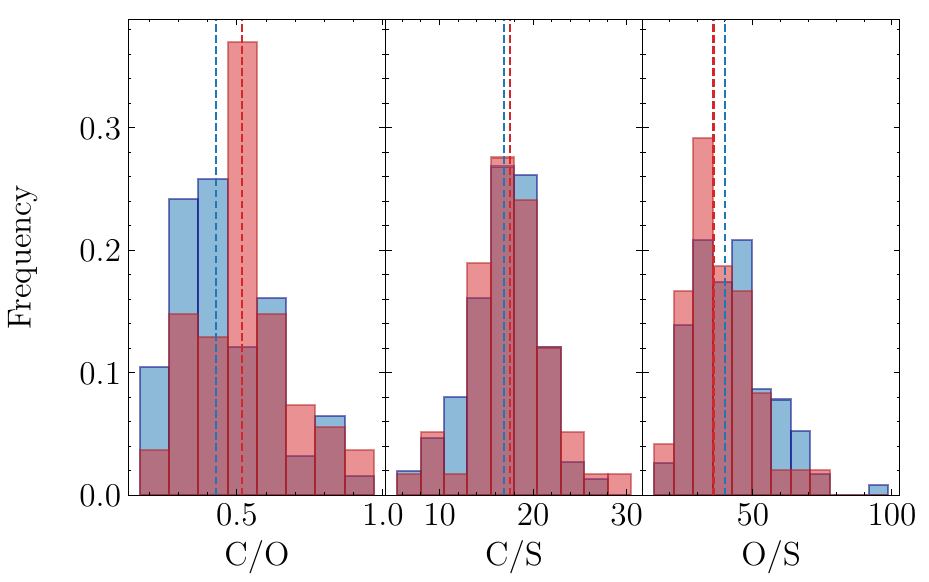}
    \caption{Histograms of $\log(R_{\text{pl}}/\text{R}_\oplus)$, $\log(P_{\text{orb}}/\text{day})$, $\log(M_{\text{pl}}/\text{M}_\oplus)$ (top panels), host star [X/H] (middle panels) and C/O, C/S and O/S abundance ratios (bottom panels). Systems having only one  confirmed exoplanet (singles) are shown in blue and systems with more than one confirmed exoplanet in red (multis). The dashed lines indicate the median values of each parameter and the colors match the ones from the histogram bars. The median differences ($\text{singles}-\text{multis}$) are 0.00, 0.01, 0.09, 0.03, 0.00, -0.09 ($(\Delta\text{C/O})/(\text{C/O}_{\text{singles}})\sim21\%$), -0.62 ($\sim4\%$) and 4.15 ($\sim10\%$) for [Fe/H], [C/H], [O/H], [S/H], [Ni/H], C/O, C/S and O/S, respectively.}
    \label{fig:histograms-single-multi}
\end{figure}

We divided our sample into stars having only one confirmed exoplanet (singles) and those having more than one confirmed exoplanet (multis). In the top row of Figure \ref{fig:histograms-single-multi}, we show the histograms of the distributions of $\log(R_{\text{pl}}/\text{R}_\oplus)$, $\log(P_{\text{orb}}/\text{day})$ and $\log(M_{\text{pl}}/\text{M}_\oplus)$. We can see from the top left and right panels that small and low-mass exoplanets are more frequent in multi-planetary systems, with median differences ($\text{singles}-\text{multis}$) of 9.05 for $\log(R_{\text{pl}}/\text{R}_\oplus)$ and 1.29 for $\log(M_{\text{pl}}/\text{M}_\oplus)$. For $\log(P_{\text{orb}}/\text{day})$ (top middle panel), the median difference between singles and multis is -0.32, meaning that exoplanets in multi-planetary systems tend to have larger orbital distances than exoplanets in single systems. We performed K-S tests between the distributions of $\log(R_{\text{pl}}/\text{R}_\oplus)$, $\log(P_{\text{orb}}/\text{day})$ and $\log(M_{\text{pl}}/\text{M}_\oplus)$ of the two groups (singles and multis) and found percentages of $p_{\text{KS}}<0.001$ of 100\% for the three of them, i.e., the distributions of radii, orbital periods and masses of exoplanets in single and multi-planetary systems are significantly different. Similar results were found for $P_{\text{orb}}$ (e.g., \citealt{Wright2009}) and $R_{\text{pl}}$ (e.g., \citealt{Muresan2026}). \cite{Muresan2026} also investigated the $R_{\text{pl}}$ distributions of singles and multis and found that, when removing the hot Jupiters from their sample, the $R_{\text{pl}}$ distributions were indistinguishable. Here, we still find significant differences between the distributions of $R_{\text{pl}}$ of singles and multis after removing the hot Jupiters (percentage of $p_{\text{KS}}<0.001$ of 70\%).

In the middle and bottom panels of Figure \ref{fig:histograms-single-multi}, we show the distribution of [Fe/H], [C/H], [O/H], [S/H], [Ni/H], C/O, C/S and O/S ratios, we note that the abundance distributions for the singles and multis are very similar for all elements, except for C/O -- for which there is an excess of exoplanets in single systems around stars with $\text{C/O}\lesssim0.50$. These results for [Fe/H] are in agreement with other studies in the literature (e.g., \citealt{Weiss2018}; \citealt{Ghezzi2021}; \citealt{Loaiza-Tacuri2025}). 
We performed K-S tests between abundance distributions of the two groups (singles and multis) and found percentages of $p_{\text{KS}}<0.001$ of 1\%, 0\%, 32\%, 5\%, 3\%, 61\%, 1\% and 9\% for [Fe/H], [C/H], [O/H], [S/H], [Ni/H], C/O, C/S and O/S, respectively, i.e., finding significant differences only between the distributions of C/O ratios of stars hosting single and multi-planetary systems.

The scarcity of massive exoplanets in multi-planetary systems might be caused by dynamical instabilities triggered by their presence, which increases collisions, ejections and planet-planet scattering, possibly reducing the number of planets in the system (e.g., \citealt{Izidoro2017}; \citealt{Pan2025}). Additionally, massive exoplanets are more frequent around metal-rich stars (e.g., \citealt{Fischer2005}; \citealt{Sousa2011}; \citealt{Mortier2013}; \citealt{Ghezzi2018}), which agrees with what we see in Figures \ref{fig:mass_ab_ratios} and \ref{fig:histograms-single-multi}. \cite{Pan2025} performed simulations of multi-planetary systems and classified the final stage of the system as ``observed'' singles or multis and found that the mean eccentricity and inclination of the planetary orbits increase significantly with host star [Fe/H]. Even though more than one planet might survive from this dynamically hot environment, their final orbits are highly misaligned and these systems might be observed as single-transit -- compared to lower metallicity counterparts in which the planets can maintain nearly circular and coplanar orbits (see their Figure 4).

\subsection{Distribution of Elemental Abundances of Stars Hosting Small and Giant Exoplanets}
\label{sec:giant_vs_small}
Recent studies have shown that the metallicity distribution of stars hosting sub-Saturn or Jupiter-like exoplanets is different from that of stars hosting smaller exoplanets, such as sub-Earths, super-Earths or sub-Neptunes. There is a dependence of giant exoplanet occurrence rate with metallicity, but this is not observed for stars hosting small exoplanets -- where a wider range of metallicities is observed for stars hosting super-Earths and sub-Neptunes (e.g., \citealt{Udry2006}; \citealt{Ghezzi2010a}; \citealt{Buchhave2012}; \citealt{Adibekyan2012b}; \citealt{Adibekyan2019}; \citealt{Ghezzi2021}; \citealt{Ghezzi2026}). However, there is a dependence with metallicity for stars hosting hot small exoplanets (e.g., \citealt{Petigura2018b}; \citealt{Wilson2022}; \citealt{Mulders2016}; \citealt{Wanderley2025}; \citealt{Ghezzi2026}). In this study, we divided the stars into two groups based on their exoplanet size classes: stars hosting giant exoplanets (sS or J) or small exoplanets (sE or SE or sN). If a star has at least one giant exoplanet, we define it as a giant exoplanet host. Throughout this and the following analysis, in order to better account for possible fluctuations and to improve the robustness of the statistical analysis, we bootstrapped the two groups 1000 times. We also performed two-sample Kolmogorov–Smirnov test\footnote{The two-sample Kolmogorov-Smirnov (K-S) tests were performed using \texttt{ks\_2samp} function from Python \textit{scipy} package.} for each bootstrapped sample. For each bootstrapped sample, we calculate its median abundance and K-S $p$-value (herein, $p_{\text{KS}}$), obtaining a distribution of median abundances and $p_{\text{KS}}$. In this work, we consider a result to be significant if the $p_{\text{KS}}<$0.001.

\begin{figure}[!ht]
    \centering
    \includegraphics[width=\linewidth]{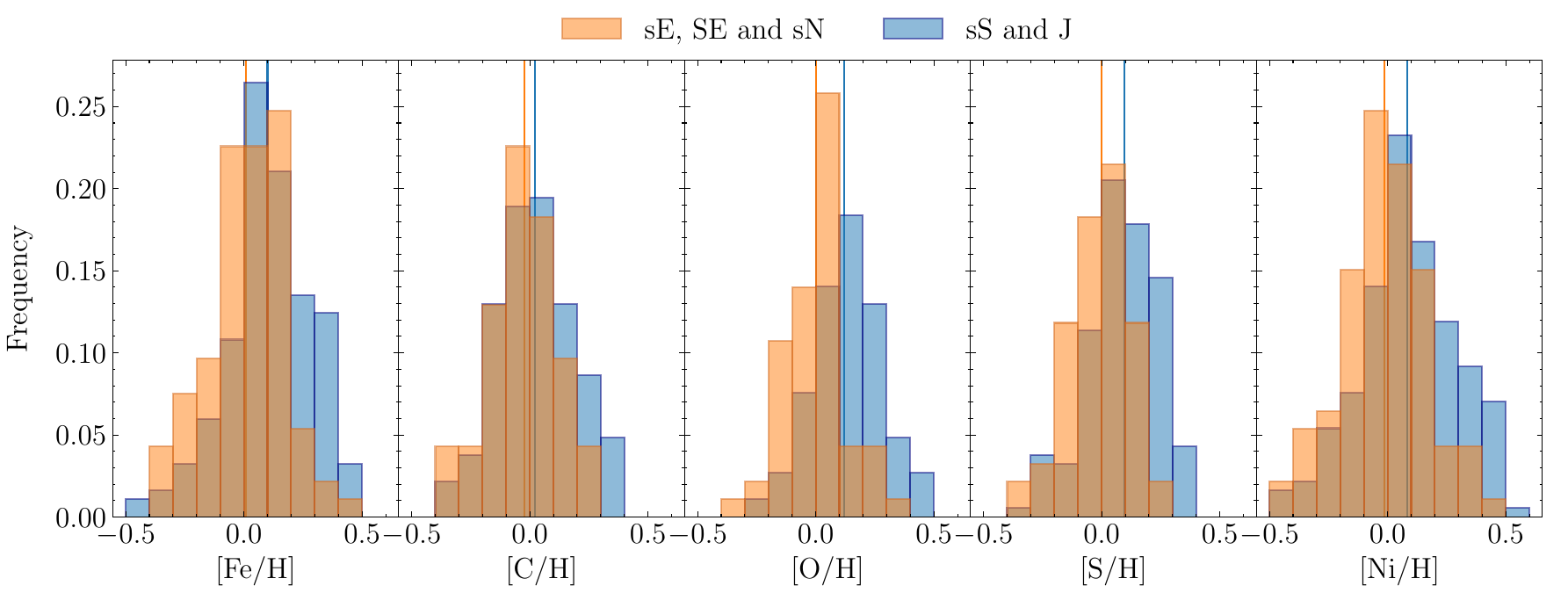}
    \vspace{0cm}
    \includegraphics[width=.6\linewidth]{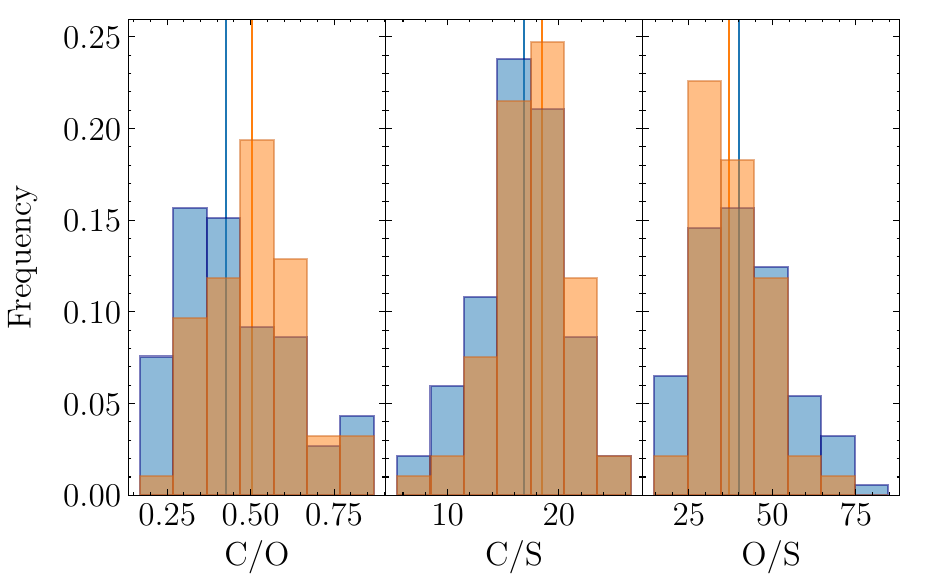}
    \caption{Distribution of elemental abundances for stars hosting giant exoplanets (sS and J, blue bars) and small exoplanets (sE, SE and sN, orange bars). In the first row, we show [Fe/H], [C/H], [O/H], [S/H] and [Ni/H]. In the second row, we show C/O, C/S and O/S. The vertical lines represent the median values for the 1000 bootstrapped samples, with colors matching the respective histograms.}
    \label{fig:06_EarthxGas}
\end{figure}

In the first row of Figure \ref{fig:06_EarthxGas}, we show that the abundance distributions for hosts of small exoplanets (orange histograms) have lower median values of [X/H] compared to those of giant exoplanets hosts (blue histograms). We can see from the comparison of the histograms that there is an excess of higher abundances in the distributions of giant exoplanet hosts for all elements. The median [X/H] differences (giant - small) are 0.09 dex, 0.04 dex, 0.12 dex, 0.10 dex and 0.09 dex for Fe, C, O, S and Ni, respectively. Similar results were found in the literature for Fe (e.g., \citealt{Sousa2011}; \citealt{Wang2015}; \citealt{Ghezzi2021}), Ni (e.g., \citealt{Adibekyan2012b}; \citealt{Ghezzi2026}) and, also, Mg (e.g., \citealt{Adibekyan2012b}) and Si (e.g., \citealt{Adibekyan2012b}; \citealt{Ghezzi2026}), which are $\alpha$-elements as O and S. In the second row of Figure \ref{fig:06_EarthxGas}, we show the distribution of the abundance ratios of C/O, C/S and O/S\footnote{$\text{X}/\text{Y}=10^{\text{A(X)}-\text{A(Y)}}$, where A(X) and A(Y) are the absolute abundances of X and Y, respectively.} ratios for the two populations. The distributions of C/O in stars hosting small and large exoplanets are clearly different. The differences in the median abundance ratios (giant - small) for C/O, C/S and O/S are, respectively, -0.08 ($(\Delta\text{C/O})/(\text{C/O}_{\text{giants}})\sim18\%$), -1.60 (9\%) and 3.01 (7\%). Such results suggest that stars hosting only small planets tend to have larger C/O ratios than those hosting at least one giant exoplanet, the differences for the C/S and O/S are not deemed to be significant (higher than $\sim$12\%). 

For the K-S tests, we find $p_{\text{KS}}<$0.001 for a large fraction of the bootstrapped samples for all elements (indicating possible differences in the distribution between giant and small exoplanet hosts), except C. For Fe, O, S, and Ni, we obtained $p_{\text{KS}}<$0.001 for approximately 65\%, 99\%, 91\% and 68\% of the bootstrapped samples, respectively. On the other hand, for C, we obtained $p_{\text{KS}}<$0.001 for only 16\%. For C/O, C/S and O/S, we obtained $p_{\text{KS}}<$0.001 for 51\%, 12\% and 7\%, respectively. 

Finally, we also performed K-S tests for [X/Fe] distributions of small and large exoplanet hosts, but they all returned $p_{\text{KS}}<$0.001 for a small fraction of the bootstrapped samples (between 2 -- 43\%) -- suggesting that the trends observed for [X/H] are reflecting the correlation between [Fe/H] and $R_{\text{pl}}$ (e.g., \citealt{Ghezzi2026}). 

\subsection{Trends with Exoplanet Radius}
\begin{figure}[!ht]
    \centering
    \begin{minipage}{.4\textwidth}
        \centering
        \includegraphics[width=\linewidth]{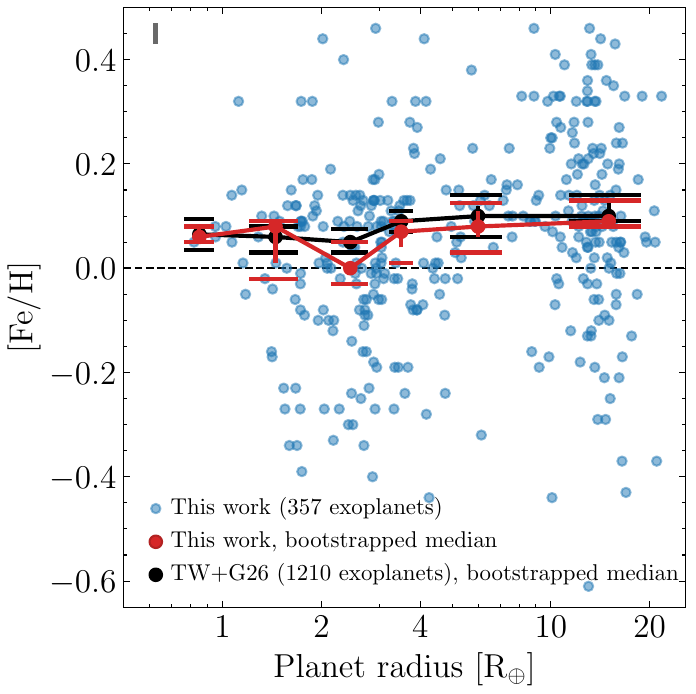}
    \end{minipage}
    \begin{minipage}{.4\textwidth}
        \centering
        \includegraphics[width=\linewidth]{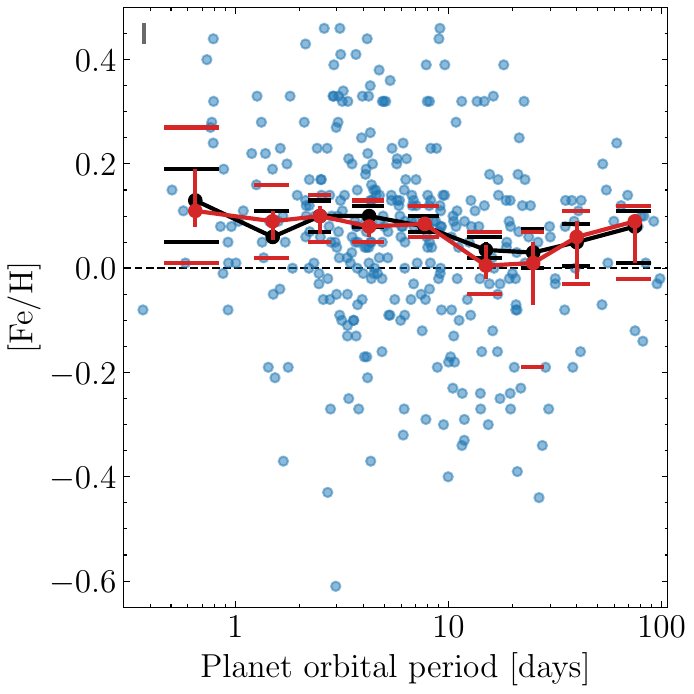}
    \end{minipage}
    \caption{Distribution of [Fe/H] of host stars with planet radius (left panel) and planet orbital period (right panel). The blue circles represent the sample from this work. The red circles represent the median abundances of the 1000 bootstrapped samples and their 68\% confidence interval range of the distribution (equivalent to a 1$\sigma$ range for a symmetric distribution) for the bins (0.7, 1], (1, 1.9], (1.9, 3], (3, 4], (4, 8] and (8, 22] R$_\oplus$ and (0.3, 1], (1, 2], (2, 3], (3, 5.5], (5.5, 10], (10, 20], (20, 30], (30, 50] and (50, 100] days. The red horizontal bars represent the 95\% confidence interval range, equivalent to a 2$\sigma$ range. The black circles and horizontal bars correspond to the same quantities derived from the combination of the sample from this work and from \citealt{Ghezzi2026}. The black dashed lines represents the solar values. In the top left of each panel, we show the median uncertainties in gray.}
    \label{fig:06_FeH_Rad_Porb}
\end{figure}

We present the distributions of the abundances of [Fe/H], [C/H], [O/H], [S/H], [Ni/H], and the abundance ratios of C/O, C/S, and O/S as a function of exoplanet radius in the different panels of Figure \ref{fig:06_FeH_Rad_Porb}.
In our sample, we clearly see a transition where there is a drastic decrease in the amount of exoplanets present approximately between 4 -- 8 R$_\oplus$. This gap falls within the sub-Saturn mass range (10 -- 100 M$_{\oplus}$), which is expected to have a lower number of planets within the core accretion formation scenario because planets beyond the ice line grow so rapidly from 10 M$_{\oplus}$ to 100 M$_{\oplus}$ that only a few are left with the intermediate mass range 10 -- 100 M$_{\oplus}$ (e.g., \citealt{Ida2004}).

For [Fe/H] (see top left panel of Figure \ref{fig:06_FeH_Rad_Porb}), the distribution of metallicity with exoplanet radius shows a clear tendency for host stars to be more metal-rich as exoplanet radius increases. We performed OLS regressions and obtained a median positive angular coefficient with median $t$-value=3.50 and $p_{\text{OLS}}<$0.001 ($p$-value of the angular coefficient determined by the OLS regression) for 58\% of the bootstrapped samples, indicating an overall increase in [Fe/H] of the host star with the exoplanet radius (e.g., \citealt{Buchhave2012}; \citeyear{Buchhave2014}; \citealt{Petigura2018b}; \citealt{Ghezzi2021}; \citealt{Wilson2022}; \citealt{Ghezzi2026}). Although the [Fe/H] distribution for stars hosting Jupiter exoplanets covers basically our entire range of metallicities, their median [Fe/H] is high (0.10 dex). Similarly to \citealt{Petigura2018b} (see their Figure 3) and \cite{Wilson2022} (see their Figure 9), we find a roughly constant mean [Fe/H] up to $\sim2$ R$_\oplus$ followed by a slight increase towards the Jupiter hosts. We also see an enhancement in [Fe/H] for stars hosting sub-Neptunes with larger radii (3 -- 4 R$_\oplus$) compared to stars hosting smaller sub-Neptunes (1.9 -- 3 R$_\oplus$), which was also observed in the APOGEE Kepler sample analyzed by \cite{Wilson2022}. The difference in the median values of the bootstrapped samples between the smaller and larger sub-Neptune hosts is 0.07 dex. However, the K-S tests resulted in a $p_{\text{KS}}<$0.001 for only 2\% of the bootstrapped samples, which indicates that there is not a statistically significant difference between the two distributions of sub-Neptune hosts.

\begin{figure}[!ht]
    \centering
    \includegraphics[width=\linewidth]{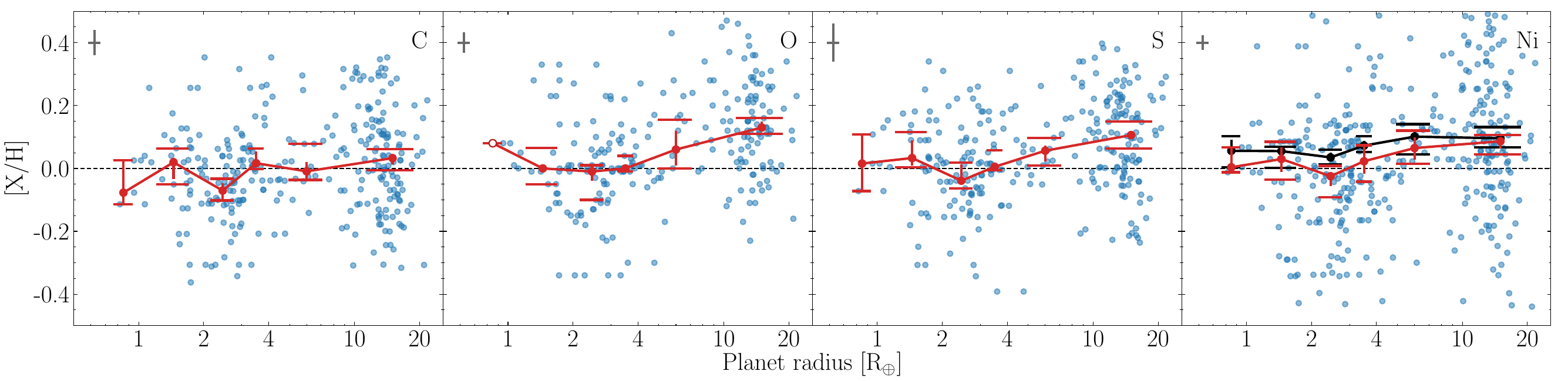}%
    \vspace{0cm}
    \includegraphics[width=\linewidth]{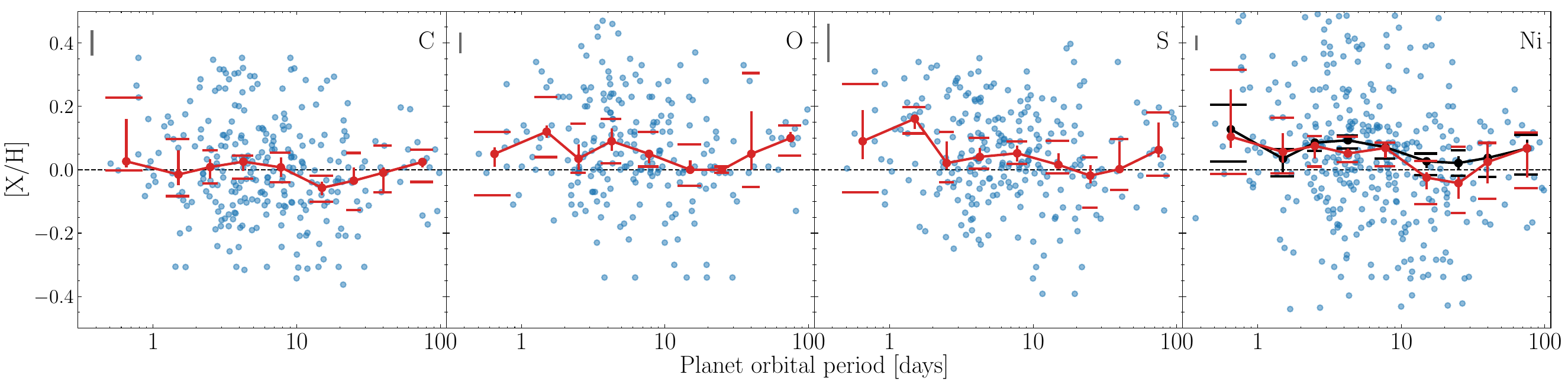}
    \caption{Distribution of [X/H] of host stars with planet radius (top row) and planet orbital period (bottom row). The symbols and lines are the same as in Figure \ref{fig:06_FeH_Rad_Porb}.}
    \label{fig:06_XH_Rad_Porb}
\end{figure}

Concerning the distribution of [X/H] as a function of exoplanet radius (see top row of Figure \ref{fig:06_XH_Rad_Porb}), we find an overall similar behavior for C, O, S, and Ni, as that for the metallicity ([Fe/H]) -- all elements generally show an increase in the abundances for large exoplanets. In particular, the distribution of the [Ni/H] abundances with $R_{\text{pl}}$ is quite similar and reminiscent of that of [Fe/H], while in the case of [C/H], the median abundances show a slight oscillation as $R_{\text{pl}}$ increases, which is not seen for [Fe/H]. The OLS regressions show significant and steeper trends for the $\alpha$ elements versus $R_{\text{pl}}$, i.e., [O/H] and [S/H] with percentage of $p_{\text{OLS}}<0.001$ of 99\% and 82\%, respectively, while the percentages are 9\% and 32\% for [C/H] and [Ni/H].

We also evaluated the [X/Fe] for our sample, finding generally flat distributions for [C/Fe], [O/Fe], [S/Fe], and [Ni/Fe] as a function of $R_{\text{pl}}$ for all elements. For [S/Fe], \cite{Wilson2022} found a small positive correlation with R$_{\text{pl}}$ -- a result that was not recovered in our data. For [C/Fe], [O/Fe], [S/Fe] and [Ni/Fe], we performed K-S tests dividing the sample between Jupiter and non-Jupiter hosts and obtained $p_{\text{KS}}<$0.001 for 27\%, 11\%, 0.5\% and 6\% of the bootstrapped samples, respectively, meaning no significant differences between the two types of host stars.

\begin{figure}[!ht]
    \centering
    \begin{minipage}{\textwidth}
    \centering
    \begin{minipage}{.3\textwidth}
        \centering
        \includegraphics[width=\linewidth]{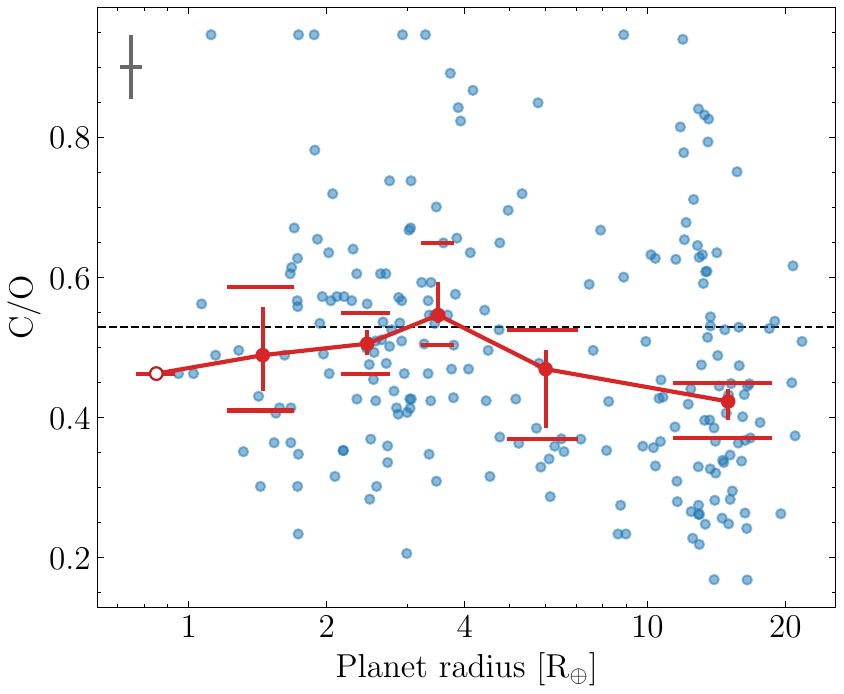}
    \end{minipage}
    \begin{minipage}{.3\textwidth}
        \centering
        \includegraphics[width=\linewidth]{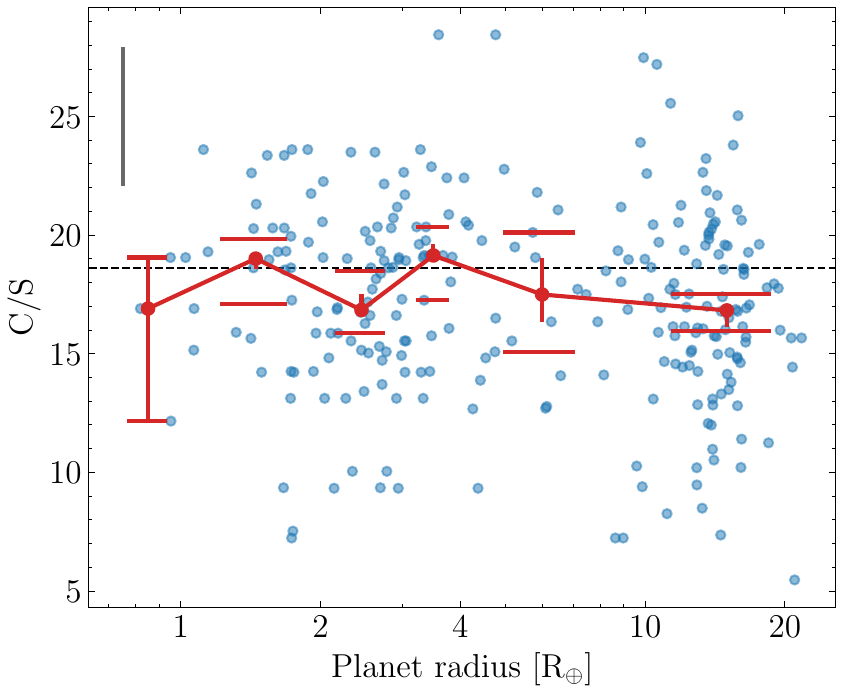}
    \end{minipage}
    \begin{minipage}{.3\textwidth}
        \centering
        \includegraphics[width=\linewidth]{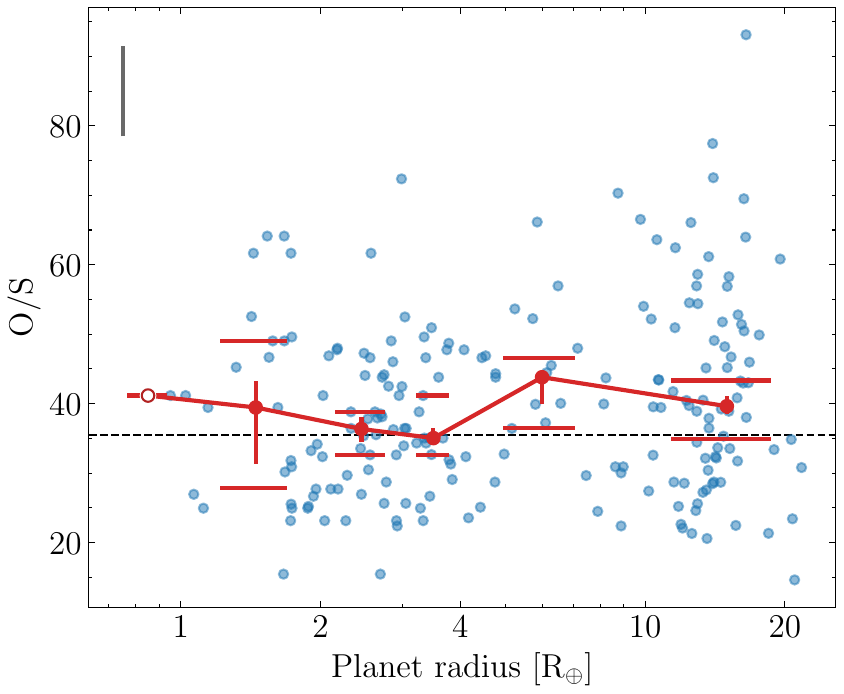}
    \end{minipage}
    \end{minipage}%
    \vspace{0cm}
    \begin{minipage}{\textwidth}
    \centering
    \begin{minipage}{.3\textwidth}
        \centering
        \includegraphics[width=\linewidth]{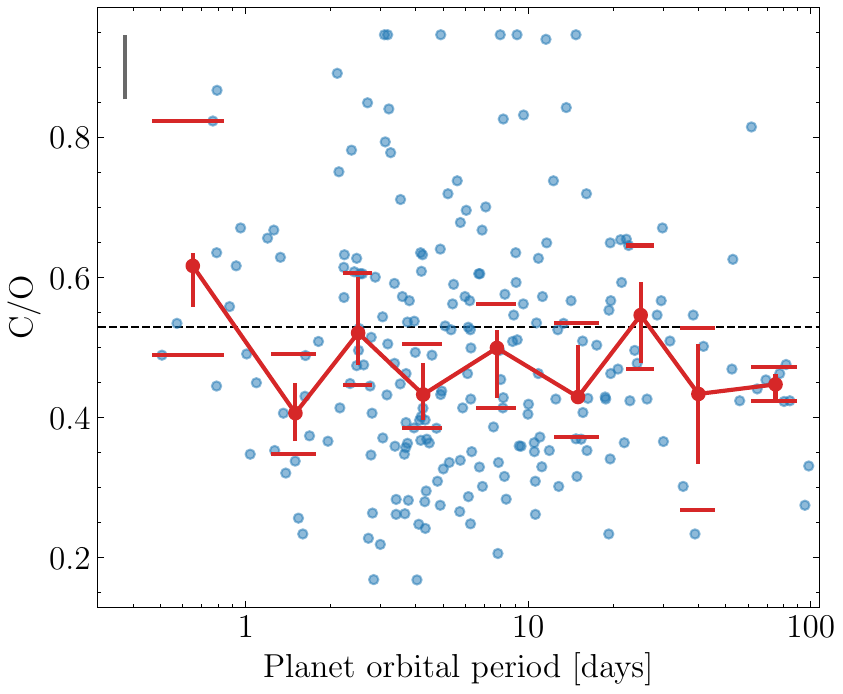}
    \end{minipage}
    \begin{minipage}{.3\textwidth}
        \centering
        \includegraphics[width=\linewidth]{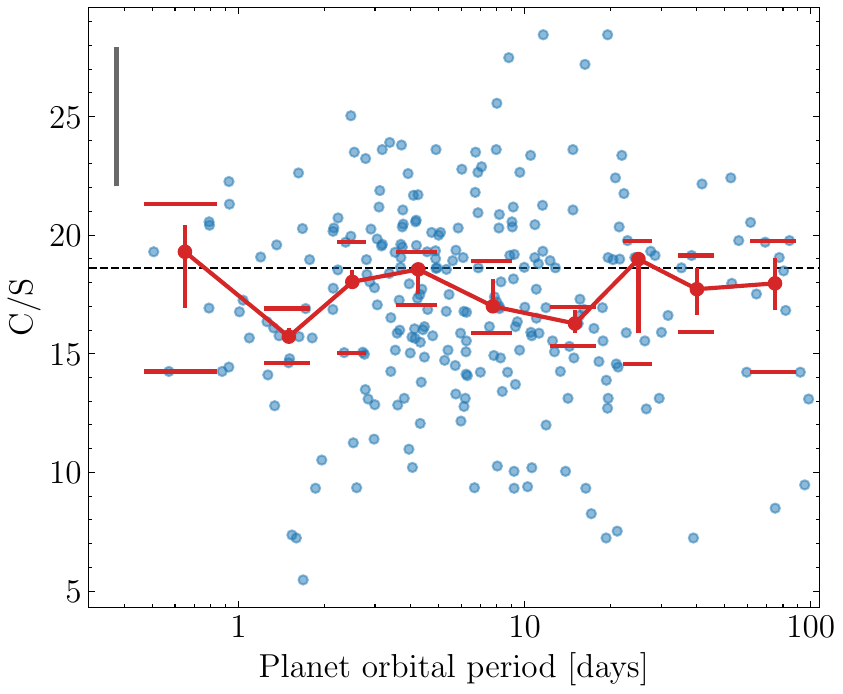}
    \end{minipage}
    \begin{minipage}{.3\textwidth}
        \centering
        \includegraphics[width=\linewidth]{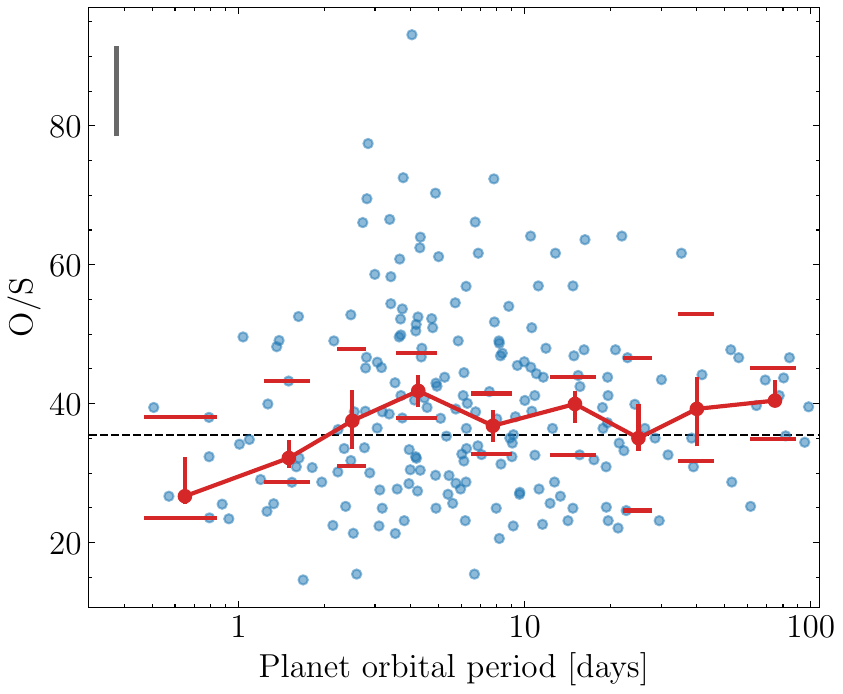}
    \end{minipage}
    \end{minipage}
    \caption{Distribution of C/O (left column), C/S (middle column) and O/S (right column) of host stars with planet radius (top row) and planet orbital period (bottom row). The symbols and lines are the same as in Figure \ref{fig:06_FeH_Rad_Porb}.}
    \label{fig:06_XX_Rad_Porb}
\end{figure}

Moving on to C/O, in the left panel of the first row of Figure \ref{fig:06_XX_Rad_Porb}, we show the distribution of host star C/O versus exoplanet radius, with median C/O ratios per R$_{\text{pl}}$ bin shown as filled red circles. The lowest median C/O value is found for Jupiter hosts (0.43$^{+0.02}_{-0.03}$) and this is lower than the solar value of 0.53, while the highest median value of C/O is 0.55$^{+0.05}_{-0.01}$ for 3 -- 4 R$_\oplus$ sub-Neptune hosts ($(\Delta\text{C/O})/(\text{C/O}_{\text{3-4 sub-Neptunes}})\approx22\%$). In comparison, 1.9-3 R$_\oplus$ sub-Neptune hosts have a median C/O ratio of 0.50$^{+0.01}_{-0.02}$ ($\Delta\approx9\%)$. We find similar results using mean and median values from bootstrap and regular mean and median values -- they agree that 3 -- 4 R$_\oplus$ sub-Neptune hosts have the highest C/O ratio and Jupiter hosts have the lowest C/O ratios. We performed K-S tests and found no significant differences between the C/O distributions of 1.9 -- 3 R$_\oplus$ and 3 -- 4 R$_\oplus$ sub-Neptune hosts ($p_{\text{KS}}<$0.001 for 11\% of the bootstrapped samples). However, we identified significant differences when comparing sub-Neptune hosts to the rest of the sample, Jupiter hosts to the rest of the sample and sub-Neptune hosts to Jupiter hosts, for which we found $p_{\text{KS}}<$0.001 for 63\%, 65\% and 78\% of the bootstrapped samples, respectively.

The distributions of host star C/S and O/S ratios as a function of exoplanet radii are presented in the middle and right panels of the first row of Figure \ref{fig:06_XX_Rad_Porb}. For C/S, the median abundances for all bins are slightly below solar (C/S$_\odot$=18.62), except for super-Earth hosts (median C/S = 19.01$^{+0.30}_{-0.40}$) and 3 -- 4 R$_\oplus$ sub-Neptune hosts (median C/S = 19.14$^{+0.47}_{-0.22}$) which are slightly above, but solar within the uncertainties. C/S ratios generally follow a similar behavior as C/O, in particular for $R_{\text{pl}}>3$ R$_\oplus$. Similarly to C/O, the lowest median C/S value is found for Jupiter hosts (16.79$^{+0.21}_{-0.67}$), and the highest median value for 3 -- 4 R$_\oplus$ sub-Neptune hosts (19.14$^{+0.47}_{-0.09}$), resulting in a significant  $\Delta\approx12.3\%$. However, at lower $R_{\text{pl}}$, there is a different trend when compared to C/O -- the 1 -- 1.9 R$_\oplus$ bin has a higher median C/S than those of the adjacent bins, while for C/O there is a smooth decline for $R_{\text{pl}}\leq3$ R$_\oplus$. For the K-S tests, we find no statistically significant differences when comparing the distributions of C/S ratios of 3 -- 4 R$_\oplus$ sub-Neptune hosts with Jupiter hosts, 3 -- 4 R$_\oplus$ sub-Neptune hosts to the rest of the sample, and Jupiter hosts to the rest of the sample, having $p_{\text{KS}}<$0.001 for 19\%, 29\% and 15\% of the bootstrapped samples.

The median abundance ratios for O/S for all bins are slightly above solar (O/S$_\odot$=35.48), except for 3 -- 4 $R_{\text{pl}}$ sub-Neptune hosts. The lowest median O/S ratio is found for 3 -- 4 R$_\oplus$ sub-Neptune hosts (median O/S = 35.04$^{+1.40}_{-0.72}$) and the highest median value for sub-Saturn hosts (median O/S = 43.80$^{+0.66}_{-3.90}$) -- the behavior is almost reverse of that of the C/O distribution, with differences in median O/S ratios corresponding to about $(\Delta\text{O/S})/(\text{O/S}_{\text{sub-Saturn}})\sim20\%$. For the K-S tests, we find no statistically significant differences when comparing the distribution of O/S ratio of 3 -- 4 R$_\oplus$ sub-Neptune hosts with sub-Saturn hosts, 3 -- 4 R$_\oplus$ sub-Neptune hosts to the rest of the sample, or sub-Saturn hosts to the rest of the sample, having $p_{\text{KS}}<$0.001 for 5\%, 4\% and 5\% of the bootstrapped samples.

\subsection{Trends with Exoplanet Orbital Period}
The distribution of host star [Fe/H] as a function of exoplanet orbital period for our sample is shown in the right panel of Figure \ref{fig:06_FeH_Rad_Porb}. Before we discuss these results, we point out that, unlike the previous studies that investigated trends of host star metallicities with exoplanet orbital periods (e.g., \citealt{Mulders2016}; \citealt{Petigura2018b}; \citealt{Wilson2018}), our sample of FGK stars has a significant number of large ($R_{\text{pl}}\geq4\ \text{R}_\oplus$) exoplanets (151 exoplanets, or, $\sim$42\% of our sample). One important result in the mentioned studies was that stars hosting hot exoplanets with orbital periods shorter than roughly 8 -- 10 days tend to have larger metallicities when compared to exoplanets with longer orbital periods, and that this was true for samples of large and small exoplanets (\citealt{Mulders2016}). Metallicity enhancements for stars hosting short period exoplanets have also been found for M dwarfs, but, in that case, the exoplanet orbital period threshold was much shorter ($\sim$4 days; \citealt{Wanderley2025}).

In our sample, 75\% of the hot exoplanets orbit stars with $\text{[Fe/H]}\geq0.00$, for which approximately half are hot Jupiters (48\%). We already know from the planet-metallicity correlation that giant exoplanets are more frequently found around metal-rich stars (e.g., \citealt{Fischer2005}; \citealt{Mortier2013}; \citealt{Ghezzi2018}). Elevated stellar metallicity change disk density profiles by increasing the dust-to-gas ratio, possibly triggering or amplifying certain migration mechanisms (e.g., \citealt{Mulders2016}; \citealt{Petigura2018b}; \citealt{Dawson2013}). Since the origin of hot Jupiters is commonly associated with the migration of these giant planets to close-in orbits after their formation at higher orbital distances (e.g., \citealt{Ida2008}; \citealt{Dawson2013}; \citealt{Dawson2018}) -- although \textit{in situ} formation still remains under debate (e.g., \citealt{Boley2016}; \citealt{Mathur2025}) -- the higher frequency of close-in exoplanets around metal rich stars observed here and in many works (e.g., \citealt{Mulders2016}; \citealt{Petigura2018b}; \citealt{Wilson2018}) can be caused by the influence of the stellar metallicity in the dynamical instabilities of the proto-planetary disk.

The median [Fe/H] results in this study (shown as red circles in Figure \ref{fig:06_FeH_Rad_Porb}) stay moderately flat with increasing $P_{\text{orb}}$ until $\sim$8 -- 10 days, having a median [Fe/H] of 0.09 dex, and are lower for orbital periods between $\sim$10 -- 30 days, with a median metallicity of 0.01 dex. We performed K–S tests by dividing the sample at orbital periods ranging from 8 to 11 days, using a step size of 0.1 day, and found percentages of $p_{\text{KS}}<$0.001 higher than 50\% for transitions placed anywhere between 8.3 -- 10.3 days. However, a larger median difference of 0.08 dex is found when placing the orbital period transition at 9.8 days, which is also where we obtained a larger percentage of 71\% of the bootstrapped samples with $p_{\text{KS}}<$0.001 for the K-S tests. Furthermore, by testing the transition at 11 days, we obtained $p_{\text{KS}}<$0.001 for 39\% of the bootstrapped samples and this fraction only decreases with increasing orbital period transition.

Interestingly, we find that for larger orbital periods, between 30 -- 100 days, there is an increase in the median [Fe/H] of host stars. It is clear from our results that the median iron abundance does not keep declining for larger orbital periods, and it is possible that this uptrend had not been seen in previous studies because our sample has a larger number of large planets at large orbital periods. We performed OLS regressions to the [Fe/H]-$P_{\text{orb}}$ data and obtained a median negative angular coefficient with median $t$-value=-1.54 and $p_{\text{OLS}}<$0.001 for 8\% of the bootstrapped samples. However, if we remove the 30 -- 50 and 50 -- 100 days orbital period bins, the OLS regressions result in a median negative angular coefficient with median $t$-value=-4.35 and $p_{\text{OLS}}<$0.001 for 82\% of the bootstrapped samples. Thus, we do recover the trend in which we see an overall decrease in the [Fe/H] of the host star with increasing planet orbital period (e.g., \citealt{Mulders2016}; \citealt{Petigura2018b}; \citealt{Wilson2018}; \citealt{Wilson2022}), but for $P_{\text{orb}}\leq30$ days. 

To further investigate the uptrend with orbital period, we combined our exoplanet sample with the Kepler sample from \cite{Ghezzi2026}, whose methodology for the determination of host star parameters and iron abundances is the exact same as in this study -- making this a homogeneous sample. The combined sample has 1210 exoplanets and their median metallicities as a function of exoplanet orbital periods are shown as black symbols in Figure \ref{fig:06_FeH_Rad_Porb}. The results for this combined sample confirm that the metallicities do not decline past $P_{\text{orb}}\sim$30 days, having $p_{\text{OLS}}<$0.001 for 43\% and $|t_{\text{OLS}}|>3$ for 32\% of the bootstrapped samples for the whole $P_{\text{orb}}$ range. We also segregated the combined sample into subsamples having small ($R_{\text{pl}}<$ 4 R$_\oplus$) and large exoplanets ($R_{\text{pl}}\geq$ 4 R$_\oplus$) and we find similar uptrends in metallicity for exoplanets with $P_{\text{orb}}>30$ days for both the small and large exoplanet regimes, , having $p_{\text{OLS}}<$0.001 for 19\% and $|t_{\text{OLS}}|>3$ for 13\% of the bootstrapped samples for the whole $P_{\text{orb}}$ range of small exoplanets and $p_{\text{OLS}}<$0.001 for 23\% and $|t_{\text{OLS}}|>3$ for 17\% for the giant exoplanets.

The distributions of [X/H] for the studied elements with exoplanet orbital period are shown in the bottom row of Figure \ref{fig:06_XH_Rad_Porb}. We can see that, overall, the behavior of the elemental abundances is similar for all elements. For completeness, we also investigated the abundance differences for the samples of stars hosting hot ($P_{\text{orb}}\leq$ 10 days) versus warm$_{(10,30]}$($10<P_{\text{orb}}\leq$ 30 days) exoplanets for the elements [C/H], [O/H], [S/H], and [Ni/H], as well as the C/O, C/S, and O/S ratios. The median abundance differences for ``hot minus warm$_{(10,30]}$" are found in Table \ref{tab:median_abundances}. For the entire sample, besides [Fe/H], we find a significant abundance difference ($\Delta >0.05$) only for the element [Ni/H]. When segregating the sample into stars hosting only small exoplanets (sE, SE and/or sN), we find significant abundance differences between hot and warm$_{(10,30]}$ for all elements. For stars hosting at least one giant exoplanet (sS and/or J), we only see significant differences for [Fe/H], [C/H] and [Ni/H]. For systems having at least one sub-Neptune, we find significant differences for [Fe/H], [S/H] and [Ni/H] -- elements that are (Fe and Ni) or behave (S) as refractories in the protoplanetary disk. For systems having at least one sub-Saturn, we find significant differences for all elements. However, for systems having at least one Jupiter, there are no significant differences between hot and warm$_{(10,30]}$. Regarding the abundance ratios, we see significant differences ($(\Delta/\text{X/Y})/(\text{X/Y}_{\text{hot}})>1-10^{0.05}\approx12\%$) for C/O and O/S for stars hosting only small planets, O/S for stars hosting at least one sub-Neptune and C/S for stars hosting at least one sub-Saturn. Finally, we performed K-S tests between hot and warm$_{(10,30]}$ hosts and obtained high percentages of $p_{\text{KS}}<$0.001 for the bootstrapped samples only for [Fe/H] and [Ni/H] for the entire sample, [Fe/H], [C/H] and [Ni/H] for stars hosting only small planets, [Fe/H] and [Ni/H] for stars hosting at least one sub-Neptune and [Fe/H], [C/H] and [Ni/H] for stars hosting at least one sub-Saturn.

We also investigated the distribution of [X/Fe] with exoplanet orbital period for all elements and found no trends over the entire range of $P_{\text{orb}}$, with $p_{\text{OLS}}<$0.001 for 3\%, 1\%, 3\% and 0\% of the OLS regressions of the bootstrapped samples for [C/Fe], [O/Fe], [S/Fe] and [Ni/Fe], respectively -- showing there are no significant trends. For [S/Fe], \cite{Wilson2022} found a significant positive correlation ($p$-value$=1.2\times10^{-5}$), which we did not recover -- as the median [S/Fe] stays mostly constant.

\begin{table}
    \centering
    \begin{splittabular}{ccccccccBccccc}
        \hline
        \hline
        Type & N$_{\text{pl}}$ & $\Delta$[Fe/H] & $\Delta$[C/H] & $\Delta$[O/H] & $\Delta$[S/H] & $\Delta$[Ni/H] & \% $p_{\text{KS}}<$0.001 & $\Delta$C/O & $\Delta$C/S & $\Delta$O/S & \% $p_{\text{KS}}<$0.001\\
         & (hot, warm$_{(10,30]}$) &  &  &  &  &  & (Fe, C, O, S, Ni) & & & & (C/O,C/S,O/S)\\
        \hline
        All & 244, 47 & 0.11$^{+0.06}_{-0.01}$ & 0.05$^{+0.05}_{-0.01}$ & 0.05$^{+0.04}_{-0.03}$ & 0.03$^{+0.06}_{-0.03}$ & 0.11$^{+0.03}_{-0.05}$ & 83, 24, 6, 11, 79 & 0.05$^{+0.05}_{-0.03}$ & 0.47$^{+0.62}_{-0.72}$ & -3.40$^{+2.77}_{-2.64}$ & 4,6,11\\
        sE/SE/sN & 89, 20 & 0.22$^{+0.06}_{-0.04}$ & 0.13$^{+0.03}_{-0.07}$ & 0.06$^{+0.06}_{-0.01}$ & 0.11$^{+0.09}_{-0.05}$ & 0.24$^{+0.06}_{-0.06}$ & 95, 66, 6, 36, 96 & 0.13$^{+0.10}_{-0.05}$ & 1.89$^{+0.53}_{-0.63}$ & -8.53$^{+2.84}_{-2.12}$ & 29,17,19\\
        sS/J & 138, 34 & 0.07$^{+0.04}_{-0.03}$ & 0.08$^{+0.03}_{-0.02}$ & 0.04$^{+0.05}_{-0.05}$ & 0.03$^{+0.04}_{-0.05}$ & 0.09$^{+0.03}_{-0.03}$ & 43, 28, 9, 4, 47 & -0.02$^{+0.06}_{-0.02}$ & 1.27$^{+1.03}_{-0.92}$ & 1.22$^{+2.99}_{-2.82}$ & 2,20,3\\
        sN & 68, 23 & 0.09$^{+0.03}_{-0.06}$ & 0.04$^{+0.10}_{-0.05}$ & 0.04$^{+0.04}_{-0.06}$ & 0.06$^{+0.06}_{-0.08}$ & 0.10$^{+0.06}_{-0.05}$ & 60, 17, 10, 21, 52 & 0.03$^{+0.08}_{-0.10}$ & 0.82$^{+1.21}_{-1.28}$ & -7.91$^{+3.43}_{-10.70}$ & 6,5,22\\
        sS & 23, 14 & 0.16$^{+0.02}_{-0.04}$ & 0.13$^{+0.02}_{-0.04}$ & 0.14$^{+0.06}_{-0.07}$ & 0.08$^{+0.06}_{-0.07}$ & 0.17$^{+0.03}_{-0.07}$ & 91, 94, 17, 25, 90 & -0.03$^{+0.16}_{-0.12}$ & 3.30$^{+2.31}_{-1.39}$ & 0.40$^{+4.07}_{-10.58}$ & 5,12,2\\
        J & 115, 20 & 0.01$^{+0.05}_{-0.04}$ & 0.01$^{+0.09}_{-0.05}$ & 0.01$^{+0.09}_{-0.06}$ & -0.01$^{+0.07}_{-0.04}$ & 0.04$^{+0.09}_{-0.03}$ & 2, 3, 0, 3, 4 & -0.02$^{+0.20}_{-0.10}$ & 1.04$^{+1.24}_{-1.40}$ & 1.76$^{+2.91}_{-6.65}$ & 3,8,4\\
        \hline
    \end{splittabular}
    \caption{Median [Fe/H], [C/H], [O/H], [S/H], [Ni/H], C/O, C/S, and O/S abundance differences between (hot - warm$_{(10,30]\text{\ days}}$) for the samples of stars hosting exoplanets. ``All" corresponds to the entire sample, sE/SE/sN to stars hosting only small exoplanets, sS/J to stars hosting at least one giant exoplanet and, sN, sS and J to stars hosting at least one sub-Neptune, sub-Saturn and Jupiter, respectively. The last column presents the percentage of bootstrapped samples with $p_{\text{KS}}<$0.001 for the K-S tests for the distributions of [Fe/H], [C/H], [O/H], [S/H], [Ni/H], and C/O, C/S and O/S.}
    \label{tab:median_abundances}
\end{table}

In the bottom row of Figure \ref{fig:06_XX_Rad_Porb}, we show the distributions of host star C/O, C/S and O/S with orbital period. Starting with C/O, the median C/O abundance ratios show a slight oscillation as $P_{\text{orb}}$ increases, but stay moderately flat with values below solar, except for stars hosting exoplanets with $P_{\text{orb}}\leq1$ day, which have the highest median C/O value of 0.62$^{+0.02}_{-0.06}$. The median C/O for host stars with exoplanets having $20<P_{\text{orb}}\leq30$ days (0.55$^{+0.05}_{-0.07}$) is basically solar. We also see a generally flat distribution of median C/S as a function of exoplanet orbital period. The median values of C/S are fairly close or below the solar value (18.62). Here again, the highest median C/S value is found for stars hosting exoplanets $P_{\text{orb}}\leq1$ day (19.29$^{+1.12}_{-2.37}$), but this bin has a very small number of exoplanets (9 exoplanets). For O/S, there is a clear increase in the median O/S ratios versus exoplanet orbital period for $P_{\text{orb}}$ up to 5.5 days. This is followed by a slight oscillation as $P_{\text{orb}}$ increases. All median O/S values having $P_{\text{orb}}>5.5$ days are above or near solar (34.48). The lowest value of 26.67$^{+5.69}_{-1.14}$ is found for stars hosting exoplanets with $P_{\text{orb}}\leq1$ day -- unlike C/O and C/S distributions for which this bin had the highest median value. 

\subsection{Trends with Exoplanet Mass}
\label{sec:trends-mpl}
\begin{figure}[!ht]
    \centering
    \begin{minipage}{\linewidth}
    \centering
    \begin{minipage}{.32\textwidth}
        \centering
        \includegraphics[width=\linewidth]{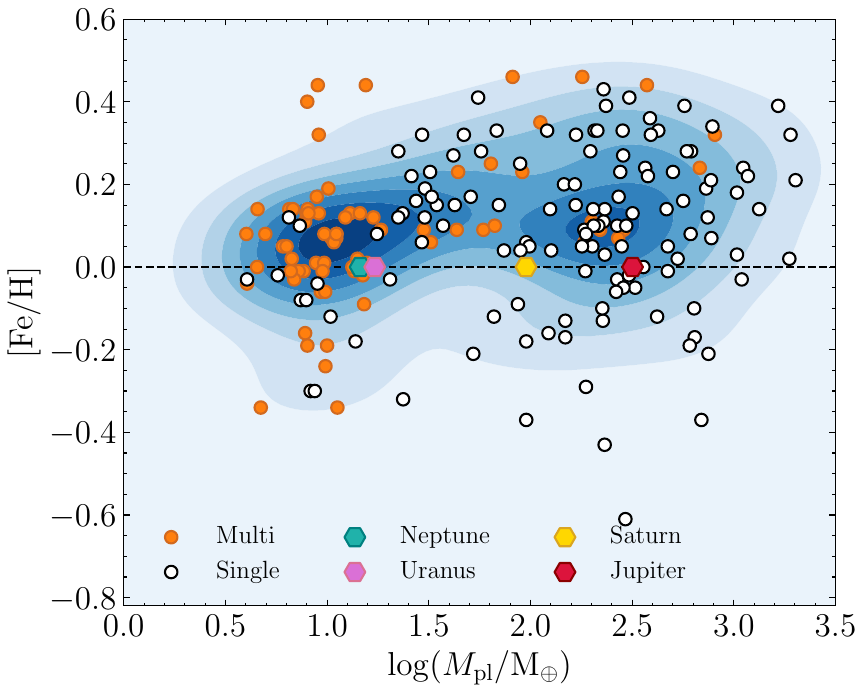}
    \end{minipage}
    \begin{minipage}{.32\textwidth}
        \centering
        \includegraphics[width=\linewidth]{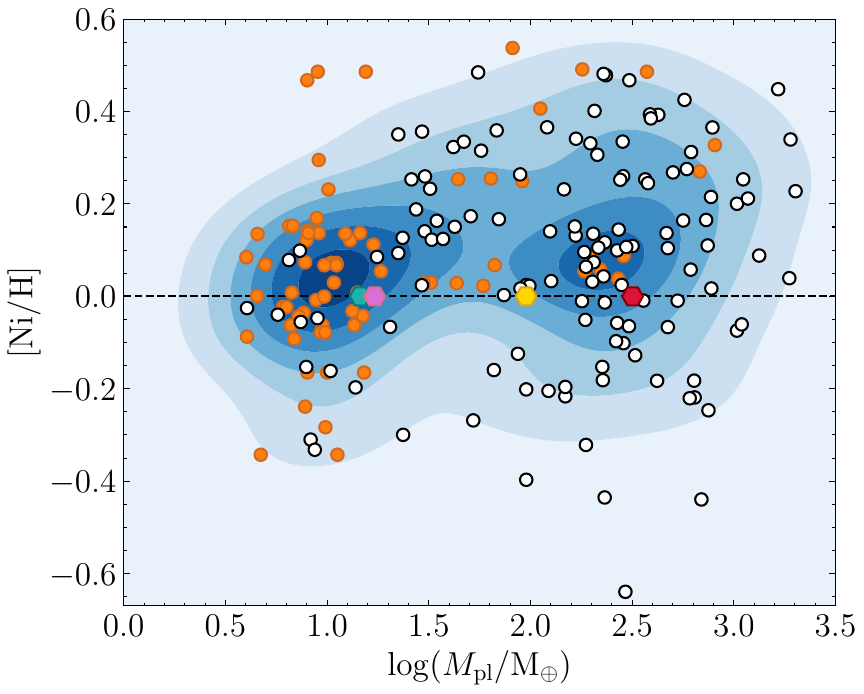}
    \end{minipage}
    \end{minipage}
    \vspace{0cm}
    \begin{minipage}{\linewidth}
    \centering
    \begin{minipage}{.32\textwidth}
        \centering
        \includegraphics[width=\linewidth]{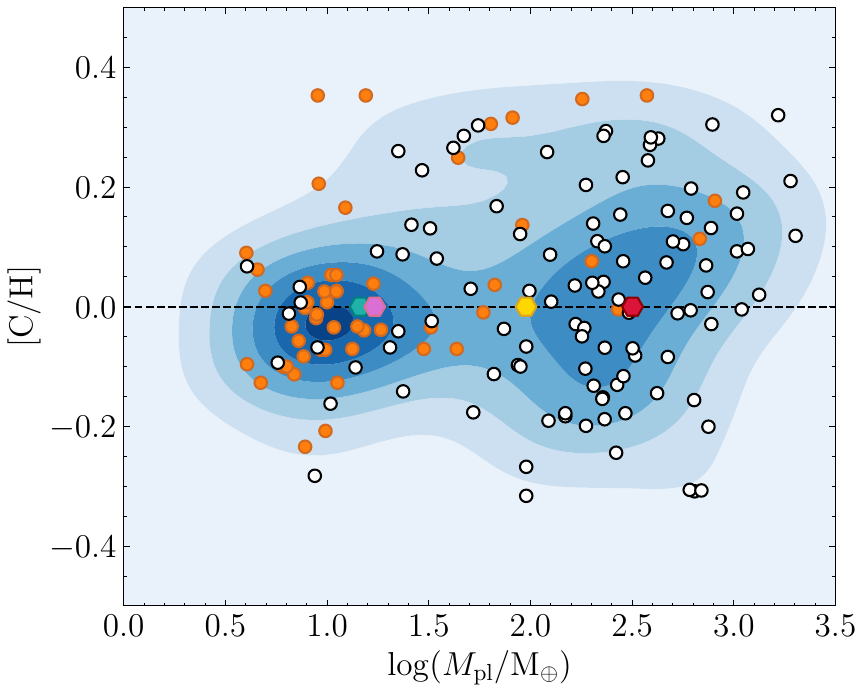}
    \end{minipage}
    \begin{minipage}{.32\textwidth}
        \centering
        \includegraphics[width=\linewidth]{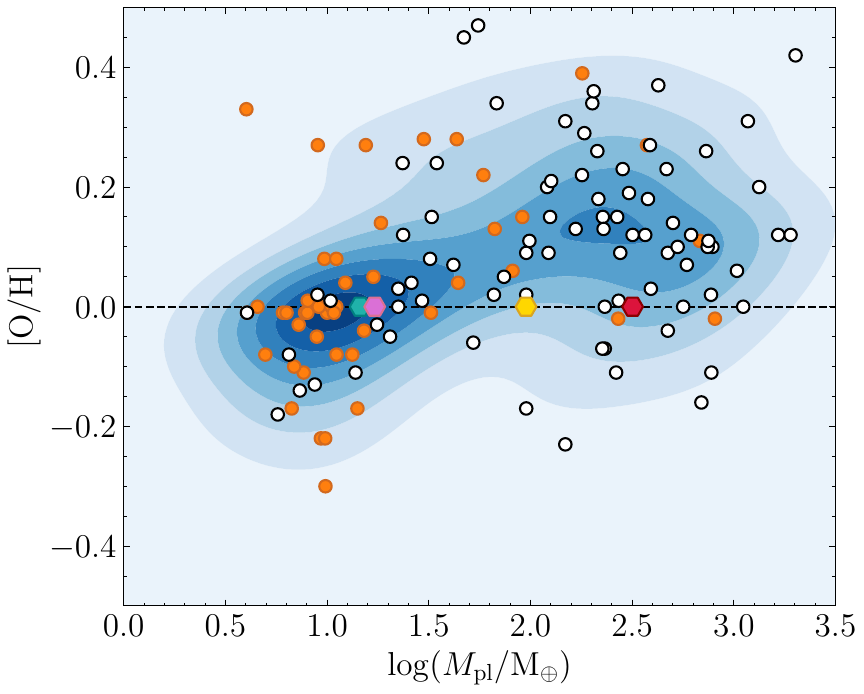}
    \end{minipage}
    \begin{minipage}{.32\textwidth}
        \centering
        \includegraphics[width=\linewidth]{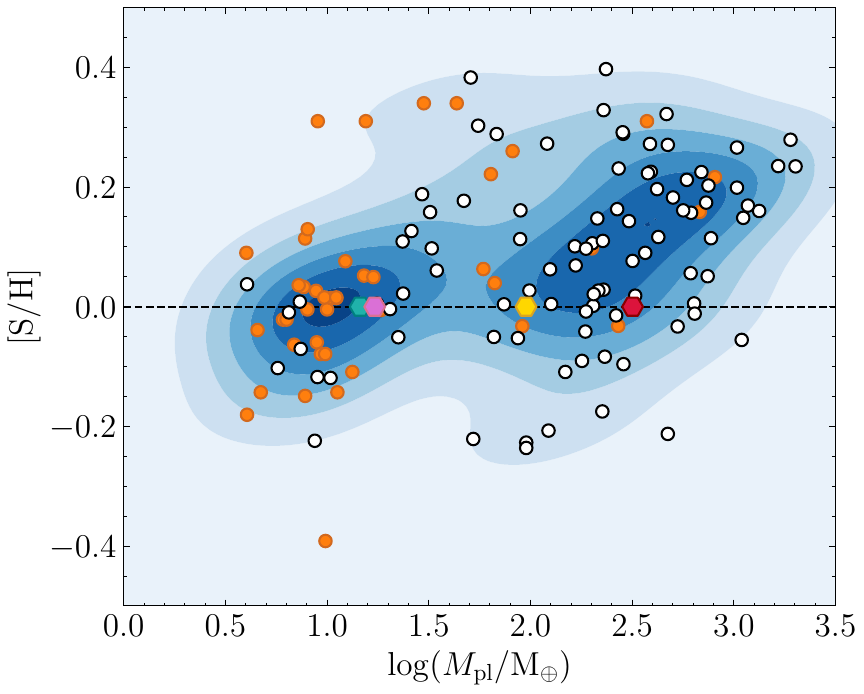}
    \end{minipage}
    \end{minipage}
    \vspace{0cm}
    \begin{minipage}{\linewidth}
    \centering
        \begin{minipage}{.32\textwidth}
            \includegraphics[width=\linewidth]{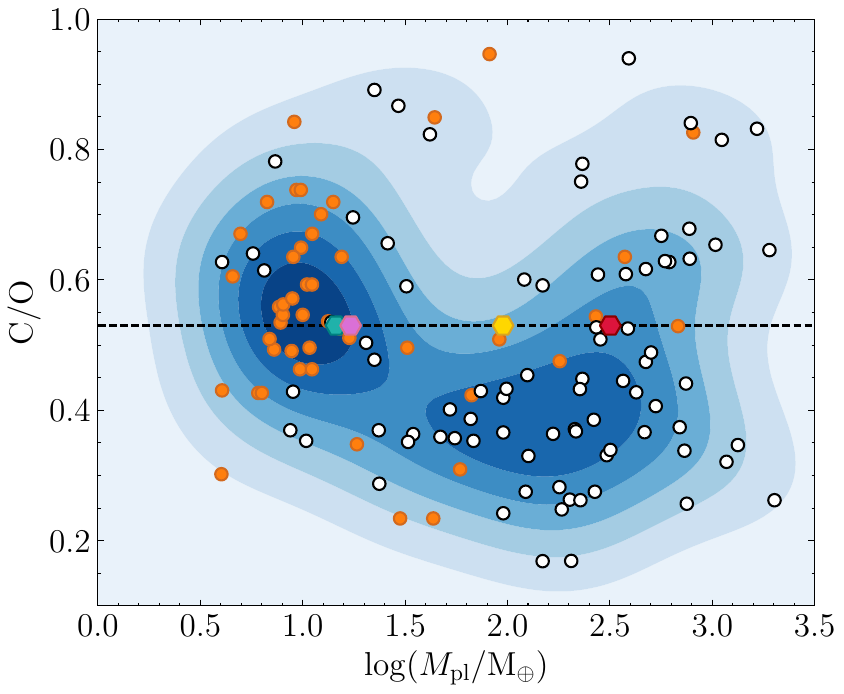}
        \end{minipage}
        \vspace{0cm}
        \begin{minipage}{.32\textwidth}
            \includegraphics[width=\linewidth]{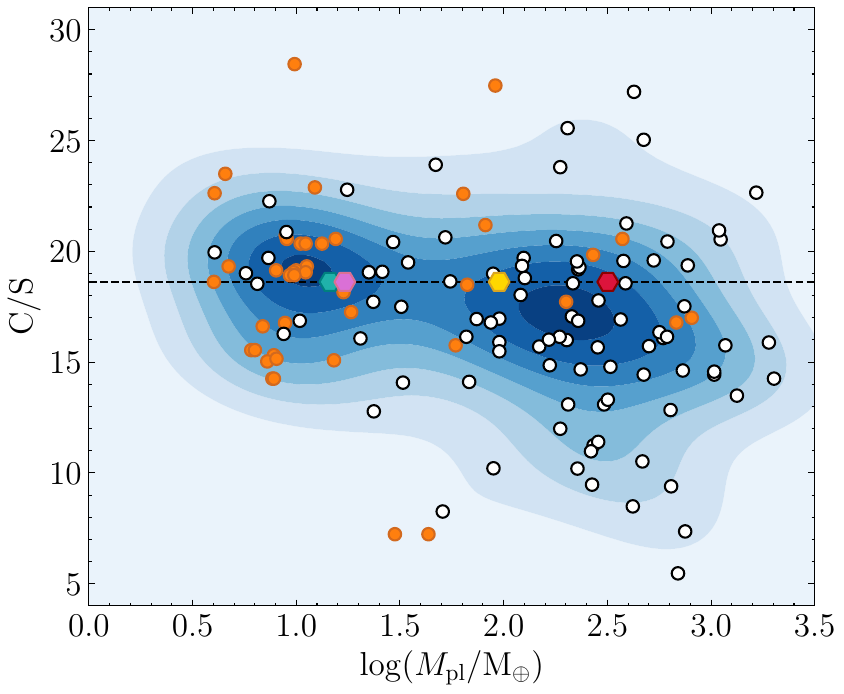}
        \end{minipage}
        \vspace{0cm}
        \begin{minipage}{.32\textwidth}
            \includegraphics[width=\linewidth]{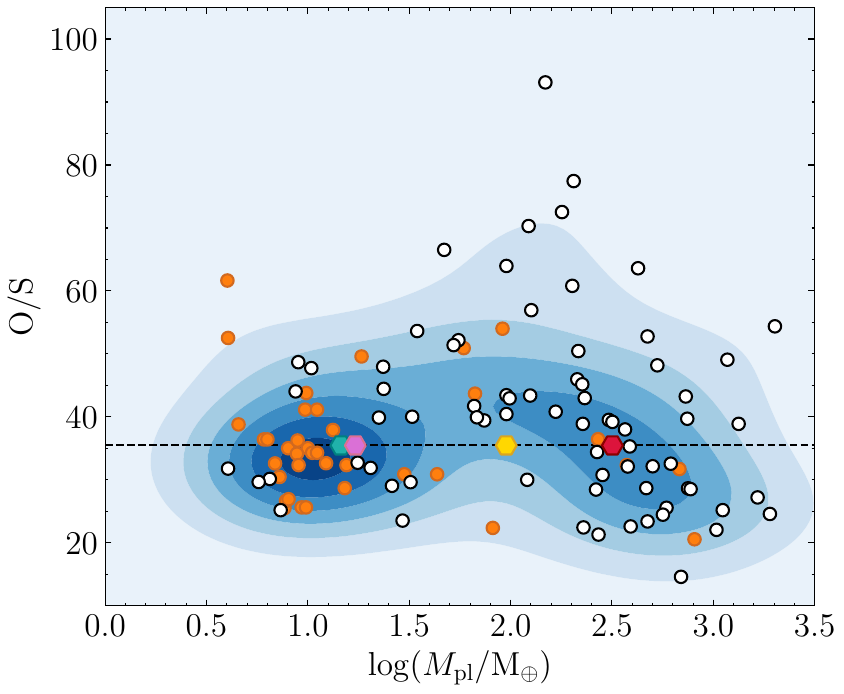}
        \end{minipage}
    \end{minipage}
    \caption{Distribution of [X/H] and abundance ratios of host stars as a function of exoplanet mass. In the first row, we show [Fe/H] (left panel) and [Ni/H] (right panel). In the second row, [C/H] (left panel), [O/H] (middle panel) and [S/H] (right panel). In the last row, C/O (left panel), C/S (middle panel), and O/S (left panel) ratios. Stars having only one exoplanet confirmed are presented in white circles (single) and, having more than one exoplanet confirmed, in orange circles (multi). Jupiter, Saturn, Uranus and Neptune are represented by the red, yellow, pink and green hexagons, respectively. The blue heat map represents the distribution of all exoplanets, independently of their status as single or multi. The black dashed lines represent the solar values.}
    \label{fig:mass_ab_ratios}
\end{figure}

Examining planetary masses adds another fundamental property with which to probe possible correlations with stellar chemistry. The masses of the exoplanets in this study come from non-homogeneous literature sources (see Table \ref{tab:planet_radii}). Although we have masses determined for 270 exoplanets, here, we will consider only exoplanets with masses having uncertainties below 20\%, resulting in a sample with 199 exoplanets. Figure \ref{fig:mass_ab_ratios} shows stellar [X/H] and abundance ratios versus exoplanetary mass. From left to right, we show Fe and Ni in the first row, C, O and S in the second row and C/O, C/S and O/S in the last row. In each panel, the solar ratios are indicated by the black horizontal dashed lines and the masses of Jupiter, Saturn, Uranus and Neptune are shown as the red, yellow, pink and green hexagons, respectively. We note a gap in all of the panels, dividing the distributions into two mass groups. These groups are separated by a gap covering a mass range of $\sim$20 -- 100 M$_\oplus$ ($\log(M_{\text{pl}}/\text{M}_\oplus)\sim$1.3 -- 2). This mass range is roughly where the sub-Saturn desert is located (e.g., \citealt{Ida2004}). The ``mass gap'' is most accentuated in [O/H], [S/H], C/O and O/S, while [Fe/H], [Ni/H], [C/H] and C/S do not yield such a clean segregation. However, this might be associated with the reduced number of available [O/H] and [S/H] abundances for planet-hosting stars compared to [Fe/H] and [Ni/H].

To identify if these gaps are statistically significant and their locations, we performed K–S tests by dividing the sample at $\log(M_{\text{pl}}/\text{M}_\oplus)$ ranging from 1.3 to 2.3, using a step size of 0.1. All the tests pointed to a gap at $\sim$1.4 ($\sim$25 M$_{\oplus}$) with a percentage of $p_{\text{KS}}<0.001$ of 94\%, 55\%, 100\%, 99\% and 86\% for [Fe/H], [C/H], [O/H], [S/H] and [Ni/H], respectively, and 74\%, 42\% and 22\% for C/O, C/S and O/S. These results corroborate with the existence of an assumed vertical gap (zero slope) for all elements and abundance ratios, except for [C/H], C/S and O/S. The previous work by \cite{DelgadoMena2021} did not find a significant difference between the distribution of the C/O ratios of stars hosting exoplanets with $M_{\text{pl}}$ below and above 30 $\text{M}_\oplus$. In their work, \cite{DelgadoMena2021} used the oxygen results from \cite{BertrandeLis2015} for both the 6158 \AA~ O I line and the 6300 \AA~ [O I] line to obtain C/O ratios. To further investigate the signature in the C/O ratios, we combined our sample with their C/O$_{6300}$ sample, resulting in a total sample of 121 host stars, and found a percentage of $p_{\text{KS}}<0.001$ of 68\% for the K-S tests between the distribution of C/O ratios of stars hosting exoplanets with $M_{\text{pl}}$ below and above 30 $\text{M}_\oplus$ -- still finding a significant difference.

We note that stars hosting low-mass exoplanets ($\log(M_{\text{pl}}/\text{M}_\oplus)<1.4$) are closely clustered around solar abundances with median abundances of 0.04, -0.02, -0.01, 0.00 and 0.01 for [Fe/H], [C/H], [O/H], [S/H], [Ni/H], respectively, and 0.56, 19.06 and 34.32 for C/O, C/S and O/S, while massive exoplanet hosts are, in general, more scattered. There is a small cluster of exoplanets around 2.4 and [X/H]$\sim0.08$ in the middle of [Fe/H] and [Ni/H] distributions but they represent a small fraction of massive exoplanets. The median abundances of massive exoplanet ($\log(M_{\text{pl}}/\text{M}_\oplus)\geq1.4$) hosts are 0.13, 0.04, 0.12, 0.12 and 0.12, and 0.44, 16.79 and 30.40, i.e., iron-peak and $\alpha$-elements above solar values, C/O and C/S below solar values and [C/H] and O/S around solar value.

In Figure \ref{fig:mass_ab_ratios}, we see that massive exoplanets are preferentially found around oxygen-enriched stars, whereas this trend is not observed for carbon abundances. This may point to the greater importance of the H$_2$O ice line compared to the CO and CO$_2$ ice lines in the formation of giant exoplanet cores within the core-accretion scenario. Since giant planet core growth must occur on a short timescale, before disk dispersal, formation at smaller orbital distances favors more frequent collisions and, consequently, more efficient accretion of solid material (e.g., \citealt{Drka2017}; \citealt{Leemker2026}). In this context, the H$_2$O ice line is located significantly closer to the star than the CO and CO$_2$ ice lines. A similar trend is observed for sulfur abundances, but this may be related to the fact that both O and S are $\alpha$-elements. To test this hypothesis, it is necessary to determine the abundances of additional $\alpha$-elements (e.g., Mg, and Si) to assess whether this pattern holds. \cite{Robinson2006}, using the SPOCS data of \cite{Fischer2005}, observed that stars hosting giant exoplanets have enhanced [Si/Fe] and [Ni/Fe] abundances compared to field stars with the same [Fe/H], which may arise primarily from the oxygen and silicon correlation. For the Kepler sample, \cite{Ghezzi2026} found statistically significant differences between the [X/H] abundances of stars hosting only small exoplanets and only giant exoplanets for the $\alpha$-elements Si and Ti, but not for Mg and Ca. Finally, for the HARPS GTO sample, \cite{CostaSilva2020} observed a higher frequency of exoplanets around stars with solar [S/Fe] ($-0.10<\text{[S/Fe]}<0.10$). This is also found in our work, in which we have a total of 86 exoplanets around stars with solar [S/Fe] compared to 51 exoplanets around stars with non-solar [S/Fe].

We performed OLS regressions and obtained percentages of $p_{\text{OLS}}<0.001$ of 30\%, 7\%, 90\%, 95\%, 23\%, 2\%, 36\% and 0\% for [Fe/H], [C/H], [O/H], [S/H], [Ni/H], C/O, C/S, and O/S, respectively, and positive angular coefficients for all elements and negative for all abundance ratios. Additionally, to account for different detection biases, we restrained our sample to exoplanets detected through planetary transit and having mass uncertainties below 20\% and found percentages of $p_{\text{OLS}}<0.001$ of 88\% and 68\% for [O/H] and [S/H], respectively, and below 14\% for the other abundances. These results are in agreement with what we found for $R_{\text{pl}}$, in which $\alpha$-elements have steeper trends. We also investigated the distribution of [X/Fe] with $\log(M_{\text{pl}}/\text{M}_\oplus)$ and found no trends for all elements, with percentages of $p_{\text{OLS}}<0.001$ below 28\% for all elements.

We divided our sample into hot ($P_{\text{orb}}\leq10$ days) and warm exoplanets ($10<P_{\text{orb}}\leq100$) and performed OLS regressions. For the hot exoplanets, we found no trends for all abundances (percentages of $p_{\text{OLS}}<0.001$ below 45\%). For the warm exoplanets, we found significant trends with [O/H] and [S/H] (percentages of $p_{\text{OLS}}<0.001$ of 56\% and 83\%, respectively), while the percentages were below 42\% for the other abundances. Considering that the hot exoplanets did not form in their current orbits, but rather migrated closer to the host star (e.g., \citealt{Ida2008}; \citealt{Dawson2013}), the absence of trends between the $\alpha$-elements and $\log(M_{\text{pl}}/\text{M}_\oplus)$ for hot exoplanets may indicate that these trends are possibly associated to the exoplanets actual formation locations, assuming that most of the warm exoplanets preserve their orbital distances.

Regarding the distribution the planets of the Solar system for [X/H], we note that Uranus (pink hexagon) and Neptune (green hexagon) fall, for all elements, close to the border of the over-density of low-mass exoplanets. Saturn (yellow hexagon) falls near the lower-mass border of the over-density of massive exoplanets for all elements, except [O/H]. Jupiter (red hexagon) falls at the intermediate-mass border of the over-density of giant exoplanets for [Fe/H], [Ni/H] and [S/H], in the middle for [C/H] and outside for [O/H].

Focusing on the C/O ratios, Figure \ref{fig:mass_ab_ratios} suggests that the underlying stellar C/O ratio plays a role in the position of the mass gap. The gap is not as pronounced at lower values of C/O ($\lesssim$0.35), with the lower-mass limit of gas giant planets (which form by runaway gas accretion) increasing with larger C/O ratios. The increasing low-mass limit for gas giant planets may point to enhanced C/O ratios driving a more efficient (or faster) build-up of giant planet solid cores to the critical mass at which runaway accretion of gas from the disk occurs. We note that at the solar value of $\text{C/O}=0.53$, not only does Jupiter (red hexagon) fall to the right of this mass gap, but Saturn (yellow hexagon) falls near the lower-mass limit edge for a gas giant planet, while both Uranus (pink hexagon) and Neptune (green hexagon) fall at masses well below the gap. In the case of the ratios of O/S, the mass gap spans the same mass range and, while the behavior of the mass limits seem to behave in the opposite sense from that of C/O, this results from the use of oxygen in the numerator instead of the denominator. Taken together, the three panels of Figure \ref{fig:mass_ab_ratios} reveal the importance of the stellar abundances of C, O, and S in influencing exoplanetary architectures.

In addition to the gap in mass, increasing host star C/O ratios seem to result in larger fractions of lower-mass exoplanets relative to giant exoplanets. There seems to be a slope in the trend of C/O versus exoplanet mass in the sense that, at low $M_{\text{pl}}$, the lower envelope of the distribution has higher C/O ratios ($\text{C/O}\sim0.4$) than the high $M_{\text{pl}}$ regime ($\text{C/O}\sim0.2$).

In order to investigate how sensitive the trends with $\log(M_{\text{pl}}/\text{M}_\oplus)$ are to the precision of [X/H], abundance ratios and $\log(M_{\text{pl}}/\text{M}_\oplus)$, we performed OLS regressions through 1000 Monte Carlo resamplings, considering a normal distribution for the uncertainties of exoplanetary masses, [X/H] and abundance ratios uncertainties, and found percentages of $p_{\text{OLS}}<0.001$ of 91\% and 84\% for [O/H] and [S/H], respectively. The other elements and abundance ratios remain with percentages below 40\%, i.e., not having significant trends with $\log(M_{\text{pl}}/\text{M}_\oplus)$. Summarizing, the trends between $\log(M_{\text{pl}}/\text{M}_\oplus)$ and $\alpha$-elements are robust.

\subsubsection{Singles versus Multis}
\label{sec:mass-single-multi}
In Figure \ref{fig:mass_ab_ratios}, we show the distribution of [X/H] and C/O, C/S and O/S ratios with $\log(M_{\text{pl}}/\text{M}_\oplus)$ for stars with single systems (white circles) and stars with multi-planetary systems (orange circles). As discussed in Section \ref{sec:single-vs-multi}, low-mass exoplanets are preferentially found in multi-planetary systems, while massive exoplanets are mostly found in single systems. We performed OLS regressions for the distributions of abundances with $\log(M_{\text{pl}}/\text{M}_\oplus)$, finding for singles $p_{\text{OLS}}<0.001$ of 71\% for [S/H] and below 27\% for the other abundances, and, for multis,  72\%, 64\% and 60\% for [Fe/H], [C/H] and [Ni/H], respectively, and below 40\% for the other abundances. In summary, for singles, we find a significant trend with [S/H], while for multis, we find significant trends with [Fe/H], [C/H] and [Ni/H].

\subsubsection{Populations of Low-[$\alpha$/Fe] versus High-[$\alpha$/Fe] Stars}
\begin{figure}[!ht]
    \centering
    \begin{minipage}{\linewidth}
    \centering
    \begin{minipage}{.32\textwidth}
        \centering
        \includegraphics[width=.95\linewidth]{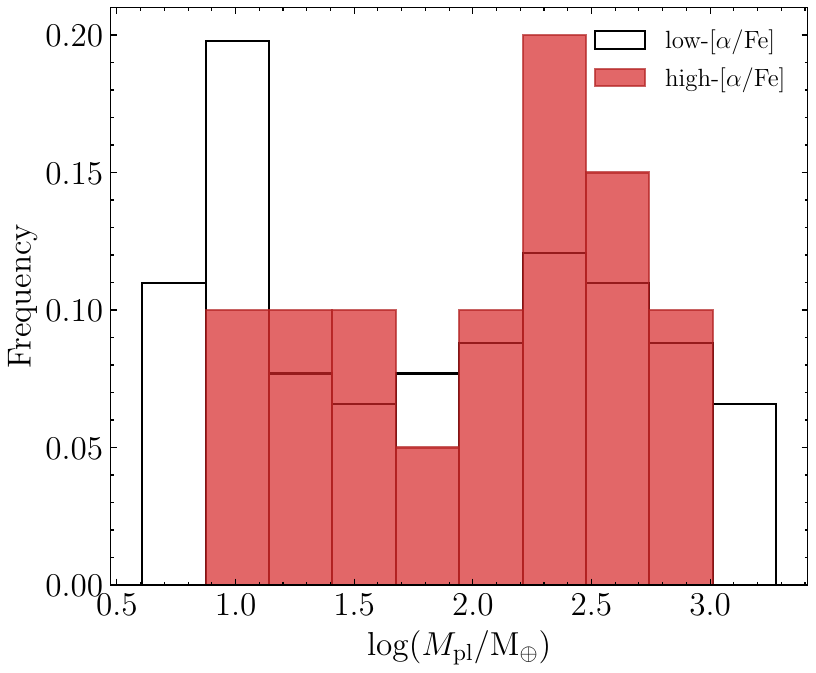}
    \end{minipage}
    \begin{minipage}{.32\textwidth}
        \centering
        \includegraphics[width=\linewidth]{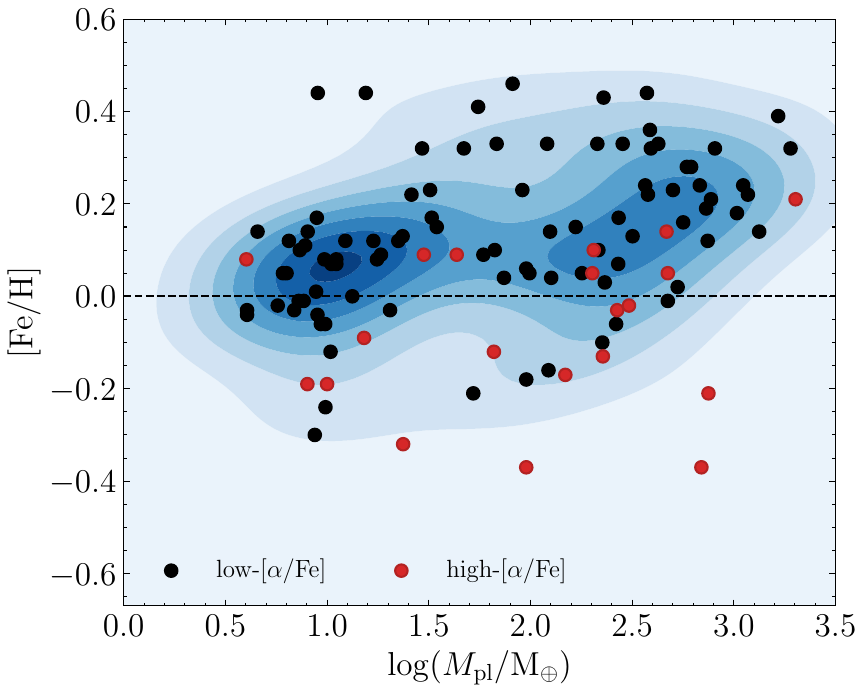}
    \end{minipage}
    \begin{minipage}{.32\textwidth}
        \centering
        \includegraphics[width=\linewidth]{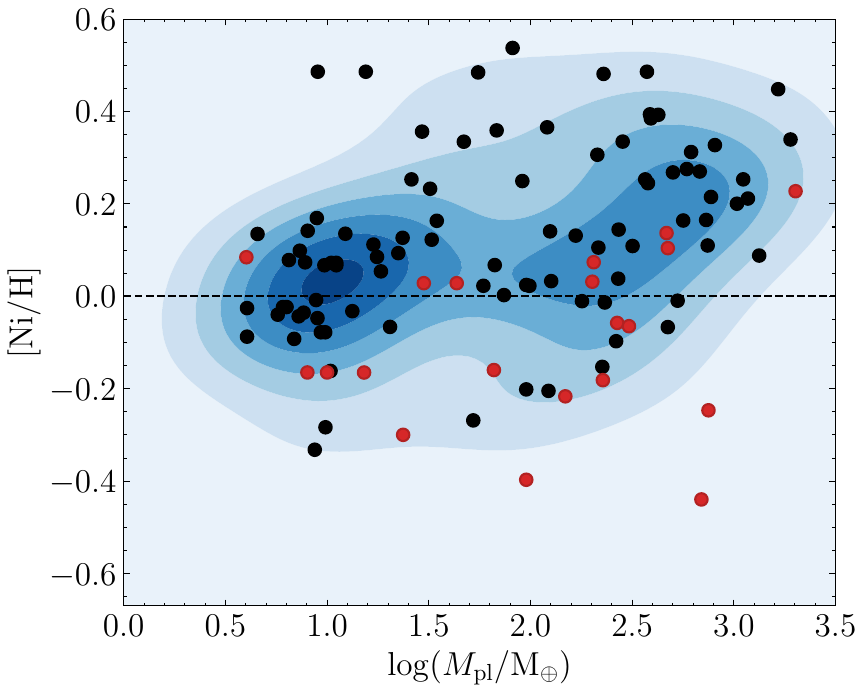}
    \end{minipage}
    \end{minipage}
    \vspace{0cm}
    \begin{minipage}{\linewidth}
    \centering
    \begin{minipage}{.32\textwidth}
        \centering
        \includegraphics[width=\linewidth]{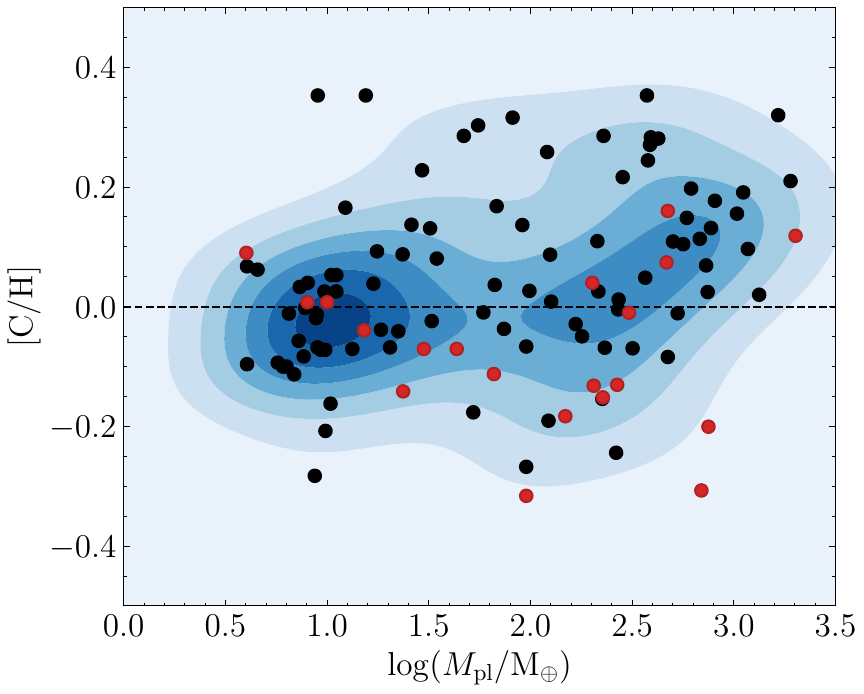}
    \end{minipage}
    \begin{minipage}{.32\textwidth}
        \centering
        \includegraphics[width=\linewidth]{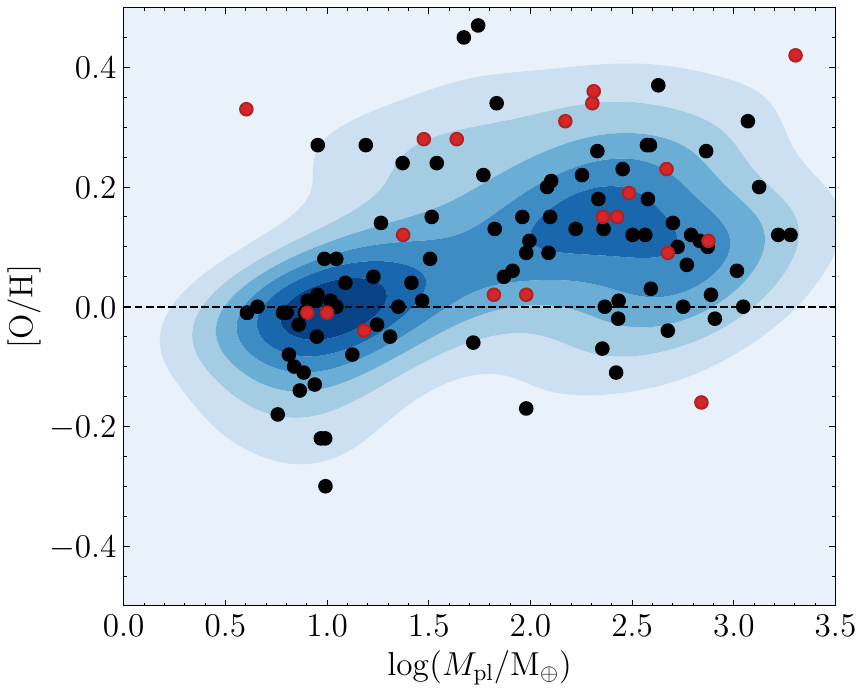}
    \end{minipage}
    \begin{minipage}{.32\textwidth}
        \centering
        \includegraphics[width=\linewidth]{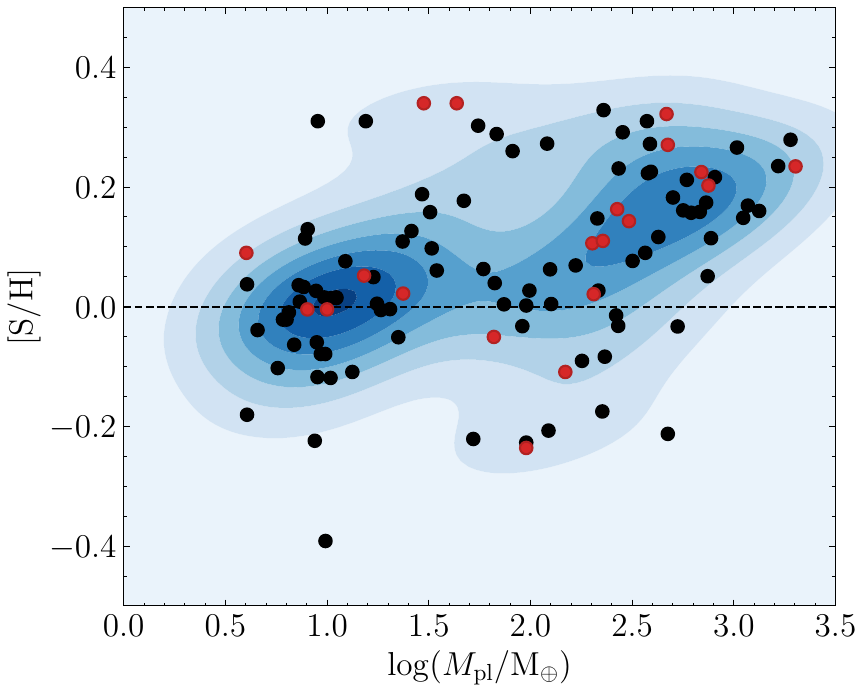}
    \end{minipage}
    \end{minipage}
    \vspace{0cm}
    \begin{minipage}{\linewidth}
    \centering
        \begin{minipage}{.32\textwidth}
            \includegraphics[width=\linewidth]{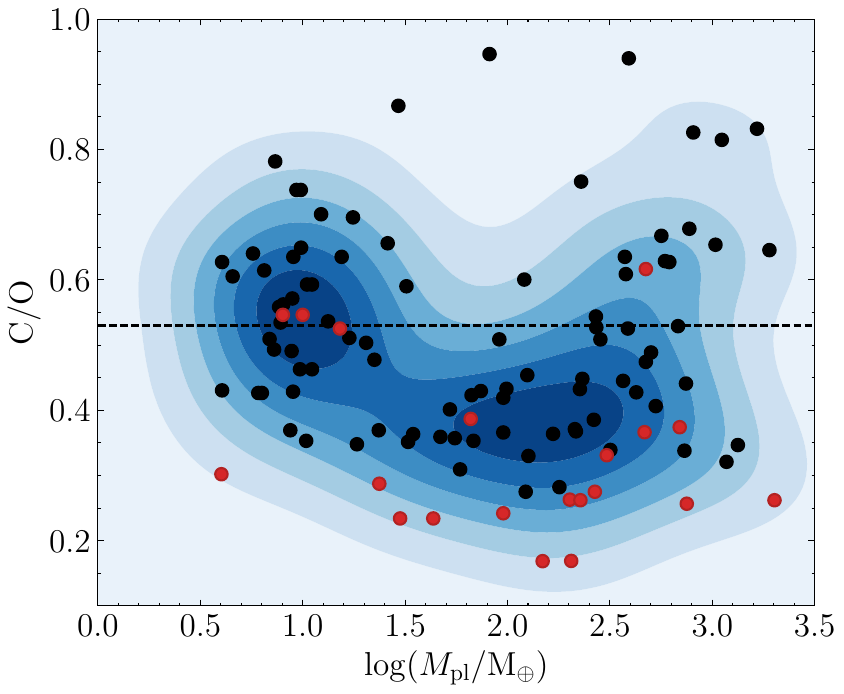}
        \end{minipage}
        \vspace{0cm}
        \begin{minipage}{.32\textwidth}
            \includegraphics[width=\linewidth]{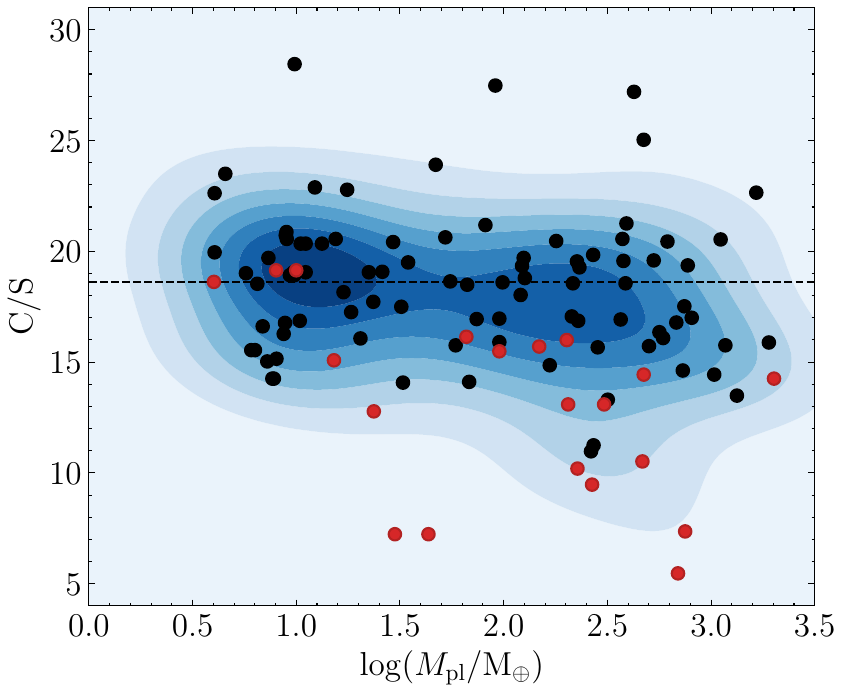}
        \end{minipage}
        \vspace{0cm}
        \begin{minipage}{.32\textwidth}
            \includegraphics[width=\linewidth]{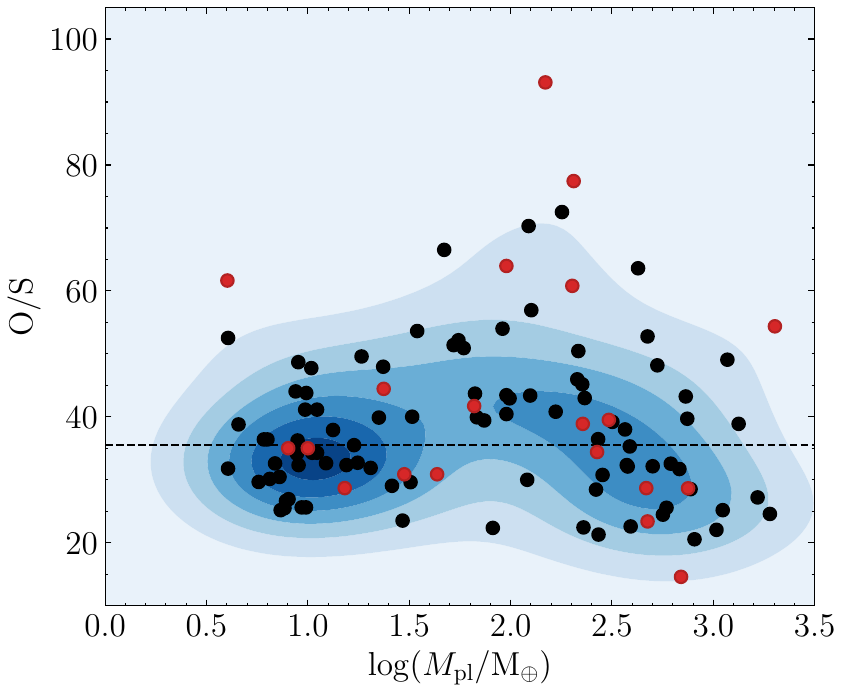}
        \end{minipage}
    \end{minipage}
    \caption{In the top left panel, we show the mass distributions of exoplanets around stars having low-[$\alpha$/Fe] (white) and high-[$\alpha$/Fe] (red). In the other panels, we show the distribution of [X/H] and abundance ratios of host stars as a function of exoplanet mass. In the top row, we show [Fe/H] (middle panel) and [Ni/H] (right panel). In the second row, [C/H] (left panel), [O/H] (middle panel) and [S/H] (right panel). In the last row, C/O (left panel), C/S (middle panel), and O/S (left panel) ratios. Stars on the low-[$\alpha$/Fe] are presented in black circles and, on the high-[$\alpha$/Fe], in red circles. The blue heat map represents the distribution of all exoplanets, independently of their status as low- or high-[$\alpha$/Fe]. The black dashed lines represent the solar values.}
    \label{fig:mpl-high-low-alpha}
\end{figure}

In order to investigate whether the previously found trends between the host star abundances of [O/H] and [S/H] with exoplanetary mass are sensitive to the host stars belonging to different Galactic populations (thin versus thick disk), we first compare, in the top left panel of Figure \ref{fig:mpl-high-low-alpha}, the histograms of $\log(M_{\text{pl}}/\text{M}_\oplus)$ for exoplanets orbiting low-[$\alpha$/Fe] (shown in white) and high-[$\alpha$/Fe] (shown in red) host stars. The two histograms cover the same range in $M_{\text{pl}}$ between 4 -- 2000 M$_\oplus$, but they have peaks at different masses: roughly at $10\ \text{M}_\oplus$ and $200\ \text{M}_\oplus$, respectively, for the low-[$\alpha$/Fe] and high-[$\alpha$/Fe] distributions. The K-S tests show that their $\log(M_{\text{pl}}/\text{M}_\oplus)$ distributions are indistinguishable (percentage of $p_{\text{KS}}<0.001$ of 2\%). 
The distributions of host star [X/H] and C/O, C/S and O/S with exoplanetary mass are shown in the other panels of Figure \ref{fig:mpl-high-low-alpha}, which are similar to those shown in Figure \ref{fig:mass_ab_ratios}, but now the different colors segregate between the high-[$\alpha$/Fe] (red circles) and low-[$\alpha$/Fe] stars (black circles).

We find statistically significant trends for host star [X/H] versus $\log(M_{\text{pl}}/\text{M}_\oplus)$ in the low-[$\alpha$/Fe] population for all elements studied. In contrast, we find no trends for the high-[$\alpha$/Fe] host star population 
(the percentages of $p_{\text{OLS}}<0.001$ are 85\%, 56\%, 81\%, 84\%, 80\%, 0\%, 2\%, and 0\% for [Fe/H], [C/H], [O/H], [S/H], [Ni/H], C/O, C/S and O/S, respectively, for the low-[$\alpha$/Fe] population and below 14\% for all abundances for the high-[$\alpha$/Fe] population). We acknowledge that it is possible that stars on the high-[$\alpha$/Fe] sequence have significant trends between [X/H] and $\log(M_{\text{pl}}/\text{M}_\oplus)$ -- which may not have been detected here due to low-number statistics (20 exoplanets).

Recall that although our sample has thick disk stars, it is composed mostly of thin disk stars and, as discussed in Section \ref{sec:trends-mpl}, for the full sample we found statistically significant trends between host star [X/H] versus $\log(M_{\text{pl}}/\text{M}_\oplus)$ only for [O/H] and [S/H] (the $\alpha$-elements studied here), and not for [Fe/H], [C/H] and [Ni/H]. In contrast, for the thin disk sample only, we now find trends for all elements. The possible explanation for this different signature is that there is a higher proportion of massive exoplanets (75\%) compared to low-mass exoplanets (25\%) around high-[$\alpha$/Fe] stars (top left panel of Figure \ref{fig:mpl-high-low-alpha}). The higher proportion of these high-[$\alpha$/Fe] massive exoplanets tends to flatten the trends between [Fe/H] and [Ni/H] versus $\log(M_{\text{pl}}/\text{M}_\oplus)$. (There are 35 low-mass exoplanets (38\%) and 56 massive exoplanets (62\%) around low-[$\alpha$/Fe] stars.) Figure \ref{fig:mpl-high-low-alpha} shows that, for [Fe/H] and [Ni/H], as expected, the exoplanets from the high-[$\alpha$/Fe] sequence (red circles) orbit stars with [X/H]$\lesssim$0.1. In our sample, below roughly solar metallicity, exoplanets from the low-[$\alpha$/Fe] sequence (black circles) are scarce.

For the host star C/O, C/S and O/S ratios, as expected, the exoplanets around high-[$\alpha$/Fe] are mostly found around stars with C/O and C/S below solar. For O/S, they are indistinguishable from the low-[$\alpha$/Fe] population. For the complete sample, we do not find statistically significant trends between host star abundance ratios and exoplanetary mass. When segregating into two groups, stars on the low- and high-[$\alpha$/Fe] sequences, we find similar results. Finally, it would be interesting to confirm these results with other $\alpha$-elements.

\subsection{Elemental Abundance Ratios in Hosting Stars and Giant Exoplanets}
Probing the elemental abundance ratios in exoplanet host stars is important due to the constraints that they can provide with respect to the formation and distribution of chemical species in the protoplanetary disk (e.g., \citealt{Thiabaud2015a}, \citeyear{Thiabaud2015b}). The distribution of the elements in the disk is mainly governed by the proximity to the proto-star over time, so present-day exoplanet elemental ratio abundances serve as probes of the relative location of planet formation to the ice lines of a variety of C/N/O-bearing (e.g., \citealt{"O2011}; \citealt{Madhusudhan2019}; \citealt{Ohno2023}) and S-bearing molecules (\citealt{Crossfield2023}) considering the standard core accretion model for planet formation (\citealt{Pollack1996}).

\subsubsection{C/O ratios}
\label{sec:co_ratio}
Besides H and He, C and O are the most abundant elements in the Universe and are present in the primordial molecular cloud and in the protoplanetary disk -- having been the first elemental abundance ratio proposed to possibly trace the formation pathway of giant planets (e.g., \citealt{Seager2005}). Different locations of the ice lines of H$_2$O, CO and CO$_2$ governs the availability of C and O and, consequently, C/O in the gas and solid phases throughout the disk (e.g., \citealt{"O2011}; \citealt{Espinoza2017}), which will contribute to the final composition of the exoplanet thought the accretion of these materials.

\begin{figure}[!ht]
    \centering
    \begin{minipage}{.4\textwidth}
        \centering
        \includegraphics[width=\linewidth]{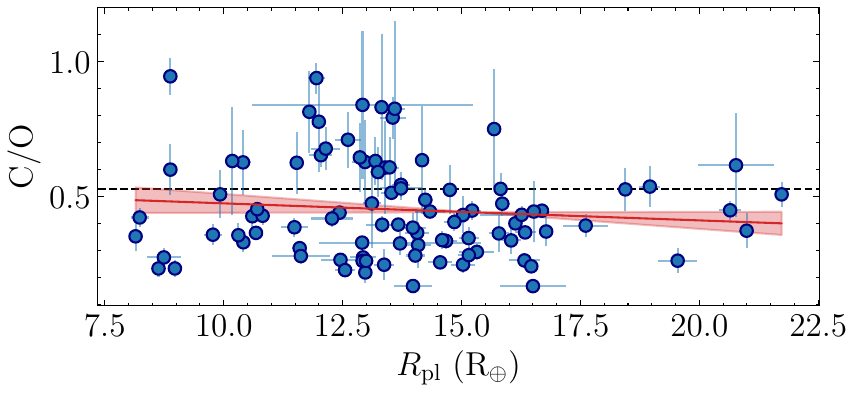}
    \end{minipage}
    \begin{minipage}{.4\textwidth}
        \centering
        \includegraphics[width=\linewidth]{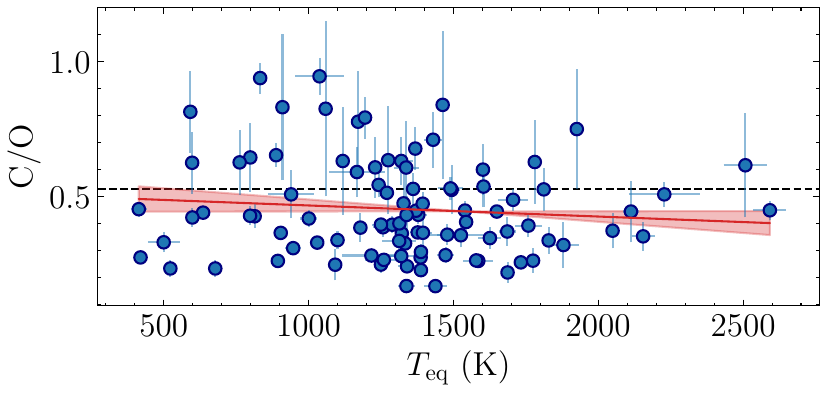}
    \end{minipage}
    \vspace{0cm}
    \begin{minipage}{.4\textwidth}
        \centering
        \includegraphics[width=\linewidth]{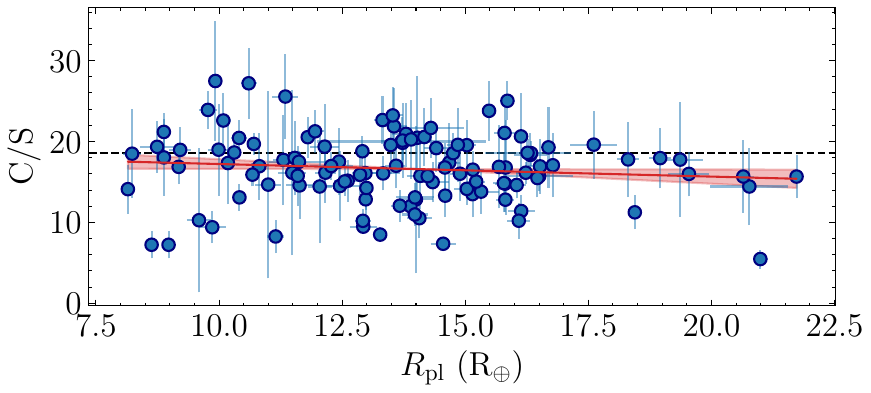}
    \end{minipage}
    \begin{minipage}{.4\textwidth}
        \centering
        \includegraphics[width=\linewidth]{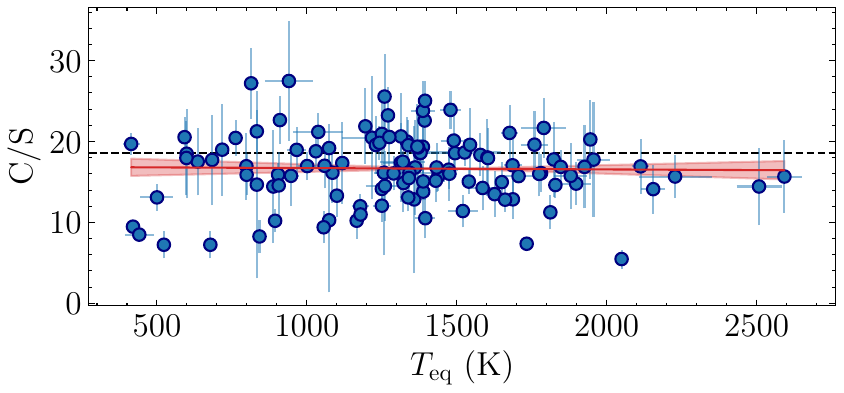}
    \end{minipage}
    \vspace{0cm}
    \begin{minipage}{.4\textwidth}
        \centering
        \includegraphics[width=\linewidth]{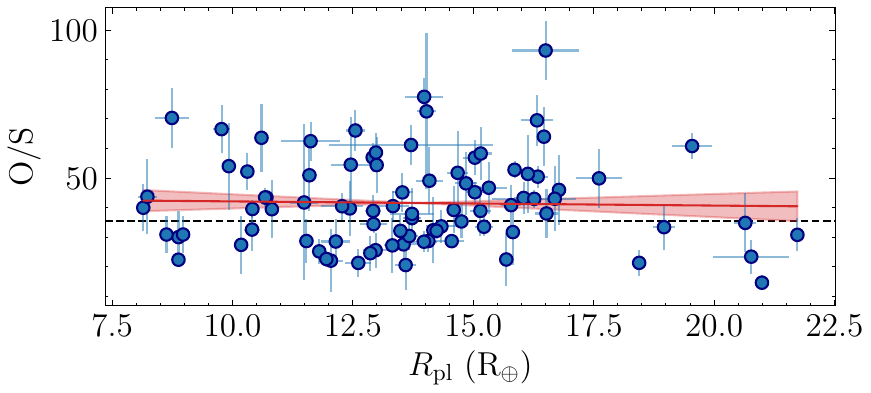}
    \end{minipage}
    \begin{minipage}{.4\textwidth}
        \centering
        \includegraphics[width=\linewidth]{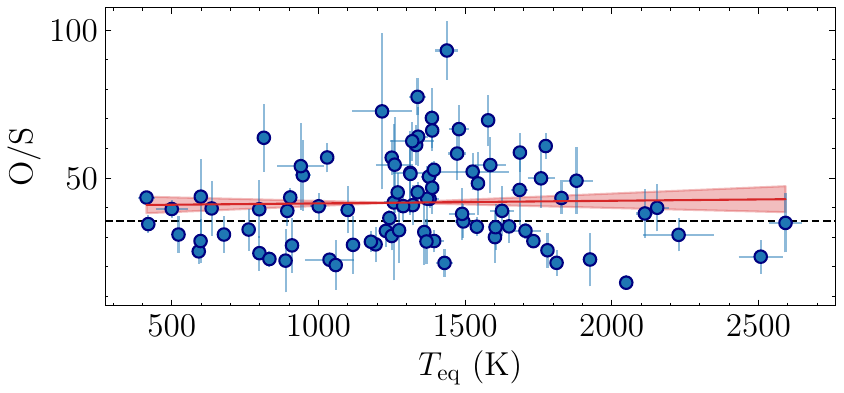}
    \end{minipage}
    \caption{Distribution of C/O (top row), C/S (middle row) and O/S (bottom row) of stars hosting hot Jupiters with planet radius (left panel) and planet equilibrium temperature (right panel). The red line is the OLS linear regression and the painted part represents the errors associated with the regression. The black dashed line represents the solar value.}
    \label{fig:06_teske_CO}
\end{figure}

The study by \cite{Teske2014} determined the carbon and oxygen abundances of a stellar sample hosting 16 Jupiter exoplanets and found no significant trends between host star C/O ratio and exoplanet equilibrium temperature ($T_{\text{eq}}$) and radius. In Figure \ref{fig:06_teske_CO} (top row), we present our C/O results for our sample containing 75 hot Jupiters. We collected the exoplanet equilibrium temperature values from the NASA Exoplanet Archive and performed OLS regressions for the host star C/O as a function of exoplanet $R_{\text{pl}}$ and $T_{\text{eq}}$. For C/O versus $T_{\text{eq}}$, we find no trend ($p_{\text{KS}}<$0.001 for 0\% of the bootstrapped samples) for a sample of 95 hot Jupiters -- in general agreement with \cite{Teske2014} results. For C/O versus $R_{\text{pl}}$, we see no trend with the OLS regressions (0\%). 

Our sample allows for investigating the behavior of the host star C/O ratio versus $R_{\text{pl}}$ and $T_{\text{eq}}$ for all exoplanet classes in this study, segregating the exoplanets samples as before: all, sE/SE/sN, sS/J, sN, sS, J, hot, warm$_{(10,30]}$ and warm$_{(10,100]}$). We performed OLS regressions of host star C/O versus $R_{\text{pl}}$ and $T_{\text{eq}}$ for all combinations and found a statistically significant trend only for host star C/O versus $R_{\text{pl}}$ for stars hosting hot exoplanets (156 exoplanets), for which we obtained median R$^2=0.07^{+0.04}_{-0.03}$, a negative angular coefficient with median $t$-value=-3.44$^{+0.87}_{-0.96}$ and $p_{\text{KS}}<$0.001 for 54\% of the bootstrapped samples. 

\subsubsection*{Comparisons with Exoplanet C/O ratios}
Over 40 giant exoplanets have their atmospheric C/O abundance ratios determined and published in the literature to date (e.g., \citealt{Bonnefoy2018}; \citealt{Changeat2022}; \citealt{WeinerMansfield2024}) and these measured C/O ratios can vary considerably between exoplanets from the same system and, also, from the host star C/O value (e.g., \citealt{Madhusudhan2012}; \citealt{Konopacky2013}; \citealt{Moses2013}). Recently, \cite{Baburaj2025} determined stellar C/O ratios of five stars with directly imaged exoplanets and found a similar stellar-planet C/O ratio (within 1$\sigma$) for seven exoplanets of their sample composed of eight giant exoplanets ($1\ \text{M}_{\text{J}}< M_{\text{pl}}<62\ \text{M}_{\text{J}}$), suggesting that these were formed via gravitational instability process. The only exception was the exoplanet HD 206893 b, which showed C/O$_{\text{pl}}>\text{C/O}_\star$. Jupiter serves an excellent comparison between stellar and giant planet C/O.  It is generally accepted that it formed through core accretion (\citealt{Pollack1996}).  Although the C/O ratio can potentially tell us much about Jupiter's formation (\citealt{Lodders2004}), its atmospheric C/O is extremely challenging to determine because water clouds form in the deep atmosphere and deplete the gas in water vapor, the main oxygen carrier. Recent determinations using data from the Juno mission vary depending on the temperature structure used in the models to describe its atmosphere, having values ranging from a solar C/O ratio (\citealt{Li2024}) to $\sim$2.9 (\citealt{Yang2026}).

\begin{figure}[!ht]
    \centering
    \begin{minipage}{.45\textwidth}
        \centering
        \includegraphics[width=\linewidth]{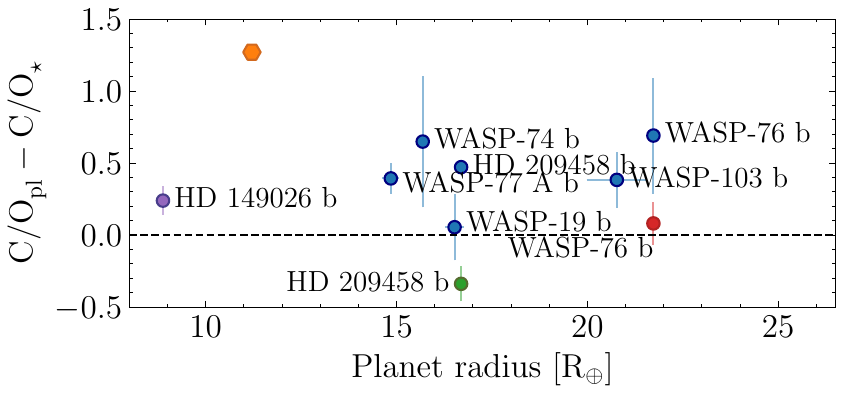}
    \end{minipage}
    \begin{minipage}{.45\textwidth}
        \centering
        \includegraphics[width=\linewidth]{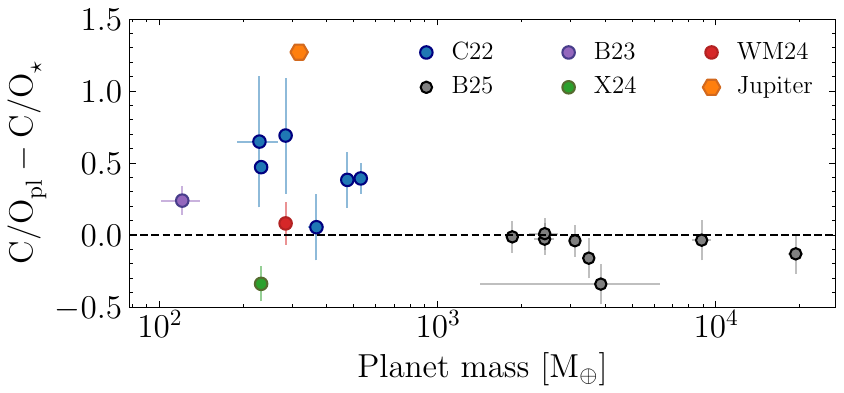}
    \end{minipage}
    \caption{Distributions of C/O$_{\text{pl}}-\text{C/O}_{\star}$ with $R_{\text{pl}}$ determined in this work (left panel) and $M_{\text{pl}}$ taken from the NASA Exoplanet Archive (right panel). The black dashed lines represent the solar values, blue, violet, green, red and gray points represent atmospheric C/O data from \cite{Changeat2022} (C22), \cite{Bean2023} (B23), \cite{Xue2024} (X24), \cite{WeinerMansfield2024} (WM24) and \cite{Baburaj2025} (B25), and the orange hexagon represents Jupiter (\citealt{Lodders2004}) for comparison.}
    \label{fig:CO_exoplanet}
\end{figure}

For our sample, we have six exoplanets in common with the study of \cite{Changeat2022}, which provide atmospheric C/O ratios recovered from their free-chemistry retrievals of Hubble Space Telescope (HST) and Spitzer data. Our derived solar C/O ratio is 0.53 while, for \cite{Changeat2022}, the solar ratio is 0.55. In Figure \ref{fig:CO_exoplanet}, we show the results for $\Delta$C/O (exoplanet-host star) as functions of exoplanet radius and mass. The host star C/O ratios obtained here range between 0.41 and 0.75, and this sample of exoplanets has masses between $\sim$228 -- 530 M$_\oplus$ or 0.72 -- 1.67 M$_{\text{J}}$ (\citealt{Gillon2014}; \citealt{Stassun2017}; \citealt{Corte2020}; \citealt{Ehrenreich2020}; \citealt{Noguer2024}). The exoplanet C/O ratios determined by \cite{Changeat2022} are found to be larger than the stellar ones (C/O$_{\text{pl}}>\text{C/O}_\star$) for all exoplanets, having a median difference of 0.43$\pm$0.13, which is higher than the median difference of -0.04$\pm$0.04 found for the directly imaged exoplanets from \cite{Baburaj2025}, represented by gray points in the right panel of the second row in Figure \ref{fig:CO_exoplanet}.

We have two exoplanets in our sample with atmospheric C/O ratios determined using transmission spectra obtained with the James Webb Space Telescope (JWST, \citealt{Gardner2006}, \citeyear{Gardner2023}), HD 149026 b (C/O$_{\text{pl}}=0.84\pm0.03$, \citealt{Bean2023}) and HD 209458 b (C/O$_{\text{pl}}=0.11^{+0.12}_{-0.06}$, \citealt{Xue2024}), and one with the Immersion GRating INfrared Spectrometer (IGRINS) on the Gemini-S telescope, WASP-76 b (C/O$_{\text{pl}}=0.90^{+0.13}_{-0.14}$, \citealt{WeinerMansfield2024}). They have stellar C/O ratios determined in our work, 0.60$\pm$0.09, 0.45$\pm$0.02 and 0.51$\pm$0.05 for HD 149026, HD 209458 and WASP-76, respectively, and exoplanetary masses from the NASA Exoplanet Archive, 120.78$\pm$19.07 M$_\oplus$ (\citealt{Stassun2017}), 232.02$\pm$12.71 M$_\oplus$ (\citealt{Stassun2017}) and 284.14$^{+4.45}_{-4.13}$ M$_\oplus$ (\citealt{Ehrenreich2020}). The median difference C/O$_{\text{pl}} - \text{C/O}_\star$ is 0.08$\pm$0.16, which is more compatible with the exoplanets from \cite{Baburaj2025} instead of \cite{Changeat2022}, as can also be seen in Figure \ref{fig:06_teske_CO} (second row, right panel). This might be due to the fact that, as stressed by \cite{Changeat2022}, HST and Spitzer data alone are not sensitive enough to precisely constrain atmospheric C/O of exoplanets. However, we also note that the exoplanets from \cite{Baburaj2025} are more massive and orbit further away from their host stars. This spread we observed for differences between the planetary and stellar C/O ratios reinforces the importance of keep pursuing the characterization of exoplanet atmospheres with high-quality data, such as from JWST, to increase the number of accurate atmospheric C/O determinations to better constrain exoplanet formation mechanisms.

Our results do not suggest trends between $\Delta$C/O ($\text{exoplanet}-\text{host star}$) versus exoplanet radius ($p\text{-value}=0.56$). However, concerning the behavior of $\Delta$C/O versus exoplanet mass, there is a hint of a possible anti-correlation when considering the $\Delta$C/O for the lower mass exoplanets from \cite{Changeat2022} and the more massive exoplanets from \cite{Baburaj2025}. There seems to be a possible threshold in exoplanet mass (from this plot, possibly around $\sim10^3$ M$_\oplus$ or $\sim3.15\ \text{M}_{\text{J}}$), although this is, at this point, a speculation) below which the C/O ratios of exoplanets tend to be larger than those of their host stars, while this is not the case for the more massive exoplanets, where their C/O ratios are roughly the same or lower than the host stars. In this context, \cite{Hoch2023} analyzed the C/O ratios of nine directly imaged planets and 25 transiting exoplanets from \cite{Changeat2022} and found a trend between exoplanet C/O ratios and their masses (M$_{\text{J}}$), with a break around 4 M$_{\text{J}}$ in which exoplanets below this mass show a wide spread of C/O ratios ($\sim$0.3 -- 1.6) and more massive exoplanets are restricted to values around 0.7$\pm$0.2 -- which aligns with our possible threshold around 3 M$_{\text{J}}$. The two groups of exoplanets are different according to K-S and Anderson-Darling tests ($p_{\text{KS}}=0.0007$ and $p_{\text{AD}}=0.0009$), indicating possibly two different formation pathways. However, considering only the higher-quality data from JWST and IGRINS, we see no significant difference ($p_{\text{KS}}=0.19$ and $p_{\text{AD}}=0.14$). We also looked for possible trends between $\Delta$C/O and $R_{\text{pl}}$, $P_{\text{orb}}$, $M_{\text{pl}}$, [X/H] and [X/Fe] and found none.

In conclusion, exoplanet formation and evolution are stochastic processes that involve many mechanisms, such as possible exoplanet migration during its formation, inhomogeneities in the disk, non-static disk temperature profile, fraction of gas and solids accreted, vertical mixing and others (e.g., \citealt{Espinoza2017}; \citealt{Madhusudhan2019}). In this regard, C/O is important but, for a more complete overview of exoplanets' formation history, C/O needs to be analyzed together with a wider variety of elemental ratios due to the high volatility of both C and O (e.g., \citealt{Pacetti2022}).

\subsubsection{C/S and O/S ratios}
\label{sec:c/s}
In contrast with C/N/O-bearing molecules, which have condensation temperatures below 200 K (\citealt{Lodders2025}), S condenses into troilite (FeS) at much higher temperatures ($\sim$700 K), placing its ice line closer to the star at $\sim$0.3 AU (e.g., \citealt{Oka2011}). \cite{Kama2019} estimated that $\gtrsim90\%$ of disk sulfur is in the form of ices or refractory minerals, which indicates that sulfur has a refractory behavior in the disk (e.g., \citealt{Lodders2004}). In light of that, \cite{Crossfield2023} used volatile-to-sulfur ratios (C/S and O/S) to investigate two different models of giant planet formation, through the accretion of pebbles versus planetesimals. For example, for the hot Jupiter WASP-39 b, they suggest that planetesimal accretion is more consistent with the observed SO$_2$ composition of the atmosphere. This emerges from the fact that pebble accretion models predict that gas giants become highly enriched in volatile elements compared to refractories (e.g., \citealt{Schneider2021a}, \citeyear{Schneider2021b}, \citeyear{Schneider2022}) and planetesimals accretion models predict solar C/S and O/S ratios, as well as C/O values (e.g., \citealt{Pacetti2022}).

Continuing to focus on the giant exoplanets, we now explore the distributions of host star C/S and O/S ratios with exoplanet radius and equilibrium temperature (see the middle and bottom rows of Figure \ref{fig:06_teske_CO}). C/S and O/S versus $R_{\text{pl}}$ and $T_{\text{eq}}$ have similar behaviors as the distribution of C/O ratios, showing a generally flat trend (percentages of $p$-values$<0.001$ for 0\% of the bootstrapped samples), with a slightly negative inclinations of the OLS regressions and no statistically significant trends.

\subsubsection*{Comparisons with exoplanet C/S and O/S ratios}
To the best of our knowledge, HAT-P-26 b is the only exoplanet with atmospheric C/S and O/S determined (\citealt{Gressier2025}) using JWST. In that study, they found atmospheric 
C/S ratio of 1.29$^{+2.87}_{-0.97}\times$solar and O/S ratio of 5.39$^{+6.62}_{-3.61}\times$solar, i.e., C/S ratio between 0.3 -- 4$\times$solar and O/S ratio between 2 -- 12$\times$solar. HAT-P-26 b is a low-mass giant exoplanet with $R_{\text{pl}}=6.333^{+0.807}_{-0.359}$ R$_\oplus$ and $M_{\text{pl}}=18.751\pm2.225$ M$_\oplus$ (\citealt{Hartman2011}). Its determined radius in our work is 6.68$\pm$0.11 R$_\oplus$, which is in agreement with \cite{Hartman2011} results. HAT-P-26 b is a hot exoplanet with $P_{\text{orb}}=4.24$ days (\citealt{Stassun2017}).

This exoplanet is part of our sample. We determined its host star atmospheric parameters and found \teff$=$5071 K, [Fe/H]$=$0.04 dex and \logg$=$4.51 dex, and obtained A(C)=8.69$\pm$0.11 dex and A(S)=7.42$\pm$0.03 dex, and a C/S ratio of 18.62$\pm$4.83. (Unfortunately, due to atmospheric contamination in the [O I] line, A(O) could not be determined.) However, as discussed in Section \ref{sec:abundances_teff}, in this low effective temperature regime, our derived C, and S abundances are likely affected by systematics. Here, we can use the calibration (OLS regression) from Section \ref{sec:abundances_teff} to correct the C and S abundances of HAT-P-26. The corrected absolute abundances are A(C)$_{\text{corrected}}$=8.55 dex and A(S)$_{\text{corrected}}$=7.20 dex, resulting in a host star C/S$_{\text{corrected}}$=22.39$\pm$4.83.

Using the atmospheric C/S value determined for HAT-P-26 b in \cite{Gressier2025} of C/S$_{\text{HAT-P-26 b}}=$0.3 -- 4$\times$ solar, and adopting the solar C/S ratio of 18.62$\pm$2.71 derived in this study, we obtain C/S$_{\text{HAT-P-26 b}}=$5.59 -- 74.48, which is 0.27 -- 3.46$\times$ the C/S ratio of the host star. According to \cite{Gressier2025}, C/S ratios $\sim$3 -- 20$\times$ solar are expected in pebble accretion formation scenarios, which typically take place beyond the FeS ice line. However, the current uncertainties in the exoplanet atmospheric C/S ratio limit the diagnostic power to distinguish between alternative scenarios for planet formation (e.g.,\citealt{Turrini2021}; \citealt{Crossfield2023}). Finally, in the coming years, the number of H$_2$S and SO$_2$ detections with JWST will increase, allowing for a broader picture of how sulfur impacts atmospheric processes and planet formation scenarios (e.g., \citealt{Polman2023}).

\section{Conclusions}
\label{sec:conclusion}
In this paper, we determined atmospheric parameters and chemical abundances for 290 solar-type exoplanet hosting stars selected from the NASA Exoplanet Archive. We analyzed high-resolution and high SNR publicly available spectra from HARPS-North and HARPS-South archives. The atmospheric parameters effective temperature (\teff), metallicity ([Fe/H]), surface gravity (\logg) and microturbulent velocity (\vmic), were determined using a classical spectroscopic analysis (\citealt{Ghezzi2018}, \citeyear{Ghezzi2021}). We determined stellar radius, mass, and age for the sample stars using the isochrone method with the PARAM code (\citealt{daSilva2006}). Chemical abundances of carbon, oxygen, sulfur, and nickel were derived using atomic lines and the LTE line analysis code MOOG (\citealt{Sneden1973}). Based on abundance data available in Vizier and the Hypatia catalog, as of December 2025, we report, for the first time, carbon abundances for 75 stars, oxygen abundances for 63 stars, sulfur abundances for 115 stars, and nickel abundances for 69 stars.

We investigated possible trends between the chemical composition of host stars, exoplanet radii ($R_{\text{pl}}$), exoplanet orbital periods ($P_{\text{orb}}$) and exoplanet mass ($M_{\text{pl}}$). In addition, we also discussed the similarities between atmospheric abundance ratios of exoplanet atmospheres and their hosting-star abundance ratios for a small sample of exoplanets. The main results obtained in this work are as follows:
\begin{itemize}
    \item In our sample, stars hosting giant exoplanets ($R_{\text{pl}}>4\ \text{R}_\oplus$) have a higher median [X/H] than stars hosting small exoplanets for all studied elements (see Figure \ref{fig:06_EarthxGas}). K-S tests were performed between the two samples (with and without giant exoplanets) and returned significant differences for all [X/H], except [C/H]. This suggests that there is likely a difference in the distributions of [Fe/H], [O/H], [S/H], and [Ni/H] between stars hosting giant and small exoplanets. In addition, we find a correlation between exoplanet radius and host star [Fe/H] (see Figure \ref{fig:06_FeH_Rad_Porb}). Such results are in line with previous findings from the literature (e.g., \citealt{Petigura2018b}; \citealt{Ghezzi2021}; \citealt{Wilson2022}; \citealt{Ghezzi2026}). The distributions of host star [O/H], [S/H], and [Ni/H] versus exoplanet radius are overall similar to the behavior for [Fe/H], but the slopes are steeper for the $\alpha$-elements [O/H] and [S/H];

    \item The [Fe/H] of stars hosting hot exoplanets ($P_{\text{orb}}\lesssim10$ days) are, on average, more metal rich than those hosting warm exoplanets, as previously found (e.g., \citealt{Mulders2016}). When considering only exoplanets with $P_{\text{orb}}\leq30$ days, OLS regressions show a statistically significant anti-correlation between host star [Fe/H] and $P_{\text{orb}}$ (see Figure \ref{fig:06_FeH_Rad_Porb}). One significant result from this study is that [Fe/H] does not continue to decline as the orbital period increases, but rather rises again for exoplanets with larger orbital periods. The median [Fe/H] abundances stay moderately flat around 0.09 dex for $P_{\text{orb}}\lesssim10$ days, decrease to 0.01 dex for $10\lesssim P_{\text{orb}}\lesssim30$ days and increase again to 0.09 dex for $P_{\text{orb}}>$ 30 days.  To further confirm these trends we combined our sample with the Kepler sample from \cite{Ghezzi2026}, resulting in a combined sample of 1210 exoplanets, all with host star parameters derived homogeneously. The results from this larger combined sample are very similar to the ones from our sample, confirming that the median [Fe/H] are enhanced for large orbital period regime; 

    \item When segregating the sample into stars hosting only small exoplanets (sub-Earths, super-Earths and/or sub-Neptunes), we find significant differences between the abundances of hot ($P_{\text{orb}}\leq10$ days) and warm ($10<P_{\text{orb}}\leq30$ days) exoplanet hosts for all elements. This is also observed between stars hosting hot and warm sub-Saturns. However, for systems having at least one Jupiter-size exoplanet, no differences were found between hosts of hot and warm exoplanets. The K-S tests suggest significant ($p$-value$<$0.001) differences for Fe, C and Ni for hosts of small exoplanets, Fe and Ni for sub-Neptune hosts, and Fe, C and Ni for sub-Saturn hosts. For C/S and O/S, when segregating the sample into stars hosting exoplanets of different size classes, we find significant differences between stars hosting hot and warm exoplanets for systems having at least 1 sub-Saturn for C/S and for systems hosting only small exoplanets and at least 1 sub-Neptune for O/S;
    
    \item Concerning the C/O ratios, hosts of 3 -- 4 R$_\oplus$ sub-Neptunes have the highest median C/O value of 0.55$^{+0.05}_{-0.01}$, while the lowest values of 0.43$^{+0.02}_{-0.03}$ are found for Jupiter hosts (see Figure \ref{fig:06_XX_Rad_Porb}). The K-S tests suggest that their differences are significant. There is an anti-correlation between C/O ratios versus $R_{\text{pl}}$ for stars hosting hot exoplanets. Also, stars hosting only small exoplanets ($R_{\text{pl}}\leq4\ \text{R}_\oplus$) have higher median C/O ratio than stars hosting at least one giant exoplanet. By segregating the sample, we find significant differences between stars hosting hot and warm exoplanets for systems having only small exoplanets;

    \item For our sample, there seems to be a gap in exoplanet mass between $\sim$20 -- 100 $M_\oplus$ in the distribution of host-star [X/H] as a function of exoplanet mass (see Figure \ref{fig:mass_ab_ratios}). Stars hosting low-mass exoplanets show median abundances around solar values for all elements, whereas stars hosting massive exoplanets exhibit higher median abundances, with the exception of carbon. The fact that giant exoplanets are preferentially found around oxygen-enriched stars, but not carbon-enriched stars, may point to the greater importance of the H$_2$O ice line compared to the CO and CO$_2$ ice lines in the formation of giant exoplanet cores;

    \item The distribution of [O/H] and [S/H] as a function of exoplanet mass, for our sample, shows steep trends -- the $\alpha$-elements studied here (see Figure \ref{fig:mass_ab_ratios}). However, when dividing the sample into hot and warm exoplanet hosts, these trends are only found for warm exoplanets -- which may be linked to the orbital distances in which these exoplanets are formed;

    \item We find a significant difference between the distributions of C/O ratios of stars hosting low-mass versus massive exoplanets; increasing C/O ratios seem to result in larger fractions of low-mass exoplanets relative to massive exoplanets (see Figure \ref{fig:mass_ab_ratios}). Additionally, the position of the gap in exoplanet mass changes with the C/O abundance ratios, with the lower-mass limit for giant exoplanets increasing with increasing C/O ratios.  This may indicate that enhanced C/O ratios result in a more efficient (or faster) build-up of solid cores to the critical mass that results in runaway gas accretion and the formation of a giant planets;

    \item Our sample has 20 exoplanets orbiting high-[$\alpha$/Fe] stars and 91 orbiting low-[$\alpha$/Fe] stars. There is a higher fraction of low-mass exoplanets around low-[$\alpha$/Fe] stars and, also, of massive exoplanets around high-[$\alpha$/Fe] stars (see Figure \ref{fig:mpl-high-low-alpha}). We find trends between host star [X/H] and $\log(M_{\text{pl}}/\text{M}_\oplus)$ only for exoplanets around low-[$\alpha$/Fe] stars -- including [Fe/H] and [Ni/H], for which no trends were found for the complete sample. The higher fraction of massive exoplanets orbiting high-[$\alpha$/Fe] stars, most of them around $\text{[Fe/H]}\lesssim0.00$, may be responsible for flatter trends between [Fe/H] and [Ni/H] versus $\log(M_{\text{pl}}/\text{M}_\oplus)$;
    
    \item The C/O ratios for the host stars were also compared with those ratios measured in the atmospheres of exoplanets (see Figure \ref{fig:CO_exoplanet}). We have six exoplanets in common with the sample of \cite{Changeat2022} (HST$+$Spitzer data) and we find that C/O$_{\text{pl}}>\text{C/O}_\star$ for all exoplanets. When comparing our results with the C/O ratios from the sample of eight directly imaged exoplanets from \cite{Baburaj2025}, which are more massive and have longer periods than ours, we see that, unlike our sample, seven out of eight exoplanets from \cite{Baburaj2025} had a lower C/O atmospheric ratios than their host stars. Such results combined suggest that there seems to be a possible threshold in exoplanet mass around $\sim3.15\ \text{M}_{\text{J}}$ beyond which the C/O ratios in host stars are larger than in the exoplanets (\citealt{Hoch2023}). However, by adding 3 exoplanets with atmospheric C/O ratios determined using higher-quality data from JWST and IGRINS, we see no threshold -- the $\Delta$C/O ratios are close to or lower than zero (ranging from -0.34 to 0.24) even for lower-mass planets.
\end{itemize}

\begin{acknowledgments}
\nolinenumbers
We thank the anonymous referee for insightful questions that helped to improve the paper. We thank Eve J. Lee for discussions that resulted in reshaping the paper. This study was financed in part by the Coordenação de Aperfeiçoamento de Pessoal de Nível Superior – Brasil (CAPES) – Finance Code 001. E.C.A. acknowledges financial support from Fundação Carlos Chagas Filho de Amparo à Pesquisa do Estado do Rio de Janeiro (FAPERJ) through a Ph.D. Scholarship Grade A under grant E-26/203.647/2024. L.G. acknowledges financial support from Fundação Carlos Chagas Filho de Amparo à Pesquisa do Estado do Rio de Janeiro (FAPERJ), through the ARC research grant E-26/211.386/2019.  This research has made use of the NASA Exoplanet Archive, which is operated by the California Institute of Technology, under contract with the National Aeronautics and Space Administration under the Exoplanet Exploration Program. This research used the facilities of the Italian Center for Astronomical Archive (IA2) operated by INAF at the Astronomical Observatory of Trieste. This research is based on data obtained from the ESO Science Archive Facility with DOI: \url{https://doi.org/10.18727/archive/33} (\citealt{esobib}). This work has made use of the VALD database, operated at Uppsala University, the Institute of Astronomy RAS in Moscow, and the University of Vienna. The research shown here acknowledges the use of the Hypatia Catalog Database, an online compilation of stellar abundance data as described in \cite{Hinkel2014}, which was supported by NASA's Nexus for Exoplanet System Science (NExSS) research coordination network and the Vanderbilt Initiative in Data-Intensive Astrophysics (VIDA). This research has made use of the SIMBAD astronomical database and the VizieR catalogue access tool, operated at CDS, Strasbourg, France (\citealt{Wenger2000}). The original description of the VizieR service was published in \cite{Ochsenbein2000}. This research has made use of the NASA/IPAC Infrared Science Archive, which is funded by the National Aeronautics and Space Administration and operated by the California Institute of Technology.
\end{acknowledgments}

%






\appendix
\section{List of Spectra}
We present the identifications of the spectra used in this work, obtained from the ESO Science Archive Facility (HARPS-South data) and the Italian Center for Astronomical Archive (HARPS-North data) in Table \ref{tab:coords-spectra}.

\label{ap:spectra}
\begin{table}[!ht]
    \centering
    \begin{splittabular}{lccccBcc}
    \hline\hline
    Star & RA J2000 & DEC J2000 & General & $N_{\text{general}}$ & Oxygen & $N_{\text{oxygen}}$\\
     & (deg) & (deg) & & & &\\
    \hline
    55 Cnc & 133.1468373 & 28.3298154 & HARPN.2014-01-02T01-37-59.495\_s1d\_A,HARPN.2014... & 3 & HARPN.2014-01-01T23-42-13.573\_s1d\_A,HARPN.2014... & 2\\
    BD+20 594 & 53.6511231 & 20.5990205 & ADP.2015-11-16T02:00:40.567,ADP.2015... & 25 & ADP.2015-11-16T02:00:40.567,ADP.2015... & 25\\
    CoRoT-1 & 102.0798591 & -3.1021394 & ADP.2014-09-18T12:10:34.533,ADP.2014... & 34 & - & -\\
    CoRoT-4 & 102.1946233 & -0.6727861 & ADP.2014-09-18T12:17:41.010,ADP.2014... & 25 & - & -\\
    \hline
    \end{splittabular}
    \caption{HARPS-South and HARPS-North data used. Column 1 shows the stellar identification. Columns 2 and 3 show the coordinates. Columns 4 and 5 show the original identification of the spectra obtained in the archives and the total number of spectra, respectively. Columns 6 and 7 show the spectra clean of telluric features or any contamination (see Section \ref{sec:oxygen}) and the total number of spectra, respectively.}
    \label{tab:coords-spectra}
    \tablecomments{This table is published in its entirety in the machine-readable format. A portion is shown here for guidance regarding its form and content.}
\end{table}

\section{Comparisons with Results from the Literature}
\label{ap:comparasions}

\subsection{Atmospheric Parameters}
\begin{figure}[!ht]
    \centering
    \includegraphics[width=\linewidth]{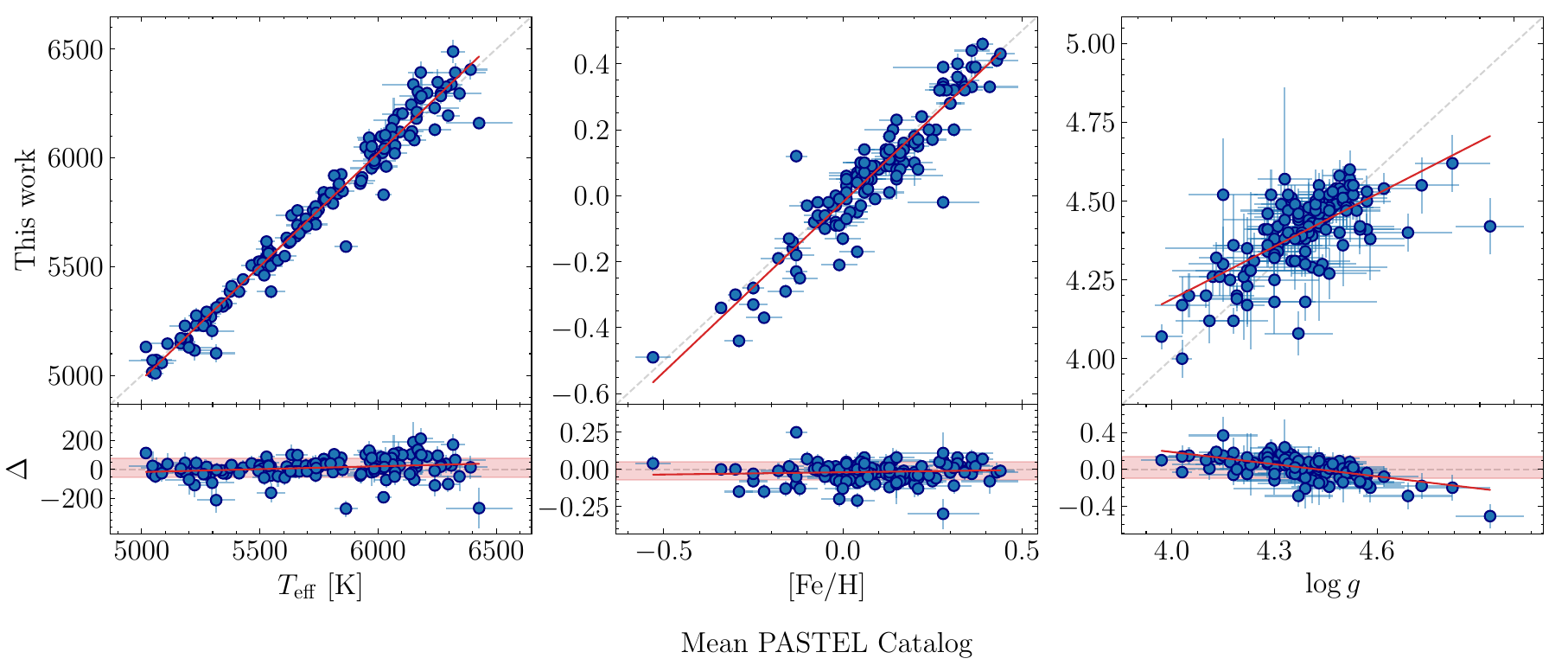}
    \caption{Comparisons between the atmospheric parameters determined in this work and from the Mean PASTEL catalog (\citealt{Soubiran2022}). \teff\ in the left column, [Fe/H] in the middle column and \logg\ in the right column. The lower panels of each row show the residuals, where $\Delta\text{par}=\text{par}_{\text{this\ work}}-\text{par}_{\text{literature}}$. The gray dashed lines represent equality, 1:1, the red lines are the linear regressions fitted to the data and the painted area is the $\pm$2MAD region around the median difference.}
    \label{fig:atm_comp}
\end{figure}

We compared our atmospheric parameters with the values from the Mean PASTEL catalog (\citealt{Soubiran2022}), which is the mean version of the PASTEL catalog (\citealt{Soubiran2016}), a compilation of atmospheric parameters determinations using high-resolution data. In Figure \ref{fig:atm_comp}, we see an excellent agreement for the 75 stars in common with the catalog, finding median differences of 12$\pm$33 K, -0.01$\pm$0.03 and 0.02$\pm$0.06 dex for \teff, [Fe/H] and \logg, respectively, and R$^2$ values of 0.04, 0.01 and 0.36 for the residuals. The R$^2$ value of \logg\ is being very influenced by the stars with \logg$>$4.65 dex in PASTEL -- without them, the value decreases to 0.22.

For the stars with the greatest differences, although still acceptable, we highlight a couple of them. For \teff, we highlight Kepler-1300, WASP-32, WASP-42 and WASP-70 A. For Kepler-1300 (5830$\pm$21 K), PASTEL has a mean value of 6023$\pm$25 K based on 3 determinations, 6021$\pm$60 K (\citealt{Petigura2017}), 6027$\pm$28 K (\citealt{Brewer2018}) and 6014$\pm$100 K (\citealt{Furlan2018}), but we find a determination more compatible with ours in the literature, 5874 K (\citealt{McQuillan2013}). For WASP-32 (6159$\pm$21 K), PASTEL has only one value of 6427$\pm$141 K (\citealt{Mortier2013}), but we find 6111$\pm$51 (\citealt{Magrini2022}) in the literature. For WASP-42 (5102$\pm$41 K), PASTEL has only one value of 5315$\pm$79 K (\citealt{Mortier2013}), but we find 5031$\pm$80 K (\citealt{Magrini2022}) and 5115 K (\citealt{Matsuno2024}) in the literature. For [Fe/H], we highlight Kepler-1300 and WASP-32. For Kepler-1300 (0.12$\pm$0.02 dex), PASTEL has a mean value of -0.13$\pm$0.03 dex. For WASP-32 (-0.02$\pm$0.02 dex), PASTEL has only one measurement of 0.28$\pm$0.10 dex (\citealt{Mortier2013}), but we find 0.03 dex (\citealt{Boettner2024}) and 0.02 dex (\citealt{Ye2025}) in the literature. For \logg, we highlight WASP-32 and WASP-71. For WASP-32 (4.42$\pm$0.09 dex), PASTEL has a mean value of 4.93$\pm$0.10 dex, but we find 4.40 dex (\citealt{Chen2021}) and 4.38 dex (\citealt{deLaverny2025})  in the literature.

\subsection{Carbon, Nickel and Sulfur}
\begin{figure}[!ht]
    \centering
    \includegraphics[width=\linewidth]{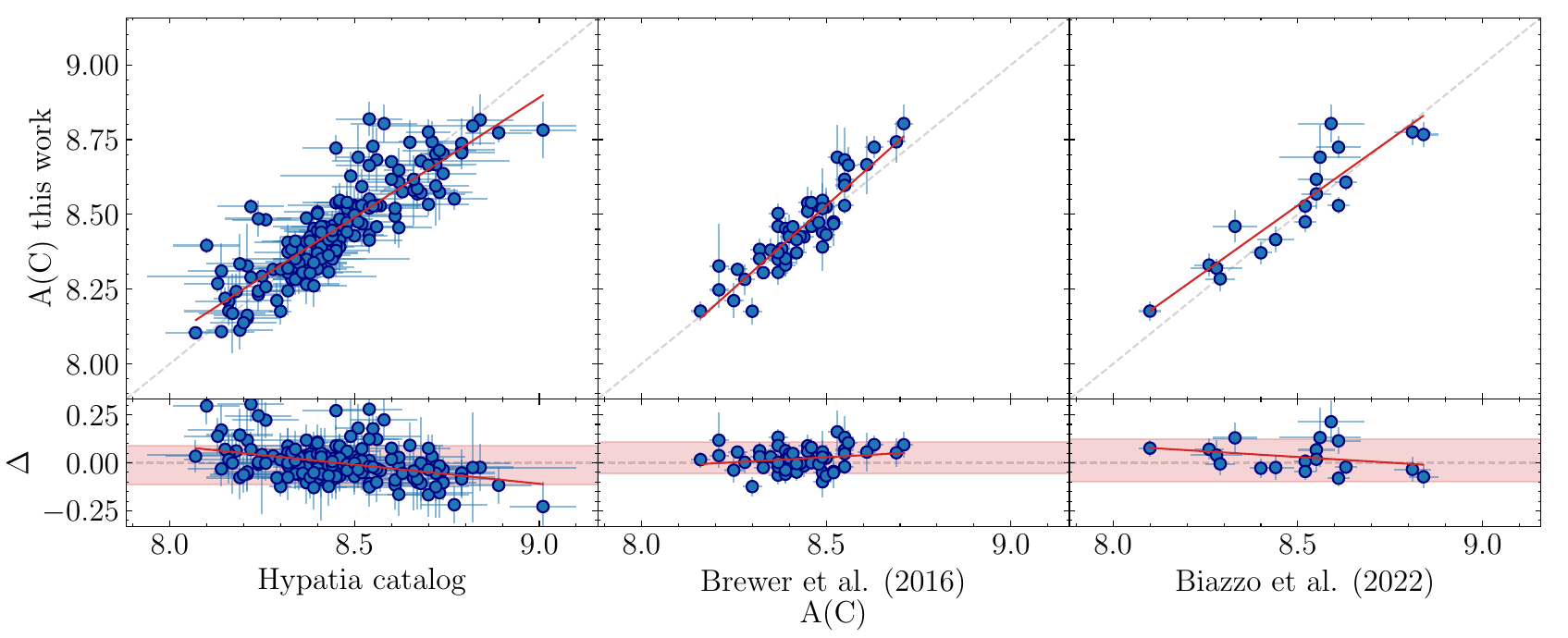}%
    \vspace{0cm}
    \includegraphics[width=\linewidth]{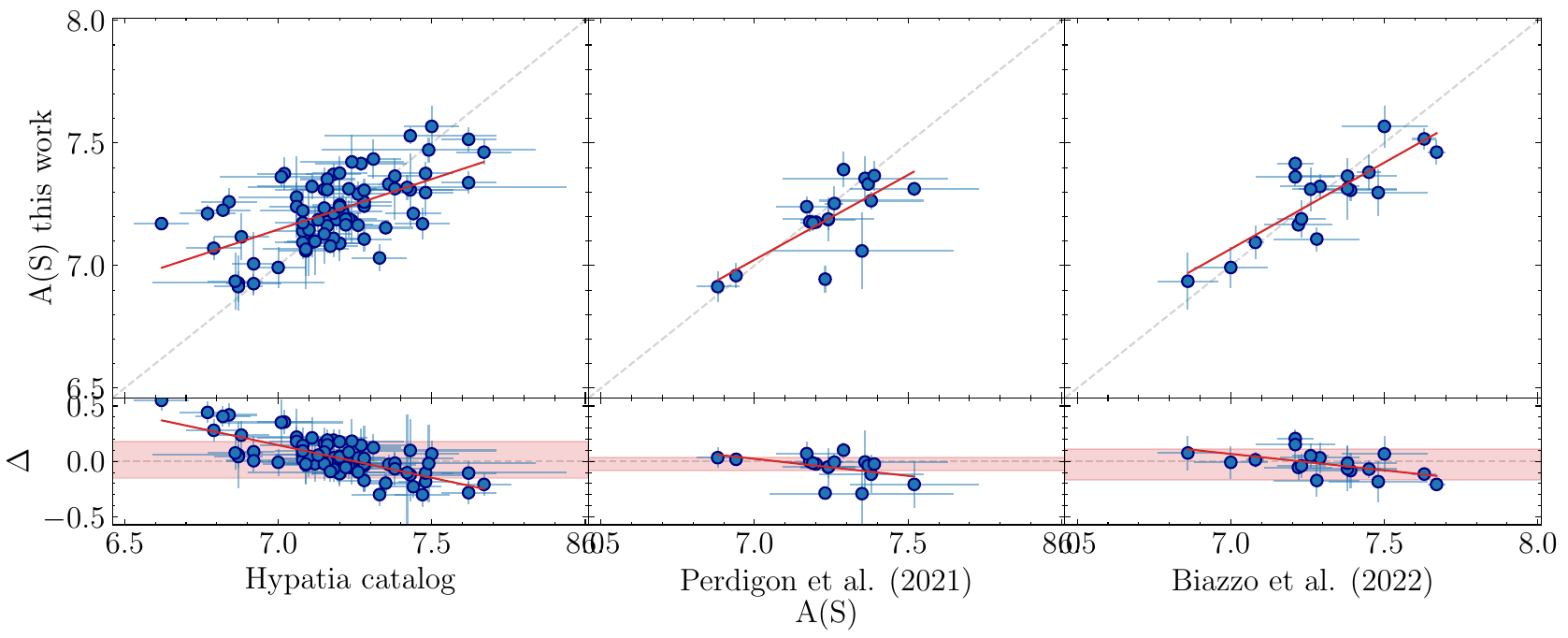}%
    \vspace{0cm}
    \includegraphics[width=\linewidth]{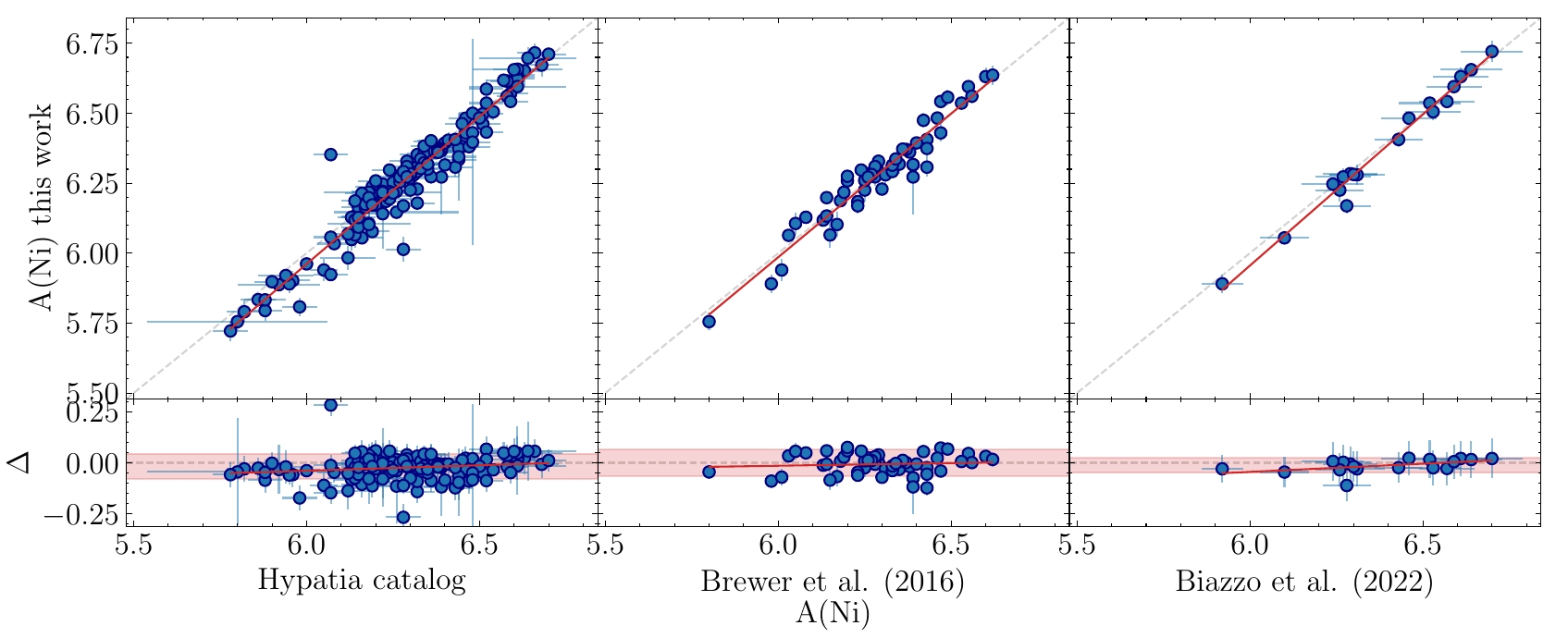}%
    \caption{Comparisons between the absolute abundances of C (top row), S (middle row) and Ni (bottom row) determined in this work and in the literature. We compare our abundances with the Hypatia catalog (H14, \citealt{Hinkel2014}), \cite{Brewer2016} (B16), \cite{Perdigon2021} (P21) and the GAPS Programme sample (BI22, \citealt{Biazzo2022}). The lines and painted areas are the same as in Figure \ref{fig:atm_comp}.}
    \label{fig:CSNi_comp}
\end{figure}

We compare our abundances determined with equivalent widths, i.e., those of C, Ni and S, with the median absolute abundances from the updated version of the Hypatia catalog (\citealt{Hinkel2014}, H14) to check if our values are compatible with a global distribution of these abundances. Additionally, we also compare our results with those from the 2022 paper of the GAPS Programme at TNG, \cite{Biazzo2022} (hereafter, BI22), for a wider appraisal of the accuracy and precision of our approach. Finally, we also compare with the abundances C and Ni abundances of \cite{Brewer2016} (hereafter, B16) and S abundances of \cite{Perdigon2021} (hereafter, P21). In Figure \ref{fig:CSNi_comp}, we compare our abundances directly.

For the Hypatia catalog, we have 59 stars in common with carbon abundances. The median difference between carbon abundances is -0.01$\pm$0.04 dex and we found a R$^2$=0.12 for the linear fit of the residuals. However, it is noteworthy that high uncertainties can be found in H14 values and this might be due to the fact that, besides having measurements of a variety of techniques, it also has different carbon indicators, mixing atomic lines and molecular band abundances together -- which, in general, do not agree in the literature (e.g., \citealt{DelgadoMena2021}). For nickel, we have an excellent agreement for the 53 stars in common, with a median difference of -0.02$\pm$0.03 dex and R$^2$=0.01 for the linear fit of the residuals. For C and Ni, we can see a clear outlier in the plots. This star is Kepler-1300, having A(C) = 8.53$\pm$0.03 dex and A(Ni) = 6.35$\pm$0.03 dex. In the catalog, their values are based on determinations by \cite{Brewer2018}, 8.22$\pm$0.09 dex and 6.07$\pm$0.05 dex, respectively. Lastly, for sulfur, we have 28 stars in common, with a median difference of 0.00$\pm$0.07 dex and R$^2$=0.51 for the linear fit of the residuals, reflecting the presence of a moderate trend. This dispersion is typical considering that the blends in sulfur lines decrease the precision of the abundance determination. There are two stars pushing the trend observed in the residuals, K2-188 (7.07$\pm$0.05 dex) and TOI-1736 (7.23$\pm$0.02 dex), having \teff\ values of 5959 K and 5753 K, respectively. In the Hypatia catalog, K2-188 and TOI-1736 values are based on determinations by \cite{Plotnikova2024}, 7.69$\pm$0.09 dex and 6.82$\pm$0.09 dex, respectively. 

For the stars in common with \cite{Brewer2016}, in general, we find an excellent agreement. For carbon, we have 43 stars in common, with a median difference of 0.03$\pm$0.04 dex and R$^2$=0.04 for the linear fit of the residuals. For nickel, we have 53 stars in common, with a median difference of 0.00$\pm$0.03 dex and R$^2$=0.01 for the linear fit of the residuals. 

For \cite{Perdigon2021} (sulfur), we have 17 stars in common, finding a median difference of -0.02$\pm$0.03 dex and R$^2$=0.19 for the linear fit of the residuals, reflecting the presence of a weak trend.

For the 18 stars in common with \cite{Biazzo2022}, we find median differences of 0.01$\pm$0.06 dex, -0.01$\pm$0.02 dex and -0.03$\pm$0.07 dex for C, Ni and S, respectively, and corresponding R$^2$ values for the linear fits of the residuals of 0.08, 0.25 and 0.30, reflecting the presence of a weak trend with Ni and S. However, the higher values of dispersion and R$^2$ found for the residuals might be due to the scarcity of stars in common.

\subsection{Oxygen}
\label{sec:cap5_oxygen}
\begin{figure}[!ht]
    \centering
    \includegraphics[width=\linewidth]{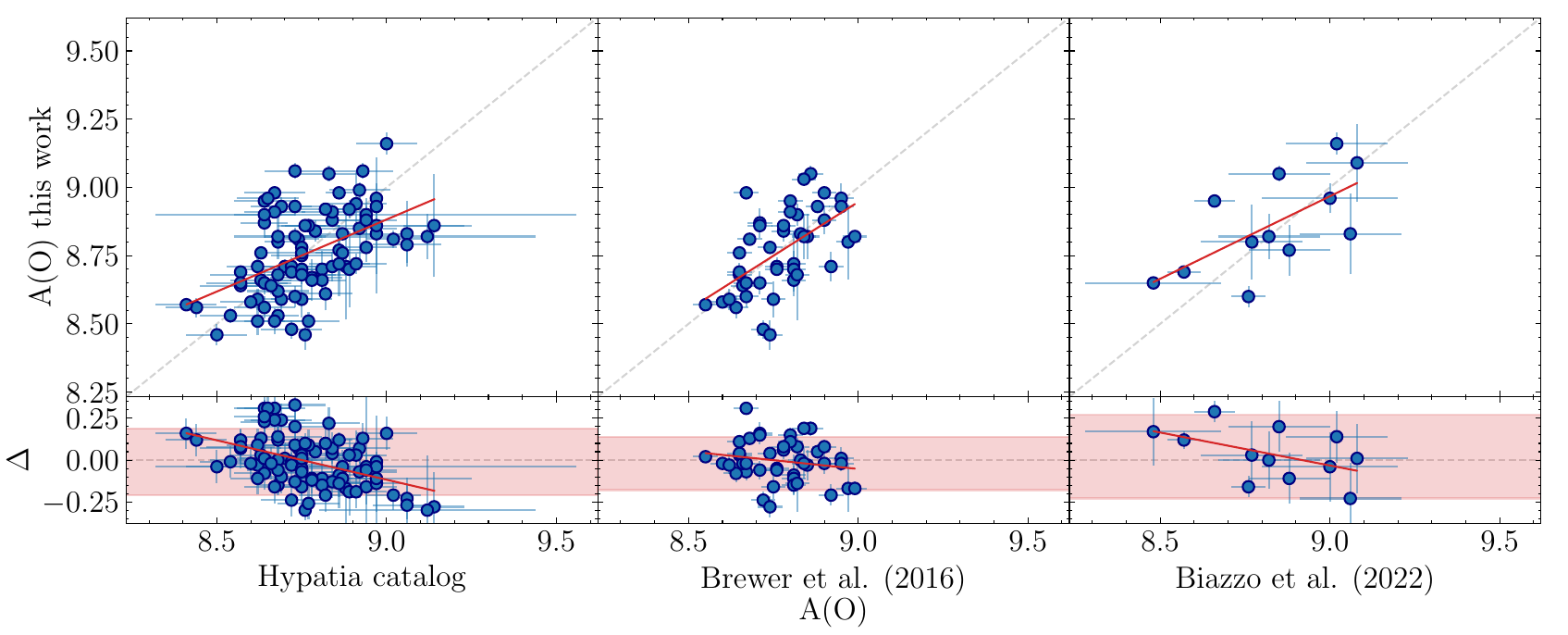}%
    \caption{Comparisons between the absolute abundances of O determined in this work and in the literature. We compare our abundances with the Hypatia catalog (H14, \citealt{Hinkel2014}), \cite{Brewer2016} (B16) and the GAPS Programme sample (BI22, \citealt{Biazzo2022}). The lines and painted areas are the same as in Figure \ref{fig:atm_comp}.}
    \label{fig:O_comp}
\end{figure}

We compare our oxygen abundances, determined using spectral synthesis, with the values of H14, B16 and BI22 (see Figure \ref{fig:O_comp}). The abundances from H14 are compiled from works in which oxygen was determined exclusively using the O I triplet at 777 nm, which are lines that suffer from NLTE effects (e.g., \citealt{Amarsi2019}). B16 determined their abundances using spectral synthesis of the O I triplet at 777 nm and some molecular lines, including OH. BI22 determined their abundances using two methods, the MOOG driver \texttt{blends} with EWs (measured with \texttt{splot} task from Tody1986) of O I triplet and NLTE corrections by \cite{Amarsi2015} and, also, spectral synthesis of the [O I] line using the MOOG driver \texttt{synth}. For the final abundances, they considered the weighted average of the values from the two methods.

Comparing with H14, B16 and BI22, we have 35, 44 and 13 stars in common, respectively, finding median differences of -0.04$\pm$0.18 dex, -0.02$\pm$0.09 dex and 0.01$\pm$0.13 dex and R$^2$ values of 0.20, 0.06 and 0.22 for the residuals, reflecting the presence of a weak trend with H14 and BI22. Again, this might be due to the low number of stars in common. However, in the literature, we can find examples of the same group using fixed atmospheric parameters and finding different oxygen abundances depending on the indicator they used. \cite{Ecuvillon2006} reported mean abundances of oxygen of a sample of 96 planet-hosting stars determined using 4 different indicators, finding 0.10$\pm$0.16 dex with near-UV OH lines, 0.12$\pm$0.16 dex with [O I] line, -0.16$\pm$0.17 dex with NLTE O I triplet and 0.15$\pm$0.17 dex with LTE O I triplet. Finally, considering the discrepancies between oxygen abundances of different indicators, adding the differences of atmospheric parameters and methodologies, these numbers represent a general good agreement between our abundances and those from the literature.

\begin{figure}[!ht]
    \centering
    \includegraphics[width=0.5\linewidth]{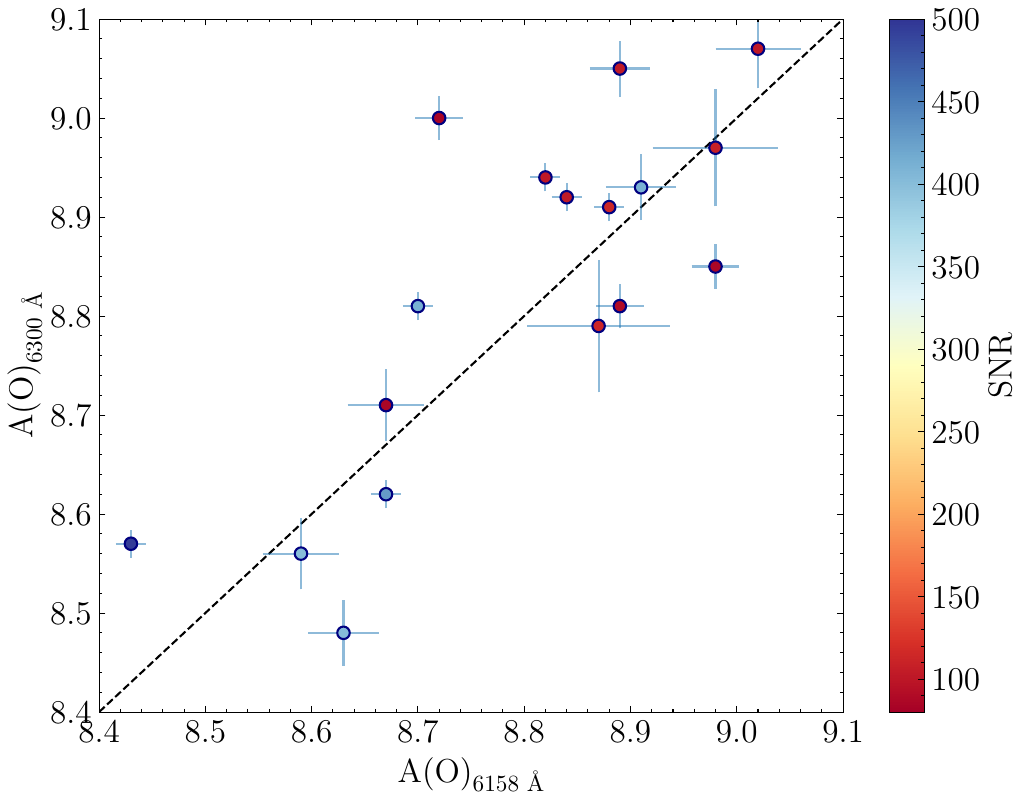}
    \caption{Comparison between abundances of oxygen determined in this work using [O I] line at 6300 \AA~and O I line at 6158 \AA. The colors represent the SNR of the spectra.}
    \label{fig:oi6158}
\end{figure}

Additionally, to the best of our knowledge, there is not a large sample of stars in common with this work having oxygen abundances determined based on the O I line at 6158 \AA. We determined oxygen abundances for 17 stars with the O I line at 6158 \AA~applying the same methodology as the one used for the [O I] line at 6300 \AA~(see Section \ref{sec:oxygen}), but manually. We obtained an excellent agreement between the oxygen abundances obtained with the two oxygen indicators, finding a median difference ($\text{A(O)}_{\text{6300}}-\text{A(O)}_{\text{6158}}$) of 0.03 dex for the entire sample, 0.04 dex for the subsample having $80<\text{SNR}<120$ and 0.00 dex for the subsample having $\text{SNR}>400$ (see Figure \ref{fig:oi6158}), and MAD of 0.08 for the tree subsamples. Finally, we obtained $\text{A(O)}=8.70\pm0.05$ dex for the Ceres 2009-02-08 solar spectrum, which is the same abundance obtained using the [O I] line at 6300 \AA.

\subsection{Stellar Radii}
\label{sec:05_StellarRadii}
We compared our stellar radii with the values from \cite{Stassun2017} (hereafter, S17), \cite{Petigura2018a} (hereafter, P18a), \cite{Kruse2019} (hereafter, K19) and \cite{Loaiza-Tacuri2024} (hereafter, LT24). S17 used \teff, [Fe/H] and \logg\ values from PASTEL catalog (\citealt{Soubiran2016}), broadband photometric data from all-sky catalogs and parallaxes from Gaia DR1 (\citealt{GaiaCollaboration2016}) to calculate stellar bolometric fluxes and angular radii. After, they determined stellar radii using the Stefan-Boltzmann law. P18a used HIRES spectra to determine atmospheric parameters for the planet-hosting stars and \texttt{isoclassify} Python package (\citealt{Huber2017}) to determine stellar masses and radii. K19 used Gaia DR2 (\citealt{GaiaCollaboration2018}) stellar radii. LT24 used HIRES spectra of K2 stars observed from campaigns zero through eight (C0-8) to determine atmospheric parameters using the $q^2$ code (\citealt{Rami2014}) for planet-hosting stars and PARAM v1.3 code (\citealt{daSilva2006}), associated with V magnitudes and Gaia DR3 parallaxes (\citealt{GaiaCollaboration2021}), to determine stellar radii.

\begin{figure}[!ht]
    \centering
    \includegraphics[width=\linewidth]{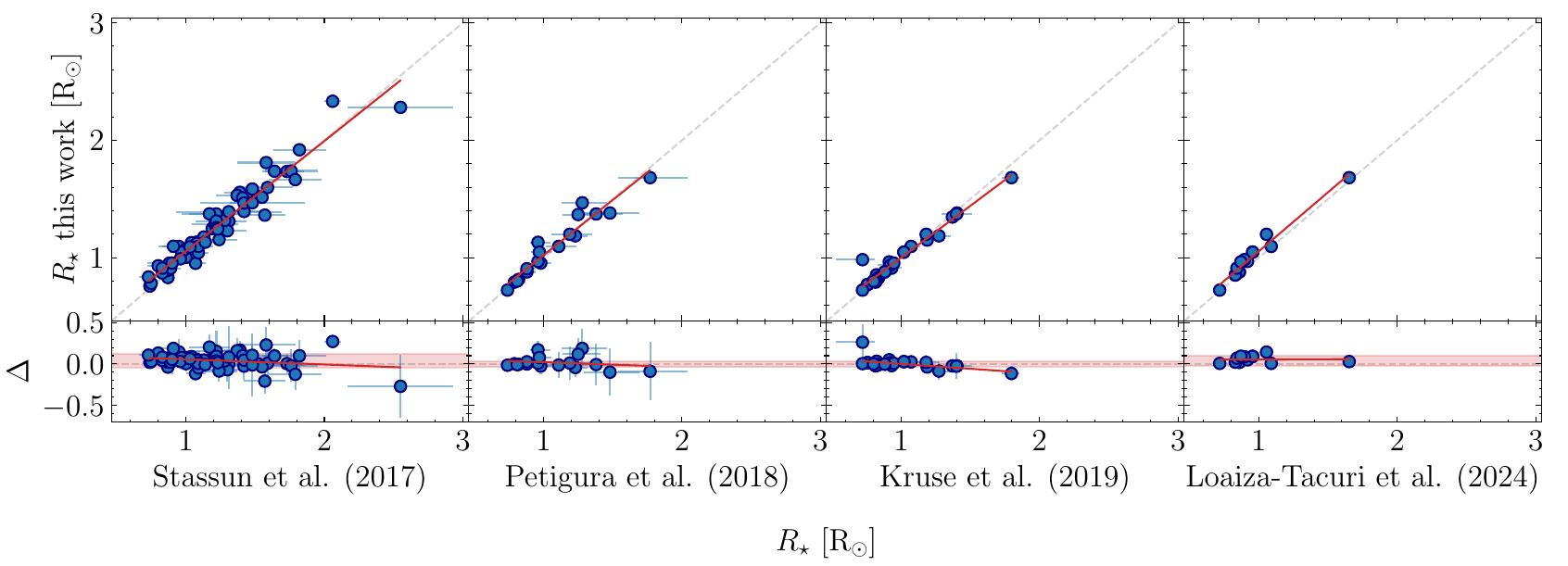}
    \caption{Comparisons between stellar radii determined in this work and in the literature. We compare with S17 (\citealt{Stassun2017}) in the first panel, P18a (\citealt{Petigura2018a}) in the second panel, K19 (\citealt{Kruse2019}) in the third panel and LT24 (\citealt{Loaiza-Tacuri2024}) in the forth panel. We show the direct comparisons in the upper panels and the residuals in the lower panels. The lines and painted areas are the same as in Figure \ref{fig:atm_comp}.}
    \label{fig:Rstar_comp}
\end{figure}

In Figure \ref{fig:Rstar_comp}, we see an excellent agreement with stellar radii values from the literature. We find a median difference of 0.06$\pm$0.03 R$_\odot$, 0.00$\pm$0.02 R$_\odot$, -0.02$\pm$0.02 R$_\odot$ and 0.04$\pm$0.02 R$_\odot$ for 21, 13, 12 and 4 stars in common with S17, P18, K19 and LT24, respectively, and R$^2$ values of 0.02, 0.06, 0.49 and 0.00 for the residuals. The high R$^2$ value for the residuals with K19 might be associated with the low number statistics since we can visually see a good agreement.

\subsection{Exoplanetary Radii}
\label{subsec:planetary_radii}

\begin{figure}[!ht]
    \centering
    \includegraphics[width=\linewidth]{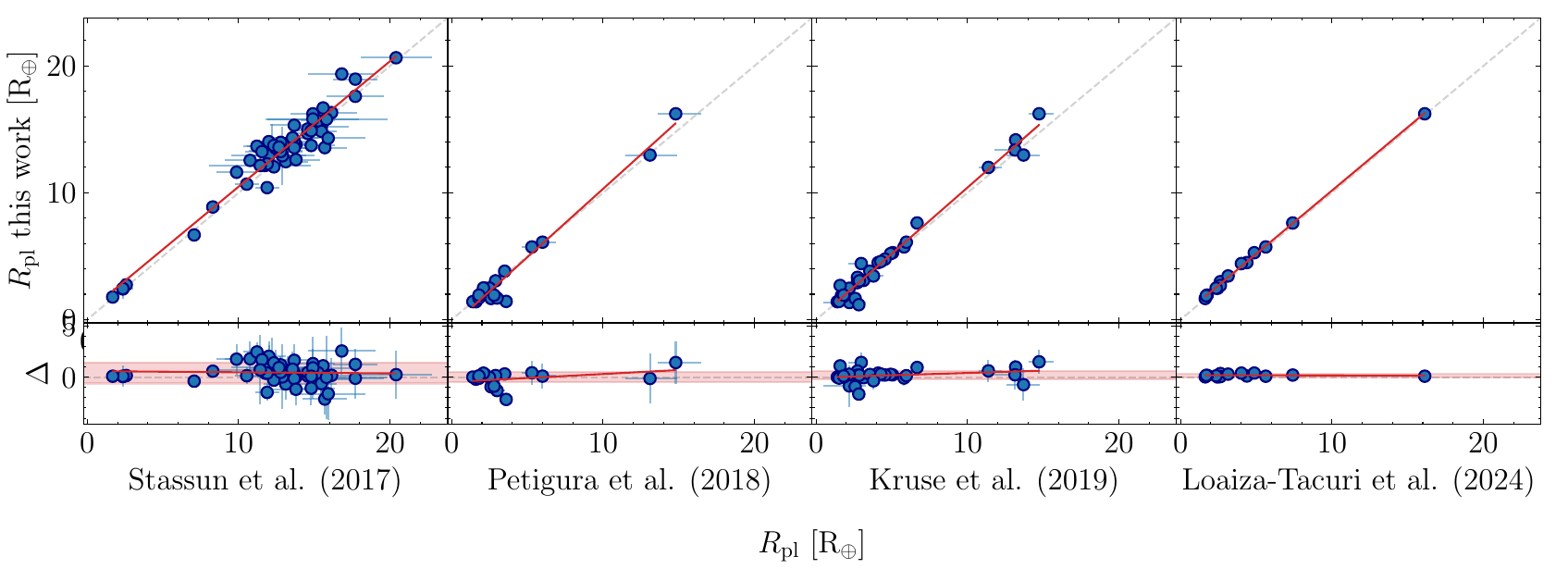}
    \caption{Comparisons between planetary radii determined in this work and in the literature. We compare with S17 (\citealt{Stassun2017}) in the first panel, P18a (\citealt{Petigura2018a}) in the second panel, K19 (\citealt{Kruse2019}) in the third panel and LT24 (\citealt{Loaiza-Tacuri2024}) in the forth panel. The lines and painted areas are the same as in Figure \ref{fig:atm_comp}.}
    \label{fig:planetary_radii}
\end{figure}

We compare our exoplanetary radii with the values calculated by the same authors mentioned in Section \ref{sec:05_StellarRadii}. With their determined stellar radii, S17 used exoplanetary parameters from \url{exoplanets.org} to determine exoplanetary radii, P18a used transit depths calculated from K2 light curves from campaigns five through eight (C5-8), K19 used transit depths calculated from K2 light curves from C0-8 and LT24 used transit depth values from NASA Exoplanet Archive. We see a good agreement with exoplanetary radii values from the literature in Figure \ref{fig:planetary_radii}. We find a median difference of 0.42$\pm$0.51 R$_\oplus$, 0.02$\pm$0.26 R$_\oplus$, 0.21$\pm$0.20 R$_\oplus$ and 0.15$\pm$0.07 R$_\oplus$ for 56, 17, 31 and 14 exoplanets in common with S17, P18, K19 and LT24, respectively, and R$^2$ values of 0.00, 0.14, 0.08 and 0.00 for the residuals. Finally, the median uncertainties for the transit depth and exoplanetary radii are 3.00\% and 2.00\%, respectively.

\bibliography{sample631}{}

@misc{ps,
doi = {10.26133/NEA12},
url = {https://catcopy.ipac.caltech.edu/dois/doi.php?id=10.26133/NEA12},
author = {{NASA Exoplanet Archive}},
title = {Planetary Systems},
publisher = {NExScI-Caltech/IPAC},
version = {Version: 2025-11-30 10:44},
year = {2025}
}

@ARTICLE{Li2024,
       author = {{Li}, Cheng and {Allison}, Michael and {Atreya}, Sushil and {Brueshaber}, Shawn and {Fletcher}, Leigh N. and {Guillot}, Tristan and {Li}, Liming and {Lunine}, Jonathan and {Miguel}, Yamila and {Orton}, Glenn and {Steffes}, Paul and {Waite}, J. Hunter and {Wong}, Michael H. and {Levin}, Steven and {Bolton}, Scott},
        title = "{Super-adiabatic temperature gradient at Jupiter's equatorial zone and implications for the water abundance}",
      journal = {\icarus},
         year = 2024,
        month = may,
       volume = {414},
          eid = {116028},
        pages = {116028},
          doi = {10.1016/j.icarus.2024.116028},
archivePrefix = {arXiv},
       eprint = {2403.05363},
 primaryClass = {astro-ph.EP},
       adsurl = {https://ui.adsabs.harvard.edu/abs/2024Icar..41416028L}
}

@ARTICLE{Robinson2006,
       author = {{Robinson}, Sarah E. and {Laughlin}, Gregory and {Bodenheimer}, Peter and {Fischer}, Debra},
        title = "{Silicon and Nickel Enrichment in Planet Host Stars: Observations and Implications for the Core Accretion Theory of Planet Formation}",
      journal = {\apj},
         year = 2006,
        month = may,
       volume = {643},
       number = {1},
        pages = {484-500},
          doi = {10.1086/502795},
archivePrefix = {arXiv},
       eprint = {astro-ph/0601656},
 primaryClass = {astro-ph},
       adsurl = {https://ui.adsabs.harvard.edu/abs/2006ApJ...643..484R}
}

@ARTICLE{Yang2026,
       author = {{Yang}, Jeehyun and {Hyder}, Ali and {Hu}, Renyu and {Lunine}, Jonathan I.},
        title = "{Coupled 1D Chemical Kinetic Transport and 2D Hydrodynamic Modeling Supports a Modest 1─1.5{\texttimes} Supersolar Oxygen Abundance in Jupiter's Atmosphere}",
      journal = {\psj},
         year = 2026,
        month = jan,
       volume = {7},
       number = {1},
          eid = {2},
        pages = {2},
          doi = {10.3847/PSJ/ae28d5},
archivePrefix = {arXiv},
       eprint = {2508.05007},
 primaryClass = {astro-ph.EP},
       adsurl = {https://ui.adsabs.harvard.edu/abs/2026PSJ.....7....2Y}
}

@ARTICLE{Espinoza2017,
       author = {{Espinoza}, N{\'e}stor and {Fortney}, Jonathan J. and {Miguel}, Yamila and {Thorngren}, Daniel and {Murray-Clay}, Ruth},
        title = "{Metal Enrichment Leads to Low Atmospheric C/O Ratios in Transiting Giant Exoplanets}",
      journal = {\apjl},
         year = 2017,
        month = mar,
       volume = {838},
       number = {1},
          eid = {L9},
        pages = {L9},
          doi = {10.3847/2041-8213/aa65ca},
archivePrefix = {arXiv},
       eprint = {1611.08616},
 primaryClass = {astro-ph.EP},
       adsurl = {https://ui.adsabs.harvard.edu/abs/2017ApJ...838L...9E}
}

@ARTICLE{Wanderley2025,
       author = {{Wanderley}, F{\'a}bio and {Cunha}, Katia and {Souto}, Diogo and {Smith}, Verne V. and {Daflon}, Simone},
        title = "{Metallicities of M Dwarf Planet Host Stars from Kepler, K2, and TESS Observed by APOGEE: Trends with Exoplanetary Radii and Orbital Periods}",
      journal = {\aj},
         year = 2025,
        month = sep,
       volume = {170},
       number = {3},
          eid = {177},
        pages = {177},
          doi = {10.3847/1538-3881/adef12},
archivePrefix = {arXiv},
       eprint = {2507.04066},
 primaryClass = {astro-ph.SR},
       adsurl = {https://ui.adsabs.harvard.edu/abs/2025AJ....170..177W}
}

@ARTICLE{Boufleur2018,
       author = {{Boufleur}, Rodrigo C. and {Emilio}, Marcelo and {Janot-Pacheco}, Eduardo and {Andrade}, Laerte and {Ferraz-Mello}, Sylvio and {do Nascimento}, Jr., Jos{\'e}-Dias and {de La Reza}, Ramiro},
        title = "{A modified CoRoT detrend algorithm and the discovery of a new planetary companion}",
      journal = {\mnras},
         year = 2018,
        month = jan,
       volume = {473},
       number = {1},
        pages = {710-720},
          doi = {10.1093/mnras/stx2187},
archivePrefix = {arXiv},
       eprint = {1709.00351},
 primaryClass = {astro-ph.EP},
       adsurl = {https://ui.adsabs.harvard.edu/abs/2018MNRAS.473..710B}
}

@ARTICLE{Vivien2024,
       author = {{Vivien}, H.~G. and {Hoyer}, S. and {Deleuil}, M. and {Sulis}, S. and {Santerne}, A. and {Christiansen}, J.~L. and {Hardegree-Ullman}, K.~K. and {Lopez}, T.~A.},
        title = "{New ephemerides and detection of transit-timing variations in the K2-138 system using high-precision CHEOPS photometry}",
      journal = {\aap},
         year = 2024,
        month = aug,
       volume = {688},
          eid = {A192},
        pages = {A192},
          doi = {10.1051/0004-6361/202348013},
archivePrefix = {arXiv},
       eprint = {2406.18267},
 primaryClass = {astro-ph.EP},
       adsurl = {https://ui.adsabs.harvard.edu/abs/2024A&A...688A.192V}
}

@ARTICLE{Pope2016,
       author = {{Pope}, Benjamin J.~S. and {Parviainen}, Hannu and {Aigrain}, Suzanne},
        title = "{Transiting exoplanet candidates from K2 Campaigns 5 and 6}",
      journal = {\mnras},
         year = 2016,
        month = oct,
       volume = {461},
       number = {4},
        pages = {3399-3409},
          doi = {10.1093/mnras/stw1373},
archivePrefix = {arXiv},
       eprint = {1606.01264},
 primaryClass = {astro-ph.EP},
       adsurl = {https://ui.adsabs.harvard.edu/abs/2016MNRAS.461.3399P}
}

@ARTICLE{Deeg2010,
       author = {{Deeg}, H.~J. and {Moutou}, C. and {Erikson}, A. and {Csizmadia}, Sz. and {Tingley}, B. and {Barge}, P. and {Bruntt}, H. and {Havel}, M. and {Aigrain}, S. and {Almenara}, J.~M. and {Alonso}, R. and {Auvergne}, M. and {Baglin}, A. and {Barbieri}, M. and {Benz}, W. and {Bonomo}, A.~S. and {Bord{\'e}}, P. and {Bouchy}, F. and {Cabrera}, J. and {Carone}, L. and {Carpano}, S. and {Ciardi}, D. and {Deleuil}, M. and {Dvorak}, R. and {Ferraz-Mello}, S. and {Fridlund}, M. and {Gandolfi}, D. and {Gazzano}, J.-C. and {Gillon}, M. and {Gondoin}, P. and {Guenther}, E. and {Guillot}, T. and {Hartog}, R. Den and {Hatzes}, A. and {Hidas}, M. and {H{\'e}brard}, G. and {Jorda}, L. and {Kabath}, P. and {Lammer}, H. and {L{\'e}ger}, A. and {Lister}, T. and {Llebaria}, A. and {Lovis}, C. and {Mayor}, M. and {Mazeh}, T. and {Ollivier}, M. and {P{\"a}tzold}, M. and {Pepe}, F. and {Pont}, F. and {Queloz}, D. and {Rabus}, M. and {Rauer}, H. and {Rouan}, D. and {Samuel}, B. and {Schneider}, J. and {Shporer}, A. and {Stecklum}, B. and {Street}, R. and {Udry}, S. and {Weingrill}, J. and {Wuchterl}, G.},
        title = "{A transiting giant planet with a temperature between 250K and 430K}",
      journal = {\nat},
         year = 2010,
        month = mar,
       volume = {464},
       number = {7287},
        pages = {384-387},
          doi = {10.1038/nature08856},
       adsurl = {https://ui.adsabs.harvard.edu/abs/2010Natur.464..384D}
}

@ARTICLE{Christiansen2017,
       author = {{Christiansen}, Jessie L. and {Vanderburg}, Andrew and {Burt}, Jennifer and {Fulton}, B.~J. and {Batygin}, Konstantin and {Benneke}, Bj{\"o}rn and {Brewer}, John M. and {Charbonneau}, David and {Ciardi}, David R. and {Collier Cameron}, Andrew and {Coughlin}, Jeffrey L. and {Crossfield}, Ian J.~M. and {Dressing}, Courtney and {Greene}, Thomas P. and {Howard}, Andrew W. and {Latham}, David W. and {Molinari}, Emilio and {Mortier}, Annelies and {Mullally}, Fergal and {Pepe}, Francesco and {Rice}, Ken and {Sinukoff}, Evan and {Sozzetti}, Alessandro and {Thompson}, Susan E. and {Udry}, St{\'e}phane and {Vogt}, Steven S. and {Barman}, Travis S. and {Batalha}, Natasha E. and {Bouchy}, Fran{\c{c}}ois and {Buchhave}, Lars A. and {Butler}, R. Paul and {Cosentino}, Rosario and {Dupuy}, Trent J. and {Ehrenreich}, David and {Fiorenzano}, Aldo and {Hansen}, Brad M.~S. and {Henning}, Thomas and {Hirsch}, Lea and {Holden}, Bradford P. and {Isaacson}, Howard T. and {Johnson}, John A. and {Knutson}, Heather A. and {Kosiarek}, Molly and {L{\'o}pez-Morales}, Mercedes and {Lovis}, Christophe and {Malavolta}, Luca and {Mayor}, Michel and {Micela}, Giuseppina and {Motalebi}, Fatemeh and {Petigura}, Erik and {Phillips}, David F. and {Piotto}, Giampaolo and {Rogers}, Leslie A. and {Sasselov}, Dimitar and {Schlieder}, Joshua E. and {S{\'e}gransan}, Damien and {Watson}, Christopher A. and {Weiss}, Lauren M.},
        title = "{Three{\textquoteright}s Company: An Additional Non-transiting Super-Earth in the Bright HD 3167 System, and Masses for All Three Planets}",
      journal = {\aj},
         year = 2017,
        month = sep,
       volume = {154},
       number = {3},
          eid = {122},
        pages = {122},
          doi = {10.3847/1538-3881/aa832d},
archivePrefix = {arXiv},
       eprint = {1706.01892},
 primaryClass = {astro-ph.EP},
       adsurl = {https://ui.adsabs.harvard.edu/abs/2017AJ....154..122C}
}

@ARTICLE{Borde2010,
       author = {{Bord{\'e}}, P. and {Bouchy}, F. and {Deleuil}, M. and {Cabrera}, J. and {Jorda}, L. and {Lovis}, C. and {Csizmadia}, S. and {Aigrain}, S. and {Almenara}, J.~M. and {Alonso}, R. and {Auvergne}, M. and {Baglin}, A. and {Barge}, P. and {Benz}, W. and {Bonomo}, A.~S. and {Bruntt}, H. and {Carone}, L. and {Carpano}, S. and {Deeg}, H. and {Dvorak}, R. and {Erikson}, A. and {Ferraz-Mello}, S. and {Fridlund}, M. and {Gandolfi}, D. and {Gazzano}, J.-C. and {Gillon}, M. and {Guenther}, E. and {Guillot}, T. and {Guterman}, P. and {Hatzes}, A. and {Havel}, M. and {H{\'e}brard}, G. and {Lammer}, H. and {L{\'e}ger}, A. and {Mayor}, M. and {Mazeh}, T. and {Moutou}, C. and {P{\"a}tzold}, M. and {Pepe}, F. and {Ollivier}, M. and {Queloz}, D. and {Rauer}, H. and {Rouan}, D. and {Samuel}, B. and {Santerne}, A. and {Schneider}, J. and {Tingley}, B. and {Udry}, S. and {Weingrill}, J. and {Wuchterl}, G.},
        title = "{Transiting exoplanets from the CoRoT space mission. XI. CoRoT-8b: a hot and dense sub-Saturn around a K1 dwarf}",
      journal = {\aap},
         year = 2010,
        month = sep,
       volume = {520},
          eid = {A66},
        pages = {A66},
          doi = {10.1051/0004-6361/201014775},
archivePrefix = {arXiv},
       eprint = {1008.0325},
 primaryClass = {astro-ph.EP},
       adsurl = {https://ui.adsabs.harvard.edu/abs/2010A&A...520A..66B}
}

@ARTICLE{Bonfanti2021,
       author = {{Bonfanti}, A. and {Delrez}, L. and {Hooton}, M.~J. and {Wilson}, T.~G. and {Fossati}, L. and {Alibert}, Y. and {Hoyer}, S. and {Mustill}, A.~J. and {Osborn}, H.~P. and {Adibekyan}, V. and {Gandolfi}, D. and {Salmon}, S. and {Sousa}, S.~G. and {Tuson}, A. and {Van Grootel}, V. and {Cabrera}, J. and {Nascimbeni}, V. and {Maxted}, P.~F.~L. and {Barros}, S.~C.~C. and {Billot}, N. and {Bonfils}, X. and {Borsato}, L. and {Broeg}, C. and {Davies}, M.~B. and {Deleuil}, M. and {Demangeon}, O.~D.~S. and {Fridlund}, M. and {Lacedelli}, G. and {Lendl}, M. and {Persson}, C. and {Santos}, N.~C. and {Scandariato}, G. and {Szab{\'o}}, Gy. M. and {Collier Cameron}, A. and {Udry}, S. and {Benz}, W. and {Beck}, M. and {Ehrenreich}, D. and {Fortier}, A. and {Isaak}, K.~G. and {Queloz}, D. and {Alonso}, R. and {Asquier}, J. and {Bandy}, T. and {B{\'a}rczy}, T. and {Barrado}, D. and {Barrag{\'a}n}, O. and {Baumjohann}, W. and {Beck}, T. and {Bekkelien}, A. and {Bergomi}, M. and {Brandeker}, A. and {Busch}, M.-D. and {Cessa}, V. and {Charnoz}, S. and {Chazelas}, B. and {Corral Van Damme}, C. and {Demory}, B.-O. and {Erikson}, A. and {Farinato}, J. and {Futyan}, D. and {Garcia Mu{\~n}oz}, A. and {Gillon}, M. and {Guedel}, M. and {Guterman}, P. and {Hasiba}, J. and {Heng}, K. and {Hernandez}, E. and {Kiss}, L. and {Kuntzer}, T. and {Laskar}, J. and {Lecavelier des Etangs}, A. and {Lovis}, C. and {Magrin}, D. and {Malvasio}, L. and {Marafatto}, L. and {Michaelis}, H. and {Munari}, M. and {Olofsson}, G. and {Ottacher}, H. and {Ottensamer}, R. and {Pagano}, I. and {Pall{\'e}}, E. and {Peter}, G. and {Piazza}, D. and {Piotto}, G. and {Pollacco}, D. and {Ragazzoni}, R. and {Rando}, N. and {Ratti}, F. and {Rauer}, H. and {Ribas}, I. and {Rieder}, M. and {Rohlfs}, R. and {Safa}, F. and {Salatti}, M. and {S{\'e}gransan}, D. and {Simon}, A.~E. and {Smith}, A.~M.~S. and {Sordet}, M. and {Steller}, M. and {Thomas}, N. and {Tschentscher}, M. and {Van Eylen}, V. and {Viotto}, V. and {Walter}, I. and {Walton}, N.~A. and {Wildi}, F. and {Wolter}, D.},
        title = "{CHEOPS observations of the HD 108236 planetary system: a fifth planet, improved ephemerides, and planetary radii}",
      journal = {\aap},
         year = 2021,
        month = feb,
       volume = {646},
          eid = {A157},
        pages = {A157},
          doi = {10.1051/0004-6361/202039608},
archivePrefix = {arXiv},
       eprint = {2101.00663},
 primaryClass = {astro-ph.EP},
       adsurl = {https://ui.adsabs.harvard.edu/abs/2021A&A...646A.157B}
}

@misc{esobib,
  doi = {10.18727/ARCHIVE/33},
  url = {https://doi.eso.org/10.18727/archive/33},
  author = {{European Southern Observatory (ESO)}},
  language = {en},
  title = {HARPS reduced data obtained by standard ESO pipeline processing},
  publisher = {European Southern Observatory (ESO)},
  year = {2014},
  copyright = {Data Access Policy for ESO Data held in the ESO Science Archive Facility}
}

@ARTICLE{Mullally2015,
       author = {{Mullally}, F. and {Coughlin}, Jeffrey L. and {Thompson}, Susan E. and {Rowe}, Jason and {Burke}, Christopher and {Latham}, David W. and {Batalha}, Natalie M. and {Bryson}, Stephen T. and {Christiansen}, Jessie and {Henze}, Christopher E. and {Ofir}, Aviv and {Quarles}, Billy and {Shporer}, Avi and {Van Eylen}, Vincent and {Van Laerhoven}, Christa and {Shah}, Yash and {Wolfgang}, Angie and {Chaplin}, W.~J. and {Xie}, Ji-Wei and {Akeson}, Rachel and {Argabright}, Vic and {Bachtell}, Eric and {Barclay}, Thomas and {Borucki}, William J. and {Caldwell}, Douglas A. and {Campbell}, Jennifer R. and {Catanzarite}, Joseph H. and {Cochran}, William D. and {Duren}, Riley M. and {Fleming}, Scott W. and {Fraquelli}, Dorothy and {Girouard}, Forrest R. and {Haas}, Michael R. and {He{\l}miniak}, Krzysztof G. and {Howell}, Steve B. and {Huber}, Daniel and {Larson}, Kipp and {Gautier}, III, Thomas N. and {Jenkins}, Jon M. and {Li}, Jie and {Lissauer}, Jack J. and {McArthur}, Scot and {Miller}, Chris and {Morris}, Robert L. and {Patil-Sabale}, Anima and {Plavchan}, Peter and {Putnam}, Dustin and {Quintana}, Elisa V. and {Ramirez}, Solange and {Silva Aguirre}, V. and {Seader}, Shawn and {Smith}, Jeffrey C. and {Steffen}, Jason H. and {Stewart}, Chris and {Stober}, Jeremy and {Still}, Martin and {Tenenbaum}, Peter and {Troeltzsch}, John and {Twicken}, Joseph D. and {Zamudio}, Khadeejah A.},
        title = "{Planetary Candidates Observed by Kepler. VI. Planet Sample from Q1--Q16 (47 Months)}",
      journal = {\apjs},
         year = 2015,
        month = apr,
       volume = {217},
       number = {2},
          eid = {31},
        pages = {31},
          doi = {10.1088/0067-0049/217/2/31},
archivePrefix = {arXiv},
       eprint = {1502.02038},
 primaryClass = {astro-ph.EP},
       adsurl = {https://ui.adsabs.harvard.edu/abs/2015ApJS..217...31M}
}

@ARTICLE{Addison2016,
       author = {{Addison}, B.~C. and {Tinney}, C.~G. and {Wright}, D.~J. and {Bayliss}, D.},
        title = "{Spin-orbit Alignment for Three Transiting Hot Jupiters: WASP-103b, WASP-87b, and WASP-66b}",
      journal = {\apj},
         year = 2016,
        month = may,
       volume = {823},
       number = {1},
          eid = {29},
        pages = {29},
          doi = {10.3847/0004-637X/823/1/29},
archivePrefix = {arXiv},
       eprint = {1603.05754},
 primaryClass = {astro-ph.EP},
       adsurl = {https://ui.adsabs.harvard.edu/abs/2016ApJ...823...29A}
}

@ARTICLE{Luck2015,
       author = {{Luck}, R. Earle},
        title = "{Abundances in the Local Region. I. G and K Giants}",
      journal = {\aj},
         year = 2015,
        month = sep,
       volume = {150},
       number = {3},
          eid = {88},
        pages = {88},
          doi = {10.1088/0004-6256/150/3/88},
archivePrefix = {arXiv},
       eprint = {1507.01466},
 primaryClass = {astro-ph.SR},
       adsurl = {https://ui.adsabs.harvard.edu/abs/2015AJ....150...88L}
}

@ARTICLE{Takeda2016,
       author = {{Takeda}, Yoichi and {Omiya}, Masashi and {Harakawa}, Hiroki and {Sato}, Bun'ei},
        title = "{Sulfur and zinc abundances of red giant stars{\textdagger}}",
      journal = {\pasj},
         year = 2016,
        month = oct,
       volume = {68},
       number = {5},
          eid = {81},
        pages = {81},
          doi = {10.1093/pasj/psw071},
archivePrefix = {arXiv},
       eprint = {1607.04385},
 primaryClass = {astro-ph.SR},
       adsurl = {https://ui.adsabs.harvard.edu/abs/2016PASJ...68...81T}
}

@ARTICLE{Takada-Hidai2002,
       author = {{Takada-Hidai}, Masahide and {Takeda}, Yoichi and {Sato}, Shizuka and {Honda}, Satoshi and {Sadakane}, Kozo and {Kawanomoto}, Satoshi and {Sargent}, Wallace L.~W. and {Lu}, Limin and {Barlow}, Thomas A.},
        title = "{Behavior of Sulfur Abundances in Metal-poor Giants and Dwarfs}",
      journal = {\apj},
         year = 2002,
        month = jul,
       volume = {573},
       number = {2},
        pages = {614-630},
          doi = {10.1086/340748},
archivePrefix = {arXiv},
       eprint = {astro-ph/0103481},
 primaryClass = {astro-ph},
       adsurl = {https://ui.adsabs.harvard.edu/abs/2002ApJ...573..614T}
}

@ARTICLE{Luck2006,
       author = {{Luck}, R. Earle and {Heiter}, Ulrike},
        title = "{Dwarfs in the Local Region}",
      journal = {\aj},
         year = 2006,
        month = jun,
       volume = {131},
       number = {6},
        pages = {3069-3092},
          doi = {10.1086/504080},
       adsurl = {https://ui.adsabs.harvard.edu/abs/2006AJ....131.3069L}
}

@ARTICLE{Gardner2006,
       author = {{Gardner}, Jonathan P. and {Mather}, John C. and {Clampin}, Mark and {Doyon}, Rene and {Greenhouse}, Matthew A. and {Hammel}, Heidi B. and {Hutchings}, John B. and {Jakobsen}, Peter and {Lilly}, Simon J. and {Long}, Knox S. and {Lunine}, Jonathan I. and {McCaughrean}, Mark J. and {Mountain}, Matt and {Nella}, John and {Rieke}, George H. and {Rieke}, Marcia J. and {Rix}, Hans-Walter and {Smith}, Eric P. and {Sonneborn}, George and {Stiavelli}, Massimo and {Stockman}, H.~S. and {Windhorst}, Rogier A. and {Wright}, Gillian S.},
        title = "{The James Webb Space Telescope}",
      journal = {\ssr},
         year = 2006,
        month = apr,
       volume = {123},
       number = {4},
        pages = {485-606},
          doi = {10.1007/s11214-006-8315-7},
archivePrefix = {arXiv},
       eprint = {astro-ph/0606175},
 primaryClass = {astro-ph},
       adsurl = {https://ui.adsabs.harvard.edu/abs/2006SSRv..123..485G}
}

@ARTICLE{Gardner2023,
       author = {{Gardner}, Jonathan P. and {Mather}, John C. and {Abbott}, Randy and {Abell}, James S. and {Abernathy}, Mark and {Abney}, Faith E. and {Abraham}, John G. and {Abraham}, Roberto and {Abul-Huda}, Yasin M. and {Acton}, Scott and {Adams}, Cynthia K. and {Adams}, Evan and {Adler}, David S. and {Adriaensen}, Maarten and {Aguilar}, Jonathan Albert and {Ahmed}, Mansoor and {Ahmed}, Nasif S. and {Ahmed}, Tanjira and {Albat}, R{\"u}deger and {Albert}, Lo{\"\i}c and {Alberts}, Stacey and {Aldridge}, David and {Allen}, Mary Marsha and {Allen}, Shaune S. and {Altenburg}, Martin and {Altunc}, Serhat and {Alvarez}, Jose Lorenzo and {{\'A}lvarez-M{\'a}rquez}, Javier and {Alves de Oliveira}, Catarina and {Ambrose}, Leslie L. and {Anandakrishnan}, Satya M. and {Andersen}, Gregory C. and {Anderson}, Harry James and {Anderson}, Jay and {Anderson}, Kristen and {Anderson}, Sara M. and {Aprea}, Julio and {Archer}, Benita J. and {Arenberg}, Jonathan W. and {Argyriou}, Ioannis and {Arribas}, Santiago and {Artigau}, {\'E}tienne and {Arvai}, Amanda Rose and {Atcheson}, Paul and {Atkinson}, Charles B. and {Averbukh}, Jesse and {Aymergen}, Cagatay and {Bacinski}, John J. and {Baggett}, Wayne E. and {Bagnasco}, Giorgio and {Baker}, Lynn L. and {Balzano}, Vicki Ann and {Banks}, Kimberly A. and {Baran}, David A. and {Barker}, Elizabeth A. and {Barrett}, Larry K. and {Barringer}, Bruce O. and {Barto}, Allison and {Bast}, William and {Baudoz}, Pierre and {Baum}, Stefi and {Beatty}, Thomas G. and {Beaulieu}, Mathilde and {Bechtold}, Kathryn and {Beck}, Tracy and {Beddard}, Megan M. and {Beichman}, Charles and {Bellagama}, Larry and {Bely}, Pierre and {Berger}, Timothy W. and {Bergeron}, Louis E. and {Bernier}, Antoine-Darveau and {Bertch}, Maria D. and {Beskow}, Charlotte and {Betz}, Laura E. and {Biagetti}, Carl P. and {Birkmann}, Stephan and {Bjorklund}, Kurt F. and {Blackwood}, James D. and {Blazek}, Ronald Paul and {Blossfeld}, Stephen and {Bluth}, Marcel and {Boccaletti}, Anthony and {Boegner}, Jr., Martin E. and {Bohlin}, Ralph C. and {Boia}, John Joseph and {B{\"o}ker}, Torsten and {Bonaventura}, N. and {Bond}, Nicholas A. and {Bosley}, Kari Ann and {Boucarut}, Rene A. and {Bouchet}, Patrice and {Bouwman}, Jeroen and {Bower}, Gary and {Bowers}, Ariel S. and {Bowers}, Charles W. and {Boyce}, Leslye A. and {Boyer}, Christine T. and {Boyer}, Martha L. and {Boyer}, Michael and {Boyer}, Robert and {Bradley}, Larry D. and {Brady}, Gregory R. and {Brandl}, Bernhard R. and {Brannen}, Judith L. and {Breda}, David and {Bremmer}, Harold G. and {Brennan}, David and {Bresnahan}, Pamela A. and {Bright}, Stacey N. and {Broiles}, Brian J. and {Bromenschenkel}, Asa and {Brooks}, Brian H. and {Brooks}, Keira J. and {Brown}, Bob and {Brown}, Bruce and {Brown}, Thomas M. and {Bruce}, Barry W. and {Bryson}, Jonathan G. and {Bujanda}, Edwin D. and {Bullock}, Blake M. and {Bunker}, A.~J. and {Bureo}, Rafael and {Burt}, Irving J. and {Bush}, James Aaron and {Bushouse}, Howard A. and {Bussman}, Marie C. and {Cabaud}, Olivier and {Cale}, Steven and {Calhoon}, Charles D. and {Calvani}, Humberto and {Canipe}, Alicia M. and {Caputo}, Francis M. and {Cara}, Mihai and {Carey}, Larkin and {Case}, Michael Eli and {Cesari}, Thaddeus and {Cetorelli}, Lee D. and {Chance}, Don R. and {Chandler}, Lynn and {Chaney}, Dave and {Chapman}, George N. and {Charlot}, S. and {Chayer}, Pierre and {Cheezum}, Jeffrey I. and {Chen}, Bin and {Chen}, Christine H. and {Cherinka}, Brian and {Chichester}, Sarah C. and {Chilton}, Zachary S. and {Chittiraibalan}, Dharini and {Clampin}, Mark and {Clark}, Charles R. and {Clark}, Kerry W. and {Clark}, Stephanie M. and {Claybrooks}, Edward E. and {Cleveland}, Keith A. and {Cohen}, Andrew L. and {Cohen}, Lester M. and {Col{\'o}n}, Knicole D. and {Coleman}, Benee L. and {Colina}, Luis and {Comber}, Brian J. and {Comeau}, Thomas M. and {Comer}, Thomas and {Conde Reis}, Alain and {Connolly}, Dennis C. and {Conroy}, Kyle E. and {Contos}, Adam R. and {Contreras}, James and {Cook}, Neil J. and {Cooper}, James L. and {Cooper}, Rachel Aviva and {Correia}, Michael F. and {Correnti}, Matteo and {Cossou}, Christophe and {Costanza}, Brian F. and {Coulais}, Alain and {Cox}, Colin R. and {Coyle}, Ray T. and {Cracraft}, Misty M. and {Crew}, Keith A. and {Curtis}, Gary J. and {Cusveller}, Bianca and {Da Costa Maciel}, Cleyciane and {Dailey}, Christopher T. and {Daugeron}, Fr{\'e}d{\'e}ric and {Davidson}, Greg S. and {Davies}, James E. and {Davis}, Katherine Anne and {Davis}, Michael S. and {Day}, Ratna and {de Chambure}, Daniel and {de Jong}, Pauline and {De Marchi}, Guido and {Dean}, Bruce H. and {Decker}, John E. and {Delisa}, Amy S. and {Dell}, Lawrence C. and {Dellagatta}, Gail},
        title = "{The James Webb Space Telescope Mission}",
      journal = {\pasp},
         year = 2023,
        month = jun,
       volume = {135},
       number = {1048},
          eid = {068001},
        pages = {068001},
          doi = {10.1088/1538-3873/acd1b5},
archivePrefix = {arXiv},
       eprint = {2304.04869},
 primaryClass = {astro-ph.IM},
       adsurl = {https://ui.adsabs.harvard.edu/abs/2023PASP..135f8001G}
}

@ARTICLE{Alderson2023,
       author = {{Alderson}, Lili and {Wakeford}, Hannah R. and {Alam}, Munazza K. and {Batalha}, Natasha E. and {Lothringer}, Joshua D. and {Adams Redai}, Jea and {Barat}, Saugata and {Brande}, Jonathan and {Damiano}, Mario and {Daylan}, Tansu and {Espinoza}, N{\'e}stor and {Flagg}, Laura and {Goyal}, Jayesh M. and {Grant}, David and {Hu}, Renyu and {Inglis}, Julie and {Lee}, Elspeth K.~H. and {Mikal-Evans}, Thomas and {Ramos-Rosado}, Lakeisha and {Roy}, Pierre-Alexis and {Wallack}, Nicole L. and {Batalha}, Natalie M. and {Bean}, Jacob L. and {Benneke}, Bj{\"o}rn and {Berta-Thompson}, Zachory K. and {Carter}, Aarynn L. and {Changeat}, Quentin and {Col{\'o}n}, Knicole D. and {Crossfield}, Ian J.~M. and {D{\'e}sert}, Jean-Michel and {Foreman-Mackey}, Daniel and {Gibson}, Neale P. and {Kreidberg}, Laura and {Line}, Michael R. and {L{\'o}pez-Morales}, Mercedes and {Molaverdikhani}, Karan and {Moran}, Sarah E. and {Morello}, Giuseppe and {Moses}, Julianne I. and {Mukherjee}, Sagnick and {Schlawin}, Everett and {Sing}, David K. and {Stevenson}, Kevin B. and {Taylor}, Jake and {Aggarwal}, Keshav and {Ahrer}, Eva-Maria and {Allen}, Natalie H. and {Barstow}, Joanna K. and {Bell}, Taylor J. and {Blecic}, Jasmina and {Casewell}, Sarah L. and {Chubb}, Katy L. and {Crouzet}, Nicolas and {Cubillos}, Patricio E. and {Decin}, Leen and {Feinstein}, Adina D. and {Fortney}, Joanthan J. and {Harrington}, Joseph and {Heng}, Kevin and {Iro}, Nicolas and {Kempton}, Eliza M. -R. and {Kirk}, James and {Knutson}, Heather A. and {Krick}, Jessica and {Leconte}, J{\'e}r{\'e}my and {Lendl}, Monika and {MacDonald}, Ryan J. and {Mancini}, Luigi and {Mansfield}, Megan and {May}, Erin M. and {Mayne}, Nathan J. and {Miguel}, Yamila and {Nikolov}, Nikolay K. and {Ohno}, Kazumasa and {Palle}, Enric and {Parmentier}, Vivien and {Petit dit de la Roche}, Dominique J.~M. and {Piaulet}, Caroline and {Powell}, Diana and {Rackham}, Benjamin V. and {Redfield}, Seth and {Rogers}, Laura K. and {Rustamkulov}, Zafar and {Tan}, Xianyu and {Tremblin}, P. and {Tsai}, Shang-Min and {Turner}, Jake D. and {de Val-Borro}, Miguel and {Venot}, Olivia and {Welbanks}, Luis and {Wheatley}, Peter J. and {Zhang}, Xi},
        title = "{Early Release Science of the exoplanet WASP-39b with JWST NIRSpec G395H}",
      journal = {\nat},
         year = 2023,
        month = feb,
       volume = {614},
       number = {7949},
        pages = {664-669},
          doi = {10.1038/s41586-022-05591-3},
archivePrefix = {arXiv},
       eprint = {2211.10488},
 primaryClass = {astro-ph.EP},
       adsurl = {https://ui.adsabs.harvard.edu/abs/2023Natur.614..664A}
}

@ARTICLE{Majewski2017,
       author = {{Majewski}, Steven R. and {Schiavon}, Ricardo P. and {Frinchaboy}, Peter M. and {Allende Prieto}, Carlos and {Barkhouser}, Robert and {Bizyaev}, Dmitry and {Blank}, Basil and {Brunner}, Sophia and {Burton}, Adam and {Carrera}, Ricardo and {Chojnowski}, S. Drew and {Cunha}, K{\'a}tia and {Epstein}, Courtney and {Fitzgerald}, Greg and {Garc{\'\i}a P{\'e}rez}, Ana E. and {Hearty}, Fred R. and {Henderson}, Chuck and {Holtzman}, Jon A. and {Johnson}, Jennifer A. and {Lam}, Charles R. and {Lawler}, James E. and {Maseman}, Paul and {M{\'e}sz{\'a}ros}, Szabolcs and {Nelson}, Matthew and {Nguyen}, Duy Coung and {Nidever}, David L. and {Pinsonneault}, Marc and {Shetrone}, Matthew and {Smee}, Stephen and {Smith}, Verne V. and {Stolberg}, Todd and {Skrutskie}, Michael F. and {Walker}, Eric and {Wilson}, John C. and {Zasowski}, Gail and {Anders}, Friedrich and {Basu}, Sarbani and {Beland}, Stephane and {Blanton}, Michael R. and {Bovy}, Jo and {Brownstein}, Joel R. and {Carlberg}, Joleen and {Chaplin}, William and {Chiappini}, Cristina and {Eisenstein}, Daniel J. and {Elsworth}, Yvonne and {Feuillet}, Diane and {Fleming}, Scott W. and {Galbraith-Frew}, Jessica and {Garc{\'\i}a}, Rafael A. and {Garc{\'\i}a-Hern{\'a}ndez}, D. An{\'\i}bal and {Gillespie}, Bruce A. and {Girardi}, L{\'e}o and {Gunn}, James E. and {Hasselquist}, Sten and {Hayden}, Michael R. and {Hekker}, Saskia and {Ivans}, Inese and {Kinemuchi}, Karen and {Klaene}, Mark and {Mahadevan}, Suvrath and {Mathur}, Savita and {Mosser}, Beno{\^\i}t and {Muna}, Demitri and {Munn}, Jeffrey A. and {Nichol}, Robert C. and {O'Connell}, Robert W. and {Parejko}, John K. and {Robin}, A.~C. and {Rocha-Pinto}, Helio and {Schultheis}, Matthias and {Serenelli}, Aldo M. and {Shane}, Neville and {Silva Aguirre}, Victor and {Sobeck}, Jennifer S. and {Thompson}, Benjamin and {Troup}, Nicholas W. and {Weinberg}, David H. and {Zamora}, Olga},
        title = "{The Apache Point Observatory Galactic Evolution Experiment (APOGEE)}",
      journal = {\aj},
         year = 2017,
        month = sep,
       volume = {154},
       number = {3},
          eid = {94},
        pages = {94},
          doi = {10.3847/1538-3881/aa784d},
archivePrefix = {arXiv},
       eprint = {1509.05420},
 primaryClass = {astro-ph.IM},
       adsurl = {https://ui.adsabs.harvard.edu/abs/2017AJ....154...94M}
}

@ARTICLE{Schneider2021a,
       author = {{Schneider}, Aaron David and {Bitsch}, Bertram},
        title = "{How drifting and evaporating pebbles shape giant planets. I. Heavy element content and atmospheric C/O}",
      journal = {\aap},
         year = {2021a},
        month = oct,
       volume = {654},
          eid = {A71},
        pages = {A71},
          doi = {10.1051/0004-6361/202039640},
archivePrefix = {arXiv},
       eprint = {2105.13267},
 primaryClass = {astro-ph.EP},
       adsurl = {https://ui.adsabs.harvard.edu/abs/2021A&A...654A..71S}
}

@ARTICLE{Schneider2021b,
       author = {{Schneider}, Aaron David and {Bitsch}, Bertram},
        title = "{How drifting and evaporating pebbles shape giant planets. II. Volatiles and refractories in atmospheres}",
      journal = {\aap},
         year = {2021b},
        month = oct,
       volume = {654},
          eid = {A72},
        pages = {A72},
          doi = {10.1051/0004-6361/202141096},
archivePrefix = {arXiv},
       eprint = {2109.03589},
 primaryClass = {astro-ph.EP},
       adsurl = {https://ui.adsabs.harvard.edu/abs/2021A&A...654A..72S}
}

@ARTICLE{Oka2011,
       author = {{Oka}, Akinori and {Nakamoto}, Taishi and {Ida}, Shigeru},
        title = "{Evolution of Snow Line in Optically Thick Protoplanetary Disks: Effects of Water Ice Opacity and Dust Grain Size}",
      journal = {\apj},
         year = 2011,
        month = sep,
       volume = {738},
       number = {2},
          eid = {141},
        pages = {141},
          doi = {10.1088/0004-637X/738/2/141},
archivePrefix = {arXiv},
       eprint = {1106.2682},
 primaryClass = {astro-ph.EP},
       adsurl = {https://ui.adsabs.harvard.edu/abs/2011ApJ...738..141O}
}

@MISC{Schneider2022,
       author = {{Schneider}, Aaron David and {Bitsch}, Bertram},
        title = "{How drifting and evaporating pebbles shape giant planets (Corrigendum)}",
 howpublished = {Astronomy \& Astrophysics, Volume 659, id.C3, 3 pp.},
         year = 2022,
        month = mar,
          eid = {C3},
        pages = {C3},
          doi = {10.1051/0004-6361/202141096e},
       adsurl = {https://ui.adsabs.harvard.edu/abs/2022A&A...659C...3S}
}

@ARTICLE{Pacetti2022,
       author = {{Pacetti}, Elenia and {Turrini}, Diego and {Schisano}, Eugenio and {Molinari}, Sergio and {Fonte}, Sergio and {Politi}, Romolo and {Hennebelle}, Patrick and {Klessen}, Ralf and {Testi}, Leonardo and {Lebreuilly}, Ugo},
        title = "{Chemical Diversity in Protoplanetary Disks and Its Impact on the Formation History of Giant Planets}",
      journal = {\apj},
         year = 2022,
        month = sep,
       volume = {937},
       number = {1},
          eid = {36},
        pages = {36},
          doi = {10.3847/1538-4357/ac8b11},
archivePrefix = {arXiv},
       eprint = {2206.14685},
 primaryClass = {astro-ph.EP},
       adsurl = {https://ui.adsabs.harvard.edu/abs/2022ApJ...937...36P}
}

@ARTICLE{Moses2013,
       author = {{Moses}, J.~I. and {Madhusudhan}, N. and {Visscher}, C. and {Freedman}, R.~S.},
        title = "{Chemical Consequences of the C/O Ratio on Hot Jupiters: Examples from WASP-12b, CoRoT-2b, XO-1b, and HD 189733b}",
      journal = {\apj},
         year = 2013,
        month = jan,
       volume = {763},
       number = {1},
          eid = {25},
        pages = {25},
          doi = {10.1088/0004-637X/763/1/25},
archivePrefix = {arXiv},
       eprint = {1211.2996},
 primaryClass = {astro-ph.EP},
       adsurl = {https://ui.adsabs.harvard.edu/abs/2013ApJ...763...25M}
}

@ARTICLE{Konopacky2013,
       author = {{Konopacky}, Quinn M. and {Barman}, Travis S. and {Macintosh}, Bruce A. and {Marois}, Christian},
        title = "{Detection of Carbon Monoxide and Water Absorption Lines in an Exoplanet Atmosphere}",
      journal = {Science},
         year = 2013,
        month = mar,
       volume = {339},
       number = {6126},
        pages = {1398-1401},
          doi = {10.1126/science.1232003},
archivePrefix = {arXiv},
       eprint = {1303.3280},
 primaryClass = {astro-ph.EP},
       adsurl = {https://ui.adsabs.harvard.edu/abs/2013Sci...339.1398K}
}

@ARTICLE{Thiabaud2015a,
       author = {{Thiabaud}, A. and {Marboeuf}, U. and {Alibert}, Y. and {Leya}, I. and {Mezger}, K.},
        title = "{Gas composition of the main volatile elements in protoplanetary discs and its implication for planet formation}",
      journal = {\aap},
         year = {2015a},
        month = feb,
       volume = {574},
          eid = {A138},
        pages = {A138},
          doi = {10.1051/0004-6361/201424868},
       adsurl = {https://ui.adsabs.harvard.edu/abs/2015A&A...574A.138T}
}

@ARTICLE{Seager2005,
       author = {{Seager}, S. and {Richardson}, L.~J. and {Hansen}, B.~M.~S. and {Menou}, K. and {Cho}, J.~Y. -K. and {Deming}, D.},
        title = "{On the Dayside Thermal Emission of Hot Jupiters}",
      journal = {\apj},
         year = 2005,
        month = oct,
       volume = {632},
       number = {2},
        pages = {1122-1131},
          doi = {10.1086/444411},
archivePrefix = {arXiv},
       eprint = {astro-ph/0504212},
 primaryClass = {astro-ph},
       adsurl = {https://ui.adsabs.harvard.edu/abs/2005ApJ...632.1122S}
}

@ARTICLE{Hartman2011,
       author = {{Hartman}, J.~D. and {Bakos}, G. {\'A}. and {Kipping}, D.~M. and {Torres}, G. and {Kov{\'a}cs}, G. and {Noyes}, R.~W. and {Latham}, D.~W. and {Howard}, A.~W. and {Fischer}, D.~A. and {Johnson}, J.~A. and {Marcy}, G.~W. and {Isaacson}, H. and {Quinn}, S.~N. and {Buchhave}, L.~A. and {B{\'e}ky}, B. and {Sasselov}, D.~D. and {Stefanik}, R.~P. and {Esquerdo}, G.~A. and {Everett}, M. and {Perumpilly}, G. and {L{\'a}z{\'a}r}, J. and {Papp}, I. and {S{\'a}ri}, P.},
        title = "{HAT-P-26b: A Low-density Neptune-mass Planet Transiting a K Star}",
      journal = {\apj},
         year = 2011,
        month = feb,
       volume = {728},
       number = {2},
          eid = {138},
        pages = {138},
          doi = {10.1088/0004-637X/728/2/138},
archivePrefix = {arXiv},
       eprint = {1010.1008},
 primaryClass = {astro-ph.EP},
       adsurl = {https://ui.adsabs.harvard.edu/abs/2011ApJ...728..138H}
}

@ARTICLE{Gressier2025,
       author = {{Gressier}, Am{\'e}lie and {Batalha}, Natasha E. and {Wogan}, Nicholas and {Alderson}, Lili and {Doud}, Dominic and {Espinoza}, N{\'e}stor and {MacDonald}, Ryan J. and {Wakeford}, Hannah R. and {Valenti}, Jeff A. and {Lewis}, Nikole K. and {Seager}, Sara and {Stevenson}, Kevin B. and {Allen}, Natalie H. and {Ca{\~n}as}, Caleb I. and {Challener}, Ryan C. and {Glidden}, Ana and {Huang}, Jingcheng and {Lin}, Zifan and {Louie}, Dana R. and {Maguire}, Cathal and {Mullens}, Elijah and {Sotzen}, Kristin and {Valentine}, Daniel and {Clampin}, Mark and {Pueyo}, Laurent and {van der Marel}, Roeland P. and {Mountain}, C. Matt},
        title = "{JWST-TST DREAMS: Sulfur Dioxide in the Atmosphere of the Neptune-mass Planet HAT-P-26 b from NIRSpec G395H Transmission Spectroscopy}",
      journal = {\aj},
         year = 2025,
        month = nov,
       volume = {170},
       number = {5},
          eid = {292},
        pages = {292},
          doi = {10.3847/1538-3881/ae0929},
archivePrefix = {arXiv},
       eprint = {2509.16082},
 primaryClass = {astro-ph.EP},
       adsurl = {https://ui.adsabs.harvard.edu/abs/2025AJ....170..292G}
}

@ARTICLE{Turrini2021,
       author = {{Turrini}, D. and {Schisano}, E. and {Fonte}, S. and {Molinari}, S. and {Politi}, R. and {Fedele}, D. and {Pani{\'c}}, O. and {Kama}, M. and {Changeat}, Q. and {Tinetti}, G.},
        title = "{Tracing the Formation History of Giant Planets in Protoplanetary Disks with Carbon, Oxygen, Nitrogen, and Sulfur}",
      journal = {\apj},
         year = 2021,
        month = mar,
       volume = {909},
       number = {1},
          eid = {40},
        pages = {40},
          doi = {10.3847/1538-4357/abd6e5},
archivePrefix = {arXiv},
       eprint = {2012.14315},
 primaryClass = {astro-ph.EP},
       adsurl = {https://ui.adsabs.harvard.edu/abs/2021ApJ...909...40T}
}

@ARTICLE{Polman2023,
       author = {{Polman}, J. and {Waters}, L.~B.~F.~M. and {Min}, M. and {Miguel}, Y. and {Khorshid}, N.},
        title = "{H$_{2}$S and SO$_{2}$ detectability in hot Jupiters. Sulphur species as indicators of metallicity and C/O ratio}",
      journal = {\aap},
         year = 2023,
        month = feb,
       volume = {670},
          eid = {A161},
        pages = {A161},
          doi = {10.1051/0004-6361/202244647},
archivePrefix = {arXiv},
       eprint = {2208.00469},
 primaryClass = {astro-ph.EP},
       adsurl = {https://ui.adsabs.harvard.edu/abs/2023A&A...670A.161P}
}

@ARTICLE{Lodders2004,
       author = {{Lodders}, Katharina},
        title = "{Jupiter Formed with More Tar than Ice}",
      journal = {\apj},
         year = 2004,
        month = aug,
       volume = {611},
       number = {1},
        pages = {587-597},
          doi = {10.1086/421970},
       adsurl = {https://ui.adsabs.harvard.edu/abs/2004ApJ...611..587L}
}

@ARTICLE{Teske2014,
       author = {{Teske}, Johanna K. and {Cunha}, Katia and {Smith}, Verne V. and {Schuler}, Simon C. and {Griffith}, Caitlin A.},
        title = "{C/O Ratios of Stars with Transiting Hot Jupiter Exoplanets}",
      journal = {\apj},
         year = 2014,
        month = jun,
       volume = {788},
       number = {1},
          eid = {39},
        pages = {39},
          doi = {10.1088/0004-637X/788/1/39},
archivePrefix = {arXiv},
       eprint = {1403.6891},
 primaryClass = {astro-ph.SR},
       adsurl = {https://ui.adsabs.harvard.edu/abs/2014ApJ...788...39T}
}

@ARTICLE{Nissen2013,
       author = {{Nissen}, P.~E.},
        title = "{The carbon-to-oxygen ratio in stars with planets}",
      journal = {\aap},
         year = 2013,
        month = apr,
       volume = {552},
          eid = {A73},
        pages = {A73},
          doi = {10.1051/0004-6361/201321234},
archivePrefix = {arXiv},
       eprint = {1303.1726},
 primaryClass = {astro-ph.SR},
       adsurl = {https://ui.adsabs.harvard.edu/abs/2013A&A...552A..73N}
}

@ARTICLE{Pollack1996,
       author = {{Pollack}, James B. and {Hubickyj}, Olenka and {Bodenheimer}, Peter and {Lissauer}, Jack J. and {Podolak}, Morris and {Greenzweig}, Yuval},
        title = "{Formation of the Giant Planets by Concurrent Accretion of Solids and Gas}",
      journal = {\icarus},
         year = 1996,
        month = nov,
       volume = {124},
       number = {1},
        pages = {62-85},
          doi = {10.1006/icar.1996.0190},
       adsurl = {https://ui.adsabs.harvard.edu/abs/1996Icar..124...62P}
}

@ARTICLE{Ohno2023,
       author = {{Ohno}, Kazumasa and {Fortney}, Jonathan J.},
        title = "{Nitrogen as a Tracer of Giant Planet Formation. II. Comprehensive Study of Nitrogen Photochemistry and Implications for Observing NH$_{3}$ and HCN in Transmission and Emission Spectra}",
      journal = {\apj},
         year = 2023,
        month = oct,
       volume = {956},
       number = {2},
          eid = {125},
        pages = {125},
          doi = {10.3847/1538-4357/ace531},
archivePrefix = {arXiv},
       eprint = {2211.16877},
 primaryClass = {astro-ph.EP},
       adsurl = {https://ui.adsabs.harvard.edu/abs/2023ApJ...956..125O}
}

@ARTICLE{Noguer2024,
       author = {{Noguer}, Federico R. and {Corley}, Suber and {Pearson}, Kyle A. and {Zellem}, Robert T. and {Simon}, Molly N. and {Burt}, Jennifer A. and {Huckabee}, Isabela and {August}, Prune C. and {Weiner Mansfield}, Megan and {Dalba}, Paul A. and {Smith}, Peter C.~B. and {Banks}, Timothy and {Bell}, Ira and {Daniel}, Dominique and {Dawson}, Lindsay and {De Mula}, Jes{\'u}s and {Deldem}, Marc and {Deligeorgopoulos}, Dimitrios and {Di Sisto}, Romina P. and {Dymock}, Roger and {Evans}, Phil and {Follero}, Giulio and {Fowler}, Martin J.~F. and {Fern{\'a}ndez-Laj{\'u}s}, Eduardo and {Hamrick}, Alex and {Iannascoli}, Nicoletta and {Kovacs}, Andre O. and {Kulh}, Denis Henrique and {Lopresti}, Claudio and {Marino}, Antonio and {Martin}, Bryan E. and {Matassa}, Paolo Arcangelo and {Napole{\~a}o}, Tasso Augusto and {Nastasi}, Alessandro and {Norris}, Anthony and {Odasso}, Alessandro and {Paschalis}, Nikolaos I. and {Pintr}, Pavel and {Postiglione}, Jake and {Randolph}, Justus and {Regembal}, Fran{\c{c}}ois and {Rousselot}, Lionel and {Gon{\c{c}}alves da Silva}, Sergio Jos{\'e} and {Smith}, Andrew and {Tomacelli}, Andrea},
        title = "{Enhancing Exoplanet Ephemerides by Leveraging Professional and Citizen Science Data: A Test Case with WASP-77 A b}",
      journal = {\pasp},
         year = 2024,
        month = jun,
       volume = {136},
       number = {6},
          eid = {064401},
        pages = {064401},
          doi = {10.1088/1538-3873/ad57f5},
archivePrefix = {arXiv},
       eprint = {2405.19615},
 primaryClass = {astro-ph.EP},
       adsurl = {https://ui.adsabs.harvard.edu/abs/2024PASP..136f4401N}
}

@ARTICLE{Ehrenreich2020,
       author = {{Ehrenreich}, David and {Lovis}, Christophe and {Allart}, Romain and {Zapatero Osorio}, Mar{\'\i}a Rosa and {Pepe}, Francesco and {Cristiani}, Stefano and {Rebolo}, Rafael and {Santos}, Nuno C. and {Borsa}, Francesco and {Demangeon}, Olivier and {Dumusque}, Xavier and {Gonz{\'a}lez Hern{\'a}ndez}, Jonay I. and {Casasayas-Barris}, N{\'u}ria and {S{\'e}gransan}, Damien and {Sousa}, S{\'e}rgio and {Abreu}, Manuel and {Adibekyan}, Vardan and {Affolter}, Michael and {Allende Prieto}, Carlos and {Alibert}, Yann and {Aliverti}, Matteo and {Alves}, David and {Amate}, Manuel and {Avila}, Gerardo and {Baldini}, Veronica and {Bandy}, Timothy and {Benz}, Willy and {Bianco}, Andrea and {Bolmont}, {\'E}meline and {Bouchy}, Fran{\c{c}}ois and {Bourrier}, Vincent and {Broeg}, Christopher and {Cabral}, Alexandre and {Calderone}, Giorgio and {Pall{\'e}}, Enric and {Cegla}, H.~M. and {Cirami}, Roberto and {Coelho}, Jo{\~a}o M.~P. and {Conconi}, Paolo and {Coretti}, Igor and {Cumani}, Claudio and {Cupani}, Guido and {Dekker}, Hans and {Delabre}, Bernard and {Deiries}, Sebastian and {D'Odorico}, Valentina and {Di Marcantonio}, Paolo and {Figueira}, Pedro and {Fragoso}, Ana and {Genolet}, Ludovic and {Genoni}, Matteo and {G{\'e}nova Santos}, Ricardo and {Hara}, Nathan and {Hughes}, Ian and {Iwert}, Olaf and {Kerber}, Florian and {Knudstrup}, Jens and {Landoni}, Marco and {Lavie}, Baptiste and {Lizon}, Jean-Louis and {Lendl}, Monika and {Lo Curto}, Gaspare and {Maire}, Charles and {Manescau}, Antonio and {Martins}, C.~J.~A.~P. and {M{\'e}gevand}, Denis and {Mehner}, Andrea and {Micela}, Giusi and {Modigliani}, Andrea and {Molaro}, Paolo and {Monteiro}, Manuel and {Monteiro}, Mario and {Moschetti}, Manuele and {M{\"u}ller}, Eric and {Nunes}, Nelson and {Oggioni}, Luca and {Oliveira}, Ant{\'o}nio and {Pariani}, Giorgio and {Pasquini}, Luca and {Poretti}, Ennio and {Rasilla}, Jos{\'e} Luis and {Redaelli}, Edoardo and {Riva}, Marco and {Santana Tschudi}, Samuel and {Santin}, Paolo and {Santos}, Pedro and {Segovia Milla}, Alex and {Seidel}, Julia V. and {Sosnowska}, Danuta and {Sozzetti}, Alessandro and {Span{\`o}}, Paolo and {Su{\'a}rez Mascare{\~n}o}, Alejandro and {Tabernero}, Hugo and {Tenegi}, Fabio and {Udry}, St{\'e}phane and {Zanutta}, Alessio and {Zerbi}, Filippo},
        title = "{Nightside condensation of iron in an ultrahot giant exoplanet}",
      journal = {\nat},
         year = 2020,
        month = apr,
       volume = {580},
       number = {7805},
        pages = {597-601},
          doi = {10.1038/s41586-020-2107-1},
archivePrefix = {arXiv},
       eprint = {2003.05528},
 primaryClass = {astro-ph.EP},
       adsurl = {https://ui.adsabs.harvard.edu/abs/2020Natur.580..597E}
}

@ARTICLE{Corte2020,
       author = {{Cort{\'e}s-Zuleta}, P{\'\i}a and {Rojo}, Patricio and {Wang}, Songhu and {Hinse}, Tobias C. and {Hoyer}, Sergio and {Sanhueza}, Bastian and {Correa-Amaro}, Patricio and {Albornoz}, Julio},
        title = "{TraMoS. V. Updated ephemeris and multi-epoch monitoring of the hot Jupiters WASP-18Ab, WASP-19b, and WASP-77Ab}",
      journal = {\aap},
         year = 2020,
        month = apr,
       volume = {636},
          eid = {A98},
        pages = {A98},
          doi = {10.1051/0004-6361/201936279},
archivePrefix = {arXiv},
       eprint = {2001.11112},
 primaryClass = {astro-ph.EP},
       adsurl = {https://ui.adsabs.harvard.edu/abs/2020A&A...636A..98C}
}

@ARTICLE{Gillon2014,
       author = {{Gillon}, M. and {Anderson}, D.~R. and {Collier-Cameron}, A. and {Delrez}, L. and {Hellier}, C. and {Jehin}, E. and {Lendl}, M. and {Maxted}, P.~F.~L. and {Pepe}, F. and {Pollacco}, D. and {Queloz}, D. and {S{\'e}gransan}, D. and {Smith}, A.~M.~S. and {Smalley}, B. and {Southworth}, J. and {Triaud}, A.~H.~M.~J. and {Udry}, S. and {Van Grootel}, V. and {West}, R.~G.},
        title = "{WASP-103 b: a new planet at the edge of tidal disruption}",
      journal = {\aap},
         year = 2014,
        month = feb,
       volume = {562},
          eid = {L3},
        pages = {L3},
          doi = {10.1051/0004-6361/201323014},
archivePrefix = {arXiv},
       eprint = {1401.2784},
 primaryClass = {astro-ph.EP},
       adsurl = {https://ui.adsabs.harvard.edu/abs/2014A&A...562L...3G}
}

@ARTICLE{Crossfield2023,
       author = {{Crossfield}, Ian J.~M.},
        title = "{Volatile-to-sulfur Ratios Can Recover a Gas Giant's Accretion History}",
      journal = {\apjl},
         year = 2023,
        month = jul,
       volume = {952},
       number = {1},
          eid = {L18},
        pages = {L18},
          doi = {10.3847/2041-8213/ace35f},
archivePrefix = {arXiv},
       eprint = {2303.17622},
 primaryClass = {astro-ph.EP},
       adsurl = {https://ui.adsabs.harvard.edu/abs/2023ApJ...952L..18C}
}

@ARTICLE{Kama2019,
       author = {{Kama}, Mihkel and {Shorttle}, Oliver and {Jermyn}, Adam S. and {Folsom}, Colin P. and {Furuya}, Kenji and {Bergin}, Edwin A. and {Walsh}, Catherine and {Keller}, Lindsay},
        title = "{Abundant Refractory Sulfur in Protoplanetary Disks}",
      journal = {\apj},
         year = 2019,
        month = nov,
       volume = {885},
       number = {2},
          eid = {114},
        pages = {114},
          doi = {10.3847/1538-4357/ab45f8},
archivePrefix = {arXiv},
       eprint = {1908.05169},
 primaryClass = {astro-ph.EP},
       adsurl = {https://ui.adsabs.harvard.edu/abs/2019ApJ...885..114K}
}

@ARTICLE{Madhusudhan2012,
       author = {{Madhusudhan}, Nikku},
        title = "{C/O Ratio as a Dimension for Characterizing Exoplanetary Atmospheres}",
      journal = {\apj},
         year = 2012,
        month = oct,
       volume = {758},
       number = {1},
          eid = {36},
        pages = {36},
          doi = {10.1088/0004-637X/758/1/36},
archivePrefix = {arXiv},
       eprint = {1209.2412},
 primaryClass = {astro-ph.EP},
       adsurl = {https://ui.adsabs.harvard.edu/abs/2012ApJ...758...36M}
}

@ARTICLE{Madhusudhan2019,
       author = {{Madhusudhan}, Nikku},
        title = "{Exoplanetary Atmospheres: Key Insights, Challenges, and Prospects}",
      journal = {\araa},
         year = 2019,
        month = aug,
       volume = {57},
        pages = {617-663},
          doi = {10.1146/annurev-astro-081817-051846},
archivePrefix = {arXiv},
       eprint = {1904.03190},
 primaryClass = {astro-ph.EP},
       adsurl = {https://ui.adsabs.harvard.edu/abs/2019ARA&A..57..617M}
}

@ARTICLE{Hoch2023,
       author = {{Hoch}, Kielan K.~W. and {Konopacky}, Quinn M. and {Theissen}, Christopher A. and {Ruffio}, Jean-Baptiste and {Barman}, Travis S. and {Rickman}, Emily L. and {Perrin}, Marshall D. and {Macintosh}, Bruce and {Marois}, Christian},
        title = "{Assessing the C/O Ratio Formation Diagnostic: A Potential Trend with Companion Mass}",
      journal = {\aj},
         year = 2023,
        month = sep,
       volume = {166},
       number = {3},
          eid = {85},
        pages = {85},
          doi = {10.3847/1538-3881/ace442},
archivePrefix = {arXiv},
       eprint = {2212.04557},
 primaryClass = {astro-ph.EP},
       adsurl = {https://ui.adsabs.harvard.edu/abs/2023AJ....166...85H}
}

@ARTICLE{Petigura2017,
       author = {{Petigura}, Erik A. and {Howard}, Andrew W. and {Marcy}, Geoffrey W. and {Johnson}, John Asher and {Isaacson}, Howard and {Cargile}, Phillip A. and {Hebb}, Leslie and {Fulton}, Benjamin J. and {Weiss}, Lauren M. and {Morton}, Timothy D. and {Winn}, Joshua N. and {Rogers}, Leslie A. and {Sinukoff}, Evan and {Hirsch}, Lea A. and {Crossfield}, Ian J.~M.},
        title = "{The California-Kepler Survey. I. High-resolution Spectroscopy of 1305 Stars Hosting Kepler Transiting Planets}",
      journal = {\aj},
         year = 2017,
        month = sep,
       volume = {154},
       number = {3},
          eid = {107},
        pages = {107},
          doi = {10.3847/1538-3881/aa80de},
archivePrefix = {arXiv},
       eprint = {1703.10400},
 primaryClass = {astro-ph.EP},
       adsurl = {https://ui.adsabs.harvard.edu/abs/2017AJ....154..107P}
}

@ARTICLE{Lodders2025,
       author = {{Lodders}, K. and {Bergemann}, M. and {Palme}, H.},
        title = "{Solar System Elemental Abundances from the Solar Photosphere and CI-Chondrites}",
      journal = {\ssr},
         year = 2025,
        month = mar,
       volume = {221},
       number = {2},
          eid = {23},
        pages = {23},
          doi = {10.1007/s11214-025-01146-w},
archivePrefix = {arXiv},
       eprint = {2502.10575},
 primaryClass = {astro-ph.SR},
       adsurl = {https://ui.adsabs.harvard.edu/abs/2025SSRv..221...23L}
}

@ARTICLE{Baburaj2025,
       author = {{Baburaj}, Aneesh and {Konopacky}, Quinn M. and {Theissen}, Christopher A. and {Peacock}, Sarah and {Huseby}, Lori and {Fulton}, Benjamin J. and {Gerasimov}, Roman and {Barman}, Travis S. and {Hoch}, Kielan K.~W.},
        title = "{A High-resolution Spectroscopic Survey of Directly Imaged Companion Hosts. I. Determination of Diagnostic Stellar Abundances for Planet Formation and Composition}",
      journal = {\aj},
         year = 2025,
        month = feb,
       volume = {169},
       number = {2},
          eid = {55},
        pages = {55},
          doi = {10.3847/1538-3881/ad8dfc},
archivePrefix = {arXiv},
       eprint = {2409.14239},
 primaryClass = {astro-ph.EP},
       adsurl = {https://ui.adsabs.harvard.edu/abs/2025AJ....169...55B}
}

@ARTICLE{Bonnefoy2018,
       author = {{Bonnefoy}, M. and {Perraut}, K. and {Lagrange}, A. -M. and {Delorme}, P. and {Vigan}, A. and {Line}, M. and {Rodet}, L. and {Ginski}, C. and {Mourard}, D. and {Marleau}, G. -D. and {Samland}, M. and {Tremblin}, P. and {Ligi}, R. and {Cantalloube}, F. and {Molli{\`e}re}, P. and {Charnay}, B. and {Kuzuhara}, M. and {Janson}, M. and {Morley}, C. and {Homeier}, D. and {D'Orazi}, V. and {Klahr}, H. and {Mordasini}, C. and {Lavie}, B. and {Baudino}, J. -L. and {Beust}, H. and {Peretti}, S. and {Musso Bartucci}, A. and {Mesa}, D. and {B{\'e}zard}, B. and {Boccaletti}, A. and {Galicher}, R. and {Hagelberg}, J. and {Desidera}, S. and {Biller}, B. and {Maire}, A. -L. and {Allard}, F. and {Borgniet}, S. and {Lannier}, J. and {Meunier}, N. and {Desort}, M. and {Alecian}, E. and {Chauvin}, G. and {Langlois}, M. and {Henning}, T. and {Mugnier}, L. and {Mouillet}, D. and {Gratton}, R. and {Brandt}, T. and {Mc Elwain}, M. and {Beuzit}, J. -L. and {Tamura}, M. and {Hori}, Y. and {Brandner}, W. and {Buenzli}, E. and {Cheetham}, A. and {Cudel}, M. and {Feldt}, M. and {Kasper}, M. and {Keppler}, M. and {Kopytova}, T. and {Meyer}, M. and {Perrot}, C. and {Rouan}, D. and {Salter}, G. and {Schmidt}, T. and {Sissa}, E. and {Zurlo}, A. and {Wildi}, F. and {Blanchard}, P. and {De Caprio}, V. and {Delboulb{\'e}}, A. and {Maurel}, D. and {Moulin}, T. and {Pavlov}, A. and {Rabou}, P. and {Ramos}, J. and {Roelfsema}, R. and {Rousset}, G. and {Stadler}, E. and {Rigal}, F. and {Weber}, L.},
        title = "{The GJ 504 system revisited. Combining interferometric, radial velocity, and high contrast imaging data}",
      journal = {\aap},
         year = 2018,
        month = oct,
       volume = {618},
          eid = {A63},
        pages = {A63},
          doi = {10.1051/0004-6361/201832942},
archivePrefix = {arXiv},
       eprint = {1807.00657},
 primaryClass = {astro-ph.EP},
       adsurl = {https://ui.adsabs.harvard.edu/abs/2018A&A...618A..63B}
}

@ARTICLE{Santos2000,
       author = {{Santos}, N.~C. and {Israelian}, G. and {Mayor}, M.},
        title = "{Chemical analysis of 8 recently discovered extra-solar planet host stars}",
      journal = {\aap},
         year = 2000,
        month = nov,
       volume = {363},
        pages = {228-238},
          doi = {10.48550/arXiv.astro-ph/0009182},
archivePrefix = {arXiv},
       eprint = {astro-ph/0009182},
 primaryClass = {astro-ph},
       adsurl = {https://ui.adsabs.harvard.edu/abs/2000A&A...363..228S}
}

@ARTICLE{Sharma2024,
       author = {{Sharma}, A. and {Stonkut{\.{e}}}, E. and {Drazdauska}, A. and {Minkevi{\v{c}}i{\={u}}t{\.{e}}}, R. and {Mikolaitis}, {\v{S}}. and {Tautvai{\v{s}}ien{\.{e}}}, G. and {Narbuntas}, T.},
        title = "{Chemical composition of planetary hosts: C, N, and {\ensuremath{\alpha}}-element abundances}",
      journal = {\aap},
         year = 2024,
        month = nov,
       volume = {691},
          eid = {A160},
        pages = {A160},
          doi = {10.1051/0004-6361/202451889},
archivePrefix = {arXiv},
       eprint = {2410.07100},
 primaryClass = {astro-ph.SR},
       adsurl = {https://ui.adsabs.harvard.edu/abs/2024A&A...691A.160S}
}

@ARTICLE{Sua2018,
       author = {{Su{\'a}rez-Andr{\'e}s}, L. and {Israelian}, G. and {Gonz{\'a}lez Hern{\'a}ndez}, J.~I. and {Adibekyan}, V. Zh. and {Delgado Mena}, E. and {Santos}, N.~C. and {Sousa}, S.~G.},
        title = "{C/O vs. Mg/Si ratios in solar type stars: The HARPS sample}",
      journal = {\aap},
         year = 2018,
        month = jun,
       volume = {614},
          eid = {A84},
        pages = {A84},
          doi = {10.1051/0004-6361/201730743},
archivePrefix = {arXiv},
       eprint = {1801.09474},
 primaryClass = {astro-ph.EP},
       adsurl = {https://ui.adsabs.harvard.edu/abs/2018A&A...614A..84S}
}

@ARTICLE{Changeat2022,
       author = {{Changeat}, Q. and {Edwards}, B. and {Al-Refaie}, A.~F. and {Tsiaras}, A. and {Skinner}, J.~W. and {Cho}, J.~Y.~K. and {Yip}, K.~H. and {Anisman}, L. and {Ikoma}, M. and {Bieger}, M.~F. and {Venot}, O. and {Shibata}, S. and {Waldmann}, I.~P. and {Tinetti}, G.},
        title = "{Five Key Exoplanet Questions Answered via the Analysis of 25 Hot-Jupiter Atmospheres in Eclipse}",
      journal = {\apjs},
         year = 2022,
        month = may,
       volume = {260},
       number = {1},
          eid = {3},
        pages = {3},
          doi = {10.3847/1538-4365/ac5cc2},
archivePrefix = {arXiv},
       eprint = {2204.11729},
 primaryClass = {astro-ph.EP},
       adsurl = {https://ui.adsabs.harvard.edu/abs/2022ApJS..260....3C}
}

@ARTICLE{WeinerMansfield2024,
       author = {{Weiner Mansfield}, Megan and {Line}, Michael R. and {Wardenier}, Joost P. and {Brogi}, Matteo and {Bean}, Jacob L. and {Beltz}, Hayley and {Smith}, Peter and {Zalesky}, Joseph A. and {Batalha}, Natasha and {Kempton}, Eliza M.-R. and {Montet}, Benjamin T. and {Owen}, James E. and {Plavchan}, Peter and {Rauscher}, Emily},
        title = "{The Metallicity and Carbon-to-oxygen Ratio of the Ultrahot Jupiter WASP-76b from Gemini-S/IGRINS}",
      journal = {\aj},
         year = 2024,
        month = jul,
       volume = {168},
       number = {1},
          eid = {14},
        pages = {14},
          doi = {10.3847/1538-3881/ad4a5f},
archivePrefix = {arXiv},
       eprint = {2405.09769},
 primaryClass = {astro-ph.EP},
       adsurl = {https://ui.adsabs.harvard.edu/abs/2024AJ....168...14W}
}

@ARTICLE{Xue2024,
       author = {{Xue}, Qiao and {Bean}, Jacob L. and {Zhang}, Michael and {Welbanks}, Luis and {Lunine}, Jonathan and {August}, Prune},
        title = "{JWST Transmission Spectroscopy of HD 209458b: A Supersolar Metallicity, a Very Low C/O, and No Evidence of CH$_{4}$, HCN, or C$_{2}$H$_{2}$}",
      journal = {\apjl},
         year = 2024,
        month = mar,
       volume = {963},
       number = {1},
          eid = {L5},
        pages = {L5},
          doi = {10.3847/2041-8213/ad2682},
archivePrefix = {arXiv},
       eprint = {2310.03245},
 primaryClass = {astro-ph.EP},
       adsurl = {https://ui.adsabs.harvard.edu/abs/2024ApJ...963L...5X}
}

@ARTICLE{Bean2023,
       author = {{Bean}, Jacob L. and {Xue}, Qiao and {August}, Prune C. and {Lunine}, Jonathan and {Zhang}, Michael and {Thorngren}, Daniel and {Tsai}, Shang-Min and {Stassun}, Keivan G. and {Schlawin}, Everett and {Ahrer}, Eva-Maria and {Ih}, Jegug and {Mansfield}, Megan},
        title = "{High atmospheric metal enrichment for a Saturn-mass planet}",
      journal = {\nat},
         year = 2023,
        month = jun,
       volume = {618},
       number = {7963},
        pages = {43-46},
          doi = {10.1038/s41586-023-05984-y},
archivePrefix = {arXiv},
       eprint = {2303.14206},
 primaryClass = {astro-ph.EP},
       adsurl = {https://ui.adsabs.harvard.edu/abs/2023Natur.618...43B}
}

@ARTICLE{Wenger2000,
       author = {{Wenger}, M. and {Ochsenbein}, F. and {Egret}, D. and {Dubois}, P. and {Bonnarel}, F. and {Borde}, S. and {Genova}, F. and {Jasniewicz}, G. and {Lalo{\"e}}, S. and {Lesteven}, S. and {Monier}, R.},
        title = "{The SIMBAD astronomical database. The CDS reference database for astronomical objects}",
      journal = {\aaps},
         year = 2000,
        month = apr,
       volume = {143},
        pages = {9-22},
          doi = {10.1051/aas:2000332},
archivePrefix = {arXiv},
       eprint = {astro-ph/0002110},
 primaryClass = {astro-ph},
       adsurl = {https://ui.adsabs.harvard.edu/abs/2000A&AS..143....9W}
}

@ARTICLE{Maciejewski2023,
       author = {{Maciejewski}, G. and {Golonka}, J. and {{\L}oboda}, W. and {Ohlert}, J. and {Fern{\'a}ndez}, M. and {Aceituno}, F.},
        title = "{A hot super-Earth planet in the WASP-84 planetary system}",
      journal = {\mnras},
         year = 2023,
        month = oct,
       volume = {525},
       number = {1},
        pages = {L43-L49},
          doi = {10.1093/mnrasl/slad078},
archivePrefix = {arXiv},
       eprint = {2305.09177},
 primaryClass = {astro-ph.EP},
       adsurl = {https://ui.adsabs.harvard.edu/abs/2023MNRAS.525L..43M}
}

@ARTICLE{Rami2014,
       author = {{Ram{\'\i}rez}, I. and {Mel{\'e}ndez}, J. and {Bean}, J. and {Asplund}, M. and {Bedell}, M. and {Monroe}, T. and {Casagrande}, L. and {Schirbel}, L. and {Dreizler}, S. and {Teske}, J. and {Tucci Maia}, M. and {Alves-Brito}, A. and {Baumann}, P.},
        title = "{The Solar Twin Planet Search. I. Fundamental parameters of the stellar sample}",
      journal = {\aap},
         year = 2014,
        month = dec,
       volume = {572},
          eid = {A48},
        pages = {A48},
          doi = {10.1051/0004-6361/201424244},
archivePrefix = {arXiv},
       eprint = {1408.4130},
 primaryClass = {astro-ph.SR},
       adsurl = {https://ui.adsabs.harvard.edu/abs/2014A&A...572A..48R}
}

@ARTICLE{Becker2019,
       author = {{Becker}, Juliette C. and {Vanderburg}, Andrew and {Rodriguez}, Joseph E. and {Omohundro}, Mark and {Adams}, Fred C. and {Stassun}, Keivan G. and {Yao}, Xinyu and {Hartman}, Joel and {Pepper}, Joshua and {Bakos}, Gaspar and {Barentsen}, Geert and {Beatty}, Thomas G. and {Bhatti}, Waqas and {Chontos}, Ashley and {Collier Cameron}, Andrew and {Hellier}, Coel and {Huber}, Daniel and {James}, David and {Kuhn}, Rudolf B. and {Lund}, Michael B. and {Pollacco}, Don and {Siverd}, Robert J. and {Stevens}, Daniel J. and {Cardoso}, Jos{\'e} Vin{\'\i}cius de Miranda and {West}, Richard},
        title = "{A Discrete Set of Possible Transit Ephemerides for Two Long-period Gas Giants Orbiting HIP 41378}",
      journal = {\aj},
         year = 2019,
        month = jan,
       volume = {157},
       number = {1},
          eid = {19},
        pages = {19},
          doi = {10.3847/1538-3881/aaf0a2},
archivePrefix = {arXiv},
       eprint = {1809.10688},
 primaryClass = {astro-ph.EP},
       adsurl = {https://ui.adsabs.harvard.edu/abs/2019AJ....157...19B}
}

@ARTICLE{Vanderburg2016,
       author = {{Vanderburg}, Andrew and {Latham}, David W. and {Buchhave}, Lars A. and {Bieryla}, Allyson and {Berlind}, Perry and {Calkins}, Michael L. and {Esquerdo}, Gilbert A. and {Welsh}, Sophie and {Johnson}, John Asher},
        title = "{Planetary Candidates from the First Year of the K2 Mission}",
      journal = {\apjs},
         year = 2016,
        month = jan,
       volume = {222},
       number = {1},
          eid = {14},
        pages = {14},
          doi = {10.3847/0067-0049/222/1/14},
archivePrefix = {arXiv},
       eprint = {1511.07820},
 primaryClass = {astro-ph.EP},
       adsurl = {https://ui.adsabs.harvard.edu/abs/2016ApJS..222...14V}
}

@ARTICLE{Barros2016,
       author = {{Barros}, S.~C.~C. and {Demangeon}, O. and {Deleuil}, M.},
        title = "{New planetary and eclipsing binary candidates from campaigns 1-6 of the K2 mission}",
      journal = {\aap},
         year = 2016,
        month = oct,
       volume = {594},
          eid = {A100},
        pages = {A100},
          doi = {10.1051/0004-6361/201628902},
archivePrefix = {arXiv},
       eprint = {1607.02339},
 primaryClass = {astro-ph.EP},
       adsurl = {https://ui.adsabs.harvard.edu/abs/2016A&A...594A.100B}
}

@ARTICLE{Yu2018,
       author = {{Yu}, Liang and {Crossfield}, Ian J.~M. and {Schlieder}, Joshua E. and {Kosiarek}, Molly R. and {Feinstein}, Adina D. and {Livingston}, John H. and {Howard}, Andrew W. and {Benneke}, Bj{\"o}rn and {Petigura}, Erik A. and {Bristow}, Makennah and {Christiansen}, Jessie L. and {Ciardi}, David R. and {Crepp}, Justin R. and {Dressing}, Courtney D. and {Fulton}, Benjamin J. and {Gonzales}, Erica J. and {Hardegree-Ullman}, Kevin K. and {Henning}, Thomas and {Isaacson}, Howard and {L{\'e}pine}, S{\'e}bastien and {Martinez}, Arturo O. and {Morales}, Farisa Y. and {Sinukoff}, Evan},
        title = "{Planetary Candidates from K2 Campaign 16}",
      journal = {\aj},
         year = 2018,
        month = jul,
       volume = {156},
       number = {1},
          eid = {22},
        pages = {22},
          doi = {10.3847/1538-3881/aac6e6},
archivePrefix = {arXiv},
       eprint = {1803.04091},
 primaryClass = {astro-ph.EP},
       adsurl = {https://ui.adsabs.harvard.edu/abs/2018AJ....156...22Y}
}

@ARTICLE{Thompson2018,
       author = {{Thompson}, Susan E. and {Coughlin}, Jeffrey L. and {Hoffman}, Kelsey and {Mullally}, Fergal and {Christiansen}, Jessie L. and {Burke}, Christopher J. and {Bryson}, Steve and {Batalha}, Natalie and {Haas}, Michael R. and {Catanzarite}, Joseph and {Rowe}, Jason F. and {Barentsen}, Geert and {Caldwell}, Douglas A. and {Clarke}, Bruce D. and {Jenkins}, Jon M. and {Li}, Jie and {Latham}, David W. and {Lissauer}, Jack J. and {Mathur}, Savita and {Morris}, Robert L. and {Seader}, Shawn E. and {Smith}, Jeffrey C. and {Klaus}, Todd C. and {Twicken}, Joseph D. and {Van Cleve}, Jeffrey E. and {Wohler}, Bill and {Akeson}, Rachel and {Ciardi}, David R. and {Cochran}, William D. and {Henze}, Christopher E. and {Howell}, Steve B. and {Huber}, Daniel and {Pr{\v{s}}a}, Andrej and {Ram{\'\i}rez}, Solange V. and {Morton}, Timothy D. and {Barclay}, Thomas and {Campbell}, Jennifer R. and {Chaplin}, William J. and {Charbonneau}, David and {Christensen-Dalsgaard}, J{\o}rgen and {Dotson}, Jessie L. and {Doyle}, Laurance and {Dunham}, Edward W. and {Dupree}, Andrea K. and {Ford}, Eric B. and {Geary}, John C. and {Girouard}, Forrest R. and {Isaacson}, Howard and {Kjeldsen}, Hans and {Quintana}, Elisa V. and {Ragozzine}, Darin and {Shabram}, Megan and {Shporer}, Avi and {Silva Aguirre}, Victor and {Steffen}, Jason H. and {Still}, Martin and {Tenenbaum}, Peter and {Welsh}, William F. and {Wolfgang}, Angie and {Zamudio}, Khadeejah A. and {Koch}, David G. and {Borucki}, William J.},
        title = "{Planetary Candidates Observed by Kepler. VIII. A Fully Automated Catalog with Measured Completeness and Reliability Based on Data Release 25}",
      journal = {\apjs},
         year = 2018,
        month = apr,
       volume = {235},
       number = {2},
          eid = {38},
        pages = {38},
          doi = {10.3847/1538-4365/aab4f9},
archivePrefix = {arXiv},
       eprint = {1710.06758},
 primaryClass = {astro-ph.EP},
       adsurl = {https://ui.adsabs.harvard.edu/abs/2018ApJS..235...38T}
}

@ARTICLE{AzevedoSilva2022,
       author = {{Azevedo Silva}, T. and {Demangeon}, O.~D.~S. and {Barros}, S.~C.~C. and {Armstrong}, D.~J. and {Otegi}, J.~F. and {Bossini}, D. and {Delgado Mena}, E. and {Sousa}, S.~G. and {Adibekyan}, V. and {Nielsen}, L.~D. and {Dorn}, C. and {Lillo-Box}, J. and {Santos}, N.~C. and {Hoyer}, S. and {Stassun}, K.~G. and {Almenara}, J.~M. and {Bayliss}, D. and {Barrado}, D. and {Boisse}, I. and {Brown}, D.~J.~A. and {D{\'\i}az}, R.~F. and {Dumusque}, X. and {Figueira}, P. and {Hadjigeorghiou}, A. and {Hojjatpanah}, S. and {Mousis}, O. and {Osborn}, A. and {Santerne}, A. and {Str{\o}m}, P.~A. and {Udry}, S. and {Wheatley}, P.~J.},
        title = "{The HD 137496 system: A dense, hot super-Mercury and a cold Jupiter}",
      journal = {\aap},
         year = 2022,
        month = jan,
       volume = {657},
          eid = {A68},
        pages = {A68},
          doi = {10.1051/0004-6361/202141520},
archivePrefix = {arXiv},
       eprint = {2111.08764},
 primaryClass = {astro-ph.EP},
       adsurl = {https://ui.adsabs.harvard.edu/abs/2022A&A...657A..68A}
}

@ARTICLE{Seager2003,
       author = {{Seager}, S. and {Mall{\'e}n-Ornelas}, G.},
        title = "{A Unique Solution of Planet and Star Parameters from an Extrasolar Planet Transit Light Curve}",
      journal = {\apj},
         year = 2003,
        month = mar,
       volume = {585},
       number = {2},
        pages = {1038-1055},
          doi = {10.1086/346105},
archivePrefix = {arXiv},
       eprint = {astro-ph/0206228},
 primaryClass = {astro-ph},
       adsurl = {https://ui.adsabs.harvard.edu/abs/2003ApJ...585.1038S}
}

@ARTICLE{Huber2017,
       author = {{Huber}, Daniel and {Zinn}, Joel and {Bojsen-Hansen}, Mathias and {Pinsonneault}, Marc and {Sahlholdt}, Christian and {Serenelli}, Aldo and {Silva Aguirre}, Victor and {Stassun}, Keivan and {Stello}, Dennis and {Tayar}, Jamie and {Bastien}, Fabienne and {Bedding}, Timothy R. and {Buchhave}, Lars A. and {Chaplin}, William J. and {Davies}, Guy R. and {Garc{\'\i}a}, Rafael A. and {Latham}, David W. and {Mathur}, Savita and {Mosser}, Benoit and {Sharma}, Sanjib},
        title = "{Asteroseismology and Gaia: Testing Scaling Relations Using 2200 Kepler  Stars with TGAS Parallaxes}",
      journal = {\apj},
         year = 2017,
        month = aug,
       volume = {844},
       number = {2},
          eid = {102},
        pages = {102},
          doi = {10.3847/1538-4357/aa75ca},
archivePrefix = {arXiv},
       eprint = {1705.04697},
 primaryClass = {astro-ph.SR},
       adsurl = {https://ui.adsabs.harvard.edu/abs/2017ApJ...844..102H}
}

@ARTICLE{Ida2008,
       author = {{Ida}, S. and {Lin}, D.~N.~C.},
        title = "{Toward a Deterministic Model of Planetary Formation. IV. Effects of Type I Migration}",
      journal = {\apj},
         year = 2008,
        month = jan,
       volume = {673},
       number = {1},
        pages = {487-501},
          doi = {10.1086/523754},
archivePrefix = {arXiv},
       eprint = {0802.1114},
 primaryClass = {astro-ph},
       adsurl = {https://ui.adsabs.harvard.edu/abs/2008ApJ...673..487I}
}

@ARTICLE{Dawson2013,
       author = {{Dawson}, Rebekah I. and {Murray-Clay}, Ruth A.},
        title = "{Giant Planets Orbiting Metal-rich Stars Show Signatures of Planet-Planet Interactions}",
      journal = {\apjl},
         year = 2013,
        month = apr,
       volume = {767},
       number = {2},
          eid = {L24},
        pages = {L24},
          doi = {10.1088/2041-8205/767/2/L24},
archivePrefix = {arXiv},
       eprint = {1302.6244},
 primaryClass = {astro-ph.EP},
       adsurl = {https://ui.adsabs.harvard.edu/abs/2013ApJ...767L..24D}
}

@ARTICLE{Boley2016,
       author = {{Boley}, A.~C. and {Granados Contreras}, A.~P. and {Gladman}, B.},
        title = "{The In Situ Formation of Giant Planets at Short Orbital Periods}",
      journal = {\apjl},
         year = 2016,
        month = feb,
       volume = {817},
       number = {2},
          eid = {L17},
        pages = {L17},
          doi = {10.3847/2041-8205/817/2/L17},
archivePrefix = {arXiv},
       eprint = {1510.04276},
 primaryClass = {astro-ph.EP},
       adsurl = {https://ui.adsabs.harvard.edu/abs/2016ApJ...817L..17B}
}

@ARTICLE{Petigura2018b,
       author = {{Petigura}, Erik A. and {Marcy}, Geoffrey W. and {Winn}, Joshua N. and {Weiss}, Lauren M. and {Fulton}, Benjamin J. and {Howard}, Andrew W. and {Sinukoff}, Evan and {Isaacson}, Howard and {Morton}, Timothy D. and {Johnson}, John Asher},
        title = "{The California-Kepler Survey. IV. Metal-rich Stars Host a Greater Diversity of Planets}",
      journal = {\aj},
         year = {2018b},
        month = feb,
       volume = {155},
       number = {2},
          eid = {89},
        pages = {89},
          doi = {10.3847/1538-3881/aaa54c},
archivePrefix = {arXiv},
       eprint = {1712.04042},
 primaryClass = {astro-ph.EP},
       adsurl = {https://ui.adsabs.harvard.edu/abs/2018AJ....155...89P}
}

@ARTICLE{Petigura2018a,
       author = {{Petigura}, Erik A. and {Crossfield}, Ian J.~M. and {Isaacson}, Howard and {Beichman}, Charles A. and {Christiansen}, Jessie L. and {Dressing}, Courtney D. and {Fulton}, Benjamin J. and {Howard}, Andrew W. and {Kosiarek}, Molly R. and {L{\'e}pine}, S{\'e}bastien and {Schlieder}, Joshua E. and {Sinukoff}, Evan and {Yee}, Samuel W.},
        title = "{Planet Candidates from K2 Campaigns 5-8 and Follow-up Optical Spectroscopy}",
      journal = {\aj},
         year = {2018a},
        month = jan,
       volume = {155},
       number = {1},
          eid = {21},
        pages = {21},
          doi = {10.3847/1538-3881/aa9b83},
archivePrefix = {arXiv},
       eprint = {1711.06377},
 primaryClass = {astro-ph.EP},
       adsurl = {https://ui.adsabs.harvard.edu/abs/2018AJ....155...21P}
}

@ARTICLE{Kruse2019,
       author = {{Kruse}, Ethan and {Agol}, Eric and {Luger}, Rodrigo and {Foreman-Mackey}, Daniel},
        title = "{Detection of Hundreds of New Planet Candidates and Eclipsing Binaries in K2 Campaigns 0-8}",
      journal = {\apjs},
         year = 2019,
        month = sep,
       volume = {244},
       number = {1},
          eid = {11},
        pages = {11},
          doi = {10.3847/1538-4365/ab346b},
archivePrefix = {arXiv},
       eprint = {1907.10806},
 primaryClass = {astro-ph.EP},
       adsurl = {https://ui.adsabs.harvard.edu/abs/2019ApJS..244...11K}
}

@ARTICLE{Stassun2017,
       author = {{Stassun}, Keivan G. and {Collins}, Karen A. and {Gaudi}, B. Scott},
        title = "{Accurate Empirical Radii and Masses of Planets and Their Host Stars with Gaia Parallaxes}",
      journal = {\aj},
         year = 2017,
        month = mar,
       volume = {153},
       number = {3},
          eid = {136},
        pages = {136},
          doi = {10.3847/1538-3881/aa5df3},
archivePrefix = {arXiv},
       eprint = {1609.04389},
 primaryClass = {astro-ph.EP},
       adsurl = {https://ui.adsabs.harvard.edu/abs/2017AJ....153..136S}
}

@ARTICLE{Ecuvillon2006,
       author = {{Ecuvillon}, A. and {Israelian}, G. and {Santos}, N.~C. and {Shchukina}, N.~G. and {Mayor}, M. and {Rebolo}, R.},
        title = "{Oxygen abundances in planet-harbouring stars. Comparison of different abundance indicators}",
      journal = {\aap},
         year = 2006,
        month = jan,
       volume = {445},
       number = {2},
        pages = {633-645},
          doi = {10.1051/0004-6361:20053469},
archivePrefix = {arXiv},
       eprint = {astro-ph/0509326},
 primaryClass = {astro-ph},
       adsurl = {https://ui.adsabs.harvard.edu/abs/2006A&A...445..633E}
}

@ARTICLE{GaiaCollaboration2018,
       author = {{Gaia Collaboration} and {Brown}, A.~G.~A. and {Vallenari}, A. and {Prusti}, T. and {de Bruijne}, J.~H.~J. and {Babusiaux}, C. and {Bailer-Jones}, C.~A.~L. and {Biermann}, M. and {Evans}, D.~W. and {Eyer}, L. and {Jansen}, F. and {Jordi}, C. and {Klioner}, S.~A. and {Lammers}, U. and {Lindegren}, L. and {Luri}, X. and {Mignard}, F. and {Panem}, C. and {Pourbaix}, D. and {Randich}, S. and {Sartoretti}, P. and {Siddiqui}, H.~I. and {Soubiran}, C. and {van Leeuwen}, F. and {Walton}, N.~A. and {Arenou}, F. and {Bastian}, U. and {Cropper}, M. and {Drimmel}, R. and {Katz}, D. and {Lattanzi}, M.~G. and {Bakker}, J. and {Cacciari}, C. and {Casta{\~n}eda}, J. and {Chaoul}, L. and {Cheek}, N. and {De Angeli}, F. and {Fabricius}, C. and {Guerra}, R. and {Holl}, B. and {Masana}, E. and {Messineo}, R. and {Mowlavi}, N. and {Nienartowicz}, K. and {Panuzzo}, P. and {Portell}, J. and {Riello}, M. and {Seabroke}, G.~M. and {Tanga}, P. and {Th{\'e}venin}, F. and {Gracia-Abril}, G. and {Comoretto}, G. and {Garcia-Reinaldos}, M. and {Teyssier}, D. and {Altmann}, M. and {Andrae}, R. and {Audard}, M. and {Bellas-Velidis}, I. and {Benson}, K. and {Berthier}, J. and {Blomme}, R. and {Burgess}, P. and {Busso}, G. and {Carry}, B. and {Cellino}, A. and {Clementini}, G. and {Clotet}, M. and {Creevey}, O. and {Davidson}, M. and {De Ridder}, J. and {Delchambre}, L. and {Dell'Oro}, A. and {Ducourant}, C. and {Fern{\'a}ndez-Hern{\'a}ndez}, J. and {Fouesneau}, M. and {Fr{\'e}mat}, Y. and {Galluccio}, L. and {Garc{\'\i}a-Torres}, M. and {Gonz{\'a}lez-N{\'u}{\~n}ez}, J. and {Gonz{\'a}lez-Vidal}, J.~J. and {Gosset}, E. and {Guy}, L.~P. and {Halbwachs}, J. -L. and {Hambly}, N.~C. and {Harrison}, D.~L. and {Hern{\'a}ndez}, J. and {Hestroffer}, D. and {Hodgkin}, S.~T. and {Hutton}, A. and {Jasniewicz}, G. and {Jean-Antoine-Piccolo}, A. and {Jordan}, S. and {Korn}, A.~J. and {Krone-Martins}, A. and {Lanzafame}, A.~C. and {Lebzelter}, T. and {L{\"o}ffler}, W. and {Manteiga}, M. and {Marrese}, P.~M. and {Mart{\'\i}n-Fleitas}, J.~M. and {Moitinho}, A. and {Mora}, A. and {Muinonen}, K. and {Osinde}, J. and {Pancino}, E. and {Pauwels}, T. and {Petit}, J. -M. and {Recio-Blanco}, A. and {Richards}, P.~J. and {Rimoldini}, L. and {Robin}, A.~C. and {Sarro}, L.~M. and {Siopis}, C. and {Smith}, M. and {Sozzetti}, A. and {S{\"u}veges}, M. and {Torra}, J. and {van Reeven}, W. and {Abbas}, U. and {Abreu Aramburu}, A. and {Accart}, S. and {Aerts}, C. and {Altavilla}, G. and {{\'A}lvarez}, M.~A. and {Alvarez}, R. and {Alves}, J. and {Anderson}, R.~I. and {Andrei}, A.~H. and {Anglada Varela}, E. and {Antiche}, E. and {Antoja}, T. and {Arcay}, B. and {Astraatmadja}, T.~L. and {Bach}, N. and {Baker}, S.~G. and {Balaguer-N{\'u}{\~n}ez}, L. and {Balm}, P. and {Barache}, C. and {Barata}, C. and {Barbato}, D. and {Barblan}, F. and {Barklem}, P.~S. and {Barrado}, D. and {Barros}, M. and {Barstow}, M.~A. and {Bartholom{\'e} Mu{\~n}oz}, S. and {Bassilana}, J. -L. and {Becciani}, U. and {Bellazzini}, M. and {Berihuete}, A. and {Bertone}, S. and {Bianchi}, L. and {Bienaym{\'e}}, O. and {Blanco-Cuaresma}, S. and {Boch}, T. and {Boeche}, C. and {Bombrun}, A. and {Borrachero}, R. and {Bossini}, D. and {Bouquillon}, S. and {Bourda}, G. and {Bragaglia}, A. and {Bramante}, L. and {Breddels}, M.~A. and {Bressan}, A. and {Brouillet}, N. and {Br{\"u}semeister}, T. and {Brugaletta}, E. and {Bucciarelli}, B. and {Burlacu}, A. and {Busonero}, D. and {Butkevich}, A.~G. and {Buzzi}, R. and {Caffau}, E. and {Cancelliere}, R. and {Cannizzaro}, G. and {Cantat-Gaudin}, T. and {Carballo}, R. and {Carlucci}, T. and {Carrasco}, J.~M. and {Casamiquela}, L. and {Castellani}, M. and {Castro-Ginard}, A. and {Charlot}, P. and {Chemin}, L. and {Chiavassa}, A. and {Cocozza}, G. and {Costigan}, G. and {Cowell}, S. and {Crifo}, F. and {Crosta}, M. and {Crowley}, C. and {Cuypers}, J. and {Dafonte}, C. and {Damerdji}, Y. and {Dapergolas}, A. and {David}, P. and {David}, M. and {de Laverny}, P. and {De Luise}, F.},
        title = "{Gaia Data Release 2. Summary of the contents and survey properties}",
      journal = {\aap},
         year = 2018,
        month = aug,
       volume = {616},
          eid = {A1},
        pages = {A1},
          doi = {10.1051/0004-6361/201833051},
archivePrefix = {arXiv},
       eprint = {1804.09365},
 primaryClass = {astro-ph.GA},
       adsurl = {https://ui.adsabs.harvard.edu/abs/2018A&A...616A...1G}
}

@ARTICLE{GaiaCollaboration2016,
       author = {{Gaia Collaboration} and {Brown}, A.~G.~A. and {Vallenari}, A. and {Prusti}, T. and {de Bruijne}, J.~H.~J. and {Mignard}, F. and {Drimmel}, R. and {Babusiaux}, C. and {Bailer-Jones}, C.~A.~L. and {Bastian}, U. and {Biermann}, M. and {Evans}, D.~W. and {Eyer}, L. and {Jansen}, F. and {Jordi}, C. and {Katz}, D. and {Klioner}, S.~A. and {Lammers}, U. and {Lindegren}, L. and {Luri}, X. and {O'Mullane}, W. and {Panem}, C. and {Pourbaix}, D. and {Randich}, S. and {Sartoretti}, P. and {Siddiqui}, H.~I. and {Soubiran}, C. and {Valette}, V. and {van Leeuwen}, F. and {Walton}, N.~A. and {Aerts}, C. and {Arenou}, F. and {Cropper}, M. and {H{\o}g}, E. and {Lattanzi}, M.~G. and {Grebel}, E.~K. and {Holland}, A.~D. and {Huc}, C. and {Passot}, X. and {Perryman}, M. and {Bramante}, L. and {Cacciari}, C. and {Casta{\~n}eda}, J. and {Chaoul}, L. and {Cheek}, N. and {De Angeli}, F. and {Fabricius}, C. and {Guerra}, R. and {Hern{\'a}ndez}, J. and {Jean-Antoine-Piccolo}, A. and {Masana}, E. and {Messineo}, R. and {Mowlavi}, N. and {Nienartowicz}, K. and {Ord{\'o}{\~n}ez-Blanco}, D. and {Panuzzo}, P. and {Portell}, J. and {Richards}, P.~J. and {Riello}, M. and {Seabroke}, G.~M. and {Tanga}, P. and {Th{\'e}venin}, F. and {Torra}, J. and {Els}, S.~G. and {Gracia-Abril}, G. and {Comoretto}, G. and {Garcia-Reinaldos}, M. and {Lock}, T. and {Mercier}, E. and {Altmann}, M. and {Andrae}, R. and {Astraatmadja}, T.~L. and {Bellas-Velidis}, I. and {Benson}, K. and {Berthier}, J. and {Blomme}, R. and {Busso}, G. and {Carry}, B. and {Cellino}, A. and {Clementini}, G. and {Cowell}, S. and {Creevey}, O. and {Cuypers}, J. and {Davidson}, M. and {De Ridder}, J. and {de Torres}, A. and {Delchambre}, L. and {Dell'Oro}, A. and {Ducourant}, C. and {Fr{\'e}mat}, Y. and {Garc{\'\i}a-Torres}, M. and {Gosset}, E. and {Halbwachs}, J. -L. and {Hambly}, N.~C. and {Harrison}, D.~L. and {Hauser}, M. and {Hestroffer}, D. and {Hodgkin}, S.~T. and {Huckle}, H.~E. and {Hutton}, A. and {Jasniewicz}, G. and {Jordan}, S. and {Kontizas}, M. and {Korn}, A.~J. and {Lanzafame}, A.~C. and {Manteiga}, M. and {Moitinho}, A. and {Muinonen}, K. and {Osinde}, J. and {Pancino}, E. and {Pauwels}, T. and {Petit}, J. -M. and {Recio-Blanco}, A. and {Robin}, A.~C. and {Sarro}, L.~M. and {Siopis}, C. and {Smith}, M. and {Smith}, K.~W. and {Sozzetti}, A. and {Thuillot}, W. and {van Reeven}, W. and {Viala}, Y. and {Abbas}, U. and {Abreu Aramburu}, A. and {Accart}, S. and {Aguado}, J.~J. and {Allan}, P.~M. and {Allasia}, W. and {Altavilla}, G. and {{\'A}lvarez}, M.~A. and {Alves}, J. and {Anderson}, R.~I. and {Andrei}, A.~H. and {Anglada Varela}, E. and {Antiche}, E. and {Antoja}, T. and {Ant{\'o}n}, S. and {Arcay}, B. and {Bach}, N. and {Baker}, S.~G. and {Balaguer-N{\'u}{\~n}ez}, L. and {Barache}, C. and {Barata}, C. and {Barbier}, A. and {Barblan}, F. and {Barrado y Navascu{\'e}s}, D. and {Barros}, M. and {Barstow}, M.~A. and {Becciani}, U. and {Bellazzini}, M. and {Bello Garc{\'\i}a}, A. and {Belokurov}, V. and {Bendjoya}, P. and {Berihuete}, A. and {Bianchi}, L. and {Bienaym{\'e}}, O. and {Billebaud}, F. and {Blagorodnova}, N. and {Blanco-Cuaresma}, S. and {Boch}, T. and {Bombrun}, A. and {Borrachero}, R. and {Bouquillon}, S. and {Bourda}, G. and {Bouy}, H. and {Bragaglia}, A. and {Breddels}, M.~A. and {Brouillet}, N. and {Br{\"u}semeister}, T. and {Bucciarelli}, B. and {Burgess}, P. and {Burgon}, R. and {Burlacu}, A. and {Busonero}, D. and {Buzzi}, R. and {Caffau}, E. and {Cambras}, J. and {Campbell}, H. and {Cancelliere}, R. and {Cantat-Gaudin}, T. and {Carlucci}, T. and {Carrasco}, J.~M. and {Castellani}, M. and {Charlot}, P. and {Charnas}, J. and {Chiavassa}, A. and {Clotet}, M. and {Cocozza}, G. and {Collins}, R.~S. and {Costigan}, G. and {Crifo}, F. and {Cross}, N.~J.~G. and {Crosta}, M. and {Crowley}, C. and {Dafonte}, C. and {Damerdji}, Y. and {Dapergolas}, A. and {David}, P. and {David}, M. and {De Cat}, P.},
        title = "{Gaia Data Release 1. Summary of the astrometric, photometric, and survey properties}",
      journal = {\aap},
         year = 2016,
        month = nov,
       volume = {595},
          eid = {A2},
        pages = {A2},
          doi = {10.1051/0004-6361/201629512},
archivePrefix = {arXiv},
       eprint = {1609.04172},
 primaryClass = {astro-ph.IM},
       adsurl = {https://ui.adsabs.harvard.edu/abs/2016A&A...595A...2G}
}

@ARTICLE{Perdigon2021,
       author = {{Perdigon}, J. and {de Laverny}, P. and {Recio-Blanco}, A. and {Fernandez-Alvar}, E. and {Santos-Peral}, P. and {Kordopatis}, G. and {{\'A}lvarez}, M.~A.},
        title = "{The AMBRE Project: Origin and evolution of sulfur in the Milky Way}",
      journal = {\aap},
         year = 2021,
        month = mar,
       volume = {647},
          eid = {A162},
        pages = {A162},
          doi = {10.1051/0004-6361/202040147},
archivePrefix = {arXiv},
       eprint = {2102.01961},
 primaryClass = {astro-ph.GA},
       adsurl = {https://ui.adsabs.harvard.edu/abs/2021A&A...647A.162P}
}

@ARTICLE{Mishurov2019,
       author = {{Mishurov}, Yu N. and {Tkachenko}, R.~V.},
        title = "{On the radial iron distribution in the Galactic disc}",
      journal = {\mnras},
         year = 2019,
        month = may,
       volume = {485},
       number = {2},
        pages = {2225-2234},
          doi = {10.1093/mnras/stz526},
       adsurl = {https://ui.adsabs.harvard.edu/abs/2019MNRAS.485.2225M}
}

@ARTICLE{Kirby2018,
       author = {{Kirby}, Evan N. and {Xie}, Justin L. and {Guo}, Rachel and {Kovalev}, Mikhail and {Bergemann}, Maria},
        title = "{Catalog of Chromium, Cobalt, and Nickel Abundances in Globular Clusters and Dwarf Galaxies}",
      journal = {\apjs},
         year = 2018,
        month = jul,
       volume = {237},
       number = {1},
          eid = {18},
        pages = {18},
          doi = {10.3847/1538-4365/aac952},
archivePrefix = {arXiv},
       eprint = {1906.08284},
 primaryClass = {astro-ph.SR},
       adsurl = {https://ui.adsabs.harvard.edu/abs/2018ApJS..237...18K}
}

@BOOK{Arnett1996,
       author = {{Arnett}, David},
        title = "{Supernovae and Nucleosynthesis: An Investigation of the History of Matter from the Big Bang to the Present}",
    publisher = {Princeton University Press},
         year = 1996,
       adsurl = {https://ui.adsabs.harvard.edu/abs/1996snih.book.....A}
}

@ARTICLE{Woosley1995,
       author = {{Woosley}, S.~E. and {Weaver}, Thomas A.},
        title = "{The Evolution and Explosion of Massive Stars. II. Explosive Hydrodynamics and Nucleosynthesis}",
      journal = {\apjs},
         year = 1995,
        month = nov,
       volume = {101},
        pages = {181},
          doi = {10.1086/192237},
       adsurl = {https://ui.adsabs.harvard.edu/abs/1995ApJS..101..181W}
}

@ARTICLE{Ochsenbein2000,
       author = {{Ochsenbein}, F. and {Bauer}, P. and {Marcout}, J.},
        title = "{The VizieR database of astronomical catalogues}",
      journal = {\aaps},
         year = 2000,
        month = apr,
       volume = {143},
        pages = {23-32},
          doi = {10.1051/aas:2000169},
archivePrefix = {arXiv},
       eprint = {astro-ph/0002122},
 primaryClass = {astro-ph},
       adsurl = {https://ui.adsabs.harvard.edu/abs/2000A&AS..143...23O}
}

@ARTICLE{Queiroz2018,
       author = {{Queiroz}, A.~B.~A. and {Anders}, F. and {Santiago}, B.~X. and {Chiappini}, C. and {Steinmetz}, M. and {Dal Ponte}, M. and {Stassun}, K.~G. and {da Costa}, L.~N. and {Maia}, M.~A.~G. and {Crestani}, J. and {Beers}, T.~C. and {Fern{\'a}ndez-Trincado}, J.~G. and {Garc{\'\i}a-Hern{\'a}ndez}, D.~A. and {Roman-Lopes}, A. and {Zamora}, O.},
        title = "{StarHorse: a Bayesian tool for determining stellar masses, ages, distances, and extinctions for field stars}",
      journal = {\mnras},
         year = 2018,
        month = may,
       volume = {476},
       number = {2},
        pages = {2556-2583},
          doi = {10.1093/mnras/sty330},
archivePrefix = {arXiv},
       eprint = {1710.09970},
 primaryClass = {astro-ph.IM},
       adsurl = {https://ui.adsabs.harvard.edu/abs/2018MNRAS.476.2556Q}
}

@ARTICLE{Drimmel2003,
       author = {{Drimmel}, R. and {Cabrera-Lavers}, A. and {L{\'o}pez-Corredoira}, M.},
        title = "{A three-dimensional Galactic extinction model}",
      journal = {\aap},
         year = 2003,
        month = oct,
       volume = {409},
        pages = {205-215},
          doi = {10.1051/0004-6361:20031070},
archivePrefix = {arXiv},
       eprint = {astro-ph/0307273},
 primaryClass = {astro-ph},
       adsurl = {https://ui.adsabs.harvard.edu/abs/2003A&A...409..205D}
}

@ARTICLE{GaiaCollaboration2021,
       author = {{Gaia Collaboration} and {Brown}, A.~G.~A. and {Vallenari}, A. and {Prusti}, T. and {de Bruijne}, J.~H.~J. and {Babusiaux}, C. and {Biermann}, M. and {Creevey}, O.~L. and {Evans}, D.~W. and {Eyer}, L. and {Hutton}, A. and {Jansen}, F. and {Jordi}, C. and {Klioner}, S.~A. and {Lammers}, U. and {Lindegren}, L. and {Luri}, X. and {Mignard}, F. and {Panem}, C. and {Pourbaix}, D. and {Randich}, S. and {Sartoretti}, P. and {Soubiran}, C. and {Walton}, N.~A. and {Arenou}, F. and {Bailer-Jones}, C.~A.~L. and {Bastian}, U. and {Cropper}, M. and {Drimmel}, R. and {Katz}, D. and {Lattanzi}, M.~G. and {van Leeuwen}, F. and {Bakker}, J. and {Cacciari}, C. and {Casta{\~n}eda}, J. and {De Angeli}, F. and {Ducourant}, C. and {Fabricius}, C. and {Fouesneau}, M. and {Fr{\'e}mat}, Y. and {Guerra}, R. and {Guerrier}, A. and {Guiraud}, J. and {Jean-Antoine Piccolo}, A. and {Masana}, E. and {Messineo}, R. and {Mowlavi}, N. and {Nicolas}, C. and {Nienartowicz}, K. and {Pailler}, F. and {Panuzzo}, P. and {Riclet}, F. and {Roux}, W. and {Seabroke}, G.~M. and {Sordo}, R. and {Tanga}, P. and {Th{\'e}venin}, F. and {Gracia-Abril}, G. and {Portell}, J. and {Teyssier}, D. and {Altmann}, M. and {Andrae}, R. and {Bellas-Velidis}, I. and {Benson}, K. and {Berthier}, J. and {Blomme}, R. and {Brugaletta}, E. and {Burgess}, P.~W. and {Busso}, G. and {Carry}, B. and {Cellino}, A. and {Cheek}, N. and {Clementini}, G. and {Damerdji}, Y. and {Davidson}, M. and {Delchambre}, L. and {Dell'Oro}, A. and {Fern{\'a}ndez-Hern{\'a}ndez}, J. and {Galluccio}, L. and {Garc{\'\i}a-Lario}, P. and {Garcia-Reinaldos}, M. and {Gonz{\'a}lez-N{\'u}{\~n}ez}, J. and {Gosset}, E. and {Haigron}, R. and {Halbwachs}, J. -L. and {Hambly}, N.~C. and {Harrison}, D.~L. and {Hatzidimitriou}, D. and {Heiter}, U. and {Hern{\'a}ndez}, J. and {Hestroffer}, D. and {Hodgkin}, S.~T. and {Holl}, B. and {Jan{\ss}en}, K. and {Jevardat de Fombelle}, G. and {Jordan}, S. and {Krone-Martins}, A. and {Lanzafame}, A.~C. and {L{\"o}ffler}, W. and {Lorca}, A. and {Manteiga}, M. and {Marchal}, O. and {Marrese}, P.~M. and {Moitinho}, A. and {Mora}, A. and {Muinonen}, K. and {Osborne}, P. and {Pancino}, E. and {Pauwels}, T. and {Petit}, J. -M. and {Recio-Blanco}, A. and {Richards}, P.~J. and {Riello}, M. and {Rimoldini}, L. and {Robin}, A.~C. and {Roegiers}, T. and {Rybizki}, J. and {Sarro}, L.~M. and {Siopis}, C. and {Smith}, M. and {Sozzetti}, A. and {Ulla}, A. and {Utrilla}, E. and {van Leeuwen}, M. and {van Reeven}, W. and {Abbas}, U. and {Abreu Aramburu}, A. and {Accart}, S. and {Aerts}, C. and {Aguado}, J.~J. and {Ajaj}, M. and {Altavilla}, G. and {{\'A}lvarez}, M.~A. and {{\'A}lvarez Cid-Fuentes}, J. and {Alves}, J. and {Anderson}, R.~I. and {Anglada Varela}, E. and {Antoja}, T. and {Audard}, M. and {Baines}, D. and {Baker}, S.~G. and {Balaguer-N{\'u}{\~n}ez}, L. and {Balbinot}, E. and {Balog}, Z. and {Barache}, C. and {Barbato}, D. and {Barros}, M. and {Barstow}, M.~A. and {Bartolom{\'e}}, S. and {Bassilana}, J. -L. and {Bauchet}, N. and {Baudesson-Stella}, A. and {Becciani}, U. and {Bellazzini}, M. and {Bernet}, M. and {Bertone}, S. and {Bianchi}, L. and {Blanco-Cuaresma}, S. and {Boch}, T. and {Bombrun}, A. and {Bossini}, D. and {Bouquillon}, S. and {Bragaglia}, A. and {Bramante}, L. and {Breedt}, E. and {Bressan}, A. and {Brouillet}, N. and {Bucciarelli}, B. and {Burlacu}, A. and {Busonero}, D. and {Butkevich}, A.~G. and {Buzzi}, R. and {Caffau}, E. and {Cancelliere}, R. and {C{\'a}novas}, H. and {Cantat-Gaudin}, T. and {Carballo}, R. and {Carlucci}, T. and {Carnerero}, M.~I. and {Carrasco}, J.~M. and {Casamiquela}, L. and {Castellani}, M. and {Castro-Ginard}, A. and {Castro Sampol}, P. and {Chaoul}, L. and {Charlot}, P. and {Chemin}, L. and {Chiavassa}, A. and {Cioni}, M. -R.~L. and {Comoretto}, G. and {Cooper}, W.~J. and {Cornez}, T. and {Cowell}, S. and {Crifo}, F. and {Crosta}, M. and {Crowley}, C. and {Dafonte}, C. and {Dapergolas}, A. and {David}, M. and {David}, P.},
        title = "{Gaia Early Data Release 3. Summary of the contents and survey properties}",
      journal = {\aap},
         year = 2021,
        month = may,
       volume = {649},
          eid = {A1},
        pages = {A1},
          doi = {10.1051/0004-6361/202039657},
archivePrefix = {arXiv},
       eprint = {2012.01533},
 primaryClass = {astro-ph.GA},
       adsurl = {https://ui.adsabs.harvard.edu/abs/2021A&A...649A...1G}
}

@ARTICLE{Martinez2019,
       author = {{Martinez}, Cintia F. and {Cunha}, Katia and {Ghezzi}, Luan and {Smith}, Verne V.},
        title = "{A Spectroscopic Analysis of the California-Kepler Survey Sample. I. Stellar Parameters, Planetary Radii, and a Slope in the Radius Gap}",
      journal = {\apj},
         year = 2019,
        month = apr,
       volume = {875},
       number = {1},
          eid = {29},
        pages = {29},
          doi = {10.3847/1538-4357/ab0d93},
archivePrefix = {arXiv},
       eprint = {1903.00174},
 primaryClass = {astro-ph.EP},
       adsurl = {https://ui.adsabs.harvard.edu/abs/2019ApJ...875...29M}
}

@ARTICLE{Weiss2018,
       author = {{Weiss}, Lauren M. and {Marcy}, Geoffrey W. and {Petigura}, Erik A. and {Fulton}, Benjamin J. and {Howard}, Andrew W. and {Winn}, Joshua N. and {Isaacson}, Howard T. and {Morton}, Timothy D. and {Hirsch}, Lea A. and {Sinukoff}, Evan J. and {Cumming}, Andrew and {Hebb}, Leslie and {Cargile}, Phillip A.},
        title = "{The California-Kepler Survey. V. Peas in a Pod: Planets in a Kepler Multi-planet System Are Similar in Size and Regularly Spaced}",
      journal = {\aj},
         year = 2018,
        month = jan,
       volume = {155},
       number = {1},
          eid = {48},
        pages = {48},
          doi = {10.3847/1538-3881/aa9ff6},
archivePrefix = {arXiv},
       eprint = {1706.06204},
 primaryClass = {astro-ph.EP},
       adsurl = {https://ui.adsabs.harvard.edu/abs/2018AJ....155...48W}
}

@ARTICLE{Loaiza-Tacuri2025,
       author = {{Loaiza-Tacuri}, V. and {Souto}, Diogo and {Quispe-Huaynasi}, F. and {Cunha}, Katia and {Daflon}, S. and {Costa-Almeida}, Ellen and {Smith}, V.~V. and {Ghezzi}, Luan},
        title = "{Stellar Characterization, Magnesium Abundances, and Chromospheric Activity Analysis of Stars with Confirmed Exoplanets from the K2 Mission}",
      journal = {\apjs},
         year = 2025,
        month = dec,
       volume = {281},
       number = {2},
          eid = {61},
        pages = {61},
          doi = {10.3847/1538-4365/ae14f9},
archivePrefix = {arXiv},
       eprint = {2510.17574},
 primaryClass = {astro-ph.SR},
       adsurl = {https://ui.adsabs.harvard.edu/abs/2025ApJS..281...61L}
}

@ARTICLE{Loaiza-Tacuri2024,
       author = {{Loaiza-Tacuri}, V. and {Cunha}, Katia and {Smith}, Verne V. and {Quispe-Huaynasi}, F. and {Costa-Almeida}, Ellen and {Ghezzi}, Luan and {Melendez}, Jorge},
        title = "{Stellar Characterization and a Chromospheric Activity Analysis of a K2 Sample of Planet-hosting Stars}",
      journal = {\apj},
         year = 2024,
        month = jul,
       volume = {970},
       number = {1},
          eid = {53},
        pages = {53},
          doi = {10.3847/1538-4357/ad4b15},
archivePrefix = {arXiv},
       eprint = {2405.08128},
 primaryClass = {astro-ph.SR},
       adsurl = {https://ui.adsabs.harvard.edu/abs/2024ApJ...970...53L}
}

@ARTICLE{Chabrier2001,
       author = {{Chabrier}, Gilles},
        title = "{The Galactic Disk Mass Budget. I. Stellar Mass Function and Density}",
      journal = {\apj},
         year = 2001,
        month = jun,
       volume = {554},
       number = {2},
        pages = {1274-1281},
          doi = {10.1086/321401},
archivePrefix = {arXiv},
       eprint = {astro-ph/0107018},
 primaryClass = {astro-ph},
       adsurl = {https://ui.adsabs.harvard.edu/abs/2001ApJ...554.1274C}
}

@ARTICLE{Bressan2012,
       author = {{Bressan}, Alessandro and {Marigo}, Paola and {Girardi}, L{\'e}o. and {Salasnich}, Bernardo and {Dal Cero}, Claudia and {Rubele}, Stefano and {Nanni}, Ambra},
        title = "{PARSEC: stellar tracks and isochrones with the PAdova and TRieste Stellar Evolution Code}",
      journal = {\mnras},
         year = 2012,
        month = nov,
       volume = {427},
       number = {1},
        pages = {127-145},
          doi = {10.1111/j.1365-2966.2012.21948.x},
archivePrefix = {arXiv},
       eprint = {1208.4498},
 primaryClass = {astro-ph.SR},
       adsurl = {https://ui.adsabs.harvard.edu/abs/2012MNRAS.427..127B}
}

@ARTICLE{Green2019,
       author = {{Green}, Gregory M. and {Schlafly}, Edward and {Zucker}, Catherine and {Speagle}, Joshua S. and {Finkbeiner}, Douglas},
        title = "{A 3D Dust Map Based on Gaia, Pan-STARRS 1, and 2MASS}",
      journal = {\apj},
         year = 2019,
        month = dec,
       volume = {887},
       number = {1},
          eid = {93},
        pages = {93},
          doi = {10.3847/1538-4357/ab5362},
archivePrefix = {arXiv},
       eprint = {1905.02734},
 primaryClass = {astro-ph.GA},
       adsurl = {https://ui.adsabs.harvard.edu/abs/2019ApJ...887...93G}
}

@ARTICLE{Anders2022,
       author = {{Anders}, F. and {Khalatyan}, A. and {Queiroz}, A.~B.~A. and {Chiappini}, C. and {Ard{\`e}vol}, J. and {Casamiquela}, L. and {Figueras}, F. and {Jim{\'e}nez-Arranz}, {\'O}. and {Jordi}, C. and {Mongui{\'o}}, M. and {Romero-G{\'o}mez}, M. and {Altamirano}, D. and {Antoja}, T. and {Assaad}, R. and {Cantat-Gaudin}, T. and {Castro-Ginard}, A. and {Enke}, H. and {Girardi}, L. and {Guiglion}, G. and {Khan}, S. and {Luri}, X. and {Miglio}, A. and {Minchev}, I. and {Ramos}, P. and {Santiago}, B.~X. and {Steinmetz}, M.},
        title = "{Photo-astrometric distances, extinctions, and astrophysical parameters for Gaia EDR3 stars brighter than G = 18.5}",
      journal = {\aap},
         year = 2022,
        month = feb,
       volume = {658},
          eid = {A91},
        pages = {A91},
          doi = {10.1051/0004-6361/202142369},
archivePrefix = {arXiv},
       eprint = {2111.01860},
 primaryClass = {astro-ph.GA},
       adsurl = {https://ui.adsabs.harvard.edu/abs/2022A&A...658A..91A}
}

@ARTICLE{Anders2019,
       author = {{Anders}, F. and {Khalatyan}, A. and {Chiappini}, C. and {Queiroz}, A.~B. and {Santiago}, B.~X. and {Jordi}, C. and {Girardi}, L. and {Brown}, A.~G.~A. and {Matijevi{\v{c}}}, G. and {Monari}, G. and {Cantat-Gaudin}, T. and {Weiler}, M. and {Khan}, S. and {Miglio}, A. and {Carrillo}, I. and {Romero-G{\'o}mez}, M. and {Minchev}, I. and {de Jong}, R.~S. and {Antoja}, T. and {Ramos}, P. and {Steinmetz}, M. and {Enke}, H.},
        title = "{Photo-astrometric distances, extinctions, and astrophysical parameters for Gaia DR2 stars brighter than G = 18}",
      journal = {\aap},
         year = 2019,
        month = aug,
       volume = {628},
          eid = {A94},
        pages = {A94},
          doi = {10.1051/0004-6361/201935765},
archivePrefix = {arXiv},
       eprint = {1904.11302},
 primaryClass = {astro-ph.GA},
       adsurl = {https://ui.adsabs.harvard.edu/abs/2019A&A...628A..94A}
}

@ARTICLE{daSilva2006,
       author = {{da Silva}, L. and {Girardi}, L. and {Pasquini}, L. and {Setiawan}, J. and {von der L{\"u}he}, O. and {de Medeiros}, J.~R. and {Hatzes}, A. and {D{\"o}llinger}, M.~P. and {Weiss}, A.},
        title = "{Basic physical parameters of a selected sample of evolved stars}",
      journal = {\aap},
         year = 2006,
        month = nov,
       volume = {458},
       number = {2},
        pages = {609-623},
          doi = {10.1051/0004-6361:20065105},
archivePrefix = {arXiv},
       eprint = {astro-ph/0608160},
 primaryClass = {astro-ph},
       adsurl = {https://ui.adsabs.harvard.edu/abs/2006A&A...458..609D}
}

@ARTICLE{Akeson2013,
       author = {{Akeson}, R.~L. and {Chen}, X. and {Ciardi}, D. and {Crane}, M. and {Good}, J. and {Harbut}, M. and {Jackson}, E. and {Kane}, S.~R. and {Laity}, A.~C. and {Leifer}, S. and {Lynn}, M. and {McElroy}, D.~L. and {Papin}, M. and {Plavchan}, P. and {Ram{\'\i}rez}, S.~V. and {Rey}, R. and {von Braun}, K. and {Wittman}, M. and {Abajian}, M. and {Ali}, B. and {Beichman}, C. and {Beekley}, A. and {Berriman}, G.~B. and {Berukoff}, S. and {Bryden}, G. and {Chan}, B. and {Groom}, S. and {Lau}, C. and {Payne}, A.~N. and {Regelson}, M. and {Saucedo}, M. and {Schmitz}, M. and {Stauffer}, J. and {Wyatt}, P. and {Zhang}, A.},
        title = "{The NASA Exoplanet Archive: Data and Tools for Exoplanet Research}",
      journal = {\pasp},
         year = 2013,
        month = aug,
       volume = {125},
       number = {930},
        pages = {989},
          doi = {10.1086/672273},
archivePrefix = {arXiv},
       eprint = {1307.2944},
 primaryClass = {astro-ph.IM},
       adsurl = {https://ui.adsabs.harvard.edu/abs/2013PASP..125..989A}
}

@ARTICLE{Mayor2003,
       author = {{Mayor}, M. and {Pepe}, F. and {Queloz}, D. and {Bouchy}, F. and {Rupprecht}, G. and {Lo Curto}, G. and {Avila}, G. and {Benz}, W. and {Bertaux}, J. -L. and {Bonfils}, X. and {Dall}, Th. and {Dekker}, H. and {Delabre}, B. and {Eckert}, W. and {Fleury}, M. and {Gilliotte}, A. and {Gojak}, D. and {Guzman}, J.~C. and {Kohler}, D. and {Lizon}, J. -L. and {Longinotti}, A. and {Lovis}, C. and {Megevand}, D. and {Pasquini}, L. and {Reyes}, J. and {Sivan}, J. -P. and {Sosnowska}, D. and {Soto}, R. and {Udry}, S. and {van Kesteren}, A. and {Weber}, L. and {Weilenmann}, U.},
        title = "{Setting New Standards with HARPS}",
      journal = {The Messenger},
         year = 2003,
        month = dec,
       volume = {114},
        pages = {20-24},
       adsurl = {https://ui.adsabs.harvard.edu/abs/2003Msngr.114...20M}
}

@ARTICLE{Ghezzi2018,
       author = {{Ghezzi}, Luan and {Montet}, Benjamin T. and {Johnson}, John Asher},
        title = "{Retired A Stars Revisited: An Updated Giant Planet Occurrence Rate as a Function of Stellar Metallicity and Mass}",
      journal = {\apj},
         year = 2018,
        month = jun,
       volume = {860},
       number = {2},
          eid = {109},
        pages = {109},
          doi = {10.3847/1538-4357/aac37c},
archivePrefix = {arXiv},
       eprint = {1804.09082},
 primaryClass = {astro-ph.SR},
       adsurl = {https://ui.adsabs.harvard.edu/abs/2018ApJ...860..109G}
}

@ARTICLE{Ghezzi2009,
       author = {{Ghezzi}, L. and {Cunha}, K. and {Smith}, V.~V. and {Margheim}, S. and {Schuler}, S. and {de Ara{\'u}jo}, F.~X. and {de la Reza}, R.},
        title = "{Measurements of the Isotopic Ratio $^{6}$Li/$^{7}$Li in Stars with Planets}",
      journal = {\apj},
         year = 2009,
        month = jun,
       volume = {698},
       number = {1},
        pages = {451-460},
          doi = {10.1088/0004-637X/698/1/451},
archivePrefix = {arXiv},
       eprint = {0903.4873},
 primaryClass = {astro-ph.SR},
       adsurl = {https://ui.adsabs.harvard.edu/abs/2009ApJ...698..451G}
}

@ARTICLE{Ghezzi2010a,
       author = {{Ghezzi}, L. and {Cunha}, K. and {Smith}, V.~V. and {de Ara{\'u}jo}, F.~X. and {Schuler}, S.~C. and {de la Reza}, R.},
        title = "{Stellar Parameters and Metallicities of Stars Hosting Jovian and Neptunian Mass Planets: A Possible Dependence of Planetary Mass on Metallicity}",
      journal = {\apj},
         year = {2010a},
        month = sep,
       volume = {720},
       number = {2},
        pages = {1290-1302},
          doi = {10.1088/0004-637X/720/2/1290},
archivePrefix = {arXiv},
       eprint = {1007.2681},
 primaryClass = {astro-ph.SR},
       adsurl = {https://ui.adsabs.harvard.edu/abs/2010ApJ...720.1290G}
}

@ARTICLE{Ghezzi2010b,
       author = {{Ghezzi}, L. and {Cunha}, K. and {Smith}, V.~V. and {de la Reza}, R.},
        title = "{Lithium Abundances in a Sample of Planet-hosting Dwarfs}",
      journal = {\apj},
         year = {2010b},
        month = nov,
       volume = {724},
       number = {1},
        pages = {154-164},
          doi = {10.1088/0004-637X/724/1/154},
archivePrefix = {arXiv},
       eprint = {1009.2130},
 primaryClass = {astro-ph.SR},
       adsurl = {https://ui.adsabs.harvard.edu/abs/2010ApJ...724..154G}
}

@INPROCEEDINGS{Castelli2003,
       author = {{Castelli}, F. and {Kurucz}, R.~L.},
        title = "{New Grids of ATLAS9 Model Atmospheres}",
    booktitle = {Modelling of Stellar Atmospheres},
         year = 2003,
       editor = {{Piskunov}, N. and {Weiss}, W.~W. and {Gray}, D.~F.},
       series = {IAU Symposium},
       volume = {210},
        month = jan,
        pages = {A20},
          doi = {10.48550/arXiv.astro-ph/0405087},
archivePrefix = {arXiv},
       eprint = {astro-ph/0405087},
 primaryClass = {astro-ph},
       adsurl = {https://ui.adsabs.harvard.edu/abs/2003IAUS..210P.A20C}
}

@PHDTHESIS{Sneden1973,
       author = {{Sneden}, Christopher Alan},
        title = "{Carbon and Nitrogen Abundances in Metal-Poor Stars.}",
       school = {University of Texas, Austin},
         year = 1973,
        month = jan,
       adsurl = {https://ui.adsabs.harvard.edu/abs/1973PhDT.......180S}
}

@ARTICLE{Sousa2015,
       author = {{Sousa}, S.~G. and {Santos}, N.~C. and {Adibekyan}, V. and {Delgado-Mena}, E. and {Israelian}, G.},
        title = "{ARES v2: new features and improved performance}",
      journal = {\aap},
         year = 2015,
        month = may,
       volume = {577},
          eid = {A67},
        pages = {A67},
          doi = {10.1051/0004-6361/201425463},
archivePrefix = {arXiv},
       eprint = {1504.02725},
 primaryClass = {astro-ph.IM},
       adsurl = {https://ui.adsabs.harvard.edu/abs/2015A&A...577A..67S}
}

@ARTICLE{Asplund2009,
       author = {{Asplund}, Martin and {Grevesse}, Nicolas and {Sauval}, A. Jacques and {Scott}, Pat},
        title = "{The Chemical Composition of the Sun}",
      journal = {\araa},
         year = 2009,
        month = sep,
       volume = {47},
       number = {1},
        pages = {481-522},
          doi = {10.1146/annurev.astro.46.060407.145222},
archivePrefix = {arXiv},
       eprint = {0909.0948},
 primaryClass = {astro-ph.SR},
       adsurl = {https://ui.adsabs.harvard.edu/abs/2009ARA&A..47..481A}
}

@ARTICLE{Ryabchikova2015,
       author = {{Ryabchikova}, T. and {Piskunov}, N. and {Kurucz}, R.~L. and {Stempels}, H.~C. and {Heiter}, U. and {Pakhomov}, Yu and {Barklem}, P.~S.},
        title = "{A major upgrade of the VALD database}",
      journal = {\physscr},
         year = 2015,
        month = may,
       volume = {90},
       number = {5},
          eid = {054005},
        pages = {054005},
          doi = {10.1088/0031-8949/90/5/054005},
       adsurl = {https://ui.adsabs.harvard.edu/abs/2015PhyS...90e4005R}
}

@ARTICLE{AllendePrieto2001,
       author = {{Allende Prieto}, Carlos and {Lambert}, David L. and {Asplund}, Martin},
        title = "{The Forbidden Abundance of Oxygen in the Sun}",
      journal = {\apjl},
         year = 2001,
        month = jul,
       volume = {556},
       number = {1},
        pages = {L63-L66},
          doi = {10.1086/322874},
archivePrefix = {arXiv},
       eprint = {astro-ph/0106360},
 primaryClass = {astro-ph},
       adsurl = {https://ui.adsabs.harvard.edu/abs/2001ApJ...556L..63A}
}

@ARTICLE{McQuillan2013,
       author = {{McQuillan}, A. and {Mazeh}, T. and {Aigrain}, S.},
        title = "{Stellar Rotation Periods of the Kepler Objects of Interest: A Dearth of Close-in Planets around Fast Rotators}",
      journal = {\apjl},
         year = 2013,
        month = sep,
       volume = {775},
       number = {1},
          eid = {L11},
        pages = {L11},
          doi = {10.1088/2041-8205/775/1/L11},
archivePrefix = {arXiv},
       eprint = {1308.1845},
 primaryClass = {astro-ph.EP},
       adsurl = {https://ui.adsabs.harvard.edu/abs/2013ApJ...775L..11M}
}

@ARTICLE{Matsuno2024,
       author = {{Matsuno}, Tadafumi and {Starkenburg}, Else and {Balbinot}, Eduardo and {Helmi}, Amina},
        title = "{Improving metallicity estimates for very metal-poor stars in the Gaia DR3 GSP-Spec catalog}",
      journal = {\aap},
         year = 2024,
        month = may,
       volume = {685},
          eid = {A59},
        pages = {A59},
          doi = {10.1051/0004-6361/202245762},
archivePrefix = {arXiv},
       eprint = {2212.11639},
 primaryClass = {astro-ph.SR},
       adsurl = {https://ui.adsabs.harvard.edu/abs/2024A&A...685A..59M}
}

@ARTICLE{Furlan2018,
       author = {{Furlan}, E. and {Ciardi}, D.~R. and {Cochran}, W.~D. and {Everett}, M.~E. and {Latham}, D.~W. and {Marcy}, G.~W. and {Buchhave}, L.~A. and {Endl}, M. and {Isaacson}, H. and {Petigura}, E.~A. and {Gautier}, III, T.~N. and {Huber}, D. and {Bieryla}, A. and {Borucki}, W.~J. and {Brugamyer}, E. and {Caldwell}, C. and {Cochran}, A. and {Howard}, A.~W. and {Howell}, S.~B. and {Johnson}, M.~C. and {MacQueen}, P.~J. and {Quinn}, S.~N. and {Robertson}, P. and {Mathur}, S. and {Batalha}, N.~M.},
        title = "{The Kepler Follow-up Observation Program. II. Stellar Parameters from Medium- and High-resolution Spectroscopy}",
      journal = {\apj},
         year = 2018,
        month = jul,
       volume = {861},
       number = {2},
          eid = {149},
        pages = {149},
          doi = {10.3847/1538-4357/aaca34},
archivePrefix = {arXiv},
       eprint = {1805.12089},
 primaryClass = {astro-ph.SR},
       adsurl = {https://ui.adsabs.harvard.edu/abs/2018ApJ...861..149F}
}

@ARTICLE{Chen2021,
       author = {{Chen}, Di-Chang and {Xie}, Ji-Wei and {Zhou}, Ji-Lin and {Dong}, Subo and {Liu}, Chao and {Wang}, Hai-Feng and {Xiang}, Mao-Sheng and {Huang}, Yang and {Luo}, Ali and {Zheng}, Zheng},
        title = "{Planets Across Space and Time (PAST). I. Characterizing the Memberships of Galactic Components and Stellar Ages: Revisiting the Kinematic Methods and Applying to Planet Host Stars}",
      journal = {\apj},
         year = 2021,
        month = mar,
       volume = {909},
       number = {2},
          eid = {115},
        pages = {115},
          doi = {10.3847/1538-4357/abd5be},
archivePrefix = {arXiv},
       eprint = {2102.09424},
 primaryClass = {astro-ph.EP},
       adsurl = {https://ui.adsabs.harvard.edu/abs/2021ApJ...909..115C}
}

@ARTICLE{deLaverny2025,
       author = {{de Laverny}, Patrick and {Ligi}, Roxanne and {Crida}, Aur{\'e}lien and {Recio-Blanco}, Alejandra and {Palicio}, Pedro A.},
        title = "{The Gaia spectroscopic catalogue of exoplanets and host stars}",
      journal = {\aap},
         year = 2025,
        month = jul,
       volume = {699},
          eid = {A100},
        pages = {A100},
          doi = {10.1051/0004-6361/202554739},
archivePrefix = {arXiv},
       eprint = {2505.22205},
 primaryClass = {astro-ph.EP},
       adsurl = {https://ui.adsabs.harvard.edu/abs/2025A&A...699A.100D}
}

@ARTICLE{Ye2025,
       author = {{Ye}, Xianhao and {Wu}, Wenbo and {Allende Prieto}, Carlos and {Aguado}, David S. and {Zhao}, Jingkun and {Gonz{\'a}lez Hern{\'a}ndez}, Jonay I. and {Rebolo}, Rafael and {Zhao}, Gang and {Li}, Zhuohan and {del Burgo}, Carlos and {Chen}, Yuqin},
        title = "{Mapping the Milky Way with Gaia Bp/Rp spectra: I. Systematic flux corrections and atmospheric parameters for 68 million stars}",
      journal = {\aap},
         year = 2025,
        month = mar,
       volume = {695},
          eid = {A75},
        pages = {A75},
          doi = {10.1051/0004-6361/202452871},
archivePrefix = {arXiv},
       eprint = {2411.19105},
 primaryClass = {astro-ph.GA},
       adsurl = {https://ui.adsabs.harvard.edu/abs/2025A&A...695A..75Y}
}

@ARTICLE{Boettner2024,
       author = {{Boettner}, C. and {Viswanathan}, A. and {Dayal}, P.},
        title = "{Exoplanets across galactic stellar populations with PLATO: Estimating exoplanet yields around FGK stars for the thin disk, thick disk, and stellar halo}",
      journal = {\aap},
         year = 2024,
        month = dec,
       volume = {692},
          eid = {A150},
        pages = {A150},
          doi = {10.1051/0004-6361/202451537},
archivePrefix = {arXiv},
       eprint = {2407.15917},
 primaryClass = {astro-ph.EP},
       adsurl = {https://ui.adsabs.harvard.edu/abs/2024A&A...692A.150B}
}

@ARTICLE{Johansson2003,
       author = {{Johansson}, S. and {Litz{\'e}n}, U. and {Lundberg}, H. and {Zhang}, Z.},
        title = "{Experimental f-Value and Isotopic Structure for the Ni I Line Blended with [O I] at 6300 {\r{A}}}",
      journal = {\apjl},
         year = 2003,
        month = feb,
       volume = {584},
       number = {2},
        pages = {L107-L110},
          doi = {10.1086/374037},
archivePrefix = {arXiv},
       eprint = {astro-ph/0301382},
 primaryClass = {astro-ph},
       adsurl = {https://ui.adsabs.harvard.edu/abs/2003ApJ...584L.107J}
}

@ARTICLE{Bensby2004,
       author = {{Bensby}, T. and {Feltzing}, S. and {Lundstr{\"o}m}, I.},
        title = "{Oxygen trends in the Galactic thin and thick disks}",
      journal = {\aap},
         year = 2004,
        month = feb,
       volume = {415},
        pages = {155-170},
          doi = {10.1051/0004-6361:20031655},
archivePrefix = {arXiv},
       eprint = {astro-ph/0310741},
 primaryClass = {astro-ph},
       adsurl = {https://ui.adsabs.harvard.edu/abs/2004A&A...415..155B}
}

@ARTICLE{Wallace2011,
       author = {{Wallace}, L. and {Hinkle}, K.~H. and {Livingston}, W.~C. and {Davis}, S.~P.},
        title = "{An Optical and Near-infrared (2958-9250 {\r{A}}) Solar Flux Atlas}",
      journal = {\apjs},
         year = 2011,
        month = jul,
       volume = {195},
       number = {1},
          eid = {6},
        pages = {6},
          doi = {10.1088/0067-0049/195/1/6},
       adsurl = {https://ui.adsabs.harvard.edu/abs/2011ApJS..195....6W}
}

@ARTICLE{Teske2013,
       author = {{Teske}, Johanna K. and {Schuler}, Simon C. and {Cunha}, Katia and {Smith}, Verne V. and {Griffith}, Caitlin A.},
        title = "{Carbon and Oxygen Abundances in the Hot Jupiter Exoplanet Host Star XO-2B and Its Binary Companion}",
      journal = {\apjl},
         year = 2013,
        month = may,
       volume = {768},
       number = {1},
          eid = {L12},
        pages = {L12},
          doi = {10.1088/2041-8205/768/1/L12},
       adsurl = {https://ui.adsabs.harvard.edu/abs/2013ApJ...768L..12T}
}

@ARTICLE{BertrandeLis2015,
       author = {{Bertran de Lis}, S. and {Delgado Mena}, E. and {Adibekyan}, V. Zh. and {Santos}, N.~C. and {Sousa}, S.~G.},
        title = "{Oxygen abundances in G- and F-type stars from HARPS. Comparison of [OI] 6300 {\r{A}} and OI 6158 {\r{A}}}",
      journal = {\aap},
         year = 2015,
        month = apr,
       volume = {576},
          eid = {A89},
        pages = {A89},
          doi = {10.1051/0004-6361/201424633},
archivePrefix = {arXiv},
       eprint = {1501.05805},
 primaryClass = {astro-ph.SR},
       adsurl = {https://ui.adsabs.harvard.edu/abs/2015A&A...576A..89B}
}

@ARTICLE{Mayor1995,
       author = {{Mayor}, Michel and {Queloz}, Didier},
        title = "{A Jupiter-mass companion to a solar-type star}",
      journal = {\nat},
         year = 1995,
        month = nov,
       volume = {378},
       number = {6555},
        pages = {355-359},
          doi = {10.1038/378355a0},
       adsurl = {https://ui.adsabs.harvard.edu/abs/1995Natur.378..355M}
}

@ARTICLE{Izidoro2017,
       author = {{Izidoro}, Andre and {Ogihara}, Masahiro and {Raymond}, Sean N. and {Morbidelli}, Alessandro and {Pierens}, Arnaud and {Bitsch}, Bertram and {Cossou}, Christophe and {Hersant}, Franck},
        title = "{Breaking the chains: hot super-Earth systems from migration and disruption of compact resonant chains}",
      journal = {\mnras},
         year = 2017,
        month = sep,
       volume = {470},
       number = {2},
        pages = {1750-1770},
          doi = {10.1093/mnras/stx1232},
archivePrefix = {arXiv},
       eprint = {1703.03634},
 primaryClass = {astro-ph.EP},
       adsurl = {https://ui.adsabs.harvard.edu/abs/2017MNRAS.470.1750I}
}

@ARTICLE{Pan2025,
       author = {{Pan}, Mengrui and {Liu}, Beibei and {Jiang}, Linjie and {Xie}, Jiwei and {Zhu}, Wei and {Ribas}, Ignasi},
        title = "{Dependence of Planet Populations on Stellar Mass and Metallicity: A Pebble-accretion-based Planet Population Synthesis Model}",
      journal = {\apj},
         year = 2025,
        month = may,
       volume = {985},
       number = {1},
          eid = {7},
        pages = {7},
          doi = {10.3847/1538-4357/adc7a9},
archivePrefix = {arXiv},
       eprint = {2504.00296},
 primaryClass = {astro-ph.EP},
       adsurl = {https://ui.adsabs.harvard.edu/abs/2025ApJ...985....7P}
}

@ARTICLE{Adibekyan2011,
       author = {{Adibekyan}, V. Zh. and {Santos}, N.~C. and {Sousa}, S.~G. and {Israelian}, G.},
        title = "{A new {\ensuremath{\alpha}}-enhanced super-solar metallicity population}",
      journal = {\aap},
         year = 2011,
        month = nov,
       volume = {535},
          eid = {L11},
        pages = {L11},
          doi = {10.1051/0004-6361/201118240},
archivePrefix = {arXiv},
       eprint = {1111.4936},
 primaryClass = {astro-ph.GA},
       adsurl = {https://ui.adsabs.harvard.edu/abs/2011A&A...535L..11A}
}

@ARTICLE{Mathur2025,
       author = {{Mathur}, Devansh and {Becker}, Juliette},
        title = "{Investigating the Formation of Planets Interior to In Situ Hot Jupiters}",
      journal = {\pasp},
         year = 2025,
        month = nov,
       volume = {137},
       number = {11},
          eid = {114402},
        pages = {114402},
          doi = {10.1088/1538-3873/ae135a},
archivePrefix = {arXiv},
       eprint = {2510.13527},
 primaryClass = {astro-ph.EP},
       adsurl = {https://ui.adsabs.harvard.edu/abs/2025PASP..137k4402M}
}

@ARTICLE{Dawson2018,
       author = {{Dawson}, Rebekah I. and {Johnson}, John Asher},
        title = "{Origins of Hot Jupiters}",
      journal = {\araa},
         year = 2018,
        month = sep,
       volume = {56},
        pages = {175-221},
          doi = {10.1146/annurev-astro-081817-051853},
archivePrefix = {arXiv},
       eprint = {1801.06117},
 primaryClass = {astro-ph.EP},
       adsurl = {https://ui.adsabs.harvard.edu/abs/2018ARA&A..56..175D}
}

@ARTICLE{Adibekyan2012b,
       author = {{Adibekyan}, V. Zh. and {Sousa}, S.~G. and {Santos}, N.~C. and {Delgado Mena}, E. and {Gonz{\'a}lez Hern{\'a}ndez}, J.~I. and {Israelian}, G. and {Mayor}, M. and {Khachatryan}, G.},
        title = "{Chemical abundances of 1111 FGK stars from the HARPS GTO planet search program. Galactic stellar populations and planets}",
      journal = {\aap},
         year = {2012b},
        month = sep,
       volume = {545},
          eid = {A32},
        pages = {A32},
          doi = {10.1051/0004-6361/201219401},
archivePrefix = {arXiv},
       eprint = {1207.2388},
 primaryClass = {astro-ph.EP},
       adsurl = {https://ui.adsabs.harvard.edu/abs/2012A&A...545A..32A}
}

@ARTICLE{Fischer2005,
       author = {{Fischer}, Debra A. and {Valenti}, Jeff},
        title = "{The Planet-Metallicity Correlation}",
      journal = {\apj},
         year = 2005,
        month = apr,
       volume = {622},
       number = {2},
        pages = {1102-1117},
          doi = {10.1086/428383},
       adsurl = {https://ui.adsabs.harvard.edu/abs/2005ApJ...622.1102F}
}

@ARTICLE{Hinkel2018,
       author = {{Hinkel}, Natalie R. and {Unterborn}, Cayman T.},
        title = "{The Star-Planet Connection. I. Using Stellar Composition to Observationally Constrain Planetary Mineralogy for the 10 Closest Stars}",
      journal = {\apj},
         year = 2018,
        month = jan,
       volume = {853},
       number = {1},
          eid = {83},
        pages = {83},
          doi = {10.3847/1538-4357/aaa5b4},
archivePrefix = {arXiv},
       eprint = {1709.08630},
 primaryClass = {astro-ph.EP},
       adsurl = {https://ui.adsabs.harvard.edu/abs/2018ApJ...853...83H}
}

@ARTICLE{Thiabaud2015b,
       author = {{Thiabaud}, Amaury and {Marboeuf}, Ulysse and {Alibert}, Yann and {Leya}, Ingo and {Mezger}, Klaus},
        title = "{Elemental ratios in stars vs planets}",
      journal = {\aap},
         year = {2015b},
        month = aug,
       volume = {580},
          eid = {A30},
        pages = {A30},
          doi = {10.1051/0004-6361/201525963},
archivePrefix = {arXiv},
       eprint = {1507.01343},
 primaryClass = {astro-ph.EP},
       adsurl = {https://ui.adsabs.harvard.edu/abs/2015A&A...580A..30T}
}

@ARTICLE{Buchhave2014,
       author = {{Buchhave}, Lars A. and {Bizzarro}, Martin and {Latham}, David W. and {Sasselov}, Dimitar and {Cochran}, William D. and {Endl}, Michael and {Isaacson}, Howard and {Juncher}, Diana and {Marcy}, Geoffrey W.},
        title = "{Three regimes of extrasolar planet radius inferred from host star metallicities}",
      journal = {\nat},
         year = 2014,
        month = may,
       volume = {509},
       number = {7502},
        pages = {593-595},
          doi = {10.1038/nature13254},
archivePrefix = {arXiv},
       eprint = {1405.7695},
 primaryClass = {astro-ph.EP},
       adsurl = {https://ui.adsabs.harvard.edu/abs/2014Natur.509..593B}
}

@ARTICLE{Wang2015,
       author = {{Wang}, Ji and {Fischer}, Debra A.},
        title = "{Revealing a Universal Planet-Metallicity Correlation for Planets of Different Sizes Around Solar-type Stars}",
      journal = {\aj},
         year = 2015,
        month = jan,
       volume = {149},
       number = {1},
          eid = {14},
        pages = {14},
          doi = {10.1088/0004-6256/149/1/14},
archivePrefix = {arXiv},
       eprint = {1310.7830},
 primaryClass = {astro-ph.EP},
       adsurl = {https://ui.adsabs.harvard.edu/abs/2015AJ....149...14W}
}

@ARTICLE{Wilson2022,
       author = {{Wilson}, Robert F. and {Ca{\~n}as}, Caleb I. and {Majewski}, Steven R. and {Cunha}, Katia and {Smith}, Verne V. and {Bender}, Chad F. and {Mahadevan}, Suvrath and {Fleming}, Scott W. and {Teske}, Johanna and {Ghezzi}, Luan and {J{\"o}nsson}, Henrik and {Beaton}, Rachael L. and {Hasselquist}, Sten and {Stassun}, Keivan and {Nitschelm}, Christian and {Garc{\'\i}a-Hern{\'a}ndez}, D.~A. and {Hayes}, Christian R. and {Tayar}, Jamie},
        title = "{The Influence of 10 Unique Chemical Elements in Shaping the Distribution of Kepler Planets}",
      journal = {\aj},
         year = 2022,
        month = mar,
       volume = {163},
       number = {3},
          eid = {128},
        pages = {128},
          doi = {10.3847/1538-3881/ac3a06},
archivePrefix = {arXiv},
       eprint = {2111.01753},
 primaryClass = {astro-ph.EP},
       adsurl = {https://ui.adsabs.harvard.edu/abs/2022AJ....163..128W}
}

@INPROCEEDINGS{Matteucci2016,
       author = {{Matteucci}, Francesca},
        title = "{Introduction to Galactic Chemical Evolution}",
    booktitle = {Journal of Physics Conference Series},
         year = 2016,
       series = {Journal of Physics Conference Series},
       volume = {703},
        month = apr,
          eid = {012004},
        pages = {012004},
          doi = {10.1088/1742-6596/703/1/012004},
archivePrefix = {arXiv},
       eprint = {1602.01004},
 primaryClass = {astro-ph.GA},
       adsurl = {https://ui.adsabs.harvard.edu/abs/2016JPhCS.703a2004M}
}

@ARTICLE{Hinkel2014,
       author = {{Hinkel}, Natalie R. and {Timmes}, F.~X. and {Young}, Patrick A. and {Pagano}, Michael D. and {Turnbull}, Margaret C.},
        title = "{Stellar Abundances in the Solar Neighborhood: The Hypatia Catalog}",
      journal = {\aj},
         year = 2014,
        month = sep,
       volume = {148},
       number = {3},
          eid = {54},
        pages = {54},
          doi = {10.1088/0004-6256/148/3/54},
archivePrefix = {arXiv},
       eprint = {1405.6719},
 primaryClass = {astro-ph.SR},
       adsurl = {https://ui.adsabs.harvard.edu/abs/2014AJ....148...54H}
}

@ARTICLE{Wilson2018,
       author = {{Wilson}, Robert F. and {Teske}, Johanna and {Majewski}, Steven R. and {Cunha}, Katia and {Smith}, Verne and {Souto}, Diogo and {Bender}, Chad and {Mahadevan}, Suvrath and {Troup}, Nicholas and {Allende Prieto}, Carlos and {Stassun}, Keivan G. and {Skrutskie}, Michael F. and {Almeida}, Andr{\'e}s and {Garc{\'\i}a-Hern{\'a}ndez}, D.~A. and {Zamora}, Olga and {Brinkmann}, Jonathan},
        title = "{Elemental Abundances of Kepler Objects of Interest in APOGEE. I. Two Distinct Orbital Period Regimes Inferred from Host Star Iron Abundances}",
      journal = {\aj},
         year = 2018,
        month = feb,
       volume = {155},
       number = {2},
          eid = {68},
        pages = {68},
          doi = {10.3847/1538-3881/aa9f27},
archivePrefix = {arXiv},
       eprint = {1712.01198},
 primaryClass = {astro-ph.EP},
       adsurl = {https://ui.adsabs.harvard.edu/abs/2018AJ....155...68W}
}

@ARTICLE{Buchhave2012,
       author = {{Buchhave}, Lars A. and {Latham}, David W. and {Johansen}, Anders and {Bizzarro}, Martin and {Torres}, Guillermo and {Rowe}, Jason F. and {Batalha}, Natalie M. and {Borucki}, William J. and {Brugamyer}, Erik and {Caldwell}, Caroline and {Bryson}, Stephen T. and {Ciardi}, David R. and {Cochran}, William D. and {Endl}, Michael and {Esquerdo}, Gilbert A. and {Ford}, Eric B. and {Geary}, John C. and {Gilliland}, Ronald L. and {Hansen}, Terese and {Isaacson}, Howard and {Laird}, John B. and {Lucas}, Philip W. and {Marcy}, Geoffrey W. and {Morse}, Jon A. and {Robertson}, Paul and {Shporer}, Avi and {Stefanik}, Robert P. and {Still}, Martin and {Quinn}, Samuel N.},
        title = "{An abundance of small exoplanets around stars with a wide range of metallicities}",
      journal = {\nat},
         year = 2012,
        month = jun,
       volume = {486},
       number = {7403},
        pages = {375-377},
          doi = {10.1038/nature11121},
       adsurl = {https://ui.adsabs.harvard.edu/abs/2012Natur.486..375B}
}

@ARTICLE{Udry2006,
       author = {{Udry}, S. and {Mayor}, M. and {Benz}, W. and {Bertaux}, J. -L. and {Bouchy}, F. and {Lovis}, C. and {Mordasini}, C. and {Pepe}, F. and {Queloz}, D. and {Sivan}, J. -P.},
        title = "{The HARPS search for southern extra-solar planets. V. A 14 Earth-masses planet orbiting HD{\,}4308}",
      journal = {\aap},
         year = 2006,
        month = feb,
       volume = {447},
       number = {1},
        pages = {361-367},
          doi = {10.1051/0004-6361:20054084},
archivePrefix = {arXiv},
       eprint = {astro-ph/0510354},
 primaryClass = {astro-ph},
       adsurl = {https://ui.adsabs.harvard.edu/abs/2006A&A...447..361U}
}

@ARTICLE{DelgadoMena2021,
       author = {{Delgado Mena}, E. and {Adibekyan}, V. and {Santos}, N.~C. and {Tsantaki}, M. and {Gonz{\'a}lez Hern{\'a}ndez}, J.~I. and {Sousa}, S.~G. and {Bertr{\'a}n de Lis}, S.},
        title = "{Chemical abundances of 1111 FGK stars from the HARPS GTO planet search program. IV. Carbon and C/O ratios for Galactic stellar populations and planet hosts}",
      journal = {\aap},
         year = 2021,
        month = nov,
       volume = {655},
          eid = {A99},
        pages = {A99},
          doi = {10.1051/0004-6361/202141588},
archivePrefix = {arXiv},
       eprint = {2109.04844},
 primaryClass = {astro-ph.SR},
       adsurl = {https://ui.adsabs.harvard.edu/abs/2021A&A...655A..99D}
}

@ARTICLE{Amarsi2019,
       author = {{Amarsi}, A.~M. and {Nissen}, P.~E. and {Sk{\'u}lad{\'o}ttir}, {\'A}.},
        title = "{Carbon, oxygen, and iron abundances in disk and halo stars. Implications of 3D non-LTE spectral line formation}",
      journal = {\aap},
         year = 2019,
        month = oct,
       volume = {630},
          eid = {A104},
        pages = {A104},
          doi = {10.1051/0004-6361/201936265},
archivePrefix = {arXiv},
       eprint = {1908.10319},
 primaryClass = {astro-ph.SR},
       adsurl = {https://ui.adsabs.harvard.edu/abs/2019A&A...630A.104A}
}

@ARTICLE{Bergemann2021,
      author = {{Bergemann}, Maria and {Hoppe}, Richard and {Semenova}, Ekaterina and {Carlsson}, Mats and {Yakovleva}, Svetlana A. and {Voronov}, Yaroslav V. and {Bautista}, Manuel and {Nemer}, Ahmad and {Belyaev}, Andrey K. and {Leenaarts}, Jorrit and {Mashonkina}, Lyudmila and {Reiners}, Ansgar and {Ellwarth}, Monika},
        title = "{Solar oxygen abundance}",
      journal = {\mnras},
         year = 2021,
        month = dec,
      volume = {508},
      number = {2},
        pages = {2236-2253},
          doi = {10.1093/mnras/stab2160},
archivePrefix = {arXiv},
      eprint = {2109.01143},
 primaryClass = {astro-ph.SR},
      adsurl = {https://ui.adsabs.harvard.edu/abs/2021MNRAS.508.2236B}
}

@ARTICLE{Steffen2015,
       author = {{Steffen}, M. and {Prakapavi{\v{c}}ius}, D. and {Caffau}, E. and {Ludwig}, H. -G. and {Bonifacio}, P. and {Cayrel}, R. and {Ku{\v{c}}inskas}, A. and {Livingston}, W.~C.},
        title = "{The photospheric solar oxygen project. IV. 3D-NLTE investigation of the 777 nm triplet lines}",
      journal = {\aap},
         year = 2015,
        month = nov,
       volume = {583},
          eid = {A57},
        pages = {A57},
          doi = {10.1051/0004-6361/201526406},
archivePrefix = {arXiv},
       eprint = {1508.03487},
 primaryClass = {astro-ph.SR},
       adsurl = {https://ui.adsabs.harvard.edu/abs/2015A&A...583A..57S}
}

@ARTICLE{Lambert1978,
       author = {{Lambert}, D.~L.},
        title = "{The abundances of the elements in the solar photosphere - VIII. Revised abundances of carbon, nitrogen and oxygen.}",
      journal = {\mnras},
         year = 1978,
        month = jan,
       volume = {182},
        pages = {249-272},
          doi = {10.1093/mnras/182.2.249},
       adsurl = {https://ui.adsabs.harvard.edu/abs/1978MNRAS.182..249L}
}

@ARTICLE{Asplund2004,
       author = {{Asplund}, M. and {Grevesse}, N. and {Sauval}, A.~J. and {Allende Prieto}, C. and {Kiselman}, D.},
        title = "{Line formation in solar granulation. IV. [O I], O I and OH lines and the photospheric O abundance}",
      journal = {\aap},
         year = 2004,
        month = apr,
       volume = {417},
        pages = {751-768},
          doi = {10.1051/0004-6361:20034328},
archivePrefix = {arXiv},
       eprint = {astro-ph/0312290},
 primaryClass = {astro-ph},
       adsurl = {https://ui.adsabs.harvard.edu/abs/2004A&A...417..751A}
}

@ARTICLE{Caffau2008,
       author = {{Caffau}, E. and {Ludwig}, H. -G. and {Steffen}, M. and {Ayres}, T.~R. and {Bonifacio}, P. and {Cayrel}, R. and {Freytag}, B. and {Plez}, B.},
        title = "{The photospheric solar oxygen project. I. Abundance analysis of atomic lines and influence of atmospheric models}",
      journal = {\aap},
         year = 2008,
        month = sep,
       volume = {488},
       number = {3},
        pages = {1031-1046},
          doi = {10.1051/0004-6361:200809885},
archivePrefix = {arXiv},
       eprint = {0805.4398},
 primaryClass = {astro-ph},
       adsurl = {https://ui.adsabs.harvard.edu/abs/2008A&A...488.1031C}
}

@ARTICLE{CubasArmas2020,
       author = {{Cubas Armas}, M. and {Asensio Ramos}, A. and {Socas-Navarro}, H.},
        title = "{Spatially resolved measurements of the solar photospheric oxygen abundance}",
      journal = {\aap},
         year = 2020,
        month = nov,
       volume = {643},
          eid = {A142},
        pages = {A142},
          doi = {10.1051/0004-6361/202037849},
archivePrefix = {arXiv},
       eprint = {2010.02151},
 primaryClass = {astro-ph.SR},
       adsurl = {https://ui.adsabs.harvard.edu/abs/2020A&A...643A.142C}
}

@ARTICLE{Asplund2021,
       author = {{Asplund}, M. and {Amarsi}, A.~M. and {Grevesse}, N.},
        title = "{The chemical make-up of the Sun: A 2020 vision}",
      journal = {\aap},
         year = 2021,
        month = sep,
       volume = {653},
          eid = {A141},
        pages = {A141},
          doi = {10.1051/0004-6361/202140445},
archivePrefix = {arXiv},
       eprint = {2105.01661},
 primaryClass = {astro-ph.SR},
       adsurl = {https://ui.adsabs.harvard.edu/abs/2021A&A...653A.141A}
}

@ARTICLE{Caffau2005,
       author = {{Caffau}, E. and {Bonifacio}, P. and {Faraggiana}, R. and {Fran{\c{c}}ois}, P. and {Gratton}, R.~G. and {Barbieri}, M.},
        title = "{Sulphur abundance in Galactic stars}",
      journal = {\aap},
         year = 2005,
        month = oct,
       volume = {441},
       number = {2},
        pages = {533-548},
          doi = {10.1051/0004-6361:20052905},
archivePrefix = {arXiv},
       eprint = {astro-ph/0507030},
 primaryClass = {astro-ph},
       adsurl = {https://ui.adsabs.harvard.edu/abs/2005A&A...441..533C}
}

@ARTICLE{Spite2011,
       author = {{Spite}, M. and {Caffau}, E. and {Andrievsky}, S.~M. and {Korotin}, S.~A. and {Depagne}, E. and {Spite}, F. and {Bonifacio}, P. and {Ludwig}, H. -G. and {Cayrel}, R. and {Fran{\c{c}}ois}, P. and {Hill}, V. and {Plez}, B. and {Andersen}, J. and {Barbuy}, B. and {Beers}, T.~C. and {Molaro}, P. and {Nordstr{\"o}m}, B. and {Primas}, F.},
        title = "{First stars. XIV. Sulfur abundances in extremely metal-poor stars}",
      journal = {\aap},
         year = 2011,
        month = apr,
       volume = {528},
          eid = {A9},
        pages = {A9},
          doi = {10.1051/0004-6361/201015926},
archivePrefix = {arXiv},
       eprint = {1012.4358},
 primaryClass = {astro-ph.SR},
       adsurl = {https://ui.adsabs.harvard.edu/abs/2011A&A...528A...9S}
}

@ARTICLE{Limongi2003,
       author = {{Limongi}, M. and {Chieffi}, A.},
        title = "{Massive stars: presupernova evolution and explosive nucleosynthesis}",
      journal = {Memorie della Societa Astronomica Italiana Supplementi},
         year = 2003,
        month = jan,
       volume = {3},
        pages = {58},
       adsurl = {https://ui.adsabs.harvard.edu/abs/2003MSAIS...3...58L}
}

@ARTICLE{Matrozis2013,
       author = {{Matrozis}, E. and {Ryde}, N. and {Dupree}, A.~K.},
        title = "{Galactic chemical evolution of sulphur. Sulphur abundances from the [S i] {\ensuremath{\lambda}}1082 nm line in giants}",
      journal = {\aap},
         year = 2013,
        month = nov,
       volume = {559},
          eid = {A115},
        pages = {A115},
          doi = {10.1051/0004-6361/201322317},
archivePrefix = {arXiv},
       eprint = {1309.0114},
 primaryClass = {astro-ph.GA},
       adsurl = {https://ui.adsabs.harvard.edu/abs/2013A&A...559A.115M}
}

@ARTICLE{CostaSilva2020,
       author = {{Costa Silva}, A.~R. and {Delgado Mena}, E. and {Tsantaki}, M.},
        title = "{Chemical abundances of 1111 FGK stars from the HARPS-GTO planet search sample. III. Sulfur}",
      journal = {\aap},
         year = 2020,
        month = feb,
       volume = {634},
          eid = {A136},
        pages = {A136},
          doi = {10.1051/0004-6361/201936523},
archivePrefix = {arXiv},
       eprint = {1912.08659},
 primaryClass = {astro-ph.SR},
       adsurl = {https://ui.adsabs.harvard.edu/abs/2020A&A...634A.136C}
}

@ARTICLE{Shields2019,
       author = {{Shields}, Aomawa L.},
        title = "{The Climates of Other Worlds: A Review of the Emerging Field of Exoplanet Climatology}",
      journal = {\apjs},
         year = 2019,
        month = aug,
       volume = {243},
       number = {2},
          eid = {30},
        pages = {30},
          doi = {10.3847/1538-4365/ab2fe7},
archivePrefix = {arXiv},
       eprint = {1909.04046},
 primaryClass = {astro-ph.EP},
       adsurl = {https://ui.adsabs.harvard.edu/abs/2019ApJS..243...30S}
}

@ARTICLE{Ehlmann2016,
       author = {{Ehlmann}, B.~L. and {Anderson}, F.~S. and {Andrews-Hanna}, J. and {Catling}, D.~C. and {Christensen}, P.~R. and {Cohen}, B.~A. and {Dressing}, C.~D. and {Edwards}, C.~S. and {Elkins-Tanton}, L.~T. and {Farley}, K.~A. and {Fassett}, C.~I. and {Fischer}, W.~W. and {Fraeman}, A.~A. and {Golombek}, M.~P. and {Hamilton}, V.~E. and {Hayes}, A.~G. and {Herd}, C.~D.~K. and {Horgan}, B. and {Hu}, R. and {Jakosky}, B.~M. and {Johnson}, J.~R. and {Kasting}, J.~F. and {Kerber}, L. and {Kinch}, K.~M. and {Kite}, E.~S. and {Knutson}, H.~A. and {Lunine}, J.~I. and {Mahaffy}, P.~R. and {Mangold}, N. and {McCubbin}, F.~M. and {Mustard}, J.~F. and {Niles}, P.~B. and {Quantin-Nataf}, C. and {Rice}, M.~S. and {Stack}, K.~M. and {Stevenson}, D.~J. and {Stewart}, S.~T. and {Toplis}, M.~J. and {Usui}, T. and {Weiss}, B.~P. and {Werner}, S.~C. and {Wordsworth}, R.~D. and {Wray}, J.~J. and {Yingst}, R.~A. and {Yung}, Y.~L. and {Zahnle}, K.~J.},
        title = "{The sustainability of habitability on terrestrial planets: Insights, questions, and needed measurements from Mars for understanding the evolution of Earth-like worlds}",
      journal = {Journal of Geophysical Research (Planets)},
         year = 2016,
        month = oct,
       volume = {121},
       number = {10},
        pages = {1927-1961},
          doi = {10.1002/2016JE005134},
       adsurl = {https://ui.adsabs.harvard.edu/abs/2016JGRE..121.1927E}
}

@ARTICLE{Me2021,
       author = {{M{\'e}ndez}, Abel and {Rivera-Valent{\'\i}n}, Edgard G. and {Schulze-Makuch}, Dirk and {Filiberto}, Justin and {Ram{\'\i}rez}, Ramses M. and {Wood}, Tana E. and {D{\'a}vila}, Alfonso and {McKay}, Chris and {Ceballos}, Kevin N. Ortiz and {Jusino-Maldonado}, Marcos and {Torres-Santiago}, Nicole J. and {Nery}, Guillermo and {Heller}, Ren{\'e} and {Byrne}, Paul K. and {Malaska}, Michael J. and {Nathan}, Erica and {Sim{\~o}es}, Marta Filipa and {Antunes}, Andr{\'e} and {Mart{\'\i}nez-Fr{\'\i}as}, Jes{\'u}s and {Carone}, Ludmila and {Izenberg}, Noam R. and {Atri}, Dimitra and {Chitty}, Humberto Itic Carvajal and {Nowajewski-Barra}, Priscilla and {Rivera-Hern{\'a}ndez}, Frances and {Brown}, Corine Y. and {Lynch}, Kennda L. and {Catling}, David and {Zuluaga}, Jorge I. and {Salazar}, Juan F. and {Chen}, Howard and {Gonz{\'a}lez}, Grizelle and {Jagadeesh}, Madhu Kashyap and {Haqq-Misra}, Jacob},
        title = "{Habitability Models for Astrobiology}",
      journal = {Astrobiology},
         year = 2021,
        month = aug,
       volume = {21},
       number = {8},
        pages = {1017-1027},
          doi = {10.1089/ast.2020.2342},
archivePrefix = {arXiv},
       eprint = {2108.05417},
 primaryClass = {astro-ph.EP},
       adsurl = {https://ui.adsabs.harvard.edu/abs/2021AsBio..21.1017M}
}

@INPROCEEDINGS{Krijt2023,
       author = {{Krijt}, S. and {Kama}, M. and {McClure}, M. and {Teske}, J. and {Bergin}, E.~A. and {Shorttle}, O. and {Walsh}, K.~J. and {Raymond}, S.~N.},
        title = "{Chemical Habitability: Supply and Retention of Life's Essential Elements During Planet Formation}",
    booktitle = {Protostars and Planets VII},
         year = 2023,
       editor = {{Inutsuka}, S. and {Aikawa}, Y. and {Muto}, T. and {Tomida}, K. and {Tamura}, M.},
       series = {Astronomical Society of the Pacific Conference Series},
       volume = {534},
        month = jul,
        pages = {1031},
          doi = {10.48550/arXiv.2203.10056},
archivePrefix = {arXiv},
       eprint = {2203.10056},
 primaryClass = {astro-ph.EP},
       adsurl = {https://ui.adsabs.harvard.edu/abs/2023ASPC..534.1031K}
}

@ARTICLE{Luck2005,
       author = {{Luck}, R. Earle and {Heiter}, Ulrike},
        title = "{Stars within 15 Parsecs: Abundances for a Northern Sample}",
      journal = {\aj},
         year = 2005,
        month = feb,
       volume = {129},
       number = {2},
        pages = {1063-1083},
          doi = {10.1086/427250},
       adsurl = {https://ui.adsabs.harvard.edu/abs/2005AJ....129.1063L}
}

@INPROCEEDINGS{Cosentino2012,
       author = {{Cosentino}, Rosario and {Lovis}, Christophe and {Pepe}, Francesco and {Collier Cameron}, Andrew and {Latham}, David W. and {Molinari}, Emilio and {Udry}, Stephane and {Bezawada}, Naidu and {Black}, Martin and {Born}, Andy and {Buchschacher}, Nicolas and {Charbonneau}, Dave and {Figueira}, Pedro and {Fleury}, Michel and {Galli}, Alberto and {Gallie}, Angus and {Gao}, Xiaofeng and {Ghedina}, Adriano and {Gonzalez}, Carlos and {Gonzalez}, Manuel and {Guerra}, Jose and {Henry}, David and {Horne}, Keith and {Hughes}, Ian and {Kelly}, Dennis and {Lodi}, Marcello and {Lunney}, David and {Maire}, Charles and {Mayor}, Michel and {Micela}, Giusi and {Ordway}, Mark P. and {Peacock}, John and {Phillips}, David and {Piotto}, Giampaolo and {Pollacco}, Don and {Queloz}, Didier and {Rice}, Ken and {Riverol}, Carlos and {Riverol}, Luis and {San Juan}, Jose and {Sasselov}, Dimitar and {Segransan}, Damien and {Sozzetti}, Alessandro and {Sosnowska}, Danuta and {Stobie}, Brian and {Szentgyorgyi}, Andrew and {Vick}, Andy and {Weber}, Luc},
        title = "{Harps-N: the new planet hunter at TNG}",
    booktitle = {Ground-based and Airborne Instrumentation for Astronomy IV},
         year = 2012,
       editor = {{McLean}, Ian S. and {Ramsay}, Suzanne K. and {Takami}, Hideki},
       series = {Society of Photo-Optical Instrumentation Engineers (SPIE) Conference Series},
       volume = {8446},
        month = sep,
          eid = {84461V},
        pages = {84461V},
          doi = {10.1117/12.925738},
       adsurl = {https://ui.adsabs.harvard.edu/abs/2012SPIE.8446E..1VC}
}

@ARTICLE{Biazzo2022,
       author = {{Biazzo}, K. and {D'Orazi}, V. and {Desidera}, S. and {Turrini}, D. and {Benatti}, S. and {Gratton}, R. and {Magrini}, L. and {Sozzetti}, A. and {Baratella}, M. and {Bonomo}, A.~S. and {Borsa}, F. and {Claudi}, R. and {Covino}, E. and {Damasso}, M. and {Di Mauro}, M.~P. and {Lanza}, A.~F. and {Maggio}, A. and {Malavolta}, L. and {Maldonado}, J. and {Marzari}, F. and {Micela}, G. and {Poretti}, E. and {Vitello}, F. and {Affer}, L. and {Bignamini}, A. and {Carleo}, I. and {Cosentino}, R. and {Fiorenzano}, A.~F.~M. and {Giacobbe}, P. and {Harutyunyan}, A. and {Leto}, G. and {Mancini}, L. and {Molinari}, E. and {Molinaro}, M. and {Nardiello}, D. and {Nascimbeni}, V. and {Pagano}, I. and {Pedani}, M. and {Piotto}, G. and {Rainer}, M. and {Scandariato}, G.},
        title = "{The GAPS Programme at TNG. XXXV. Fundamental properties of transiting exoplanet host stars}",
      journal = {\aap},
         year = 2022,
        month = aug,
       volume = {664},
          eid = {A161},
        pages = {A161},
          doi = {10.1051/0004-6361/202243467},
archivePrefix = {arXiv},
       eprint = {2205.15796},
 primaryClass = {astro-ph.SR},
       adsurl = {https://ui.adsabs.harvard.edu/abs/2022A&A...664A.161B}
}

@ARTICLE{Soubiran2022,
       author = {{Soubiran}, C. and {Brouillet}, N. and {Casamiquela}, L.},
        title = "{Assessment of [Fe/H] determinations for FGK stars in spectroscopic surveys}",
      journal = {\aap},
         year = 2022,
        month = jul,
       volume = {663},
          eid = {A4},
        pages = {A4},
          doi = {10.1051/0004-6361/202142409},
archivePrefix = {arXiv},
       eprint = {2112.07545},
 primaryClass = {astro-ph.SR},
       adsurl = {https://ui.adsabs.harvard.edu/abs/2022A&A...663A...4S}
}

@ARTICLE{Sousa2007,
       author = {{Sousa}, S.~G. and {Santos}, N.~C. and {Israelian}, G. and {Mayor}, M. and {Monteiro}, M.~J.~P.~F.~G.},
        title = "{A new code for automatic determination of equivalent widths: Automatic Routine for line Equivalent widths in stellar Spectra (ARES)}",
      journal = {\aap},
         year = 2007,
        month = jul,
       volume = {469},
       number = {2},
        pages = {783-791},
          doi = {10.1051/0004-6361:20077288},
archivePrefix = {arXiv},
       eprint = {astro-ph/0703696},
 primaryClass = {astro-ph},
       adsurl = {https://ui.adsabs.harvard.edu/abs/2007A&A...469..783S}
}

@ARTICLE{Sousa2011,
       author = {{Sousa}, S.~G. and {Santos}, N.~C. and {Israelian}, G. and {Mayor}, M. and {Udry}, S.},
        title = "{Spectroscopic stellar parameters for 582 FGK stars in the HARPS volume-limited sample. Revising the metallicity-planet correlation}",
      journal = {\aap},
         year = 2011,
        month = sep,
       volume = {533},
          eid = {A141},
        pages = {A141},
          doi = {10.1051/0004-6361/201117699},
archivePrefix = {arXiv},
       eprint = {1108.5279},
 primaryClass = {astro-ph.EP},
       adsurl = {https://ui.adsabs.harvard.edu/abs/2011A&A...533A.141S}
}

@ARTICLE{Soubiran2016,
       author = {{Soubiran}, Caroline and {Le Campion}, Jean-Fran{\c{c}}ois and {Brouillet}, Nathalie and {Chemin}, Laurent},
        title = "{The PASTEL catalogue: 2016 version}",
      journal = {\aap},
         year = 2016,
        month = jun,
       volume = {591},
          eid = {A118},
        pages = {A118},
          doi = {10.1051/0004-6361/201628497},
archivePrefix = {arXiv},
       eprint = {1605.07384},
 primaryClass = {astro-ph.SR},
       adsurl = {https://ui.adsabs.harvard.edu/abs/2016A&A...591A.118S}
}

@ARTICLE{Magrini2022,
       author = {{Magrini}, L. and {Danielski}, C. and {Bossini}, D. and {Rainer}, M. and {Turrini}, D. and {Benatti}, S. and {Brucalassi}, A. and {Tsantaki}, M. and {Delgado Mena}, E. and {Sanna}, N. and {Biazzo}, K. and {Campante}, T.~L. and {Van der Swaelmen}, M. and {Sousa}, S.~G. and {He{\l}miniak}, K.~G. and {Neitzel}, A.~W. and {Adibekyan}, V. and {Bruno}, G. and {Casali}, G.},
        title = "{Ariel stellar characterisation. I. Homogeneous stellar parameters of 187 FGK planet host stars: Description and validation of the method}",
      journal = {\aap},
         year = 2022,
        month = jul,
       volume = {663},
          eid = {A161},
        pages = {A161},
          doi = {10.1051/0004-6361/202243405},
archivePrefix = {arXiv},
       eprint = {2204.08825},
 primaryClass = {astro-ph.SR},
       adsurl = {https://ui.adsabs.harvard.edu/abs/2022A&A...663A.161M}
}

@ARTICLE{Amarsi2015,
       author = {{Amarsi}, A.~M. and {Asplund}, M. and {Collet}, R. and {Leenaarts}, J.},
        title = "{The Galactic chemical evolution of oxygen inferred from 3D non-LTE spectral-line-formation calculations.}",
      journal = {\mnras},
         year = 2015,
        month = nov,
       volume = {454},
        pages = {L11-L15},
          doi = {10.1093/mnrasl/slv122},
archivePrefix = {arXiv},
       eprint = {1508.04857},
 primaryClass = {astro-ph.SR},
       adsurl = {https://ui.adsabs.harvard.edu/abs/2015MNRAS.454L..11A}
}

@ARTICLE{Teske2024,
       author = {{Teske}, Johanna K.},
        title = "{The Star{\textendash}Planet Composition Connection}",
      journal = {\araa},
         year = 2024,
        month = sep,
       volume = {62},
       number = {1},
        pages = {333-368},
          doi = {10.1146/annurev-astro-071221-053007},
       adsurl = {https://ui.adsabs.harvard.edu/abs/2024ARA&A..62..333T}
}

@ARTICLE{Gonzalez1997,
       author = {{Gonzalez}, Guillermo},
        title = "{The stellar metallicity-giant planet connection}",
      journal = {\mnras},
         year = 1997,
        month = feb,
       volume = {285},
       number = {2},
        pages = {403-412},
          doi = {10.1093/mnras/285.2.403},
       adsurl = {https://ui.adsabs.harvard.edu/abs/1997MNRAS.285..403G}
}

@ARTICLE{Mortier2013,
       author = {{Mortier}, A. and {Santos}, N.~C. and {Sousa}, S.~G. and {Adibekyan}, V. Zh. and {Delgado Mena}, E. and {Tsantaki}, M. and {Israelian}, G. and {Mayor}, M.},
        title = "{New and updated stellar parameters for 71 evolved planet hosts. On the metallicity-giant planet connection}",
      journal = {\aap},
         year = 2013,
        month = sep,
       volume = {557},
          eid = {A70},
        pages = {A70},
          doi = {10.1051/0004-6361/201321641},
archivePrefix = {arXiv},
       eprint = {1307.7870},
 primaryClass = {astro-ph.EP},
       adsurl = {https://ui.adsabs.harvard.edu/abs/2013A&A...557A..70M}
}

@ARTICLE{Ida2004,
       author = {{Ida}, S. and {Lin}, D.~N.~C.},
        title = "{Toward a Deterministic Model of Planetary Formation. I. A Desert in the Mass and Semimajor Axis Distributions of Extrasolar Planets}",
      journal = {\apj},
         year = 2004,
        month = mar,
       volume = {604},
       number = {1},
        pages = {388-413},
          doi = {10.1086/381724},
archivePrefix = {arXiv},
       eprint = {astro-ph/0312144},
 primaryClass = {astro-ph},
       adsurl = {https://ui.adsabs.harvard.edu/abs/2004ApJ...604..388I}
}

@ARTICLE{Adibekyan2019,
       author = {{Adibekyan}, Vardan},
        title = "{Heavy Metal Rules. I. Exoplanet Incidence and Metallicity}",
      journal = {Geosciences},
         year = 2019,
        month = feb,
       volume = {9},
       number = {3},
        pages = {105},
          doi = {10.3390/geosciences9030105},
archivePrefix = {arXiv},
       eprint = {1902.04493},
 primaryClass = {astro-ph.EP},
       adsurl = {https://ui.adsabs.harvard.edu/abs/2019Geosc...9..105A}
}

@ARTICLE{Mulders2016,
       author = {{Mulders}, Gijs D. and {Pascucci}, Ilaria and {Apai}, D{\'a}niel and {Frasca}, Antonio and {Molenda-{\.Z}akowicz}, Joanna},
        title = "{A Super-solar Metallicity for Stars with Hot Rocky Exoplanets}",
      journal = {\aj},
         year = 2016,
        month = dec,
       volume = {152},
       number = {6},
          eid = {187},
        pages = {187},
          doi = {10.3847/0004-6256/152/6/187},
archivePrefix = {arXiv},
       eprint = {1609.05898},
 primaryClass = {astro-ph.EP},
       adsurl = {https://ui.adsabs.harvard.edu/abs/2016AJ....152..187M}
}

@ARTICLE{Leemker2026,
       author = {{Leemker}, M. and {Facchini}, S. and {Curone}, P. and {Rampinelli}, L. and {Benisty}, M. and {Garufi}, A. and {Humphreys}, E.},
        title = "{Zooming into the water snow line: High-resolution water observations of the HL Tau disk}",
      journal = {\aap},
         year = 2026,
        month = jan,
       volume = {705},
          eid = {A193},
        pages = {A193},
          doi = {10.1051/0004-6361/202557609},
archivePrefix = {arXiv},
       eprint = {2511.16737},
 primaryClass = {astro-ph.EP},
       adsurl = {https://ui.adsabs.harvard.edu/abs/2026A&A...705A.193L}
}

@ARTICLE{Drka2017,
       author = {{Dr{\k{a}}{\.z}kowska}, J. and {Alibert}, Y.},
        title = "{Planetesimal formation starts at the snow line}",
      journal = {\aap},
         year = 2017,
        month = dec,
       volume = {608},
          eid = {A92},
        pages = {A92},
          doi = {10.1051/0004-6361/201731491},
archivePrefix = {arXiv},
       eprint = {1710.00009},
 primaryClass = {astro-ph.EP},
       adsurl = {https://ui.adsabs.harvard.edu/abs/2017A&A...608A..92D}
}

@ARTICLE{Ghezzi2026,
       author = {{Ghezzi}, Luan and {Costa-Almeida}, Ellen and {Loaiza-Tacuri}, Ver{\'o}nica and {Cunha}, Katia},
        title = "{A Comprehensive Study of the Relations between the Properties of Planetary Systems and the Chemical Compositions of Their Host Stars}",
      journal = {\apj},
         year = 2026,
        month = feb,
       volume = {998},
       number = {2},
          eid = {301},
        pages = {301},
          doi = {10.3847/1538-4357/ae317d},
archivePrefix = {arXiv},
       eprint = {2601.00962},
 primaryClass = {astro-ph.EP},
       adsurl = {https://ui.adsabs.harvard.edu/abs/2026ApJ...998..301G}
}

@ARTICLE{Ghezzi2021,
       author = {{Ghezzi}, Luan and {Martinez}, Cintia F. and {Wilson}, Robert F. and {Cunha}, Katia and {Smith}, Verne V. and {Majewski}, Steven R.},
        title = "{A Spectroscopic Analysis of the California-Kepler Survey Sample. II. Correlations of Stellar Metallicities with Planetary Architectures}",
      journal = {\apj},
         year = 2021,
        month = oct,
       volume = {920},
       number = {1},
          eid = {19},
        pages = {19},
          doi = {10.3847/1538-4357/ac14c3},
archivePrefix = {arXiv},
       eprint = {2107.04153},
 primaryClass = {astro-ph.EP},
       adsurl = {https://ui.adsabs.harvard.edu/abs/2021ApJ...920...19G}
}

@ARTICLE{Brugamyer2011,
       author = {{Brugamyer}, Erik and {Dodson-Robinson}, Sarah E. and {Cochran}, William D. and {Sneden}, Christopher},
        title = "{Silicon and Oxygen Abundances in Planet-host Stars}",
      journal = {\apj},
         year = 2011,
        month = sep,
       volume = {738},
       number = {1},
          eid = {97},
        pages = {97},
          doi = {10.1088/0004-637X/738/1/97},
archivePrefix = {arXiv},
       eprint = {1106.5509},
 primaryClass = {astro-ph.EP},
       adsurl = {https://ui.adsabs.harvard.edu/abs/2011ApJ...738...97B}
}

@ARTICLE{Jones2024,
       author = {{Jones}, Mat{\'\i}as I. and {Reinarz}, Yared and {Brahm}, Rafael and {Tala Pinto}, Marcelo and {Eberhardt}, Jan and {Rojas}, Felipe and {Triaud}, Amaury H.~M.~J. and {Gupta}, Arvind F. and {Ziegler}, Carl and {Hobson}, Melissa J. and {Jord{\'a}n}, Andr{\'e}s and {Henning}, Thomas and {Trifonov}, Trifon and {Schlecker}, Martin and {Espinoza}, N{\'e}stor and {Torres-Miranda}, Pascal and {Sarkis}, Paula and {Ulmer-Moll}, Sol{\`e}ne and {Lendl}, Monika and {Uzundag}, Murat and {Moyano}, Maximiliano and {Hesse}, Katharine and {Caldwell}, Douglas A. and {Shporer}, Avi and {Lund}, Michael B. and {Jenkins}, Jon M. and {Seager}, Sara and {Winn}, Joshua N. and {Ricker}, George R. and {Burke}, Christopher J. and {Figueira}, Pedro and {Psaridi}, Angelica and {Al Moulla}, Khaled and {Mounzer}, Dany and {Standing}, Matthew R. and {Martin}, David V. and {Dransfield}, Georgina and {Baycroft}, Thomas and {Dragomir}, Diana and {Boyle}, Gavin and {Suc}, Vincent and {Mann}, Andrew W. and {Timmermans}, Mathilde and {Ducrot}, Elsa and {Hooton}, Matthew J. and {Zu{\~n}iga-Fern{\'a}ndez}, Sebasti{\'a}n and {Sebastian}, Daniel and {Gillon}, Michael and {Queloz}, Didier and {Carson}, Joe and {Lissauer}, Jack J.},
        title = "{A long-period transiting substellar companion in the super-Jupiters to brown dwarfs mass regime and a prototypical warm-Jupiter detected by TESS}",
      journal = {\aap},
         year = 2024,
        month = mar,
       volume = {683},
          eid = {A192},
        pages = {A192},
          doi = {10.1051/0004-6361/202348147},
archivePrefix = {arXiv},
       eprint = {2401.09657},
 primaryClass = {astro-ph.EP},
       adsurl = {https://ui.adsabs.harvard.edu/abs/2024A&A...683A.192J}
}

@ARTICLE{Eberhardt2025,
       author = {{Eberhardt}, Jan and {Trifonov}, Trifon and {Henning}, Thomas and {Tala Pinto}, Marcelo and {Brahm}, Rafael and {Jord{\'a}n}, Andr{\'e}s and {Espinoza}, Nestor and {Jones}, Mat{\'\i}as I. and {Hobson}, Melissa J. and {Rojas}, Felipe I. and {Schlecker}, Martin and {Acu{\~n}a}, Lorena and {Burn}, Remo and {Boyle}, Gavin and {Leiva}, Rodrigo and {McCormac}, James and {Dunckel}, Nicholas and {Dragomir}, Diana and {Crane}, Jeffrey D. and {Shectman}, Stephen and {Teske}, Johanna K. and {Osip}, David and {Gupta}, Arvind F. and {Ulmer-Moll}, Sol{\`e}ne and {Bouchy}, Fran{\c{c}}ois and {Lendl}, Monika and {Gandolfi}, Davide and {Ricker}, George R. and {Jenkins}, Jon M. and {Seager}, Sara and {Winn}, Joshua N.},
        title = "{TOI-6695: A Pair of Near-resonant Massive Planets Observed with TESS from the WINE Survey}",
      journal = {\aj},
         year = 2025,
        month = jun,
       volume = {169},
       number = {6},
          eid = {298},
        pages = {298},
          doi = {10.3847/1538-3881/adc44e},
       adsurl = {https://ui.adsabs.harvard.edu/abs/2025AJ....169..298E}
}

@ARTICLE{Gill2024,
       author = {{Gill}, Samuel and {Bayliss}, Daniel and {Ulmer-Moll}, Sol{\`e}ne and {Wheatley}, Peter J. and {Brahm}, Rafael and {Anderson}, David R. and {Armstrong}, David and {Apergis}, Ioannis and {Alves}, Douglas R. and {Burleigh}, Matthew R. and {Butler}, R.~P. and {Bouchy}, Fran{\c{c}}ois and {Battley}, Matthew P. and {Bryant}, Edward M. and {Bieryla}, Allyson and {Crane}, Jeffrey D. and {Collins}, Karen A. and {Casewell}, Sarah L. and {Carleo}, Ilaria and {Claringbold}, Alastair B. and {Dalba}, Paul A. and {Dragomir}, Diana and {Eigm{\"u}ller}, Philipp and {Eberhardt}, Jan and {Fausnaugh}, Michael and {G{\"u}nther}, Maximilian N. and {Grieves}, Nolan and {Goad}, Michael R. and {Gillen}, Edward and {Hagelberg}, Janis and {Hobson}, Melissa and {Hedges}, Christina and {Henderson}, Beth A. and {Hawthorn}, Faith and {Henning}, Thomas and {Jones}, Mat{\'\i}as I. and {Jord{\'a}n}, Andr{\'e}s and {Jenkins}, James S. and {Kunimoto}, Michelle and {Krenn}, Andreas F. and {Kendall}, Alicia and {Lendl}, Monika and {McCormac}, James and {Moyano}, Maximiliano and {Torres-Miranda}, Pascal and {Nielsen}, Louise D. and {Osborn}, Ares and {Otegi}, Jon and {Osborn}, Hugh and {Quinn}, Samuel N. and {Rodriguez}, Joseph E. and {Ramsay}, Gavin and {Schlecker}, Martin and {Shectman}, Stephen A. and {Seager}, Sara and {Tilbrook}, Rosanna H. and {Trifonov}, Trifon and {Teske}, Johanna K. and {Udry}, Stephane and {Vines}, Jose I. and {West}, Richard R. and {Wohler}, Bill and {Winn}, Joshua N. and {Wang}, Sharon X. and {Zhou}, George and {Zivave}, Tafadzwa},
        title = "{Correction to: TOI-2447 b / NGTS-29 b: a 69-day Saturn around a Solar analogue}",
      journal = {\mnras},
         year = 2024,
        month = sep,
       volume = {533},
       number = {1},
        pages = {109-109},
          doi = {10.1093/mnras/stae1804},
archivePrefix = {arXiv},
       eprint = {2405.07367},
 primaryClass = {astro-ph.EP},
       adsurl = {https://ui.adsabs.harvard.edu/abs/2024MNRAS.533..109G}
}

@ARTICLE{Battley2024,
       author = {{Battley}, M.~P. and {Collins}, K.~A. and {Ulmer-Moll}, S. and {Quinn}, S.~N. and {Lendl}, M. and {Gill}, S. and {Brahm}, R. and {Hobson}, M.~J. and {Osborn}, H.~P. and {Deline}, A. and {Faria}, J.~P. and {Claringbold}, A.~B. and {Chakraborty}, H. and {Stassun}, K.~G. and {Hellier}, C. and {Alves}, D.~R. and {Ziegler}, C. and {Anderson}, D.~R. and {Apergis}, I. and {Armstrong}, D.~J. and {Bayliss}, D. and {Beletsky}, Y. and {Bieryla}, A. and {Bouchy}, F. and {Burleigh}, M.~R. and {Butler}, R.~P. and {Casewell}, S.~L. and {Christiansen}, J.~L. and {Crane}, J.~D. and {Dalba}, P.~A. and {Daylan}, T. and {Figueira}, P. and {Gillen}, E. and {Goad}, M.~R. and {G{\"u}nther}, M.~N. and {Henderson}, B.~A. and {Henning}, T. and {Jenkins}, J.~S. and {Jord{\'a}n}, A. and {Kanodia}, S. and {Kendall}, A. and {Kunimoto}, M. and {Latham}, D.~W. and {Levine}, A.~M. and {McCormac}, J. and {Moyano}, M. and {Osborn}, A. and {Osip}, D. and {Pritchard}, T.~A. and {Psaridi}, A. and {Rice}, M. and {Rodriguez}, J.~E. and {Saha}, S. and {Seager}, S. and {Shectman}, S.~A. and {Smith}, A.~M.~S. and {Teske}, J.~K. and {Ting}, E.~B. and {Udry}, S. and {Vines}, J.~I. and {Watson}, C.~A. and {West}, R.~G. and {Wheatley}, P.~J. and {Winn}, J.~N. and {Yee}, S.~W. and {Zhao}, Y.},
        title = "{NGTS-30b/TOI-4862b: An  1 Gyr old 98-day transiting warm Jupiter}",
      journal = {\aap},
         year = 2024,
        month = jun,
       volume = {686},
          eid = {A230},
        pages = {A230},
          doi = {10.1051/0004-6361/202449307},
archivePrefix = {arXiv},
       eprint = {2404.02974},
 primaryClass = {astro-ph.EP},
       adsurl = {https://ui.adsabs.harvard.edu/abs/2024A&A...686A.230B}
}

@ARTICLE{Howard2025,
       author = {{Howard}, Andrew W. and {Sinukoff}, Evan and {Blunt}, Sarah and {Petigura}, Erik A. and {Crossfield}, Ian J.~M. and {Isaacson}, Howard and {Kosiarek}, Molly and {Rubenzahl}, Ryan A. and {Brewer}, John M. and {Fulton}, Benjamin J. and {Dressing}, Courtney D. and {Hirsch}, Lea A. and {Knutson}, Heather and {Livingston}, John H. and {Mills}, Sean M. and {Roy}, Arpita and {Weiss}, Lauren M. and {Benneke}, Bjorn and {Ciardi}, David R. and {Christiansen}, Jessie L. and {Cochran}, William D. and {Crepp}, Justin R. and {Gonzales}, Erica and {Hansen}, Brad M.~S. and {Hardegree-Ullman}, Kevin and {Howell}, Steve B. and {L{\'e}pine}, S{\'e}bastien and {Martinez}, Arturo O. and {Rogers}, Leslie A. and {Schlieder}, Joshua E. and {Werner}, Michael and {Polanski}, Alex S. and {Angelo}, Isabel and {Beard}, Corey and {Behmard}, Aida and {Bouma}, Luke G. and {Brinkman}, Casey L. and {Chontos}, Ashley and {Dai}, Fei and {Dalba}, Paul A. and {Giacalone}, Steven and {Grunblatt}, Samuel K. and {Hill}, Michelle L. and {Kane}, Stephen R. and {Lubin}, Jack and {Mayo}, Andrew W. and {Mocnik}, Teo and {Murphy}, Joseph M. Akana and {Rice}, Malena and {Rosenthal}, Lee J. and {Tyler}, Dakotah and {Van Zandt}, Judah and {Yee}, Samuel W.},
        title = "{Planet Masses, Radii, and Orbits from NASA's K2 Mission}",
      journal = {\apjs},
         year = 2025,
        month = jun,
       volume = {278},
       number = {2},
          eid = {52},
        pages = {52},
          doi = {10.3847/1538-4365/adc5e4},
archivePrefix = {arXiv},
       eprint = {2502.04436},
 primaryClass = {astro-ph.EP},
       adsurl = {https://ui.adsabs.harvard.edu/abs/2025ApJS..278...52H}
}

@ARTICLE{Nicholson2024,
       author = {{Nicholson}, B.~A. and {Aigrain}, S. and {Eisner}, N.~L. and {Cretignier}, M. and {Barrag{\'a}n}, O. and {Kaye}, L. and {Taylor}, J. and {Owen}, J. and {Mortier}, A. and {Affer}, L. and {Boschin}, W. and {Buchhave}, L.~A. and {Cameron}, A. Collier and {Damasso}, M. and {Fabrizio}, L. Di and {DiTomasso}, V. and {Dumusque}, X. and {Ghedina}, A. and {Latham}, D.~W. and {L{\'o}pez-Morales}, M. and {Lorenzi}, V. and {Fiorenzano}, A.~F. Mart{\'\i}nez and {Molinari}, E. and {Pedani}, M. and {Pinamonti}, M. and {Rice}, K. and {Sozzetti}, A.},
        title = "{HD152843 b \& c: the masses and orbital periods of a sub-Neptune and a superpuff Neptune}",
      journal = {\mnras},
         year = 2024,
        month = aug,
       volume = {532},
       number = {4},
        pages = {4632-4644},
          doi = {10.1093/mnras/stae1821},
archivePrefix = {arXiv},
       eprint = {2310.15068},
 primaryClass = {astro-ph.EP},
       adsurl = {https://ui.adsabs.harvard.edu/abs/2024MNRAS.532.4632N}
}

@ARTICLE{Delrez2021,
       author = {{Delrez}, Laetitia and {Ehrenreich}, David and {Alibert}, Yann and {Bonfanti}, Andrea and {Borsato}, Luca and {Fossati}, Luca and {Hooton}, Matthew J. and {Hoyer}, Sergio and {Pozuelos}, Francisco J. and {Salmon}, S{\'e}bastien and {Sulis}, Sophia and {Wilson}, Thomas G. and {Adibekyan}, Vardan and {Bourrier}, Vincent and {Brandeker}, Alexis and {Charnoz}, S{\'e}bastien and {Deline}, Adrien and {Guterman}, Pascal and {Haldemann}, Jonas and {Hara}, Nathan and {Oshagh}, Mahmoudreza and {Sousa}, Sergio G. and {Van Grootel}, Val{\'e}rie and {Alonso}, Roi and {Anglada-Escud{\'e}}, Guillem and {B{\'a}rczy}, Tam{\'a}s and {Barrado}, David and {Barros}, Susana C.~C. and {Baumjohann}, Wolfgang and {Beck}, Mathias and {Bekkelien}, Anja and {Benz}, Willy and {Billot}, Nicolas and {Bonfils}, Xavier and {Broeg}, Christopher and {Cabrera}, Juan and {Collier Cameron}, Andrew and {Davies}, Melvyn B. and {Deleuil}, Magali and {Delisle}, Jean-Baptiste and {Demangeon}, Olivier D.~S. and {Demory}, Brice-Olivier and {Erikson}, Anders and {Fortier}, Andrea and {Fridlund}, Malcolm and {Futyan}, David and {Gandolfi}, Davide and {Garcia Mu{\~n}oz}, Antonio and {Gillon}, Micha{\"e}l and {Guedel}, Manuel and {Heng}, Kevin and {Kiss}, L{\'a}szl{\'o} and {Laskar}, Jacques and {Lecavelier des Etangs}, Alain and {Lendl}, Monika and {Lovis}, Christophe and {Maxted}, Pierre F.~L. and {Nascimbeni}, Valerio and {Olofsson}, G{\"o}ran and {Osborn}, Hugh P. and {Pagano}, Isabella and {Pall{\'e}}, Enric and {Piotto}, Giampaolo and {Pollacco}, Don and {Queloz}, Didier and {Rauer}, Heike and {Ragazzoni}, Roberto and {Ribas}, Ignasi and {Santos}, Nuno C. and {Scandariato}, Gaetano and {S{\'e}gransan}, Damien and {Simon}, Attila E. and {Smith}, Alexis M.~S. and {Steller}, Manfred and {Szab{\'o}}, Gyula M. and {Thomas}, Nicolas and {Udry}, St{\'e}phane and {Walton}, Nicholas A.},
        title = "{Transit detection of the long-period volatile-rich super-Earth {\ensuremath{\nu}}$^{2}$ Lupi d with CHEOPS}",
      journal = {Nature Astronomy},
         year = 2021,
        month = jun,
       volume = {5},
        pages = {775-787},
          doi = {10.1038/s41550-021-01381-5},
archivePrefix = {arXiv},
       eprint = {2106.14491},
 primaryClass = {astro-ph.EP},
       adsurl = {https://ui.adsabs.harvard.edu/abs/2021NatAs...5..775D}
}

@ARTICLE{Vanderburg2019,
       author = {{Vanderburg}, Andrew and {Huang}, Chelsea X. and {Rodriguez}, Joseph E. and {Becker}, Juliette C. and {Ricker}, George R. and {Vanderspek}, Roland K. and {Latham}, David W. and {Seager}, Sara and {Winn}, Joshua N. and {Jenkins}, Jon M. and {Addison}, Brett and {Bieryla}, Allyson and {Brice{\~n}o}, Cesar and {Bowler}, Brendan P. and {Brown}, Timothy M. and {Burke}, Christopher J. and {Burt}, Jennifer A. and {Caldwell}, Douglas A. and {Clark}, Jake T. and {Crossfield}, Ian and {Dittmann}, Jason A. and {Dynes}, Scott and {Fulton}, Benjamin J. and {Guerrero}, Natalia and {Harbeck}, Daniel and {Horner}, Jonathan and {Kane}, Stephen R. and {Kielkopf}, John and {Kraus}, Adam L. and {Kreidberg}, Laura and {Law}, Nicolas and {Mann}, Andrew W. and {Mengel}, Matthew W. and {Morton}, Timothy D. and {Okumura}, Jack and {Pearce}, Logan A. and {Plavchan}, Peter and {Quinn}, Samuel N. and {Rabus}, Markus and {Rose}, Mark E. and {Rowden}, Pam and {Shporer}, Avi and {Siverd}, Robert J. and {Smith}, Jeffrey C. and {Stassun}, Keivan and {Tinney}, C.~G. and {Wittenmyer}, Rob and {Wright}, Duncan J. and {Zhang}, Hui and {Zhou}, George and {Ziegler}, Carl A.},
        title = "{TESS Spots a Compact System of Super-Earths around the Naked-eye Star HR 858}",
      journal = {\apjl},
         year = 2019,
        month = aug,
       volume = {881},
       number = {1},
          eid = {L19},
        pages = {L19},
          doi = {10.3847/2041-8213/ab322d},
archivePrefix = {arXiv},
       eprint = {1905.05193},
 primaryClass = {astro-ph.EP},
       adsurl = {https://ui.adsabs.harvard.edu/abs/2019ApJ...881L..19V}
}

@ARTICLE{Plotnikova2024,
       author = {{Plotnikova}, A. and {Spina}, L. and {Ratcliffe}, B. and {Casali}, G. and {Carraro}, G.},
        title = "{The chemical evolution of the Milky Way thin disk using solar twins}",
      journal = {\aap},
         year = 2024,
        month = nov,
       volume = {691},
          eid = {A298},
        pages = {A298},
          doi = {10.1051/0004-6361/202451167},
archivePrefix = {arXiv},
       eprint = {2411.03067},
 primaryClass = {astro-ph.GA},
       adsurl = {https://ui.adsabs.harvard.edu/abs/2024A&A...691A.298P}
}

@ARTICLE{Brewer2018,
       author = {{Brewer}, John M. and {Fischer}, Debra A.},
        title = "{Spectral Properties of Cool Stars: Extended Abundance Analysis of Kepler Objects of Interest}",
      journal = {\apjs},
         year = 2018,
        month = aug,
       volume = {237},
       number = {2},
          eid = {38},
        pages = {38},
          doi = {10.3847/1538-4365/aad501},
archivePrefix = {arXiv},
       eprint = {1804.00673},
 primaryClass = {astro-ph.SR},
       adsurl = {https://ui.adsabs.harvard.edu/abs/2018ApJS..237...38B}
}

@ARTICLE{Brewer2016,
       author = {{Brewer}, John M. and {Fischer}, Debra A. and {Valenti}, Jeff A. and {Piskunov}, Nikolai},
        title = "{Spectral Properties of Cool Stars: Extended Abundance Analysis of 1,617 Planet-search Stars}",
      journal = {\apjs},
         year = 2016,
        month = aug,
       volume = {225},
       number = {2},
          eid = {32},
        pages = {32},
          doi = {10.3847/0067-0049/225/2/32},
archivePrefix = {arXiv},
       eprint = {1606.07929},
 primaryClass = {astro-ph.SR},
       adsurl = {https://ui.adsabs.harvard.edu/abs/2016ApJS..225...32B}
}

@ARTICLE{Moore1945,
       author = {{Moore}, Charlotte E.},
        title = "{A Multiplet Table of Astrophysical Interest. Revised Edition. Part I - Table of Multiplets}",
      journal = {Contributions from the Princeton University Observatory},
         year = 1945,
        month = jan,
       volume = {20},
        pages = {1-110},
       adsurl = {https://ui.adsabs.harvard.edu/abs/1945CoPri..20....1M}
}
\bibliographystyle{aasjournal}



\end{document}